\documentclass[twocolumn,traditabstract,longauth]{aa}
\usepackage{amsmath,amsfonts,amssymb}
\usepackage{txfonts}
\usepackage{graphicx}
\usepackage{fixltx2e}
\usepackage{natbib}
\usepackage{upgreek}
\usepackage[flushleft]{threeparttable}
\usepackage{epsfig}
\usepackage{morefloats}
\usepackage{ifthen}

\usepackage[breaklinks, colorlinks, citecolor=blue]{hyperref}

\bibpunct{(}{)}{;}{a}{}{,} 

\def\setsymbol#1#2{\expandafter\def\csname #1\endcsname{#2}}
\def\getsymbol#1{\csname #1\endcsname}

\def\Planck{\textit{Planck}}

\newbox\tablebox    \newdimen\tablewidth
\def\leaderfil{\leaders\hbox to 5pt{\hss.\hss}\hfil}
\def\endPlancktable{\tablewidth=\columnwidth 
    $$\hss\copy\tablebox\hss$$
    \vskip-\lastskip\vskip -2pt}
\def\endPlancktablewide{\tablewidth=\textwidth 
    $$\hss\copy\tablebox\hss$$
    \vskip-\lastskip\vskip -2pt}
\def\tablenote#1 #2\par{\begingroup \parindent=0.8em
    \abovedisplayshortskip=0pt\belowdisplayshortskip=0pt
    \noindent
    $$\hss\vbox{\hsize\tablewidth \hangindent=\parindent \hangafter=1 \noindent
    \hbox to \parindent{$^#1$\hss}\strut#2\strut\par}\hss$$
    \endgroup}
\def\doubleline{\vskip 3pt\hrule \vskip 1.5pt \hrule \vskip 5pt}

\def\L2{\ifmmode L_2\else $L_2$\fi}

\def\DeltaT{\ifmmode \Delta T\else $\Delta T$\fi}
\def\deltat{\ifmmode \Delta t\else $\Delta t$\fi}
\def\fknee{\ifmmode f_{\rm knee}\else $f_{\rm knee}$\fi}
\def\Fmax{\ifmmode F_{\rm max}\else $F_{\rm max}$\fi}
\def\solar{\ifmmode{\rm M}_{\mathord\odot}\else${\rm M}_{\mathord\odot}$\fi}
\def\Msolar{\ifmmode{\rm M}_{\mathord\odot}\else${\rm M}_{\mathord\odot}$\fi}
\def\Lsolar{\ifmmode{\rm L}_{\mathord\odot}\else${\rm L}_{\mathord\odot}$\fi}
\def\inv{\ifmmode^{-1}\else$^{-1}$\fi}
\def\mo{\ifmmode^{-1}\else$^{-1}$\fi}
\def\sup#1{\ifmmode ^{\rm #1}\else $^{\rm #1}$\fi}
\def\expo#1{\ifmmode \times 10^{#1}\else $\times 10^{#1}$\fi}
\def\,{\thinspace}
\def\lsim{\mathrel{\raise .4ex\hbox{\rlap{$<$}\lower 1.2ex\hbox{$\sim$}}}}
\def\gsim{\mathrel{\raise .4ex\hbox{\rlap{$>$}\lower 1.2ex\hbox{$\sim$}}}}

\def\simprop{\mathrel{\raise .4ex\hbox{\rlap{$\propto$}\lower 1.2ex\hbox{$\sim$}}}}
\def\deg{\ifmmode^\circ\else$^\circ$\fi}
\def\pdeg{\ifmmode $\setbox0=\hbox{$^{\circ}$}\rlap{\hskip.11\wd0 .}$^{\circ}
          \else \setbox0=\hbox{$^{\circ}$}\rlap{\hskip.11\wd0 .}$^{\circ}$\fi}
\def\arcs{\ifmmode {^{\scriptstyle\prime\prime}}
          \else $^{\scriptstyle\prime\prime}$\fi}
\def\arcm{\ifmmode {^{\scriptstyle\prime}}
          \else $^{\scriptstyle\prime}$\fi}
\newdimen\sa  \newdimen\sb
\def\parcs{\sa=.07em \sb=.03em
     \ifmmode \hbox{\rlap{.}}^{\scriptstyle\prime\kern -\sb\prime}\hbox{\kern -\sa}
     \else \rlap{.}$^{\scriptstyle\prime\kern -\sb\prime}$\kern -\sa\fi}
\def\parcm{\sa=.08em \sb=.03em
     \ifmmode \hbox{\rlap{.}\kern\sa}^{\scriptstyle\prime}\hbox{\kern-\sb}
     \else \rlap{.}\kern\sa$^{\scriptstyle\prime}$\kern-\sb\fi}
\def\ra[#1 #2 #3.#4]{#1\sup{h}#2\sup{m}#3\sup{s}\llap.#4}
\def\dec[#1 #2 #3.#4]{#1\deg#2\arcm#3\arcs\llap.#4}
\def\deco[#1 #2 #3]{#1\deg#2\arcm#3\arcs}
\def\rra[#1 #2]{#1\sup{h}#2\sup{m}}

\def\dots{\relax\ifmmode \ldots\else $\ldots$\fi}
\def\WHzsr{\ifmmode $W\,Hz\mo\,sr\mo$\else W\,Hz\mo\,sr\mo\fi}
\def\mHz{\ifmmode $\,mHz$\else \,mHz\fi}
\def\GHz{\ifmmode $\,GHz$\else \,GHz\fi}
\def\mKs{\ifmmode $\,mK\,s$^{1/2}\else \,mK\,s$^{1/2}$\fi}
\def\muKs{\ifmmode \,\mu$K\,s$^{1/2}\else \,$\mu$K\,s$^{1/2}$\fi}
\def\muKRJs{\ifmmode \,\mu$K$_{\rm RJ}$\,s$^{1/2}\else \,$\mu$K$_{\rm RJ}$\,s$^{1/2}$\fi}
\def\muKHz{\ifmmode \,\mu$K\,Hz$^{-1/2}\else \,$\mu$K\,Hz$^{-1/2}$\fi}
\def\MJysr{\ifmmode \,$MJy\,sr\mo$\else \,MJy\,sr\mo\fi}
\def\MJysrmK{\ifmmode \,$MJy\,sr\mo$\,mK$_{\rm CMB}\mo\else \,MJy\,sr\mo\,mK$_{\rm CMB}\mo$\fi}
\def\microns{\ifmmode \,\mu$m$\else \,$\mu$m\fi}

\def\muK{\ifmmode \,\mu$K$\else \,$\mu$\hbox{K}\fi}
\def\microK{\ifmmode \,\mu$K$\else \,$\mu$\hbox{K}\fi}
\def\muW{\ifmmode \,\mu$W$\else \,$\mu$\hbox{W}\fi}
\def\kms{\ifmmode $\,km\,s$^{-1}\else \,km\,s$^{-1}$\fi}
\def\kmsMpc{\ifmmode $\,\kms\,Mpc\mo$\else \,\kms\,Mpc\mo\fi}
\providecommand{\sorthelp}[1]{}

\newcommand{\mdg}{^{\circ}}
\newcommand{\mdgp}{^{\circ}\!\!.}

\newcommand{\planck}{\textit{Planck}}
\newcommand{\wmap}{WMAP} 
\newcommand{\WMAP}{WMAP}
\newcommand{\fermi}{\textit{Fermi}}
\newcommand{\rosat}{ROSAT}
\newcommand{\ROSAT}{ROSAT}
\newcommand{\Hipparcos}{{\sc Hipparcos}} 

\newcommand{\cc}{\mathcal{C}} 
\newcommand{\ttp}{$T$--$T$}
\newcommand{\WISE}{WISE}
\newcommand{\Wise}{\WISE}
\newcommand{\IRAS}{IRAS}
\newcommand{\iras}{\IRAS}
\newcommand{\COBE}{\textit{COBE\/}}
\newcommand{\DIRBE}{\COBE-DIRBE}
\newcommand{\Spitzer}{\textit{Spitzer\/}}

\newcommand{\ghz}{\,GHz}

\newcommand{\MHz}{\,MHz}

\newcommand{\um}{\,$\upmu$m}

\newcommand{\uK}{\,$\upmu$K}
\newcommand{\upk}{$\upmu$K}
\newcommand{\healpix}{{\tt HEALPix}}
\newcommand{\mdotyr}{\,M$_\odot$\,yr$^{-1}$}

\newcommand{\commander}{{\tt Commander}}
\newcommand{\Commander}{\commander}
\newcommand{\galprop}{{\tt GALPROP}}
\newcommand{\spdust}{{\tt SpDust2}}

\newcommand{\ha}{H$\alpha$}
\newcommand{\hi}{\ion{H}{i}}
\newcommand{\hii}{\ion{H}{ii}}
\newcommand{\lorionis}{$\lambda$ Orionis}
\newcommand{\ukmjy}{\uK\,(MJy\,sr$^{-1}$)$^{-1}$}
\newcommand{\upkmjy}{\upk\,(MJy\,sr$^{-1}$)$^{-1}$}
\newcommand{\asinh}{asinh}

\begin{document}


\title{\Planck\ 2015 results. XXV. Diffuse low-frequency Galactic foregrounds}
\authorrunning{Planck Collaboration}
\titlerunning{Planck diffuse low-frequency Galactic foregrounds}
\author{\small
Planck Collaboration: P.~A.~R.~Ade\inst{95}
\and
N.~Aghanim\inst{64}
\and
M.~I.~R.~Alves\inst{105, 11, 64}
\and
M.~Arnaud\inst{79}
\and
M.~Ashdown\inst{75, 6}
\and
J.~Aumont\inst{64}
\and
C.~Baccigalupi\inst{93}
\and
A.~J.~Banday\inst{105, 11}
\and
R.~B.~Barreiro\inst{70}
\and
J.~G.~Bartlett\inst{1, 72}
% ...
}
\institute{\small
APC, AstroParticule et Cosmologie, Universit\'{e} Paris Diderot, CNRS/IN2P3, CEA/lrfu, Observatoire de Paris, Sorbonne Paris Cit\'{e}, 10, rue Alice Domon et L\'{e}onie Duquet, 75205 Paris Cedex 13, France\goodbreak
\and
Aalto University Mets\"{a}hovi Radio Observatory and Dept of Radio Science and Engineering, P.O. Box 13000, FI-00076 AALTO, Finland\goodbreak
\and
African Institute for Mathematical Sciences, 6-8 Melrose Road, Muizenberg, Cape Town, South Africa\goodbreak
\and
Agenzia Spaziale Italiana Science Data Center, Via del Politecnico snc, 00133, Roma, Italy\goodbreak
\and
Aix Marseille Universit\'{e}, CNRS, LAM (Laboratoire d'Astrophysique de Marseille) UMR 7326, 13388, Marseille, France\goodbreak
\and
Astrophysics Group, Cavendish Laboratory, University of Cambridge, J J Thomson Avenue, Cambridge CB3 0HE, U.K.\goodbreak
\and
Astrophysics \& Cosmology Research Unit, School of Mathematics, Statistics \& Computer Science, University of KwaZulu-Natal, Westville Campus, Private Bag X54001, Durban 4000, South Africa\goodbreak
\and
Atacama Large Millimeter/submillimeter Array, ALMA Santiago Central Offices, Alonso de Cordova 3107, Vitacura, Casilla 763 0355, Santiago, Chile\goodbreak
\and
CGEE, SCS Qd 9, Lote C, Torre C, 4$^{\circ}$ andar, Ed. Parque Cidade Corporate, CEP 70308-200, Bras\'{i}lia, DF, Brazil\goodbreak
\and
CITA, University of Toronto, 60 St. George St., Toronto, ON M5S 3H8, Canada\goodbreak
\and
CNRS, IRAP, 9 Av. colonel Roche, BP 44346, F-31028 Toulouse cedex 4, France\goodbreak
\and
CRANN, Trinity College, Dublin, Ireland\goodbreak
\and
California Institute of Technology, Pasadena, California, U.S.A.\goodbreak
\and
Centre for Theoretical Cosmology, DAMTP, University of Cambridge, Wilberforce Road, Cambridge CB3 0WA, U.K.\goodbreak
\and
Centro de Estudios de F\'{i}sica del Cosmos de Arag\'{o}n (CEFCA), Plaza San Juan, 1, planta 2, E-44001, Teruel, Spain\goodbreak
\and
Computational Cosmology Center, Lawrence Berkeley National Laboratory, Berkeley, California, U.S.A.\goodbreak
\and
Consejo Superior de Investigaciones Cient\'{\i}ficas (CSIC), Madrid, Spain\goodbreak
\and
DSM/Irfu/SPP, CEA-Saclay, F-91191 Gif-sur-Yvette Cedex, France\goodbreak
\and
DTU Space, National Space Institute, Technical University of Denmark, Elektrovej 327, DK-2800 Kgs. Lyngby, Denmark\goodbreak
\and
D\'{e}partement de Physique Th\'{e}orique, Universit\'{e} de Gen\`{e}ve, 24, Quai E. Ansermet,1211 Gen\`{e}ve 4, Switzerland\goodbreak
\and
Departamento de Astrof\'{i}sica, Universidad de La Laguna (ULL), E-38206 La Laguna, Tenerife, Spain\goodbreak
\and
Departamento de F\'{\i}sica, Universidad de Oviedo, Avda. Calvo Sotelo s/n, Oviedo, Spain\goodbreak
\and
Department of Astronomy and Astrophysics, University of Toronto, 50 Saint George Street, Toronto, Ontario, Canada\goodbreak
\and
Department of Astrophysics/IMAPP, Radboud University Nijmegen, P.O. Box 9010, 6500 GL Nijmegen, The Netherlands\goodbreak
\and
Department of Physics \& Astronomy, University of British Columbia, 6224 Agricultural Road, Vancouver, British Columbia, Canada\goodbreak
\and
Department of Physics and Astronomy, Dana and David Dornsife College of Letter, Arts and Sciences, University of Southern California, Los Angeles, CA 90089, U.S.A.\goodbreak
\and
Department of Physics and Astronomy, University College London, London WC1E 6BT, U.K.\goodbreak
\and
Department of Physics, Florida State University, Keen Physics Building, 77 Chieftan Way, Tallahassee, Florida, U.S.A.\goodbreak
\and
Department of Physics, Gustaf H\"{a}llstr\"{o}min katu 2a, University of Helsinki, Helsinki, Finland\goodbreak
\and
Department of Physics, Princeton University, Princeton, New Jersey, U.S.A.\goodbreak
\and
Department of Physics, University of California, Santa Barbara, California, U.S.A.\goodbreak
\and
Department of Physics, University of Illinois at Urbana-Champaign, 1110 West Green Street, Urbana, Illinois, U.S.A.\goodbreak
\and
Dipartimento di Fisica e Astronomia G. Galilei, Universit\`{a} degli Studi di Padova, via Marzolo 8, 35131 Padova, Italy\goodbreak
\and
Dipartimento di Fisica e Scienze della Terra, Universit\`{a} di Ferrara, Via Saragat 1, 44122 Ferrara, Italy\goodbreak
\and
Dipartimento di Fisica, Universit\`{a} La Sapienza, P. le A. Moro 2, Roma, Italy\goodbreak
\and
Dipartimento di Fisica, Universit\`{a} degli Studi di Milano, Via Celoria, 16, Milano, Italy\goodbreak
\and
Dipartimento di Fisica, Universit\`{a} degli Studi di Trieste, via A. Valerio 2, Trieste, Italy\goodbreak
\and
Dipartimento di Matematica, Universit\`{a} di Roma Tor Vergata, Via della Ricerca Scientifica, 1, Roma, Italy\goodbreak
\and
Discovery Center, Niels Bohr Institute, Blegdamsvej 17, Copenhagen, Denmark\goodbreak
\and
European Southern Observatory, ESO Vitacura, Alonso de Cordova 3107, Vitacura, Casilla 19001, Santiago, Chile\goodbreak
\and
European Space Agency, ESAC, Planck Science Office, Camino bajo del Castillo, s/n, Urbanizaci\'{o}n Villafranca del Castillo, Villanueva de la Ca\~{n}ada, Madrid, Spain\goodbreak
\and
European Space Agency, ESTEC, Keplerlaan 1, 2201 AZ Noordwijk, The Netherlands\goodbreak
\and
Gran Sasso Science Institute, INFN, viale F. Crispi 7, 67100 L'Aquila, Italy\goodbreak
\and
HGSFP and University of Heidelberg, Theoretical Physics Department, Philosophenweg 16, 69120, Heidelberg, Germany\goodbreak
\and
Haverford College Astronomy Department, 370 Lancaster Avenue, Haverford, Pennsylvania, U.S.A.\goodbreak
\and
Helsinki Institute of Physics, Gustaf H\"{a}llstr\"{o}min katu 2, University of Helsinki, Helsinki, Finland\goodbreak
\and
INAF - Osservatorio Astrofisico di Catania, Via S. Sofia 78, Catania, Italy\goodbreak
\and
INAF - Osservatorio Astronomico di Padova, Vicolo dell'Osservatorio 5, Padova, Italy\goodbreak
\and
INAF - Osservatorio Astronomico di Roma, via di Frascati 33, Monte Porzio Catone, Italy\goodbreak
\and
INAF - Osservatorio Astronomico di Trieste, Via G.B. Tiepolo 11, Trieste, Italy\goodbreak
\and
INAF/IASF Bologna, Via Gobetti 101, Bologna, Italy\goodbreak
\and
INAF/IASF Milano, Via E. Bassini 15, Milano, Italy\goodbreak
\and
INFN, Sezione di Bologna, viale Berti Pichat 6/2, 40127 Bologna, Italy\goodbreak
\and
INFN, Sezione di Ferrara, Via Saragat 1, 44122 Ferrara, Italy\goodbreak
\and
INFN, Sezione di Roma 1, Universit\`{a} di Roma Sapienza, Piazzale Aldo Moro 2, 00185, Roma, Italy\goodbreak
\and
INFN, Sezione di Roma 2, Universit\`{a} di Roma Tor Vergata, Via della Ricerca Scientifica, 1, Roma, Italy\goodbreak
\and
INFN/National Institute for Nuclear Physics, Via Valerio 2, I-34127 Trieste, Italy\goodbreak
\and
IPAG: Institut de Plan\'{e}tologie et d'Astrophysique de Grenoble, Universit\'{e} Grenoble Alpes, IPAG, F-38000 Grenoble, France, CNRS, IPAG, F-38000 Grenoble, France\goodbreak
\and
IUCAA, Post Bag 4, Ganeshkhind, Pune University Campus, Pune 411 007, India\goodbreak
\and
Imperial College London, Astrophysics group, Blackett Laboratory, Prince Consort Road, London, SW7 2AZ, U.K.\goodbreak
\and
Infrared Processing and Analysis Center, California Institute of Technology, Pasadena, CA 91125, U.S.A.\goodbreak
\and
Institut N\'{e}el, CNRS, Universit\'{e} Joseph Fourier Grenoble I, 25 rue des Martyrs, Grenoble, France\goodbreak
\and
Institut Universitaire de France, 103, bd Saint-Michel, 75005, Paris, France\goodbreak
\and
Institut d'Astrophysique Spatiale, CNRS, Univ. Paris-Sud, Universit\'{e} Paris-Saclay, B\^{a}t. 121, 91405 Orsay cedex, France\goodbreak
\and
Institut d'Astrophysique de Paris, CNRS (UMR7095), 98 bis Boulevard Arago, F-75014, Paris, France\goodbreak
\and
Institut f\"ur Theoretische Teilchenphysik und Kosmologie, RWTH Aachen University, D-52056 Aachen, Germany\goodbreak
\and
Institute of Astronomy, University of Cambridge, Madingley Road, Cambridge CB3 0HA, U.K.\goodbreak
\and
Institute of Theoretical Astrophysics, University of Oslo, Blindern, Oslo, Norway\goodbreak
\and
Instituto de Astrof\'{\i}sica de Canarias, C/V\'{\i}a L\'{a}ctea s/n, La Laguna, Tenerife, Spain\goodbreak
\and
Instituto de F\'{\i}sica de Cantabria (CSIC-Universidad de Cantabria), Avda. de los Castros s/n, Santander, Spain\goodbreak
\and
Istituto Nazionale di Fisica Nucleare, Sezione di Padova, via Marzolo 8, I-35131 Padova, Italy\goodbreak
\and
Jet Propulsion Laboratory, California Institute of Technology, 4800 Oak Grove Drive, Pasadena, California, U.S.A.\goodbreak
\and
Jodrell Bank Centre for Astrophysics, Alan Turing Building, School of Physics and Astronomy, The University of Manchester, Oxford Road, Manchester, M13 9PL, U.K.\goodbreak
\and
Kavli Institute for Cosmological Physics, University of Chicago, Chicago, IL 60637, USA\goodbreak
\and
Kavli Institute for Cosmology Cambridge, Madingley Road, Cambridge, CB3 0HA, U.K.\goodbreak
\and
Kazan Federal University, 18 Kremlyovskaya St., Kazan, 420008, Russia\goodbreak
\and
LAL, Universit\'{e} Paris-Sud, CNRS/IN2P3, Orsay, France\goodbreak
\and
LERMA, CNRS, Observatoire de Paris, 61 Avenue de l'Observatoire, Paris, France\goodbreak
\and
Laboratoire AIM, IRFU/Service d'Astrophysique - CEA/DSM - CNRS - Universit\'{e} Paris Diderot, B\^{a}t. 709, CEA-Saclay, F-91191 Gif-sur-Yvette Cedex, France\goodbreak
\and
Laboratoire Traitement et Communication de l'Information, CNRS (UMR 5141) and T\'{e}l\'{e}com ParisTech, 46 rue Barrault F-75634 Paris Cedex 13, France\goodbreak
\and
Laboratoire de Physique Subatomique et Cosmologie, Universit\'{e} Grenoble-Alpes, CNRS/IN2P3, 53, rue des Martyrs, 38026 Grenoble Cedex, France\goodbreak
\and
Laboratoire de Physique Th\'{e}orique, Universit\'{e} Paris-Sud 11 \& CNRS, B\^{a}timent 210, 91405 Orsay, France\goodbreak
\and
Lawrence Berkeley National Laboratory, Berkeley, California, U.S.A.\goodbreak
\and
Lebedev Physical Institute of the Russian Academy of Sciences, Astro Space Centre, 84/32 Profsoyuznaya st., Moscow, GSP-7, 117997, Russia\goodbreak
\and
Max-Planck-Institut f\"{u}r Astrophysik, Karl-Schwarzschild-Str. 1, 85741 Garching, Germany\goodbreak
\and
Max-Planck-Institut f\"{u}r Extraterrestrische Physik, Giessenbachstra{\ss}e, 85748 Garching, Germany\goodbreak
\and
McGill Physics, Ernest Rutherford Physics Building, McGill University, 3600 rue University, Montr\'{e}al, QC, H3A 2T8, Canada\goodbreak
\and
National University of Ireland, Department of Experimental Physics, Maynooth, Co. Kildare, Ireland\goodbreak
\and
Nicolaus Copernicus Astronomical Center, Bartycka 18, 00-716 Warsaw, Poland\goodbreak
\and
Niels Bohr Institute, Blegdamsvej 17, Copenhagen, Denmark\goodbreak
\and
Nordita (Nordic Institute for Theoretical Physics), Roslagstullsbacken 23, SE-106 91 Stockholm, Sweden\goodbreak
\and
Optical Science Laboratory, University College London, Gower Street, London, U.K.\goodbreak
\and
SISSA, Astrophysics Sector, via Bonomea 265, 34136, Trieste, Italy\goodbreak
\and
SMARTEST Research Centre, Universit\`{a} degli Studi e-Campus, Via Isimbardi 10, Novedrate (CO), 22060, Italy\goodbreak
\and
School of Physics and Astronomy, Cardiff University, Queens Buildings, The Parade, Cardiff, CF24 3AA, U.K.\goodbreak
\and
School of Physics and Astronomy, University of Nottingham, Nottingham NG7 2RD, U.K.\goodbreak
\and
Sorbonne Universit\'{e}-UPMC, UMR7095, Institut d'Astrophysique de Paris, 98 bis Boulevard Arago, F-75014, Paris, France\goodbreak
\and
Space Research Institute (IKI), Russian Academy of Sciences, Profsoyuznaya Str, 84/32, Moscow, 117997, Russia\goodbreak
\and
Space Sciences Laboratory, University of California, Berkeley, California, U.S.A.\goodbreak
\and
Special Astrophysical Observatory, Russian Academy of Sciences, Nizhnij Arkhyz, Zelenchukskiy region, Karachai-Cherkessian Republic, 369167, Russia\goodbreak
\and
Sub-Department of Astrophysics, University of Oxford, Keble Road, Oxford OX1 3RH, U.K.\goodbreak
\and
The Oskar Klein Centre for Cosmoparticle Physics, Department of Physics,Stockholm University, AlbaNova, SE-106 91 Stockholm, Sweden\goodbreak
\and
Theory Division, PH-TH, CERN, CH-1211, Geneva 23, Switzerland\goodbreak
\and
UPMC Univ Paris 06, UMR7095, 98 bis Boulevard Arago, F-75014, Paris, France\goodbreak
\and
Universit\'{e} de Toulouse, UPS-OMP, IRAP, F-31028 Toulouse cedex 4, France\goodbreak
% ...
}


\abstract{We discuss the Galactic foreground emission between 20 and 100\GHz\ based on observations by \Planck\ and \WMAP. The total intensity in this part of the spectrum is dominated by free-free and spinning dust emission, whereas the polarized intensity is dominated by synchrotron emission. The \Commander\ component-separation tool has been used to separate the various astrophysical processes in total intensity. Comparison with radio recombination line templates verifies the recovery of the free-free emission along the Galactic plane. Comparison of the high-latitude \ha\ emission with our free-free map shows residuals that correlate with dust optical depth, consistent with a fraction ($\approx 30\,\%$) of \ha\ having been scattered by high-latitude dust. We highlight a number of diffuse spinning dust morphological features at high latitude. There is substantial spatial variation in the spinning dust spectrum, with the emission peak (in $I_\nu$) ranging from below 20\GHz\ to more than 50\GHz. There is a strong tendency for the spinning dust component near many prominent \hii\ regions to have a higher peak frequency, suggesting that this increase in peak frequency is associated with dust in the photo-dissociation regions around the nebulae. The emissivity of spinning dust in these diffuse regions is of the same order as previous detections in the literature. Over the entire sky, the \commander\ solution finds more anomalous microwave emission (AME) than the WMAP component maps, at the expense of synchrotron and free-free emission. This can be explained by the difficulty in separating multiple broadband components with a limited number of frequency maps. Future surveys, particularly at 5--20\GHz, will greatly improve the separation by constraining the synchrotron spectrum. We combine \Planck\ and \WMAP\ data to make the highest signal-to-noise ratio maps yet of the intensity of the all-sky polarized synchrotron emission at frequencies above a few GHz. Most of the high-latitude polarized emission is associated with distinct large-scale loops and spurs, and we re-discuss their structure. We argue that nearly all the emission at $40\deg > l > -90\deg$ is part of the Loop I structure, and show that the emission extends much further in to the southern Galactic hemisphere than previously recognised, giving \mbox{Loop I} an ovoid rather than circular outline. However, it does not continue as far as the ``\fermi\ bubble/microwave haze'', making it less probable that these are part of the same structure. We identify a number of new faint features in the polarized sky, including a dearth of polarized synchrotron emission directly correlated with a narrow, roughly $20\deg$ long filament seen in \ha\ at high Galactic latitude. Finally, we look for evidence of polarized AME, however many AME regions are significantly contaminated by polarized synchrotron emission, and we find a 2$\sigma$ upper limit of 1.6\,\% in the Perseus region.}
\keywords{Diffuse radiation -- ISM: general -- Radiation mechanisms: general -- Radio continuum: ISM -- Polarization}

\maketitle


\section{Introduction}
\label{sec:introduction}

Diffuse Galactic radio emission consists of a number of distinct components that emit via different emission mechanisms (notably synchrotron, free-free, and spinning dust). There is considerable interest in understanding these components in order to subtract foregrounds cleanly from cosmic microwave background (CMB) data and as a probe of the physics of the interstellar medium (ISM) and Galactic structure. The separation of the diffuse foregrounds into their separate constituent components is extremely difficult, since their spectra (except for free-free emission) are not well-known, and they have comparable intensities at microwave frequencies \citep{Leach2008}.

At low frequencies ($\lesssim 10$\GHz), synchrotron radiation from electrons spiralling in the Galactic magnetic field dominates the sky. At frequencies around 1\GHz\ the spectral index ($T \propto \nu^{\,\beta}$) is $\beta \approx\! -2.7$, while at higher frequencies it appears to steepen to \mbox{$\beta \approx \!-3.0$}. This broadly agrees with theoretical calculations, such as those from the \galprop\footnote{\url{http://galprop.stanford.edu};\hfill\break\url{http://sourceforge.net/projects/galprop/}} code \citep{Orlando2013}; however, a detailed comparison has not been made up to now, due to difficulties in the component separation process. Synchrotron radiation is intrinsically highly polarized and is a strong polarized CMB foreground up to around 100\GHz. Free-free emission emits over a range of radio frequencies and can be significant up to about 100\GHz\ owing to its flatter spectral index ($\beta$\,$=$\,$-2.1$). Anomalous microwave emission (AME) is an additional foreground that has been detected at frequencies 10--60\GHz\ \citep{Leitch1997,deOliveira-Costa2004,Finkbeiner2004,Davies2006,planck2011-7.2,planck2011-7.3,planck2013-XV} above the expected levels of synchrotron and free-free emission. The most plausible origin for AME is electric dipole radiation from tiny, rapidly spinning dust grains \citep{Draine1998}, which provides an excellent fit to the data, particularly from well-studied molecular clouds \citep{planck2011-7.2}. However, the diffuse emission at high latitudes has still to be definitively identified, and other possibilities exist, such as magneto-dipole radiation from thermal fluctuations of magnetized grains \citep{Draine1999,Liu2014}. The polarization of AME is of great interest to CMB cosmologists, since a significant level of polarization could hamper measurements of $B$ modes. Current measurements place upper limits on the polarization of AME at a few per cent or below \citep{Dickinson2011,Macellari2011,Rubino-Martin2012}.

In the 2013 \Planck\footnote{\Planck\ (\url{http://www.esa.int/Planck}) is a project of the European Space Agency  (ESA) with instruments provided by two scientific consortia funded by ESA member states and led by Principal Investigators from France and Italy, telescope reflectors provided through a collaboration between ESA and a scientific consortium led and funded by Denmark, and additional contributions from NASA (USA).} results based on spectral fitting, the component separation procedure did not achieve a separation of the distinct low-frequency components, instead lumping these into a single low-frequency foreground  \citep{planck2013-p06}. However, other methods have been used to do this under specific assumptions; \cite{planck2013-XII} used correlated component analysis in the southern Gould Belt region where synchrotron emission is relatively smooth and weak, while \cite{planck2014-XXII} used template fitting to separate components that were correlated with each spatial template. In the 2015 analysis \citep{planck2014-a12} we include the \wmap\ \citep{Bennett2013} and 408 MHz \citep{haslam1982} all-sky maps  to enable a separation of synchrotron, free-free, and AME. The separation is based on fitting a spectral model to each sky pixel using \commander, a Gibbs sampling code \citep{Eriksen2008}.

The current \planck\ data release also includes polarization information. At angular scales larger than 1\degr\ the polarized emission is dominated by synchrotron and thermal dust radiation. Synchrotron radiation is more important at low frequencies and dominates the polarized \planck\ maps at 30 and 44\GHz. This synchrotron emission comes mainly from the Galactic plane and also from filamentary structures that can extend over 100\deg\ across the sky. While these ``loops'' and ``spurs'' have long been known in total intensity \citep{Quigley1965,Berkhuijsen1971,Berkhuijsen1971b}, their pattern is clarified by the new polarization maps. \citet{Vidal2014a} present an analysis of the filaments using \wmap\ data. They have catalogued them, studied the polarization angle distribution and polarized spectral indices, and presented a model to explain the origin of some of them. Here, with the new \planck\ polarization data, we expand on this, identifying new polarized features and understanding better the already known ones, thanks to the improved signal-to-noise (S/N) ratio of the data.

We begin in Sect.~\ref{sec:data} by summarizing the data sets employed, including a brief discussion of some of the most important systematic effects. We then take a first look at the intensity foregrounds in Sect.~\ref{sec:ilc} by employing a constrained internal linear combination (ILC) algorithm to remove the CMB and free-free emission components whose spectra are well known. In Sect.~\ref{sec:compsep} we then discuss each of the low frequency foreground components (AME, free-free, and synchrotron emission) as determined by the \commander\ algorithm. We use these new maps to investigate the distribution and spectra of these foregrounds. In Sect.~\ref{sec:polarization} we use \WMAP/\Planck\ polarization maps to study the distribution of the low frequency polarized foregrounds. We make a high S/N ratio synchrotron map by combining the \WMAP/\Planck\ data into a single product, assuming a power-law model for synchrotron emission. We discuss in detail the large-scale features in the low frequency polarized sky including the well-known loops and spurs. Other results will be published in separate articles, including the diffuse synchrotron power spectrum and further analysis of AME polarization constraints from Galactic clouds. We conclude in Sect.~\ref{sec:conclusions}.

\section{Data}
\label{sec:data}

\begin{table*}[tb]
\begingroup
\newdimen\tblskip \tblskip=5pt
\caption{Summary of radio data sets used, with selected properties.}
\label{tab:data}
\nointerlineskip
\vskip -3mm
\footnotesize
\setbox\tablebox=\vbox{
    \newdimen\digitwidth 
    \setbox0=\hbox{\rm 0} 
    \digitwidth=\wd0 
    \catcode`*=\active 
    \def*{\kern\digitwidth}
    \newdimen\signwidth 
    \setbox0=\hbox{+} 
    \signwidth=\wd0 
    \catcode`!=\active 
    \def!{\kern\signwidth}
    \newdimen\pointwidth
    \setbox0=\hbox{{.}}
    \pointwidth=\wd0
    \catcode`?=\active
    \def?{\kern\pointwidth}
    \newdimen\notewidth
    \setbox0=\hbox{$^\mathrm{a}$}
    \notewidth=\wd0
    \catcode`@=\active
    \def@{\kern\notewidth}
    \halign{#\hfil\tabskip 10pt&
        #\hfil\tabskip 2pt&
        \hfil#\hfil\tabskip 2pt&
        \hfil#\hfil\tabskip 2pt&
        \hfil#\hfil\tabskip 2pt&
        \hfil#\hfil\tabskip 2pt&
        \hfil#\hfil\tabskip 2pt&
        \hfil#\hfil\tabskip 2pt&
        \hfil#\hfil\tabskip 0pt\cr
        \noalign{\doubleline}
        Map    & Instrument & $\nu_0^\mathrm{a}$ & $\eta_{\Delta T}(\nu_0)^\mathrm{b}$ & $\sigma_I^\mathrm{c}$ & $w_0^\mathrm{d}$ & $w_1^\mathrm{d}$ & $w_2^\mathrm{d}$ & $w_3^\mathrm{d}$\cr
        \noalign{\vskip 3pt}
               &            &  [GHz]    &  & [$\mu$K] &  & & &\cr
        \noalign{\vskip 3pt\hrule\vskip 5pt}
        Haslam  & Effelsberg/Jodrell/Parkes & **0.408 & 1.000            & $3\times 10^6$ & \dots & \dots  & \dots & \dots\cr
        K band  & \WMAP                     & *22.8** & 0.987            & 6.00@          & \dots & $!1.007$ & \dots    & $!1.549$\cr
        30\GHz  & \Planck\ LFI              & *28.4** & 0.979            & 2.49@          & $!1.013$ & \dots & $!3.585$    & $-0.485$\cr
        Ka band & \WMAP                     & *33.0** & 0.972            & 4.44@          & \dots & \dots & \dots    & $-0.788$\cr
        Q band  & \WMAP                     & *40.6** & 0.958            & 3.86@          & \dots & \dots & \dots    & $-1.342$\cr
        44\GHz  & \Planck\ LFI              & *44.1** & 0.951            & 2.80@          & \dots & \dots & $-9.571$ & $-2.457$\cr
        V band  & \WMAP                     & *60.8** & 0.910            & 4.32@          & \dots & \dots & \dots    & \dots\cr
        70\GHz  & \Planck\ LFI              & *70.4** & 0.881            & 2.22@          & \dots & \dots & \dots    & \dots\cr
        W band  & \WMAP                     & *93.5** & 0.802            & 5.14@          & \dots & \dots & \dots    & \dots\cr
        100\GHz & \Planck\ HFI              & 100?*** & 0.777            & 0.86@          & \dots & \dots & \dots    & \dots\cr
        143\GHz & \Planck\ HFI              & 143?*** & 0.592            & 0.37@          & $-1.052$ & $-1.088$ & $!6.194$ & $!3.648$\cr
        217\GHz & \Planck\ HFI              & 217?*** & 0.334            & 0.51@          & \dots & \dots & \dots    & \dots\cr
        353\GHz & \Planck\ HFI              & 353?*** & 0.075            & 1.69@          & $!0.039$ & $!0.038$ & $-0.208$ & $-0.126$\cr
        545\GHz & \Planck\ HFI              & 545?*** & 0.006            & 0.51$^\mathrm{e}$ & \dots  & \dots & \dots & \dots\cr
        857\GHz & \Planck\ HFI              & 857?*** & $6\times10^{-5}$ & 0.48$^\mathrm{e}$ & \dots  & \dots & \dots & \dots\cr
        \noalign{\vskip 5pt\hrule\vskip 3pt}
    }
}
\endPlancktablewide
\tablenote {{\rm a}} Reference frequency.\par
\tablenote {{\rm b}} Conversion factor from differential thermodynamic temperature to Rayleigh-Jeans brightness temperature.\par
\tablenote {{\rm c}} Median rms per beam for the 1\deg\ smoothed Stokes $I$ maps.\par
\tablenote {{\rm d}} Weights for the linear combination images shown in Figs.~\ref{fig:overview}(a) and \ref{fig:confusion}(a) ($w_0$), Fig.~\ref{fig:confusion}(b) ($w_1$), Fig.~\ref{fig:ff_nulled}(a) ($w_2$), and Fig.~\ref{fig:ff_nulled}(b) ($w_3$).\par
\tablenote {{\rm e}} Units MJy\,sr$^{-1}$.\par
\endgroup
\end{table*}

Table\,\ref{tab:data} lists the primary radio data sets used in our analysis, with properties that are important here. We note that brightness temperatures are in the Rayleigh-Jeans convention unless otherwise stated (\wmap\ and \planck\ data are natively in CMB thermodynamic temperature). We now briefly discuss these data sets and particular issues that are relevant to our analysis. 

\subsection{\wmap\ and \planck\ data}

The primary data used in the analysis are the 9-year \wmap\ maps \citep{Bennett2013} and the 2015 release of \planck\ maps \citep{planck2014-a01,planck2014-a03,planck2014-a09} from the Low Frequency Instrument (LFI) and High Frequency Instrument (HFI). Details of the map preparation, including smoothing to 1\deg\ full-width half-maximum (FWHM) resolution and correction of residual offsets and gain errors, are given in \citet{planck2014-a12}. The products are at \healpix\ $N_{\rm side}=256$. In addition, for polarization analysis we have prepared maps at 2\degr\ resolution, using the same methodology as for the 1\degr\ maps. Component separation depends critically on the accuracy of the input data, and so we briefly review data quality issues here.

\paragraph{Thermal noise.} This is accurately characterized in the \Planck\ and \WMAP\ data. For both missions, the noise is quite variable over the sky. We have calculated the thermal noise in our smoothed maps via propagation of errors, assuming the noise is independent between pixels at our full resolution. After smoothing, the noise in nearby pixels is strongly correlated.

\paragraph{Confusion.} Confusion noise denotes the uncertainties caused by faint compact sources within each beam. It has been estimated from source counts made at much higher angular resolution. To some extent, confusion is automatically fitted out as part of component separation. This would be exactly true if the source had a spectrum that could be well fitted by our model, and was non-variable. In fact, typical sources responsible for confusion in our maps are blazars, characterized by strong variability and ``flat'' ($\beta \approx -2$) spectra, which are similar to free-free emission, but may contain multiple peaks \citep[e.g.,][]{planck2011-6.1}. Hence a fraction of the confusion is an effective noise component in our fitting.

\paragraph{Calibration.} Uncertainties for both \WMAP\ and \Planck\ due to calibration are about 0.2\,\% for most frequencies and this contributes negligibly to our error budget. The two highest-frequency HFI channels have raw uncertainties of $\lesssim$\,10\,\% \citep{planck2014-a09}, and corrections were derived as part of the component separation fit; these uncertainties have very little impact on the low-frequency components discussed in this paper. The global calibration uncertainty of the \citet{haslam1982} map is about 10\,\%, but this only impacts our estimate for the break in the synchrotron spectrum. There is also an additional source of error that arises from the broad bandwidths of the \Planck\ and \WMAP\ detectors. We can take these into account by applying ``colour corrections'' that depend on the spectral response of the instrument and the spectrum of the sky signal \citep{Jarosik2011,planck2013-p02,planck2013-p03d,planck2014-a03,planck2014-a08}. For \planck\ LFI these corrections are typically 1--3\,\% while they are 5--10\,\% for HFI. The residual errors from such corrections are likely to be $\lesssim$\,1\,\%, especially for the LFI channels of interest here. We implement \WMAP\ colour corrections using the same approach as for the LFI, allowing for the secular drift of the centre frequency \citep{Bennett2013} and the small bandpass shifts derived in \citet{planck2014-a12}. Uncertainties due to residual beam asymmetry in the smoothed WMAP/{\it Planck} data will be, relative to other uncertainties, minimal because we have used the deconvolved WMAP data and the additional smoothing to $1^{\circ}$ FWHM resolution will symmeterize the beam.

\paragraph{Offsets.} Neither \Planck\ nor \WMAP\ measured the absolute zero level in total intensity, and so there is an arbitrary offset in each map. There is also a small but obvious residual dipole in differences between \Planck\ and \WMAP\ maps. In the analysis in Sect.~\ref{sec:ilc} we use the approach of \citet{Wehus2014} that uses \ttp\ plots to estimate the dipole and monopole amplitudes for each map (exact values were updated based on the 2015 release \Planck\ maps); the approach used for the \commander\ analysis is described in \citet{planck2014-a12}. The zero levels in the polarization maps (Stokes $Q$ and $U$) are much better defined, and are recovered with negligible error by \Planck. In contrast, two large angular scale modes in the \WMAP\ $Q$ and $U$ maps are sensitive to zero-level uncertainties in the raw data. These ``poorly constrained modes'' were down-weighted by the \WMAP\ team in their power spectrum analysis \citep{Bennett2013}, but this approach is not available when attempting to model the spectrum pixel-by-pixel. They are strong enough to seriously affect the absolute polarized brightness over most of the sky, except for the regions of strongest polarized intensity, such as the Galactic plane and the North Polar Spur.

\paragraph{Polarization leakage.} Various processes cause leakage of total intensity into the polarization signal, notably mismatch between the beamshapes for the two polarizations in each horn, and bandpass mismatch (e.g., \citealp{leahy2010}). The \WMAP\ scanning strategy allows a rather accurate correction for these effects, but a more indirect approach is needed for \Planck\ \citep{planck2014-a03} and the corrections applied are believed to be accurate to a few tenths of a percent of the total intensity. This is a significant problem along the Galactic plane, where the fractional polarization is very low and so the residual bandpass leakage error is relatively high. Beam mismatch leakage is also strong on the plane, due to the steep intensity gradients there. But for most of the sky, the residual leakage is essentially negligible. 

\paragraph{Unmodelled emission.} This includes the Sunyaev-Zeldovich (SZ) effect over most of the sky,\footnote{The largest SZ signals, due to the Coma and Virgo clusters, are explicitly fitted by \Commander, but this is only feasible because other foreground components are very faint in these regions.} most spectral lines, and the spectrum of extragalactic point sources. Hundreds of transitions are known in the part of the spectrum of interest here, but over most of the sky they are expected to be very weak compared to the continuum in our broad bands. However, they are not always negligible: rich line spectra are generated in photo-dissociation regions (PDRs) surrounding \hii\ regions, and also in the X-ray dominated region around the Galactic centre \citep{Takekawa2014}. While this spectral contamination is certainly unusual, there are no blind spectral line surveys in most of our bands that would allow us to set reliable upper limits to line contamination. The \commander\ fit actually includes line emission (CO and HCN) in four of our bands.

\subsection{Ancillary data} \label{sec:ancillary}

Below 20\GHz, several radio maps are available. \citet{Reich2004} have presented a preliminary map at 1.4\GHz, but the combination of instrumental artefacts and calibration uncertainties in the survey do not meet our stringent requirements on the error budget, which also caused us to drop a number of \Planck\ datasets as described in \citet{planck2014-a12}. We do use the 408-MHz map of \citet{haslam1982}, as re-processed by \citet{Remazeilles2014} to reduce instrumental striping and provide better point-source subtraction than previous versions. The noise in the \cite{haslam1982} map is not characterized in detail, but in any case is much less than residual scanning artefacts. We assume an rms error of 1\,K, including both effects. We also assume a $10\,\%$ uncertainty in the overall intensity scale.

The Galactic synchrotron spectrum shows significant curvature below 10\GHz, while free-free absorption sets in for bright \hii\ regions below a few hundred MHz. We therefore do not consider maps at frequencies below 408\,MHz.

We use the full-sky dust-corrected \ha\ map of \cite{Dickinson2003} to compare with our free-free solution at high latitudes where dust absorption is relatively small. The fraction of dust lying between us and the \ha-emitting gas is assumed to be $1/3$ ($f_\mathrm{d}=0.33$).  We also use the \citet{Finkbeiner2003} \ha\ map to test the results. At low latitudes, we make use of the radio recombination line (RRL) survey recovered from the HIPASS data at 1.4\GHz\ \citep{Alves2014}. This survey detects RRLs from ionized gas along the narrow Galactic plane with a resolution of 15\arcm. Although of limited sensitivity, it provides the radio equivalent of \ha\ maps without absorption along the line of sight.

We use the IRIS reprocessed far infrared (FIR) map from the IRAS survey at 100, 60, 25, and 12\um\ \citep{Miville-Deschenes2005}. The intensity scale is good to $13\,\%$ and instrumental noise is effectively insignificant after smoothing to 1\deg\ resolution. We also use the \hi\ map from the LAB survey \citep{Kalberla2005}.

We also make use of the \ROSAT\ diffuse sky survey maps. The original images provided by the Max-Planck-Institut f\"{u}r Extraterrestrische Physik (MPE)\footnote{\href{http://www.xray.mpe.mpg.de/rosat/survey/sxrb/12/ass.html}{www.xray.mpe.mpg.de/rosat/survey/sxrb/12/ass.html}. Although these web pages display GIF images including the 1997 make-up  observations, the downloadable FITS images include only the 1990--91 data.} are stored as six panels that between them cover the whole sky, with slight overlaps between adjacent panels. The pixel size is 12\arcm. The 0.1--2\,keV energy range is divided into six bands. The publicly available FITS images are derived from the analysis by \citet{Snowden1997} of the 1990--1991 \ROSAT\ survey observations, and do not include the make-up data collected in 1997 and presented by \citet{Freyberg1999}; consequently they show narrow strips of missing data. After correcting erroneous field centres in the MPE FITS headers, we found the pixel coordinates in the original images of \healpix\ \citep{Gorski2005} pixels at $N_{\rm side} = 2048$, so that the original pixels are highly oversampled. Nearest-neighbour interpolation was used for placing the intensity and uncertainty values into the \healpix\ grid. We averaged the data and uncertainties onto an $N_{\rm side} =256$ pixel grid (pixel size 13\parcm7, comparable to the original), weighting each raw pixel by the area contributed to the output \healpix\ pixel. We then merged the \healpix\ maps of the six original panels into a single all-sky map, using simple averaging for pixels covered by two or more panels.\footnote{All-sky \healpix\ maps of the six \ROSAT\ bands created as described here are available at \href{http://www.jb.man.ac.uk/research/cosmos/rosat/}{www.jb.man.ac.uk/research/cosmos/rosat/}.}

\begin{figure*}
\setlength{\unitlength}{1cm}
\begin{center}
\includegraphics[height=0.49\textwidth,angle=90]{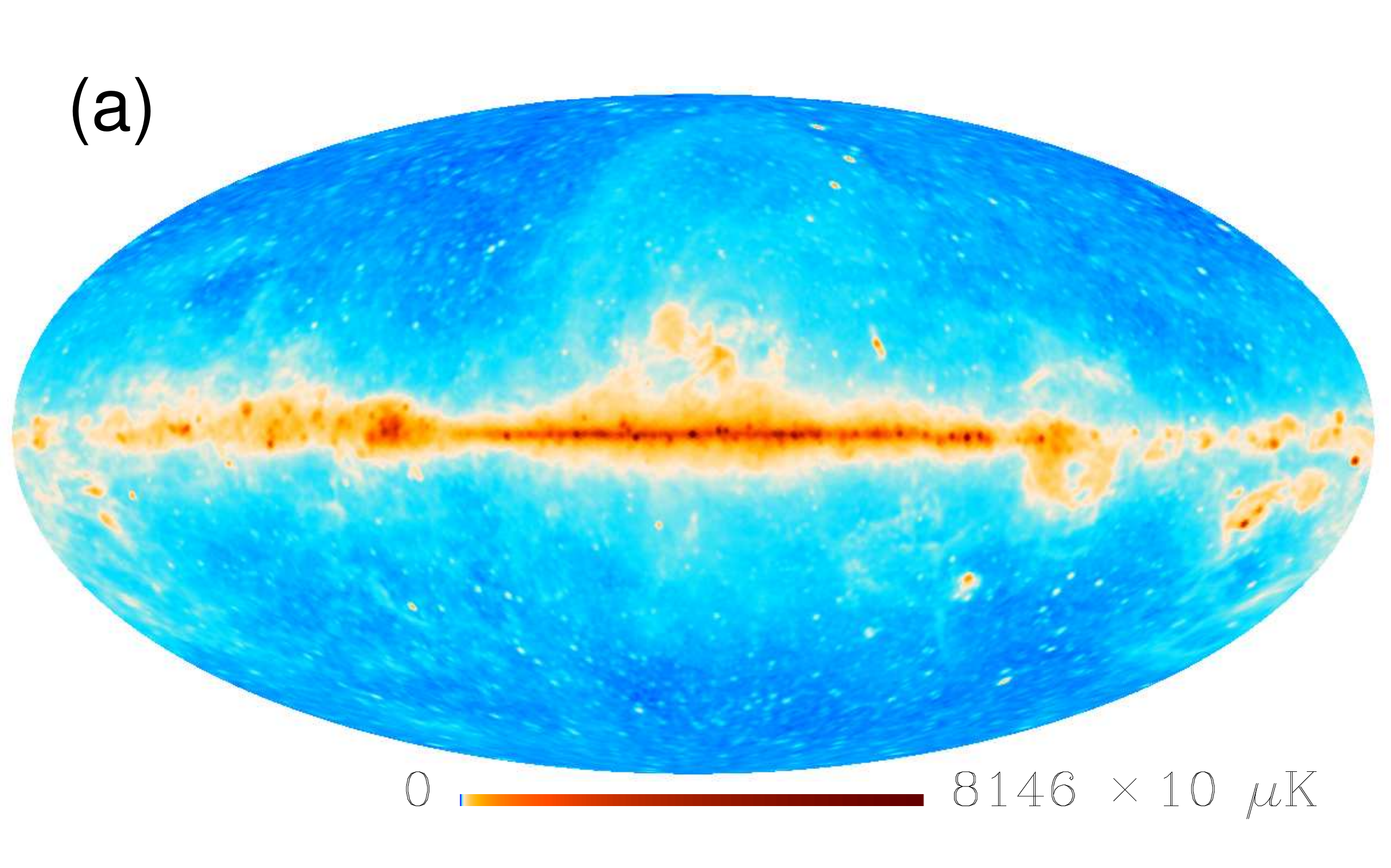}
\includegraphics[height=0.49\textwidth,angle=90]{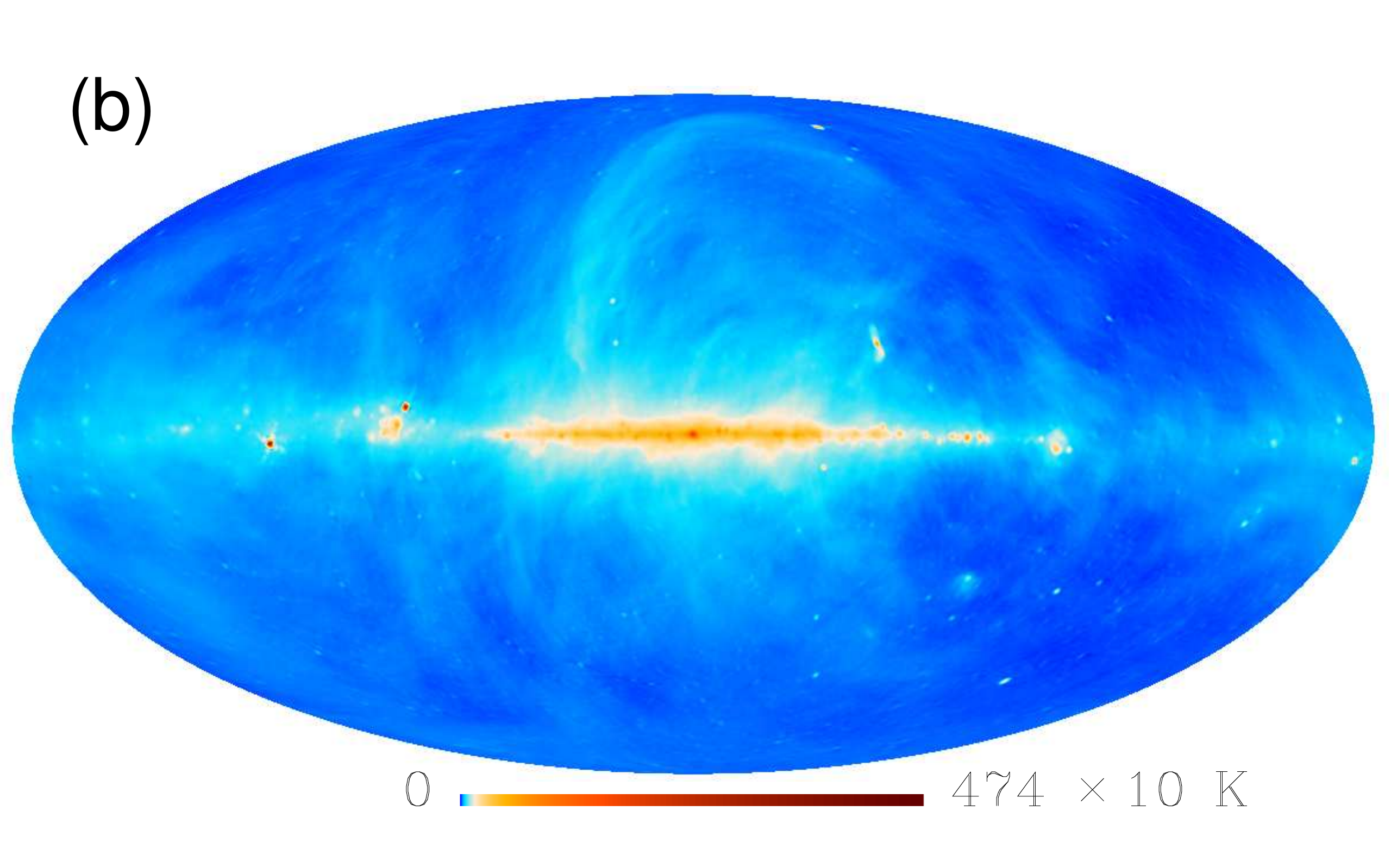}
\includegraphics[height=0.49\textwidth,angle=90]{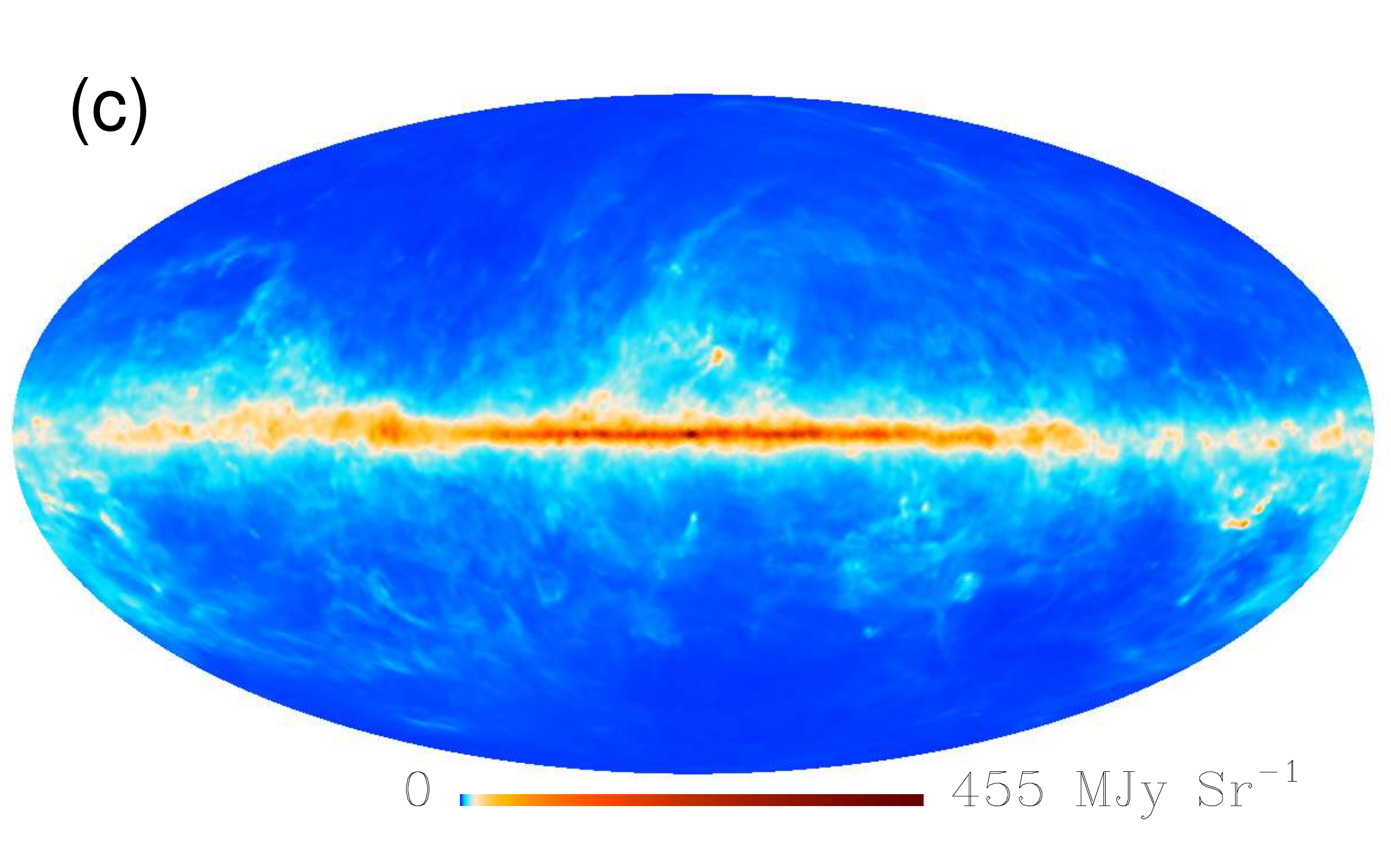}
\includegraphics[height=0.49\textwidth,angle=90]{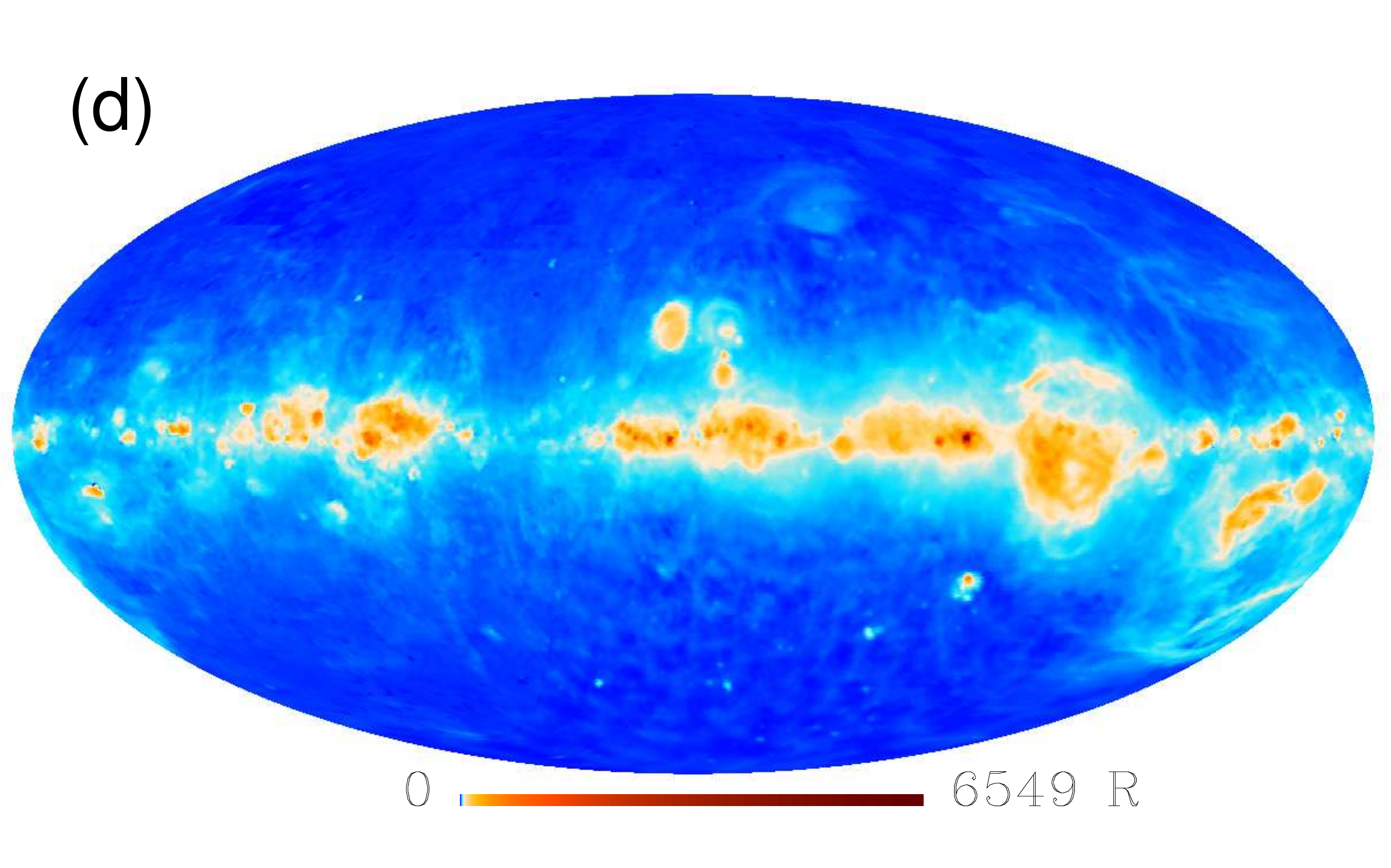}
\end{center}
\caption{(a) CMB-nulled map at 28.4\GHz, constructed using the ILC method (see text), where in this map, $\beta$\,$=$\,$-3$ emission is unchanged by construction; 
(b) 408\,MHz map \citep{Remazeilles2014}, strongly dominated by synchrotron emission;
(c) 545\GHz\ \Planck\ map, strongly dominated by thermal dust emission; and
(d) \ha\ map \citep{Dickinson2003}. 
An \asinh $(I)$ colour scale is used for all images, where $I$ is the intensity in the units indicated. The coordinate along the colour bar is linear in $I$ near zero and becomes logarithmic ($\ln(2I)$) at high intensity. All maps are at 1\deg\ resolution.}
\label{fig:overview}
\end{figure*}

\section{A first look at the low-frequency foregrounds using constrained internal linear combinations}
\label{sec:ilc}

Of all the emission mechanisms contributing to the cm-wavelength sky, the CMB is the most accurately measured, since the data are calibrated directly on the CMB dipole, and it is also the only component with a precisely known spectrum. It can therefore be straightforwardly eliminated by constructing linear combinations of the individual frequency maps. Therefore, as a first look, we apply a linear combination to the \WMAP/\Planck\ maps to investigate the range of spectral indices and morphologies at frequencies around 20--50\GHz. Our approach (Appendix~\ref{sec:ILC_method}) is similar to, but simpler and more limited than, standard ILC techniques.

Figure~\ref{fig:overview}(a) shows the total foreground emission in the \Planck\ 30\GHz\ band (reference frequency of 28.4\GHz), constructed via
\begin{equation}
T_{\rm foreground} = w_{30} T_{\rm 30} + w_{143} T_{\rm 143} + w_{353} T_{\rm 353}
\end{equation}
where the subscripts refer to channel frequencies, the weights $w_i$ are given in Table\,\ref{tab:data} and, as always in this section, are normalized to return a $\beta=-3$ power law unchanged. We use the \Planck\ HFI 143\GHz\ band as our primary model of the CMB, since it has the lowest noise and is dominated by only two components, the CMB itself and thermal dust emission. The latter is cleaned to adequate precision for display purposes using the HFI 353\GHz\ map. The dust is modelled as a uniform modified blackbody spectrum with $\beta_\mathrm{d}=1.51$ and $T_\mathrm{d}=19.6$\,K \mbox{\citep{planck2014-XXII}}. This combination fractionally diminishes the amplitude of free-free emission, but only by 0.7\,\%, which is negligible for display purposes.

\begin{figure*}
\begin{center}
\includegraphics[width=0.4\textwidth]{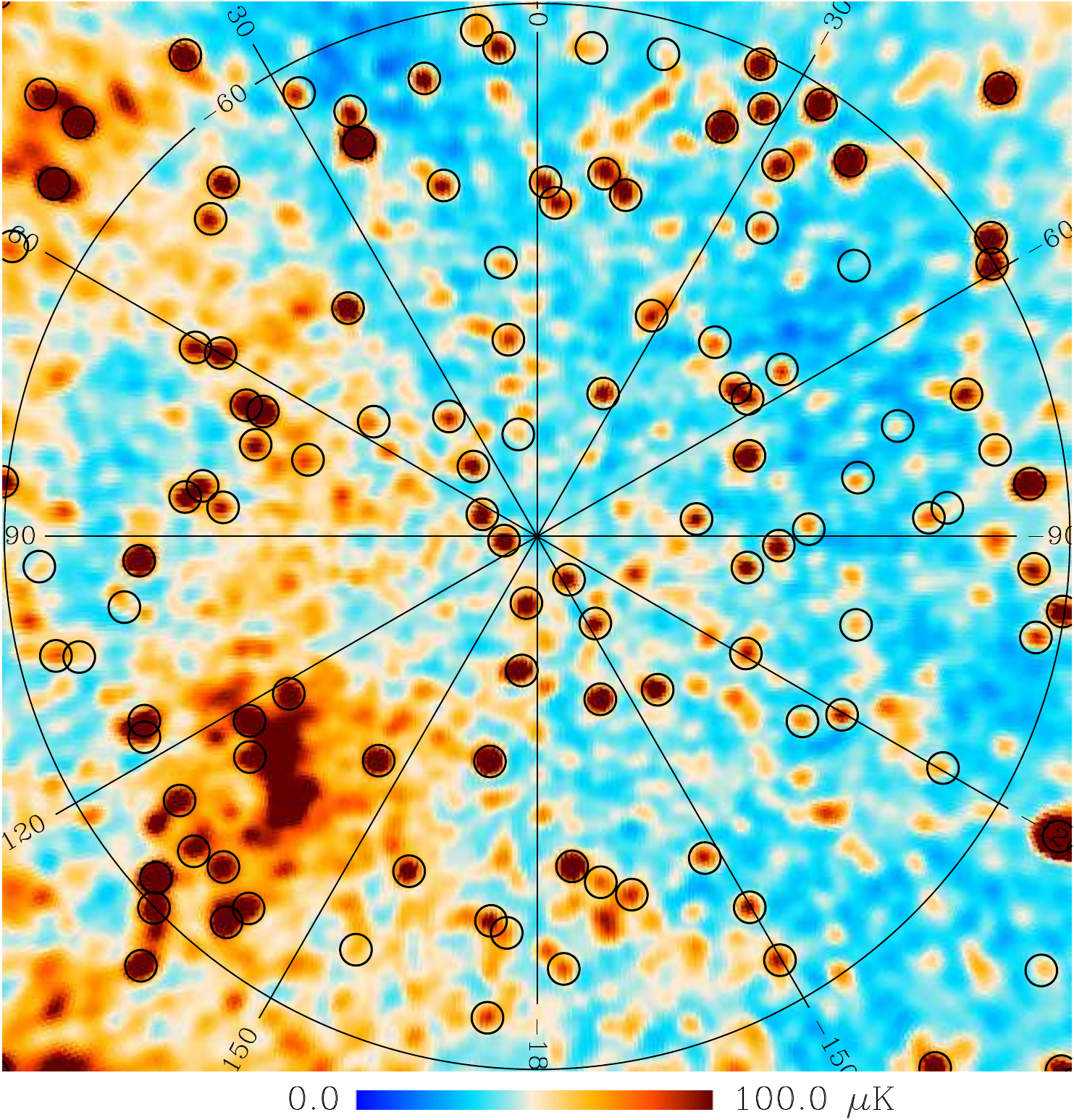}
\includegraphics[width=0.4\textwidth]{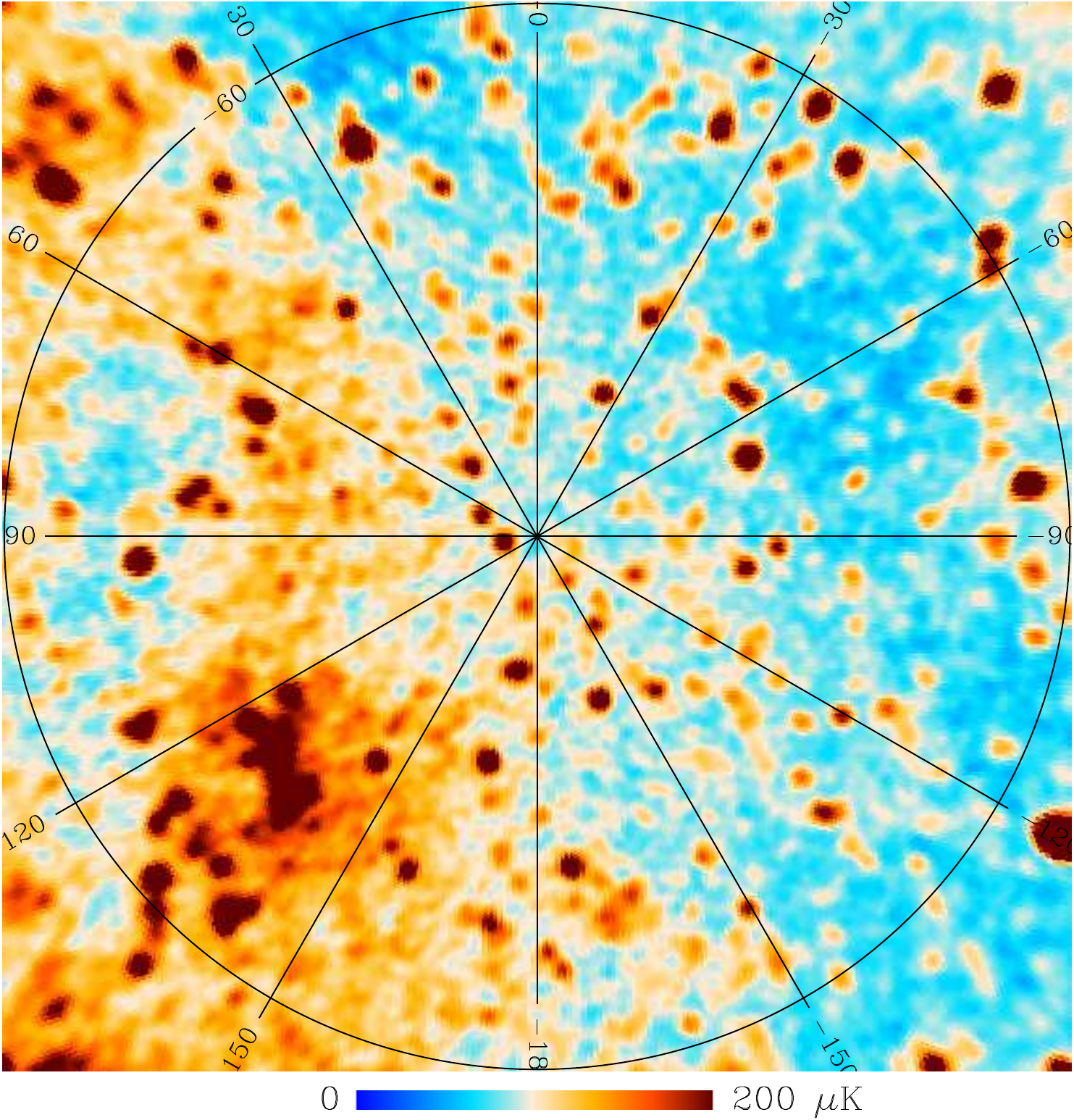}
\caption{{\it Left}: (a) South polar region of the CMB-nulled 28.4\GHz\ map. Overplotted circles are the positions of 30-GHz sources from the second \Planck\ Catalogue of Compact Sources \citep[PCCS2,][]{planck2014-a35}. The 90\,\% completeness limit for the catalogue (426\,mJy) corresponds to a peak brightness of $\DeltaT \approx 50$\uK\ at the 1\degr\ resolution of this map. {\it Right}: (b) same region in a CMB-nulled version of the \WMAP\ K-band map (22.8\GHz). The diffuse arc in the range $30\degr < l < 150\degr$ is part of Loop II. The colour scales are linear here.}
\label{fig:confusion}
\end{center}
\end{figure*}

Figure~\ref{fig:overview} also shows our ancillary maps at 408 MHz (tracing synchrotron emission), \ha\ (tracing free-free emission), and the \Planck\ 545\GHz\ map (tracing thermal dust emission). Features of all these templates appear in the 28.4\GHz\ foreground map: the brightest features along the Galactic plane are mostly free-free emission; at high latitudes in the northern hemisphere the North Polar synchrotron spur is prominent; while obvious dust-related features include the $\rho$ Oph complex, the Chameleon/Musca complex of cold clouds arcing around the South Celestial Pole, the Polaris Flare, and the comet-shaped R\,CrA molecular cloud. In addition to the diffuse components, there are thousands of point sources. In fact, at this resolution the map is strongly confusion-limited (Fig.\,\ref{fig:confusion}); the median thermal noise is 2.5\uK, but the rms in patches a few degrees across is about 10\uK, even in regions far from the plane where catalogued sources have been avoided. Consequently a signal is detected significantly in almost every pixel, and only a few SZ decrements (notably from the Coma Cluster) give a negative map intensity.

\begin{figure}
\setlength{\unitlength}{1cm}
\begin{center}
\begin{picture}(9,11)
\put(0,5.5){\includegraphics[height=9cm,angle=90]{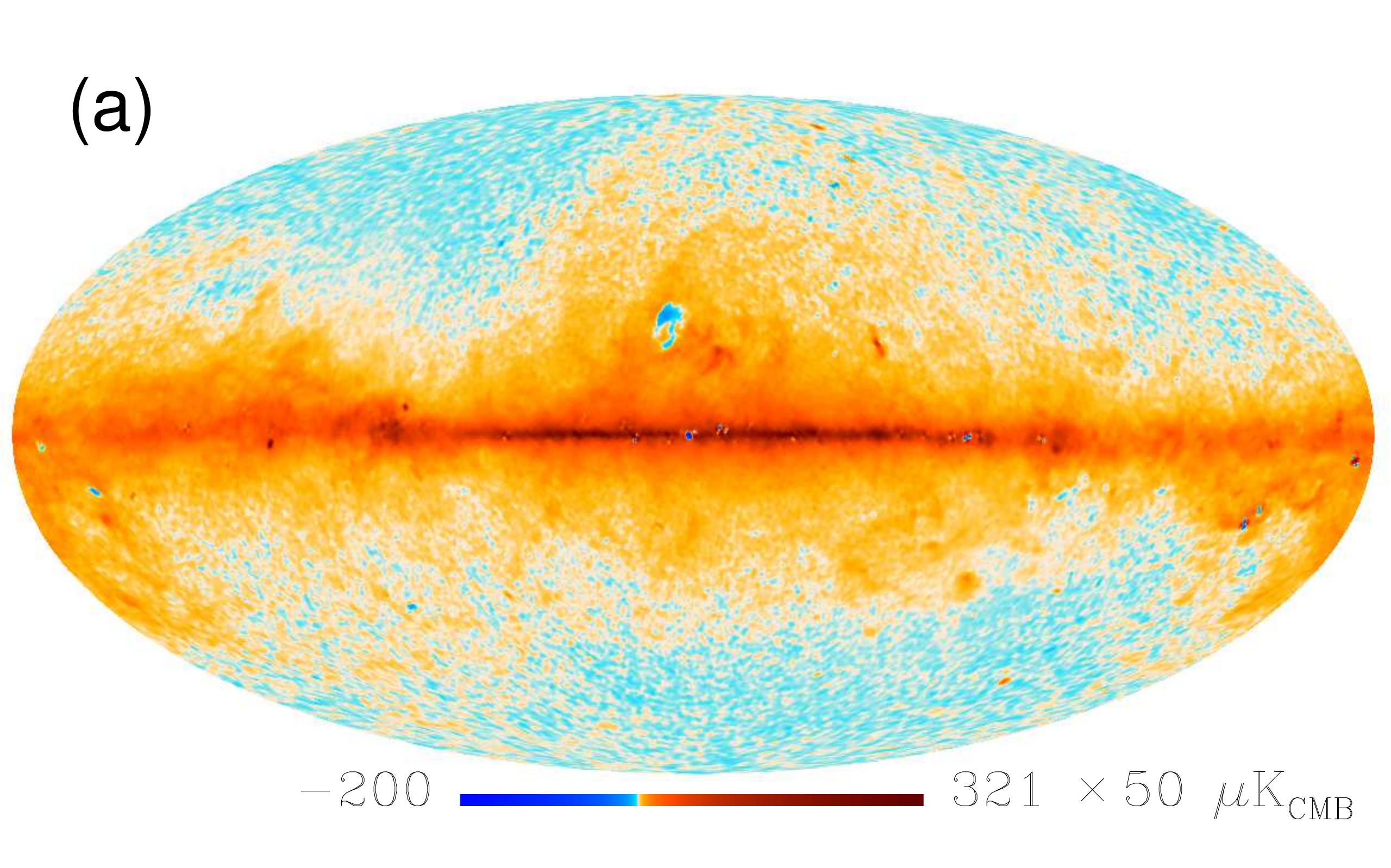}}
\put(0,0){\includegraphics[height=9cm,angle=90]{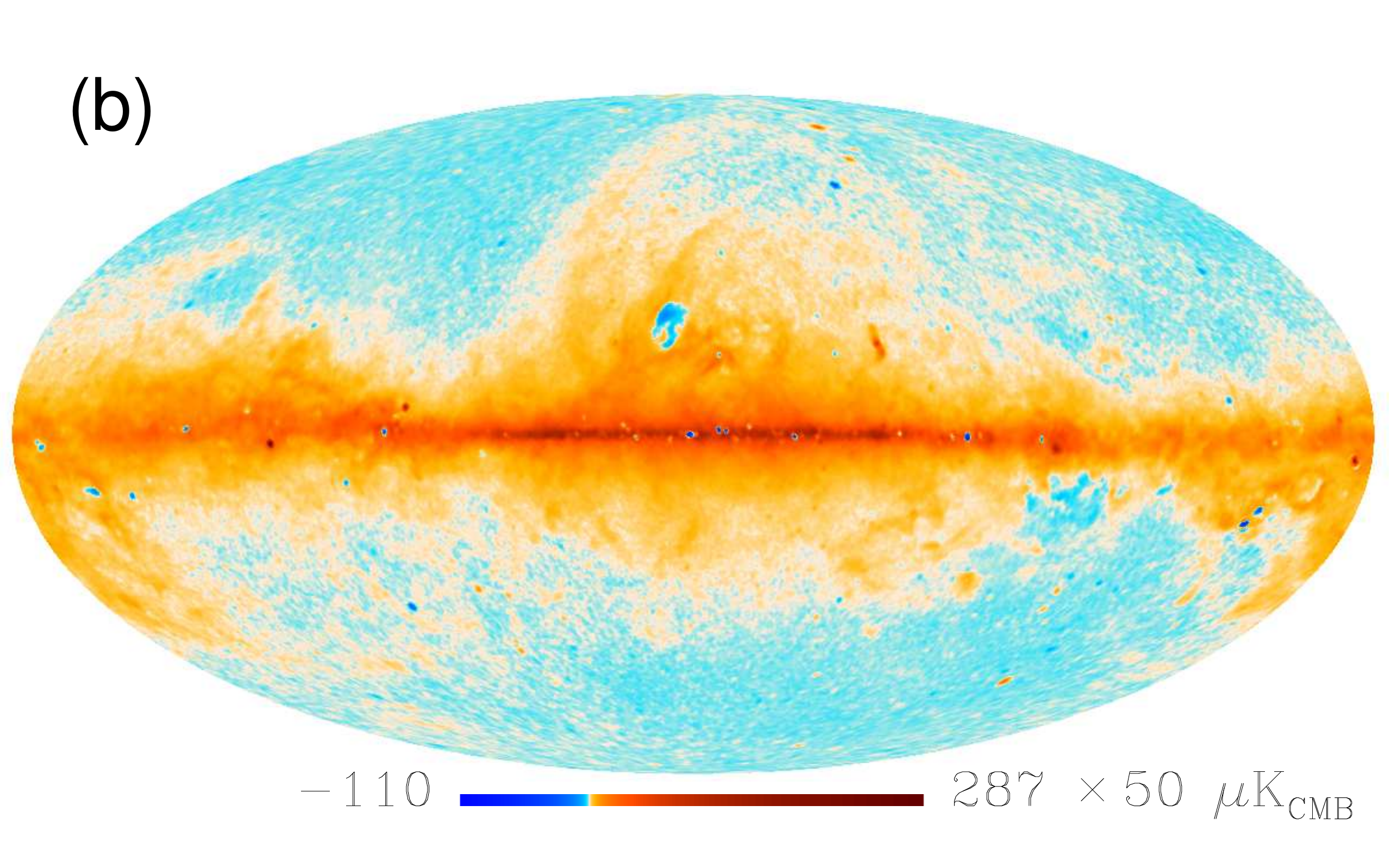}}
\end{picture}
\end{center}
\caption{Maps with CMB, free-free, and thermal dust emission nulled, scaled so that a $\beta=-3$ power-law spectrum is unchanged in amplitude.
{\it Top}: (a) \Planck-only map (28.4, 44.1, 143, 353\GHz), scaled to 28.4\GHz. 
{\it Bottom}: (b) combined \Planck\ and \WMAP. Weights used are given in Table\,\ref{tab:data}. An \asinh\ colour scale is used and an offset of $80$\,$\mu$K$_\mathrm{CMB}$ has been subtracted to enhance the contrast.}
\label{fig:ff_nulled}
\end{figure}

After the CMB, the next best determined spectrum is free-free emission. Assuming local thermodynamic equilibrium, the spectral shape depends, very weakly, on a single parameter, the electron temperature $T_\mathrm{e}$; we use the approximation of \citet{Draine_book} detailed in Sect.~\ref{sec:ff} below. For plausible values, $3000 < T_\mathrm{e} < 15000$\,K, the ratio of observed brightness between our 30 and 44\GHz\ bands varies only between 2.52 and 2.49 (including the effect of colour corrections). We can therefore null this component, taking $T_\mathrm{e} = 7500$\,K, by differencing CMB-corrected maps at 28.4 and 44.1\GHz. Figure\,\ref{fig:ff_nulled}(a) shows the result.

The drawback of this procedure is that the thermal noise is amplified by a factor of 11, partly because the differencing introduces noise from the 44\GHz\ image, and partly due to the renormalization. The situation can be considerably improved by including \WMAP\ data: Fig.\,\ref{fig:ff_nulled}(b) shows a similar analysis that includes the three lowest-frequency \WMAP\ bands, where we maximize the S/N ratio of our reference $\beta=-3$ spectrum as well as eliminating CMB, free-free, and thermal dust. In this case the S/N ratio is 5.5 times worse than in the original 28.4\GHz\ map; the better performance is partly due to the larger frequency ratio ($22.8/44.1$ vs. $28.4/44.1$), which reduces the unwanted cancellation of the synchrotron emission.

Emission with a power-law spectrum steeper than free-free, i.e., $\beta < -2.1$, will show up as positive in the free-free nulled maps, and these show (albeit with higher noise) the characteristic synchrotron and AME features noted in our description of Fig.~\ref{fig:overview} above. Most of the extragalactic sources are nearly nulled, since their spectrum is similar to free-free; those remaining are mostly optically-thin synchrotron sources, but sources that varied between the \WMAP\ and \Planck\ mean epochs also show up (as negative residuals if they were brighter during the \Planck\ observations).

In this double difference image the assumed thermal dust spectrum begins to have a significant impact; if we use $\beta_{\rm d} = 1.66$ and $T_{\rm d} = 19.0$\,K \citep{planck2013-XIV} the inner Galactic plane becomes about 20\,\% brighter. Given that $T_{\rm d}$ is known to vary significantly between the inner and outer Galaxy \citep[e.g.,][]{planck2013-p06b}, more sophisticated modelling is required to obtain the most from the data.

Nevertheless, the ILC approach does reveal an important complexity of the low-frequency foreground spectrum in a straightforward way, namely as prominent extended regions of negative intensity, notably coincident with the Sh2-27 \hii\ region at \mbox{$(l,b)=(6\deg, 24\deg)$}, and in the southern part of the Gum nebula, around $(l,b)=(257\deg,-14\deg)$. These are regions with excess flux compared to the free-free spectrum near 40\GHz. They can be fitted with a spinning dust spectrum shifted to peak at higher frequencies than standard models, but still within the range of plausible interstellar parameters \citep[e.g.,][]{planck2011-7.2,planck2013-XV}.The Sh2-27 spinning dust is likely to be associated with the famous translucent cloud in which interstellar molecules were first discovered, as absorption lines in the spectrum of the ionizing star $\zeta$ Oph \citep{Adams1941}. In fact, many bright \hii\ regions show negative residuals, e.g., the California nebula $(l,b)=(161\deg, -13\deg)$, Orion A and B $(l,b)=(210\deg,-19\deg)$ and $(207\deg,-17\deg)$, and many of the bright nebulae along the Galactic plane.

These residuals are not due to errors in our assumed $T_\mathrm{e}$ values; pushing $T_\mathrm{e}$ as high as 15000\,K makes almost no difference to the spectrum and so leaves all these features in place. They are also insensitive to changes in estimated colour corrections between \citet{planck2013-p02} and \citet{planck2014-a03}. They become less prominent for steeper assumed thermal dust spectra ($\beta_{\rm d}$), but this is mainly due to the impact of $\beta_{\rm d}$ on the underlying Galactic ridge emission noted above.

The systematic negative residuals suggest that high-frequency spinning dust may often be associated with the PDRs around \hii\ regions. \cite{Dobler2008b} also identified a potential high-frequency spinning dust component in \WMAP\ data that was correlated with H$\alpha$ from the Warm Ionized Medium. An alternative explanation in at least some cases (in particular the Galactic centre) might be significant line emission in the  39--46\GHz\ region common to \WMAP\ Q and the LFI 44-GHz bands; however well-known lines in this band, such as SiO (1--0) and methanol masers, do not seem to be bright enough to cause the effect \citep[][Jordan priv. comm.]{Jordan2015}. The implications of the wide range in AME peak frequency are discussed in Sect.~\ref{sec:ame}.

From this preliminary analysis we can draw two further lessons. First, separating components with similar spectra makes great demands on sensitivity; even separating free-free emission ($\beta = -2.1$) from synchrotron ($\beta\approx -3$) has dramatically reduced our effective S/N ratio. The synchrotron spectral index varies by only a few tenths across the sky, and to map it with useful precision we need a substantial gain in sensitivity. This has been achieved mainly by working at low resolution \mbox{\citep{Fuskeland2014,Vidal2014a}}; in the current paper we use a fixed template for the synchrotron spectrum (Sect.~\ref{sec:synchrotron} below).

Second, low-frequency foregrounds have complicated spectra: at a minimum, we must solve for the amplitudes of free-free emission, synchrotron emission, and AME, and for the latter the peak frequency and spectral width. However, even including the \WMAP\ data, we have only five frequency maps between 22 and 44\GHz\ to solve for these five parameters. With no spare degrees of freedom, we would be susceptible to both errors in the assumed spectral model and any artefacts in the data. We therefore need to use maps at higher and lower frequencies. 

Above 44\GHz\ we have the \Planck\ maps at 70 and 100\GHz\ and \WMAP\ V- and W-bands (60 and 94\GHz). There is significant line contamination in the W- and 100\GHz-bands, which has to be included in the component separation. In all these maps the strongest foreground component is the low-frequency tail of the thermal dust emission, for which it is necessary to fit at least the temperature $T_{\rm d}$ and probably the emissivity index $\beta_{\rm d}$ as well. As a result, including these higher frequency maps gives us more data points but also require us to solve for more spectral parameters. Moreover, since the foreground emission is weakest at these frequencies, the maps have relatively poor S/N ratio. The \commander\ approach discussed in the next section is our best attempt to date to handle these complexities.

\section{Total intensity foregrounds with \Commander}
\label{sec:compsep}

The analysis of Sect.~\ref{sec:ilc} demonstrated the complexity of the low-frequency foreground spectrum, but for that very reason was unable to separate the various diffuse foreground components that contribute to the maps. A more sophisticated method is required, and numerous algorithms exist that utilize the spectral and/or morphological information in the data. One of the most powerful algorithms is parametric fitting via Gibbs sampling, which has been implemented in the \commander\ code \citep{Eriksen2008}. The details of the specific implementation to these data are described in \cite{planck2014-a12}. In this section, we discuss the models for the low-frequency components used by \Commander, and compare the derived foreground products with previous results such as \WMAP\ component-separation products. We also compare the results with expectations, either based on theory or extrapolations from ancillary data.
\subsection{Free-free}
\label{sec:ff}
\newcommand{\ffHaGum}{8.9}
\newcommand{\errffHaGum}{0.9}
\newcommand{\ffHaFull}{8.0}
\newcommand{\errffHaFull}{0.8}
\newcommand{\ffHaTheo}{11.1} 
\newcommand{\errffHaTheo}{0.9} 
\newcommand{\scattFrac}{28}  
\newcommand{\errscattFrac}{12}

\begin{figure}[tb]
  \begin{center}
 \includegraphics[width=0.49\textwidth]{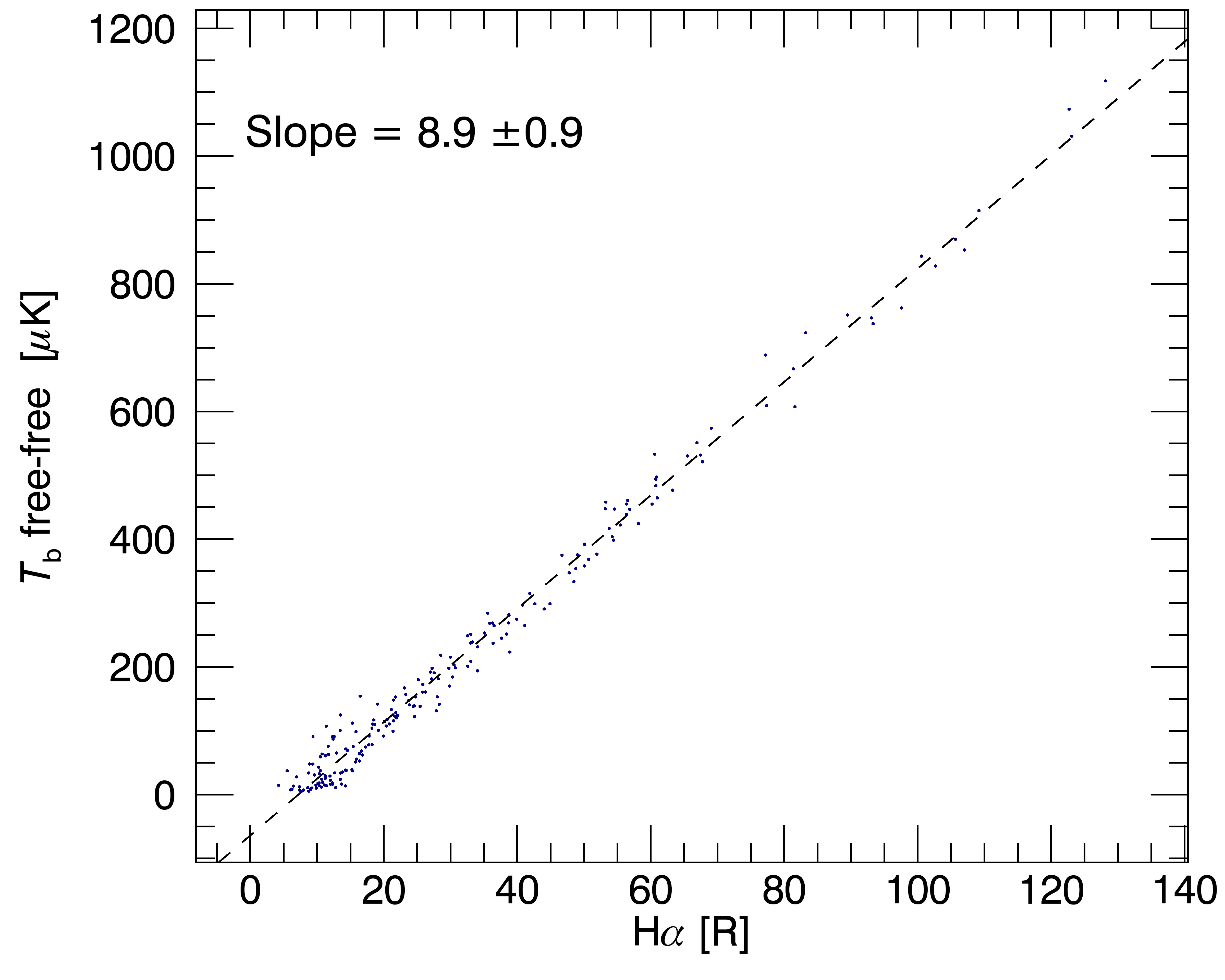}
    \includegraphics[width=0.49\textwidth]{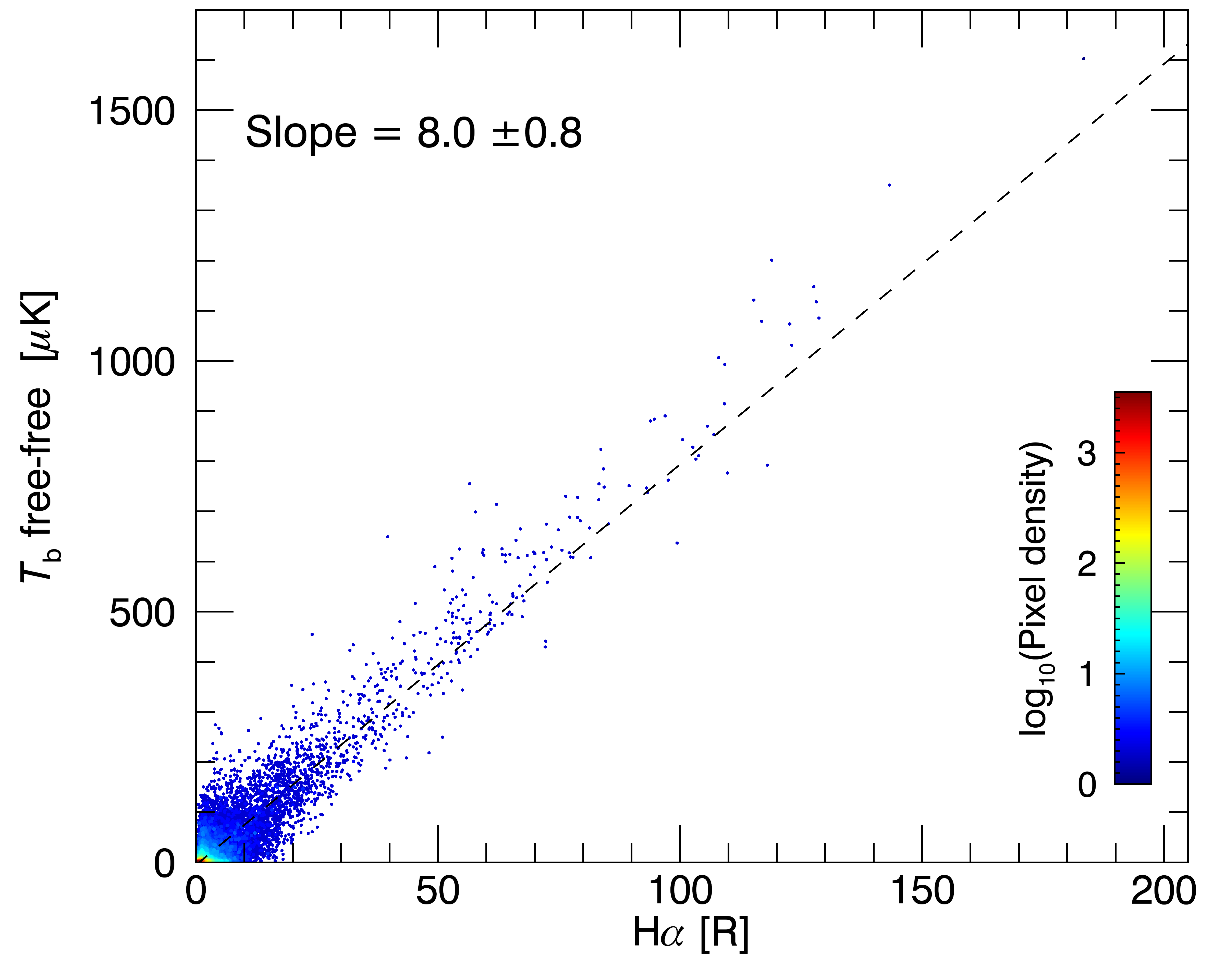}
    \caption{{\it Top}: \ttp\ plot of the free-free amplitude at 22.8\GHz\ against the \ha\ intensity [R] in the Gum Nebula (see Fig.~\ref{fig:gum_box} for the location of the region). {\it Bottom}: \ttp\ plot for 85\,\% of the sky. Regions that show large dust absorption have been masked out, mostly within $|b|\lesssim7\deg$.}
    \label{fig:TTplot_halpha}
  \end{center}
\end{figure}

The free-free radio continuum brightness can be estimated by the use of recombination lines. The most widely used is the \ha\ line ($\lambda$656.28\,nm), which traces warm ionized gas and is proportional to the emission measure (EM), in the same way as free-free radio continuum \citep{Dickinson2003}. Thus \ha\ is a reliable estimator provided that the gas is in local thermodynamic equilibrium (LTE), that there is no significant dust absorption along the line-of-sight, that there is no significant scattered component of \ha\ light, and that the electron temperature is known. The relationship between the free-free brightness $T_{\rm b}^{\rm ff}$ [$\mu$K] and \ha\ intensity $I_{{\rm H}\alpha}$ [R] for the optically thin limit is given by
\begin{equation}
  \frac{T_{\rm b}^{\rm ff}}{I_{{\rm H}\alpha}} = 1512 \, T_{4}^{0.517} \, 10^{0.029/T_{4}}  \, \nu_{\rm GHz}^{-2.0} \, g_{\rm ff}~,
\end{equation}
where $T_4$ is the electron temperature in units of $10^4$\,K, $\nu_{\rm GHz}$ the frequency in GHz, and $g_{\rm ff}$ is the Gaunt factor, which takes into account quantum mechanical effects. In the pre-factor we have included a factor of 1.08 to account for the \ion{He}{ii} contribution that adds to the free-free continuum.\!\footnote{We note that in \cite{planck2014-a12}, the equation for the free-free continuum brightness does not take into account this additional contribution from helium.} \citet{Draine_book} provides an approximation to $g_{\rm ff}$ that is accurate to within about 1\,\% in the \planck\ frequency range:
\begin{equation}
 g_{\rm ff}(\nu, T_\mathrm{e}) = \ln \left\{ \exp \left[5.960 - \frac{\sqrt{3}}{\pi} \ln(\nu_{\rm GHz}  T_{4}^{-3/2})  \right] + e \right\}~.
\end{equation}

At high Galactic latitudes, the absorption by dust is a small effect ($\lesssim 0.1$\,mag), and can be corrected for to first-order. The electron temperature is known to vary throughout the Galaxy, but has typical values of ($7500 \pm 1000$)\,K in the local diffuse ISM \citep{Shaver1983,Paladini2004,Alves2012}. This corresponds to a theoretical radio-to-\ha\ ratio of ($\ffHaTheo \pm \errffHaTheo$)\,$\mu$K\,R$^{-1}$ at 22.8\GHz, assuming $T_\mathrm{e}=(7500\pm1000)$\,K. However, previous template fitting results have indicated that the measured ratio is lower than this value, at $\approx 7$--$10\,\mu$K\,R$^{-1}$ \citep{Davies2006,Ghosh2012}.

This apparent discrepancy can be resolved with a lower electron temperature ($T_\mathrm{e} \approx 4000$\,K), although this is outside the range expected for our position in the Galaxy. Dust absorption cannot explain it since it has the opposite effect: it would increase the derived \ha\ emissivity resulting in an even lower ratio. The most popular explanation is that a fraction of the observed \ha\ intensity is from light scattered by dust grains (assuming the gas is close to being in LTE). Previous estimates of this fraction suggested relatively low values (around 10\,\%), while more recent analyses indicate that as much as half of the high latitude \ha\ intensity could be scattered \citep{Witt2010}. Other analyses give values around the 10--20\,\% level \citep{Wood1999,Brandt2012,Barnes2014}. We attempt to estimate the scattered fraction in Sect.~\ref{sec:scattering}. We begin by comparing our derived free-free amplitude with the observed \ha. We note that there are essentially no constraints on $T_\mathrm{e}$ from the \Commander\ solution, which is based on variations in the free-free spectral index, except in the inner Galactic plane (see Sect.~\ref{sec:rrl}).

\begin{figure}[tb]
\begin{center}
  \includegraphics[width=0.49\textwidth]{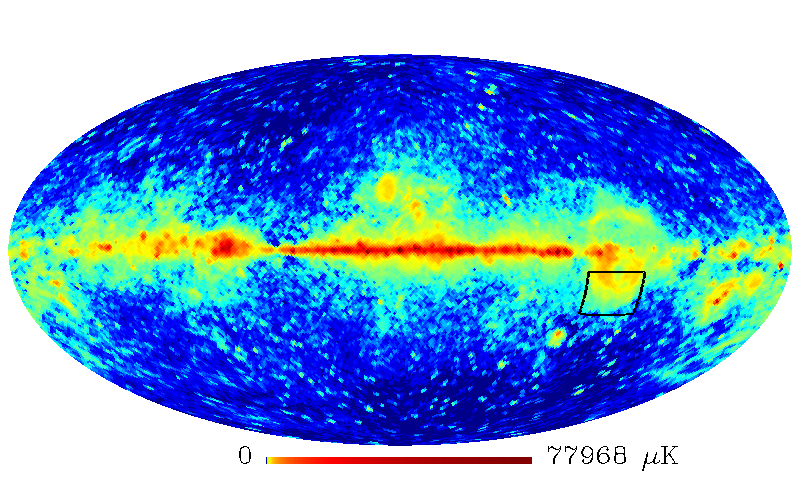}
  \caption{\commander\ free-free full-sky map at 22.8\GHz. The box shows the region we selected in the Gum nebula for the scatter plot shown in Fig.~\ref{fig:TTplot_halpha}.
    \label{fig:gum_box} }
\end{center} 
\end{figure}

Figure~\ref{fig:TTplot_halpha} (top panel) shows a \ttp\ plot of the free-free amplitude against the \ha\ intensity for the Gum nebula region (which is known to be dominated by free-free emission), as defined in Fig.~\ref{fig:gum_box}.  A very good correlation can be seen, which indicates the robustness of the \commander\ free-free amplitude. The best-fitting slope is ($\ffHaGum \pm \errffHaGum$)\,$\mu$K\,R$^{-1}$. This corresponds to an electron temperature of ($5200\pm900$)\,K, which is consistent for the temperature measured around the same region using Hydrogen RRL by \citet{Woermann2000}. We also measure the free-free-\ha\ correlation using a full-sky map, masking the more dusty regions (Fig.\,\ref{fig:TTplot_halpha}, bottom panel). The mask was defined by ignoring all pixels with an optical depth at 353\GHz, $\tau_\mathrm{353}$, larger than $1.5\times10^{-5}$ \citep[this corresponds to an extinction at the \ha\ wavelength of A(\ha)=0.54\,mag, assuming a reddening value $R_V=3.1$;][]{planck2013-p06b}, which results in a masked area of approximately 15\,\% of the sky, mostly on the Galactic plane. In the region studied, we find a best-fitting slope between the \commander\ free-free map and the \ha\ map of ($\ffHaFull \pm\errffHaFull$)\,$\mu$K\,R$^{-1}$. This is close to the value for the Gum nebula, and is consistent with previous analyses of the free-free-to-\ha\ ratio at this frequency \citep{Banday2003,Davies2006,Ghosh2012}.  This gives us additional confidence that the component separation has worked relatively well for the free-free component. Assuming the theory is correct, this lower value for the free-free to \ha\ ratio suggests a lower value for $T_\mathrm{e}$, of around 4500\,K. However, it is likely that a fraction of the \ha\ intensity is from scattering of \ha\ light from the Galactic ridge by dust grains.

\subsubsection{High latitude \ha\ scattering}
\label{sec:scattering}

\begin{figure}[tb]
\begin{center}
  \includegraphics[width=0.49\textwidth]{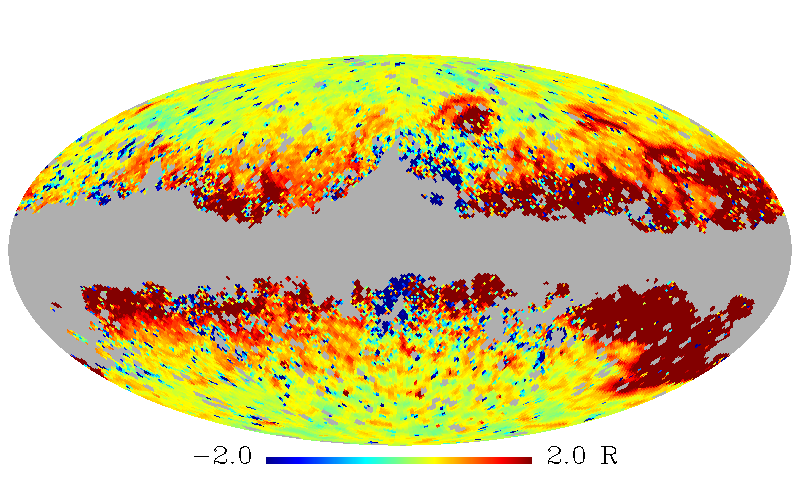}
  \includegraphics[width=0.49\textwidth]{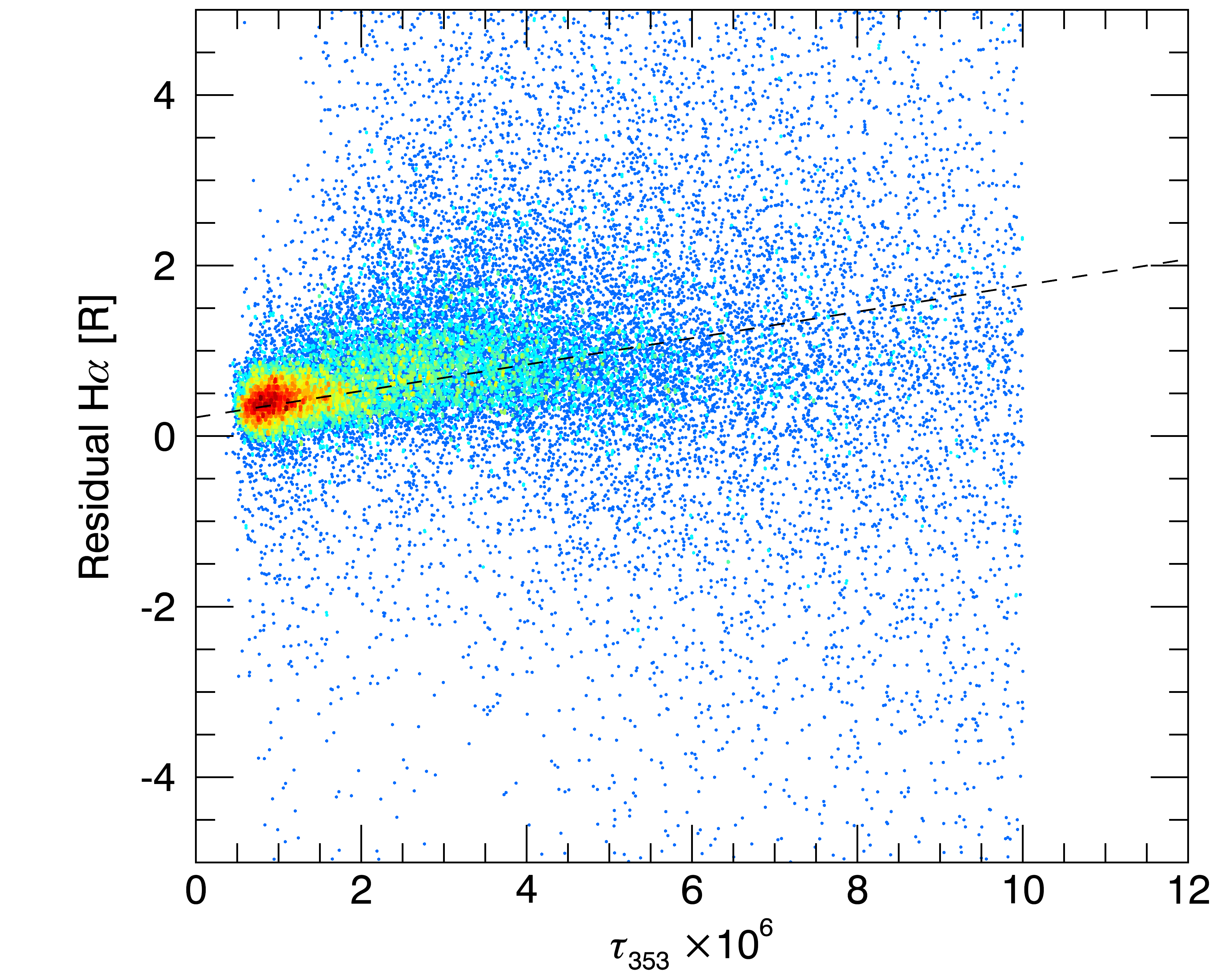}
  \caption{{\it Top:} residual \ha\ emission after subtraction of the scaled free-free component. {\it Bottom:} scatter plot of the residual \ha\ intensity against the thermal dust optical depth at 353\GHz\ outside the mask shown above, corresponding to $A({\rm H}\alpha)\lesssim 0.53$ ($\tau_{353}<1\times10^{-5}$). The best-fitting linear relation is shown as a dashed line. \label{fig:res_ha_dust}}
\end{center} 
\end{figure}

In this section we estimate the fraction, $f_\mathrm{scatt}$, of \ha\ light that is scattered by dust grains. This is believed to be responsible for the low free-free to \ha\ ratio discussed in the previous section, and could reduce the intrinsic \ha\ intensity by up to $\approx 50\,\%$ \citep{Witt2010}. To do this, we must constrain $T_\mathrm{e}$, since there is a one-to-one degeneracy between $T_\mathrm{e}$ and $f_\mathrm{scatt}$. For high latitude sight-lines away from the Galactic plane, we can assume that the bulk of the gas is at the local value $T_\mathrm{e}=(7500 \pm 1000)$\,K, thus breaking the degeneracy. We initially ignore variations in dust absorption, since we have masked out the regions that are most affected. We test the effect of this assumption by making a nominal correction for dust absorption.

First, we use the theoretical proportionality constant that relates the \ha\ intensity with the \commander\ free-free emission, ($\ffHaTheo \pm\errffHaTheo$)\,$\mu$K\,R$^{-1}$, to scale the free-free template to Rayleigh units. We note that the uncertainty accounts for a $\pm 1000$\,K uncertainty in $T_\mathrm{e}$. We subtract this template from the \ha\ map to obtain the \ha\ emission that does not come from in-situ recombination, i.e., originating from \ha\ light scattered by dust grains. The top panel in Fig.~\ref{fig:res_ha_dust} shows the residual map, with the pixels masked out shown in grey and the colour scale being truncated to $\pm 2\,$R, which saturates the brightest regions closer to the Galactic plane and the Gould Belt. A feature of this residual map is the blue central region, which resembles the microwave haze noted by \citet{Finkbeiner2004}, \citet{Dobler2008} and \citet{planck2012-IX}. This is not unexpected, since the haze was discovered by assuming that \ha\ is an accurate free-free template. This is an indication that component separation is ambiguous in this complex region, since mixed emission from the Gould Belt system makes the separation between AME, free-free, and synchrotron particularly difficult. There may also be component separation issues in other regions of the sky at a lower level, which could have a significant impact on the these residuals.

However, we can still set an upper limit for the amount of scattered \ha\ light, based on the value of the free-free to \ha\ ratio measured over the sky. The mean value at 22.8\GHz\ that we measure over most of the sky is ($\ffHaFull \pm\errffHaFull)$\,$\mu$K\,R$^{-1}$. From this, if we assume that the difference relative to the theoretical value of ($\ffHaTheo\pm\errffHaTheo)$\,$\mu$K\,R$^{-1}$ is due to an excess of \ha\ emission from scattering, we can estimate the scattering fraction to be $f_{\rm scatt}=(\scattFrac \pm \errscattFrac)$\,\%. This is an average value for the high latitude sky, so individual regions with different electron temperature might be present a level higher than this. It is consistent with previous estimates of around 20\,\% of scattered light on average at high latitudes \citep{Wood1999,Witt2010,Dong2011,Brandt2012,Barnes2014}. If we repeat the calculation using a dust-corrected \ha\ map, assuming that $1/3$ of the dust lies in front of the \ha-emitting gas (see \citealt{Dickinson2003}), we find $f_{\rm scatt}=(36\pm12)\,\%$.

Our constraints are clearly not very tight, and are dependent on our assumption on the mean electron temperature and on the level of dust absorption. Given the relatively large uncertainties, an electron temperature of about 5000\,K would be enough to bring the ratios within the 1$\sigma$ uncertainty of the observed range.

We now explore the correlation between the residual \ha\ emission and the dust optical depth at 353\GHz. Scattering by dust grains should produce a positive correlation between these two maps. We plot this in the bottom panel of Fig. \ref{fig:res_ha_dust}, where for low values of the opacity, $\tau_{353}\lesssim 1\times10^{-5}$, the correlation is indeed positive. The large dispersion of the points in the figure is a consequence of the noise and variations of electron temperature at high Galactic latitude. In spite of that, a significant correlation is observed. This suggests that the effect of scattered \ha\ light might be real and that its effects must be taken into account to determine precise values of $T_\mathrm{e}$. But we remind the reader that this could also partly be explained by the regions corresponding to $\approx 0.5$--1\,R (at high Galactic latitudes, $|b| \gtrsim 40\deg$, where the positive correlation is observed) having a lower than average value for $T_\mathrm{e}$.

We do not find strong evidence for high latitude regions with a scattered light fraction of around 50\,\%. We also do not see an anti-correlation of the free-free-to-\ha\ ratio with the thermal dust optical depth (a proxy for column density of dust), as would be expected if there were a significant fraction of scattered light. We hypothesize that the electron temperature at our exact Galactic position could be slightly lower than expected (say 6000\,K), which would reduce $f_{\rm scatt}$ to $\approx 10\,\%$.

\subsubsection{Free-free emission in the plane}
\label{sec:rrl}

\begin{figure}
\centering
\includegraphics[scale=0.35]{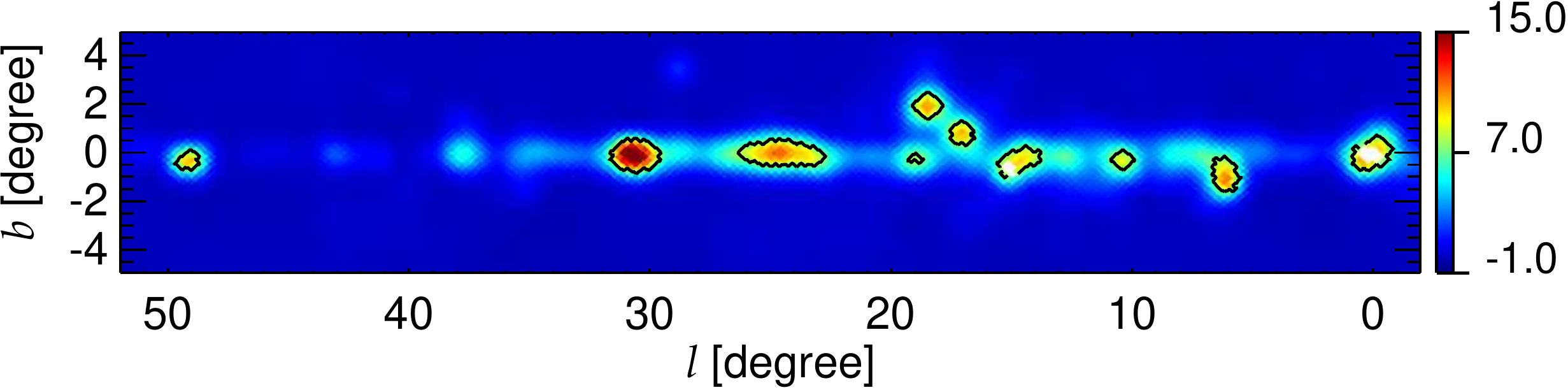}
\includegraphics[scale=0.35]{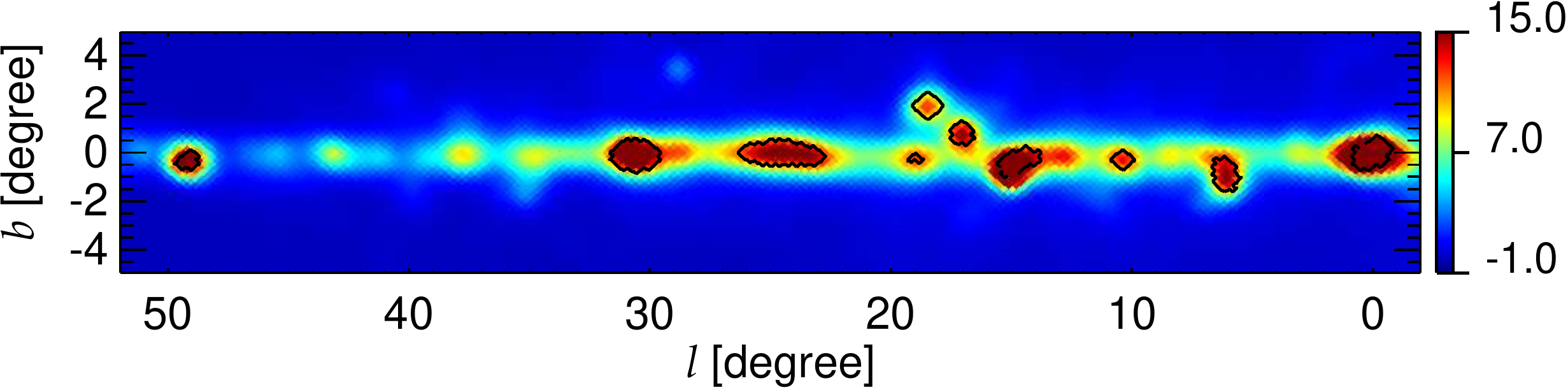}
\includegraphics[scale=0.35]{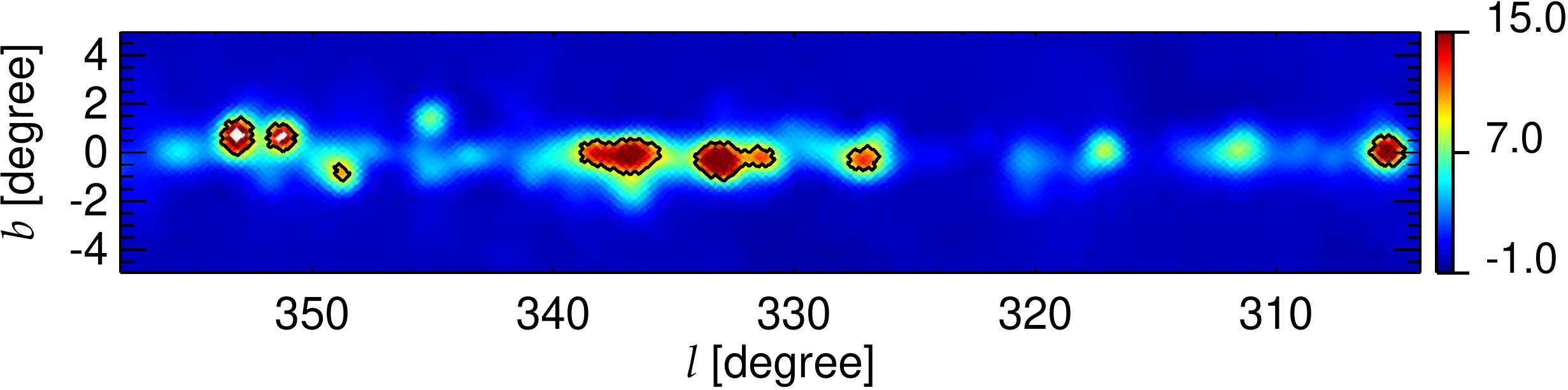}
\includegraphics[scale=0.35]{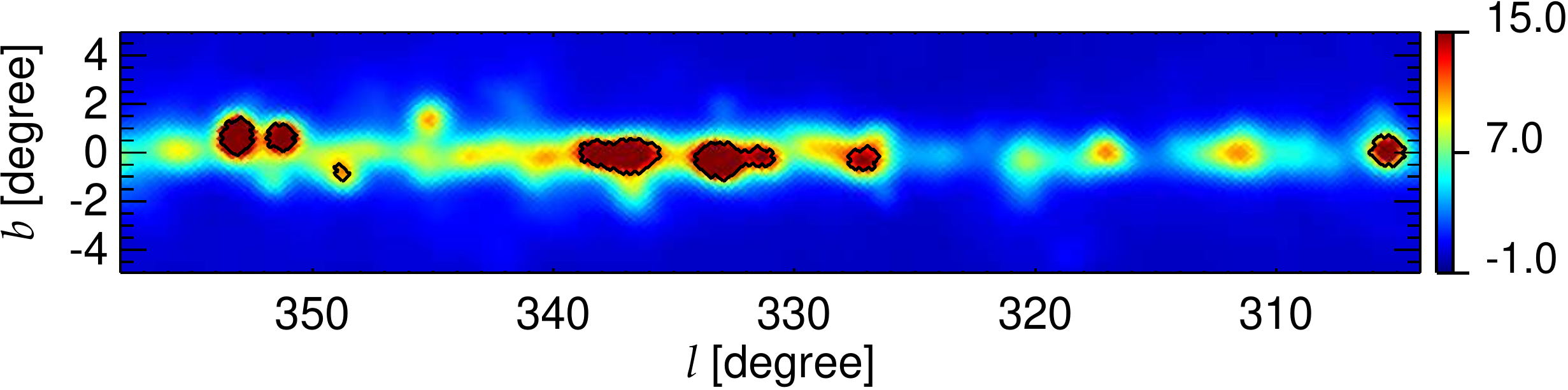}
\includegraphics[scale=0.35]{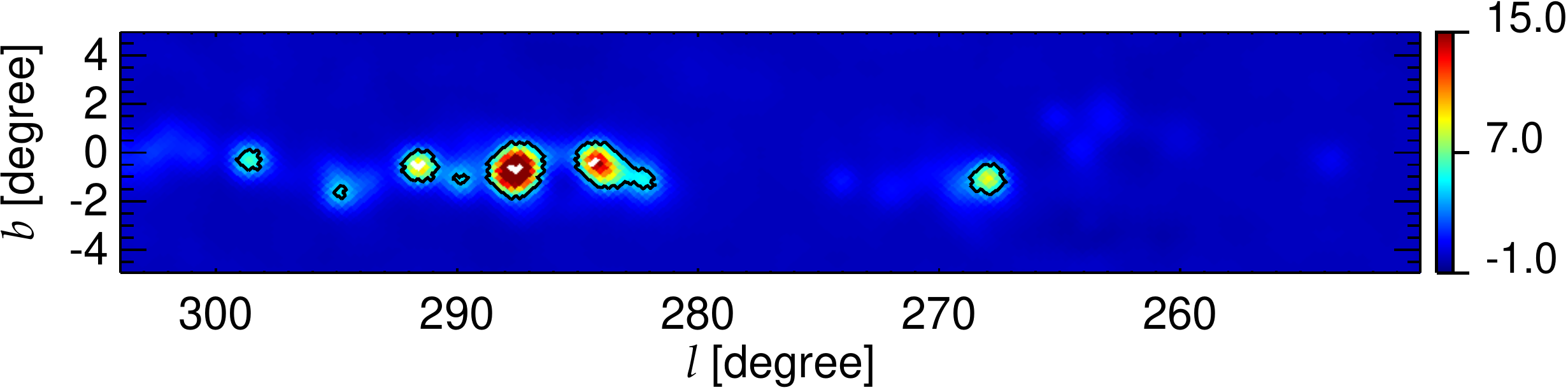}
\includegraphics[scale=0.35]{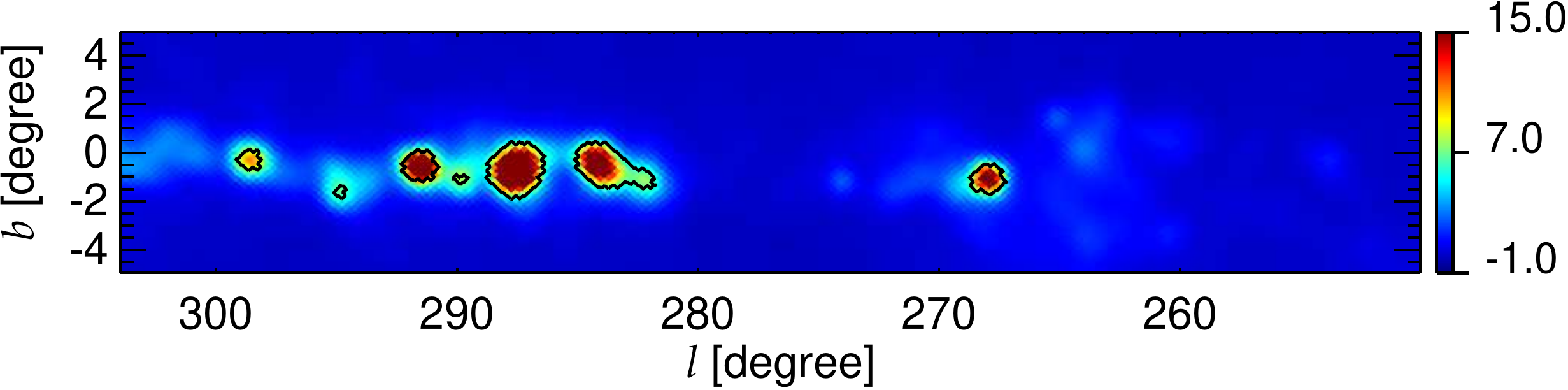}
\includegraphics[scale=0.35]{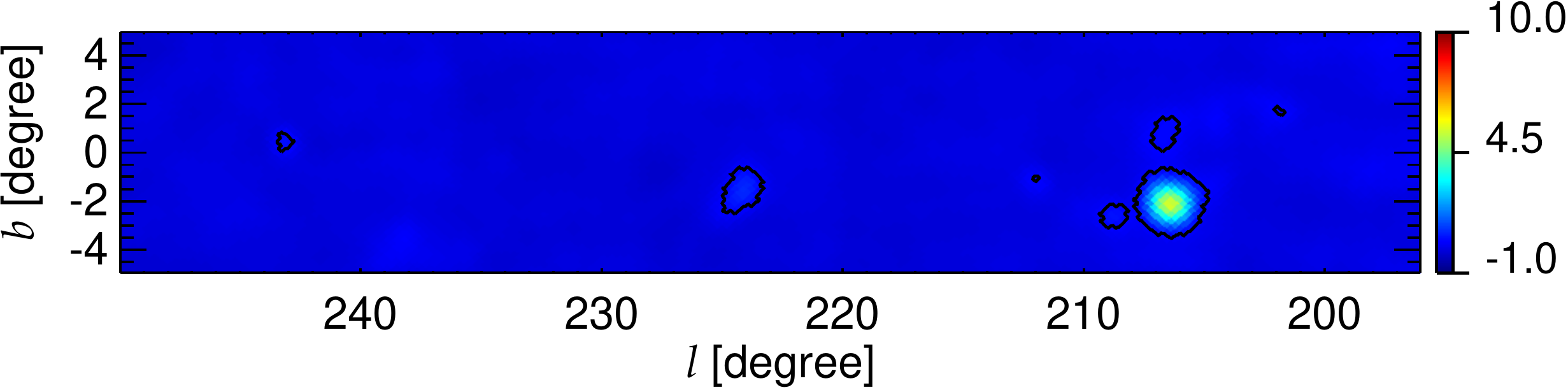}
\includegraphics[scale=0.35]{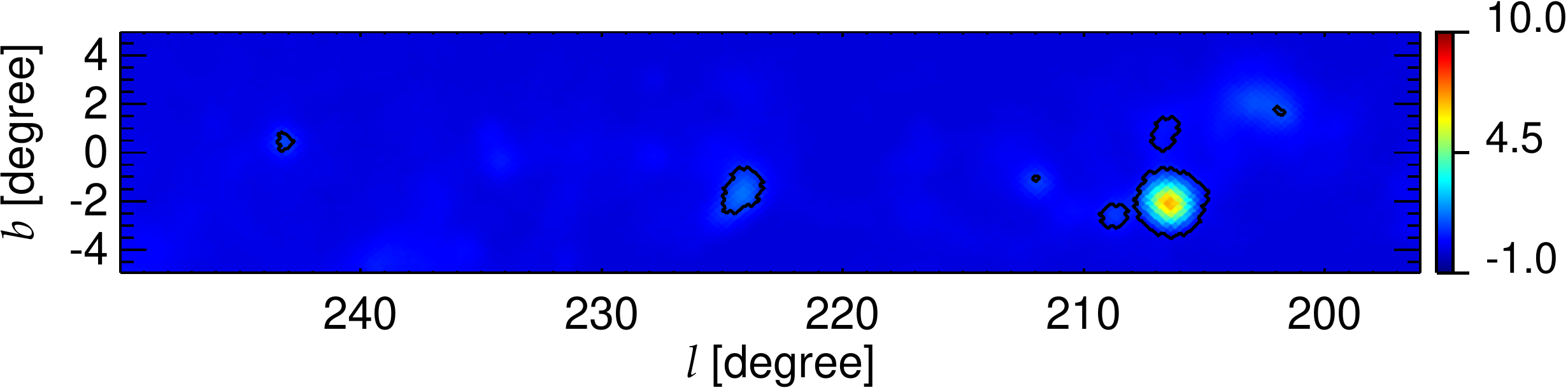}
\caption{Maps of the free-free emission along the Galactic plane, separated into four different longitude ranges. For each $l$-range, the RRL and \commander\ estimates are shown in the top and bottom panels, respectively. The maps are in units of K at 1.4\GHz\ and at 1\deg\ resolution. There is a single contour for each pair of panels, set at a value corresponding to the minimum temperature of the 2\,\% brightest pixels of the RRL map, corresponding to 7.4, 9.6, 4.4, and 0.4\,K, from top to bottom, respectively. The white pixels at the location of some bright \hii\ regions in the maps of the first, third and fifth rows, correspond to saturated pixels in the RRL
  survey.} 
\label{fig1}
\end{figure}

\begin{figure*}
\centering
\includegraphics[width=0.33\textwidth]{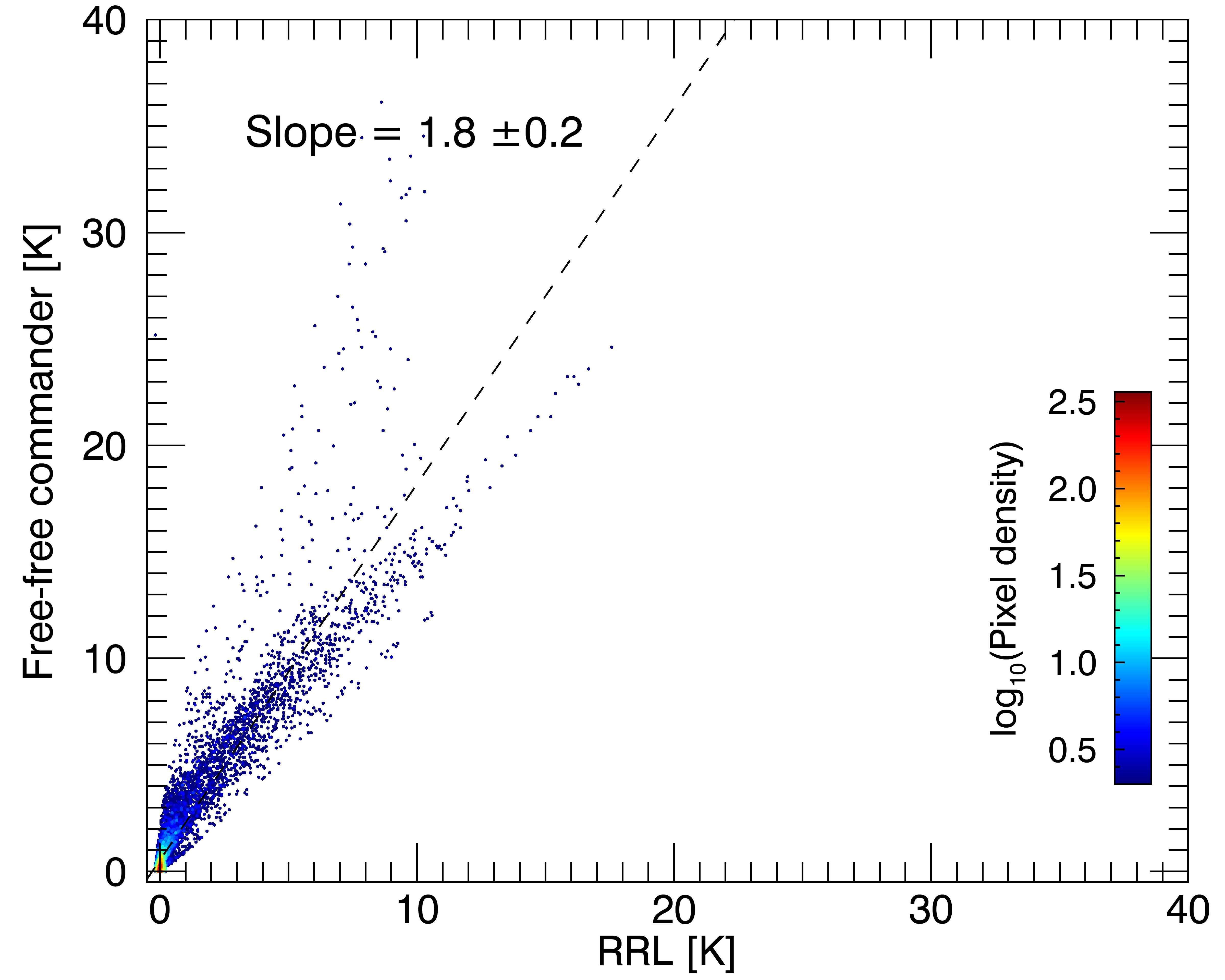}
\includegraphics[width=0.33\textwidth]{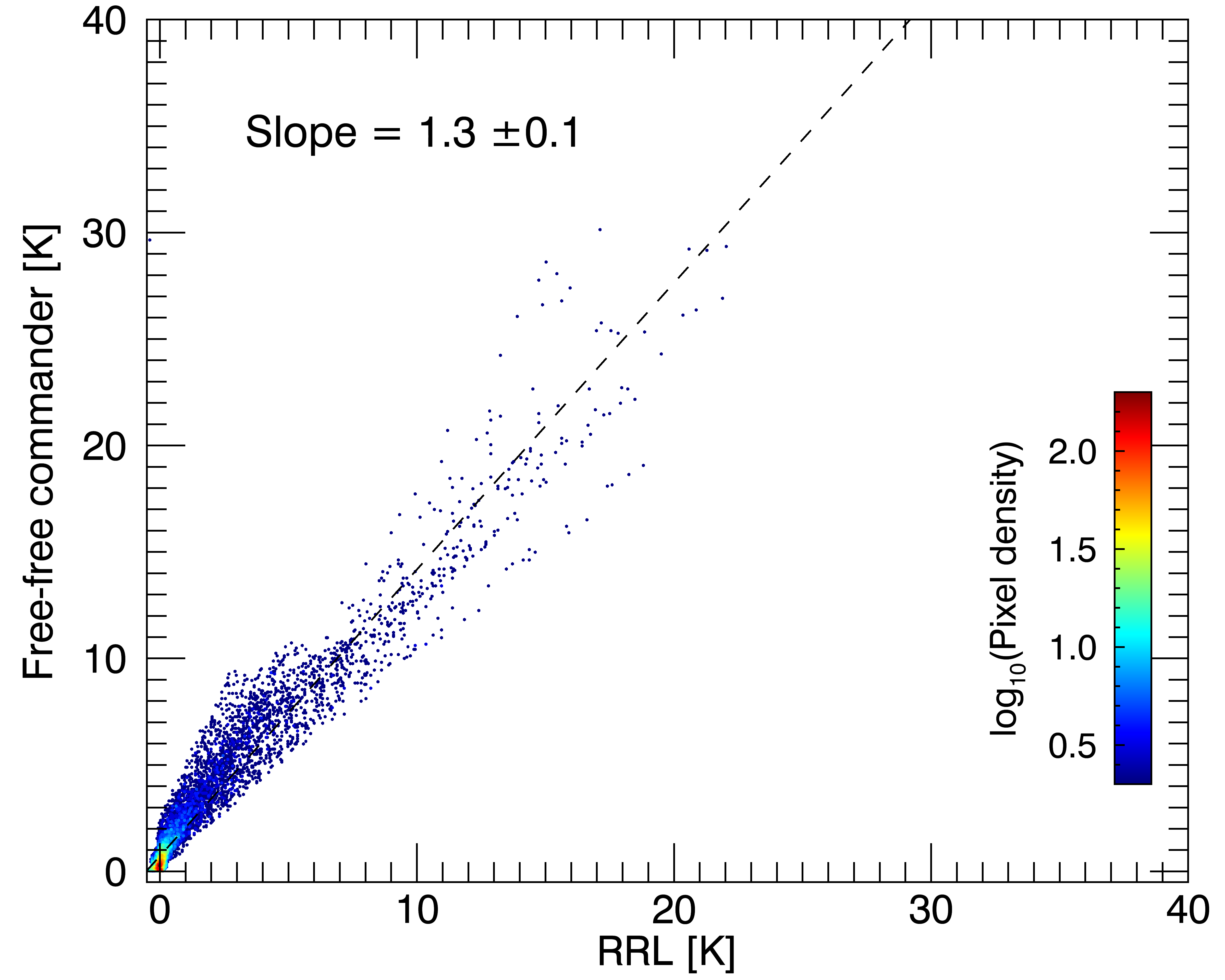}
\includegraphics[width=0.33\textwidth]{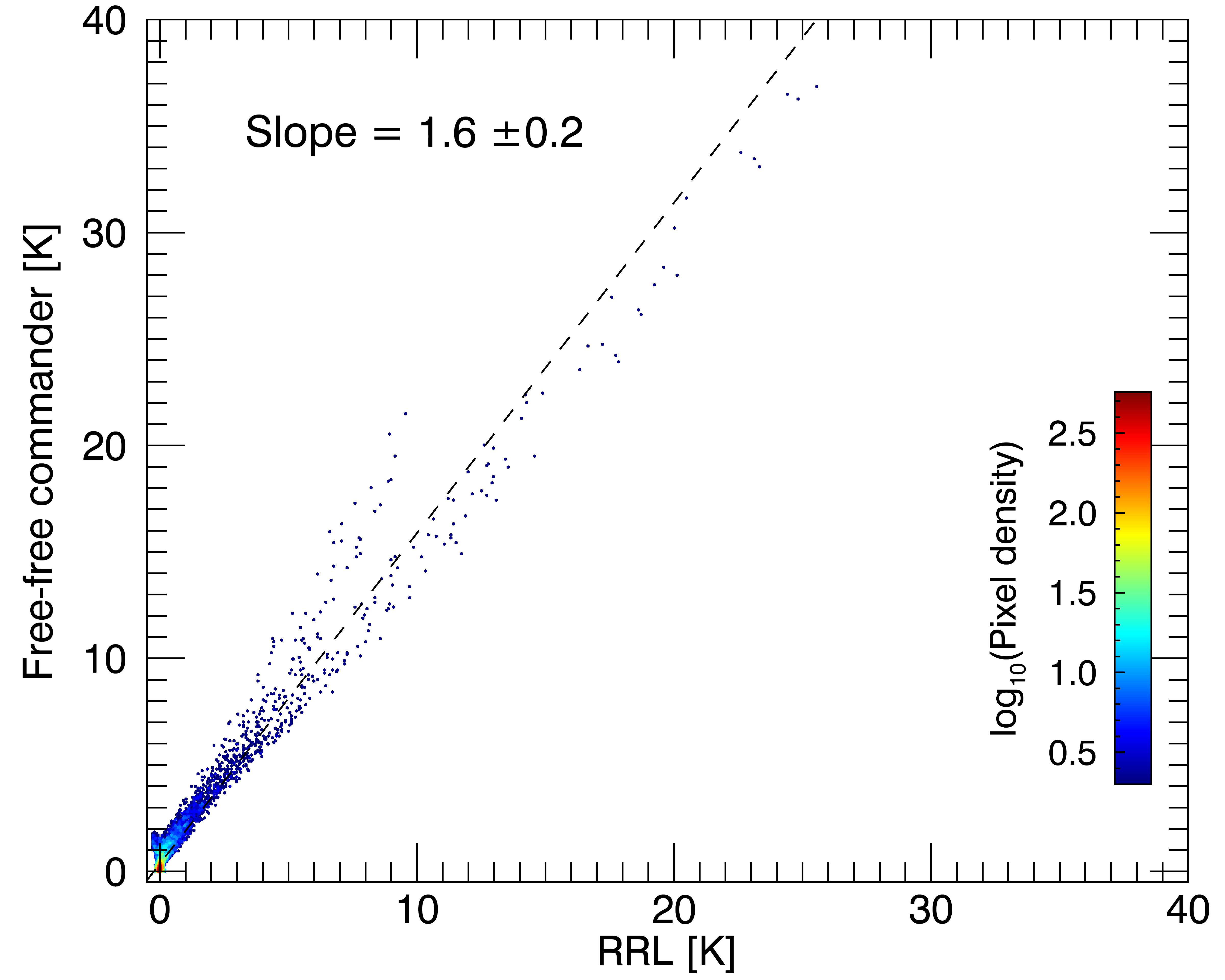}
\caption{Scatter plots of the \commander\ versus the RRL free-free map for three regions of the Galactic plane (the first three regions shown in Fig.\,\ref{fig1}. The slope values given in each panel result from a linear fit to all the data points.}
\label{fig2}
\end{figure*}

\begin{figure*}
\centering
\includegraphics[scale=0.35]{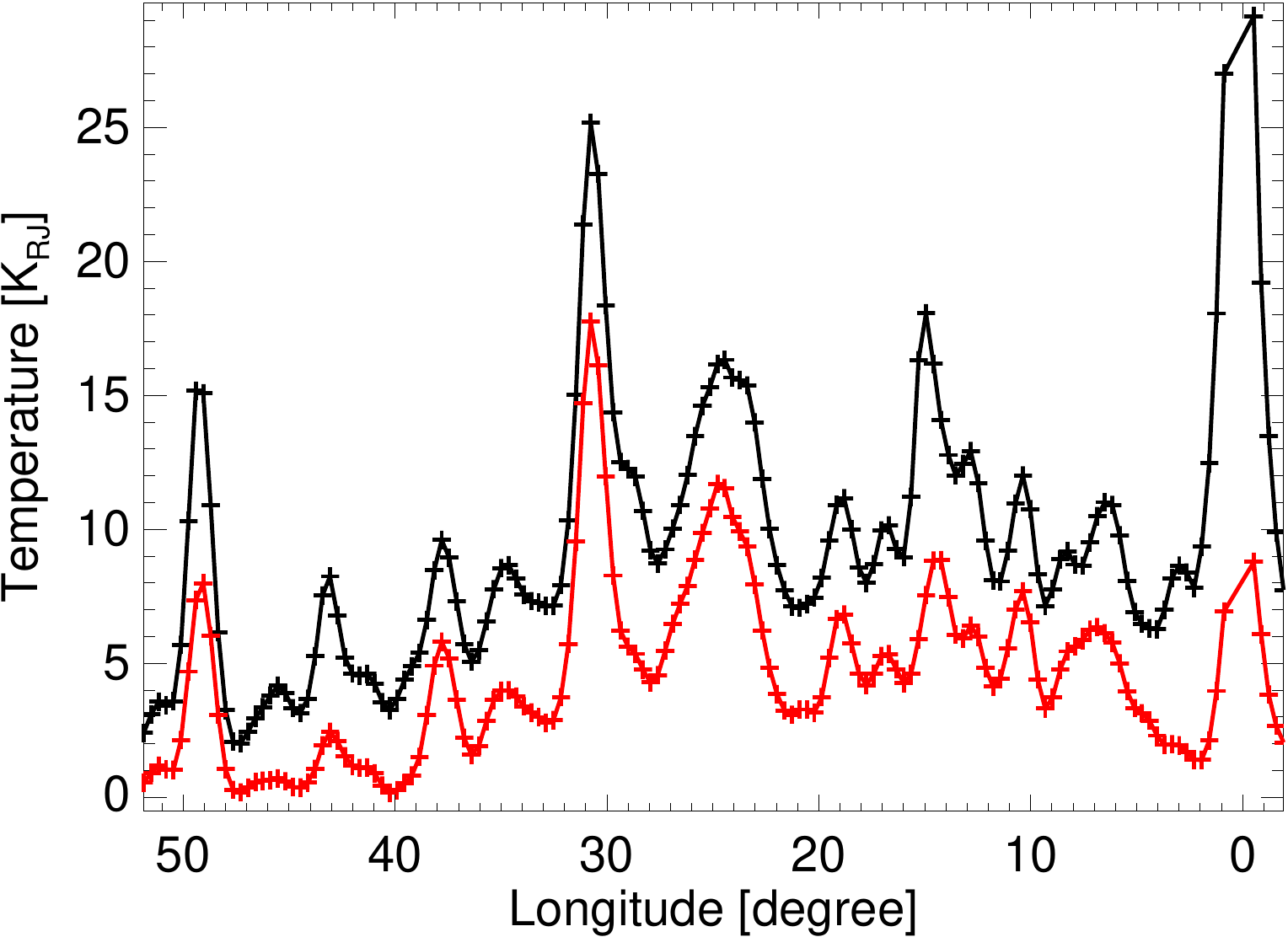}
\includegraphics[scale=0.35]{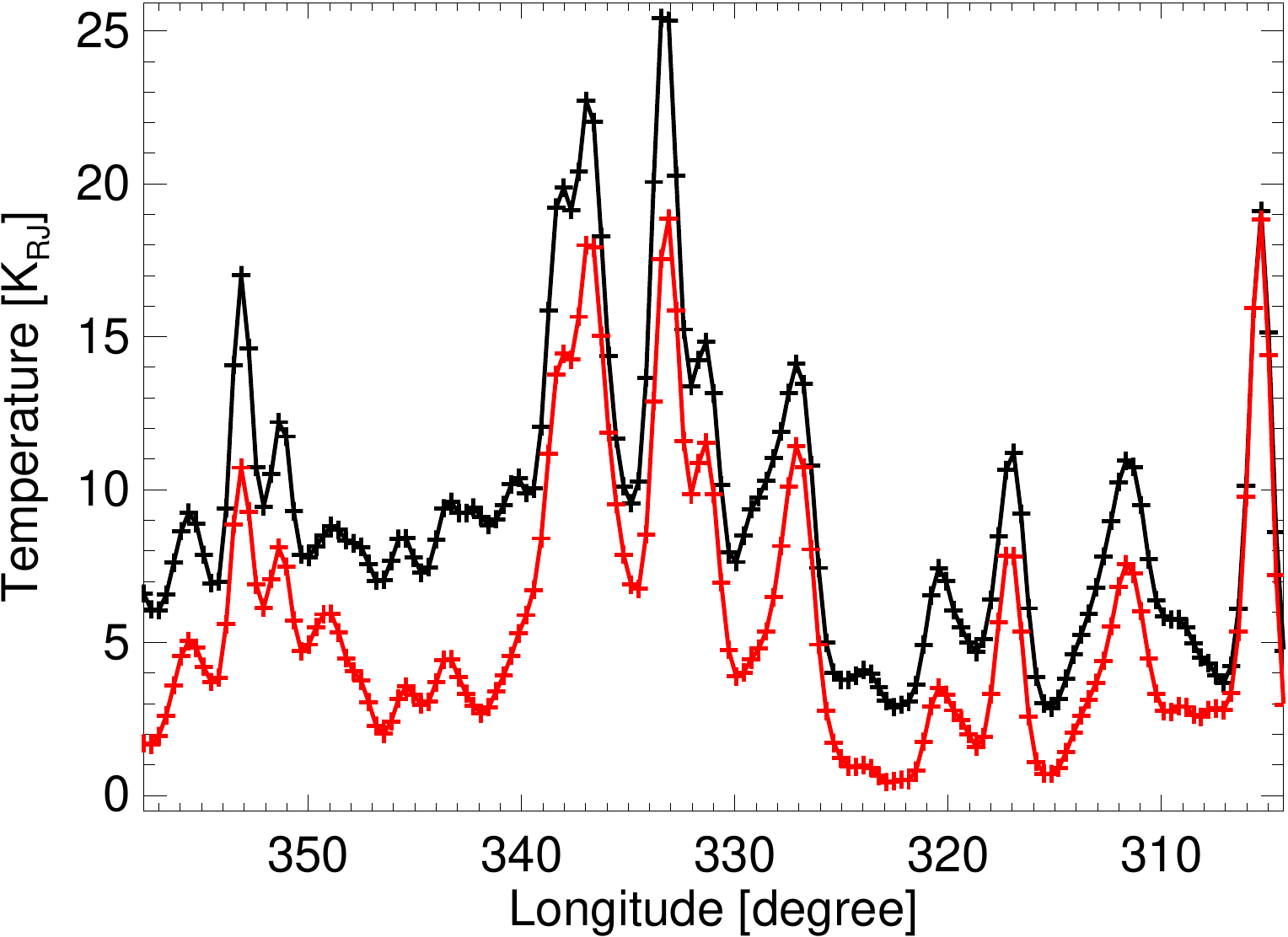}
\includegraphics[scale=0.35]{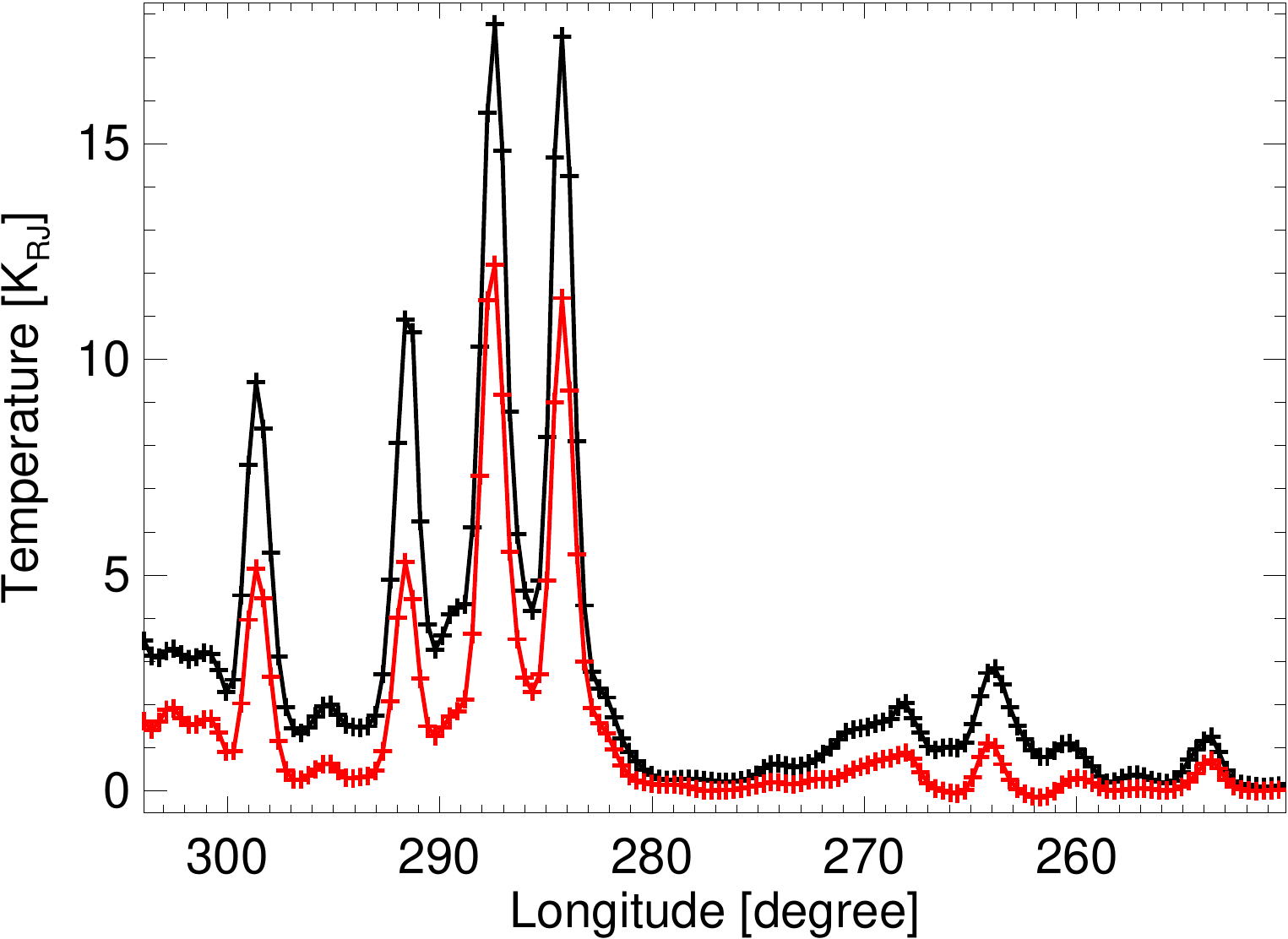}
\caption{Longitude profiles of the \commander\ (black) and RRL (red) free-free maps, averaged within $|b|\leq1\deg$, for three regions of the Galactic plane (the first three regions shown in Fig.\,\ref{fig1}). The comparison is for the bright compact regions that show as peaks on top of a broad baseline level due to extended emission.}
\label{fig3}
\end{figure*}

Along the Galactic plane, the \ha\ line is highly absorbed by dust and thus fails to provide a reliable measure of the free-free emission \citep{Dickinson2003}. In this region of the sky, we can use RRLs to estimate the thermal continuum emission and compare with the \Commander\ free-free solution. The integrated RRL brightness is proportional to the free-free continuum in exactly the same way as for the optical \ha\ recombination line. 

We use the data from the \hi\ Parkes All-Sky Survey (HIPASS; \citealt{Barnes2001}) of the southern sky, which have been re-analysed to extract RRLs at 1.4\GHz\ to produce the first continguous RRL survey of the Galactic plane \citep{Alves2014}. The full survey covers the longitude range $196\deg \rightarrow 0\deg \rightarrow 52\deg$ and $|b| \leq 5\deg$, with an angular resolution of 14\parcm4. The intensity calibration is good to better than 10\,\%.

Figure~\ref{fig1} shows maps of the free-free emission from the \commander\ fit and from the RRLs. The free-free amplitude is converted from the RRL temperature assuming a simple model for $T_\mathrm{e}$, based on an assumed temperature gradient of \mbox{($500\pm 100$)\,K\,kpc$^{-1}$} with distance from the Galactic centre \citep{Alves2014}. The Galactocentric distance at each longitude is estimated based on the \cite{Fich1989} rotation curve and using the central velocity of the RRLs. Obviously this is a crude approximation, since the RRL is a line-of-sight integral weighted by $n_\mathrm{e}^2$ and individual lines of sight could have a range of values. We estimate that the electron temperatures for a given line-of-sight are uncertain at the $\pm 1000$\,K level, which corresponds to an uncertainty in the free-free amplitude of up to 20\,\%. Thus the overall uncertainty on the free-free amplitude from the RRL data is 25\,\% at most. Nevertheless, there is an overall good level of agreement between the two estimates, both in terms of morphology and amplitude. A combination of bright compact sources (\hii\ regions) and diffuse emission within about $\pm 1\degr$ of the plane contribute, particularly in the inner Galaxy ($l=300\degr \rightarrow 0\degr \rightarrow 60\degr$). In the outer Galaxy ($l<300\degr$) the free-free emission is much weaker and only a few moderately bright \hii\ regions can be seen.

The scatter plots of Fig.~\ref{fig2} indicate that the two data sets are highly correlated, which suggests that the component separation of the free-free emission from the other components has been relatively successful. However, there are clearly changes in slope from region to region.  In general, the slopes are greater than 1, indicating that the \Commander\ free-free map is, on average, brighter that the predicted RRL map. The best-fitting slopes are typically 50\,\% higher than the predicted values.

Figure~\ref{fig3} shows longitudinal slices across the Galactic plane ($b=0\degr$) at $1\degr$ resolution. The bright \hii\ regions are clearly visible as bright peaks. Virtually all the bright peaks in the \Commander\ free-free solution have a counterpart in the RRL prediction. These peaks sit on diffuse emission that should be subtracted before a comparison of peak temperatures is made; large-scale ($\gtrsim 4\degr$) emission will not be reliable in the RRL data \citep{Alves2014}.  Many of the peaks have comparable amplitudes, while some of them are discrepant by up to a factor of 2. We note that the Galactic centre ($l=0\degr$) is low in the RRL map due to receiver saturation artefacts. The ratio of free-free to RRL brightness for the brightest 6 peaks in the first region ($l=358\degr \rightarrow 0\degr \rightarrow 52\degr$, excluding the $l=0\degr$ peak) is $1.14\pm0.04$, while the next 10 brightest peaks have a ratio of $1.36 \pm 0.08$ (the uncertainty is the rms scatter of the ratios).

The discrepancy between the two free-free estimates on bright \hii\ regions, which has been noted before \citep{Alves2010,planck2013-XIV,planck2014-XXIII}, is presumably due to a combination of effects: (i) differences in the electron temperature used to estimate the free-free brightness; (ii) beam effects in the RRL map. The RRL data are calibrated in the full beam scale, such that the flux density of compact sources is underestimated by 10--30\,\% depending on their size, although the longitude profiles of Fig.~\ref{fig3} suggest that the discrepancy between the \commander\ and RRL maps across individual sources does not depend on their extent; and (iii) residual emission in the \commander\ map, e.g., at $(l,b)=(46\pdeg8,-0\pdeg3)$ associated with the HC30 supernova remnant.

Compared to the brightest sources in each region, the weaker diffuse emission appears to have an even steeper slope, approaching 2, i.e., the discrepancy between the RRL and \commander\ solution is worse for very extended emission. This is likely to be due to excess synchrotron radiation at higher frequencies that has been accounted for in the simple \Commander\ model by the free-free component. In particular, the synchrotron component is effectively modelled as a power law in frequency with a fixed spectral index, while there is considerable evidence for spectral flattening of synchrotron emission at low latitudes (\citealp{Kogut2007}; \citealp{deOliveiraCosta2008}; \citealp{gold2010}; \citealp{planck2014-XXIII}). This will contribute to some level of excess emission at frequencies 20--100\GHz, which can result in an apparent increase in the free-free and/or AME amplitude in the \Commander\ fits, an effect that can be seen as a broad background (zero-level) in the longitude plots of Fig.~\ref{fig3}. We also note that the RRL data will not reliably trace large-scale ($\gtrsim 4\degr$) emission, since the observations were made in $8\degr \times 8\degr$ patches of sky. 

In summary, the \commander\ free-free map appears to be a reliable tracer of the brightest \hii\ regions, with an accuracy of around 20\,\%. Weaker emission regions appears to be over-estimated by up to a factor of 2, relative to estimates using RRL data. As mentioned in \cite{planck2014-a12}, including new data in the frequency range 2--20\GHz\ will yield improved component separation products, particularly for synchrotron and AME, which in turn will improve the accuracy of the free-free solution.

\subsection{Anomalous microwave emission} \label{sec:ame}

\begin{figure*}[tbh]
\begin{center}
\includegraphics[width=0.6\textwidth,angle=90]{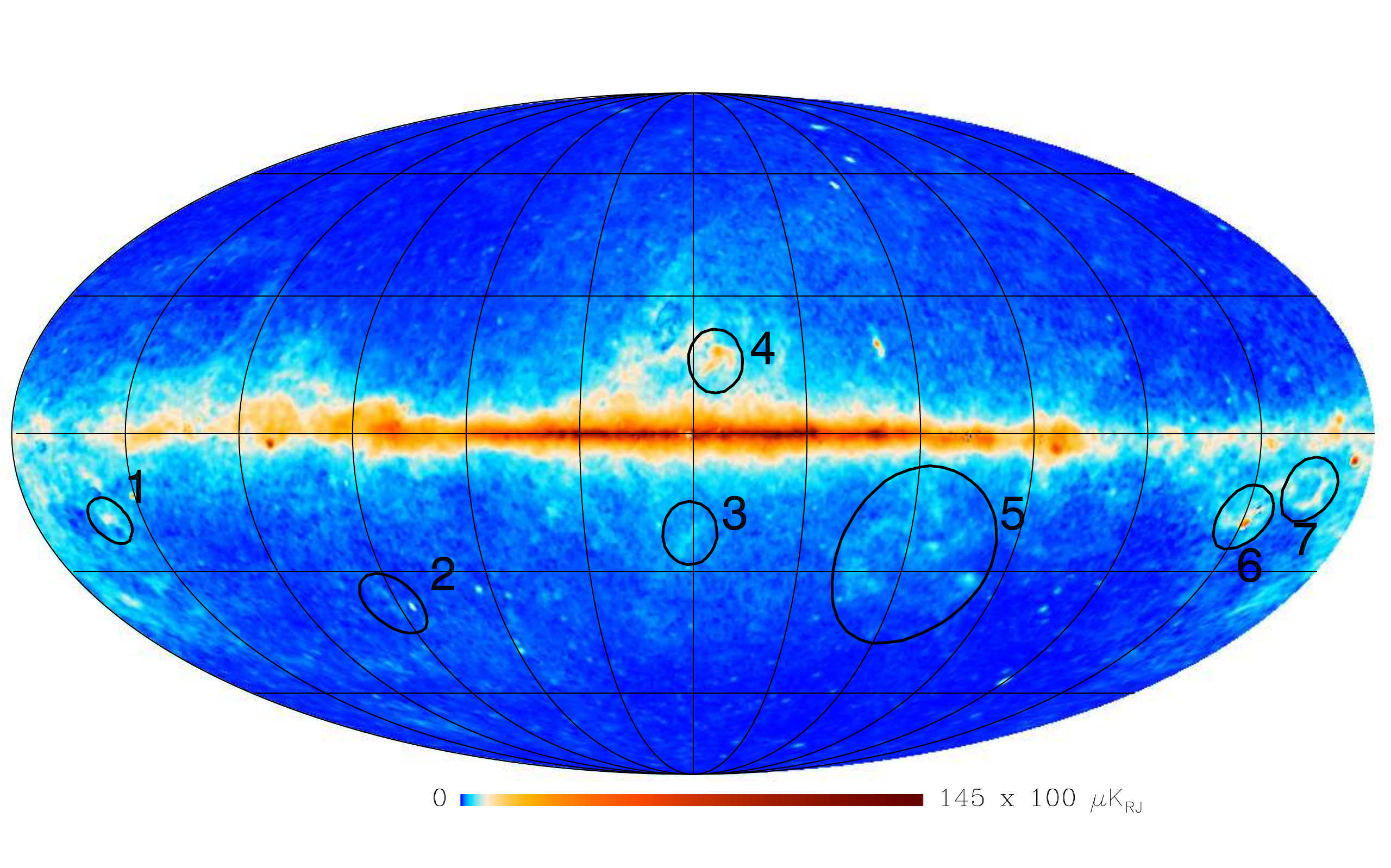}
\caption{All-sky map of AME from \Commander\ at 22.8\GHz\ plotted using a Mollweide projection, with a 30\deg\ graticule and an \asinh\ colour scheme. Seven regions of diffuse AME have been highlighted, and are discussed in the text.}
\label{fig:ame_fullsky}
\end{center}
\end{figure*}

\begin{figure}[tbh]
\newcommand{\widthfig}{0.49} \newcommand{\angfig}{0}
\begin{center}
  \includegraphics[angle=\angfig,width=\widthfig\textwidth]{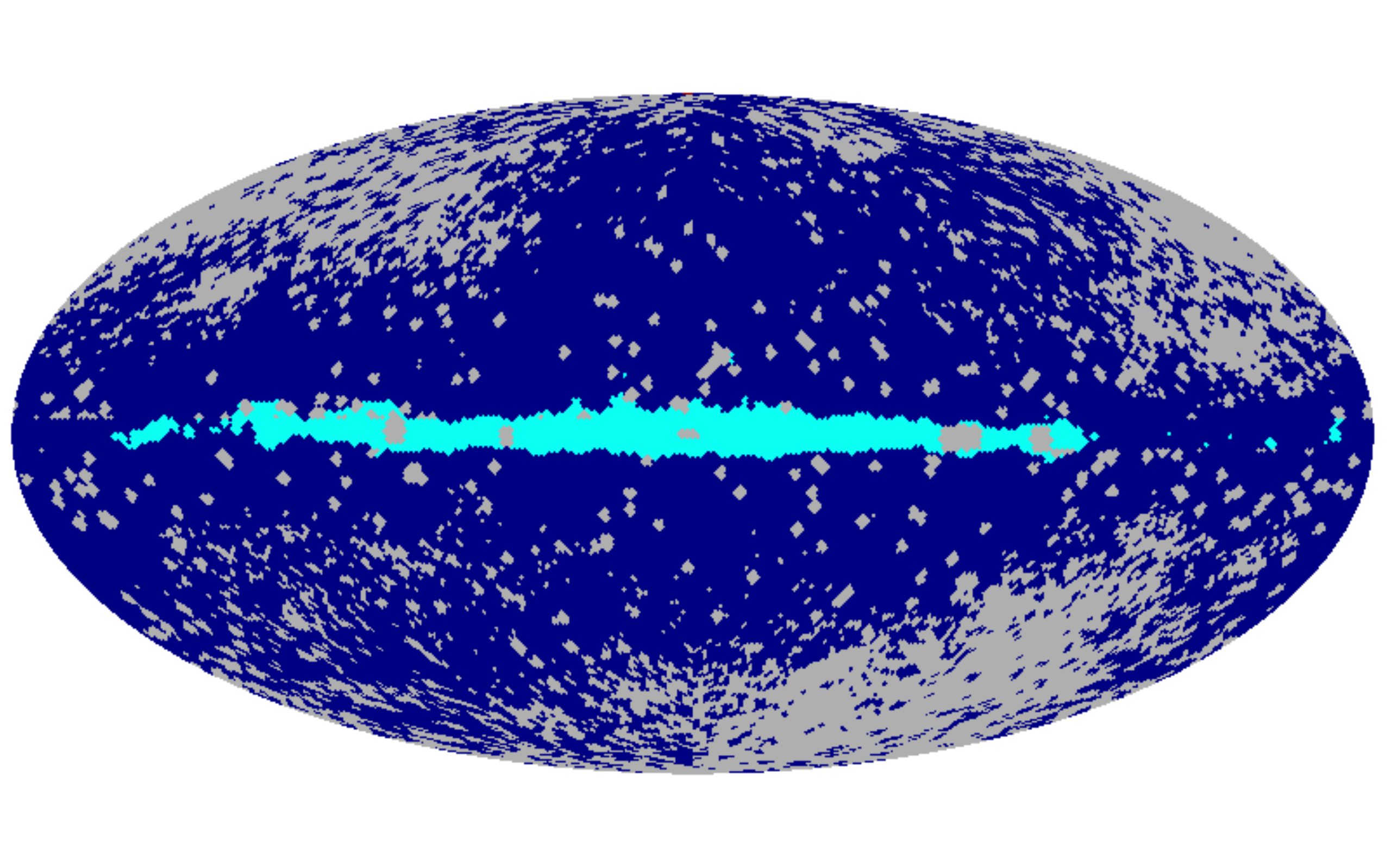}
  \caption{Mask used for the AME \ttp\ plots. The regions in grey are masked in the \ttp\ plots. The region in light blue shows where the AME at 22.8\GHz\ is brighter than 1\,mK; this is also masked when determining the best-fitting emissivity. We also mask the ecliptic plane ($|\beta|<20\deg$) when analysing the \iras\ and \Wise\ data.}
\label{fig:ame_mask}
\end{center}
\end{figure}

\begin{figure*}[tbh]
\newcommand{\widthfig}{0.32} \newcommand{\angfig}{0}
\begin{center}
  \includegraphics[angle=\angfig,width=\widthfig\textwidth]{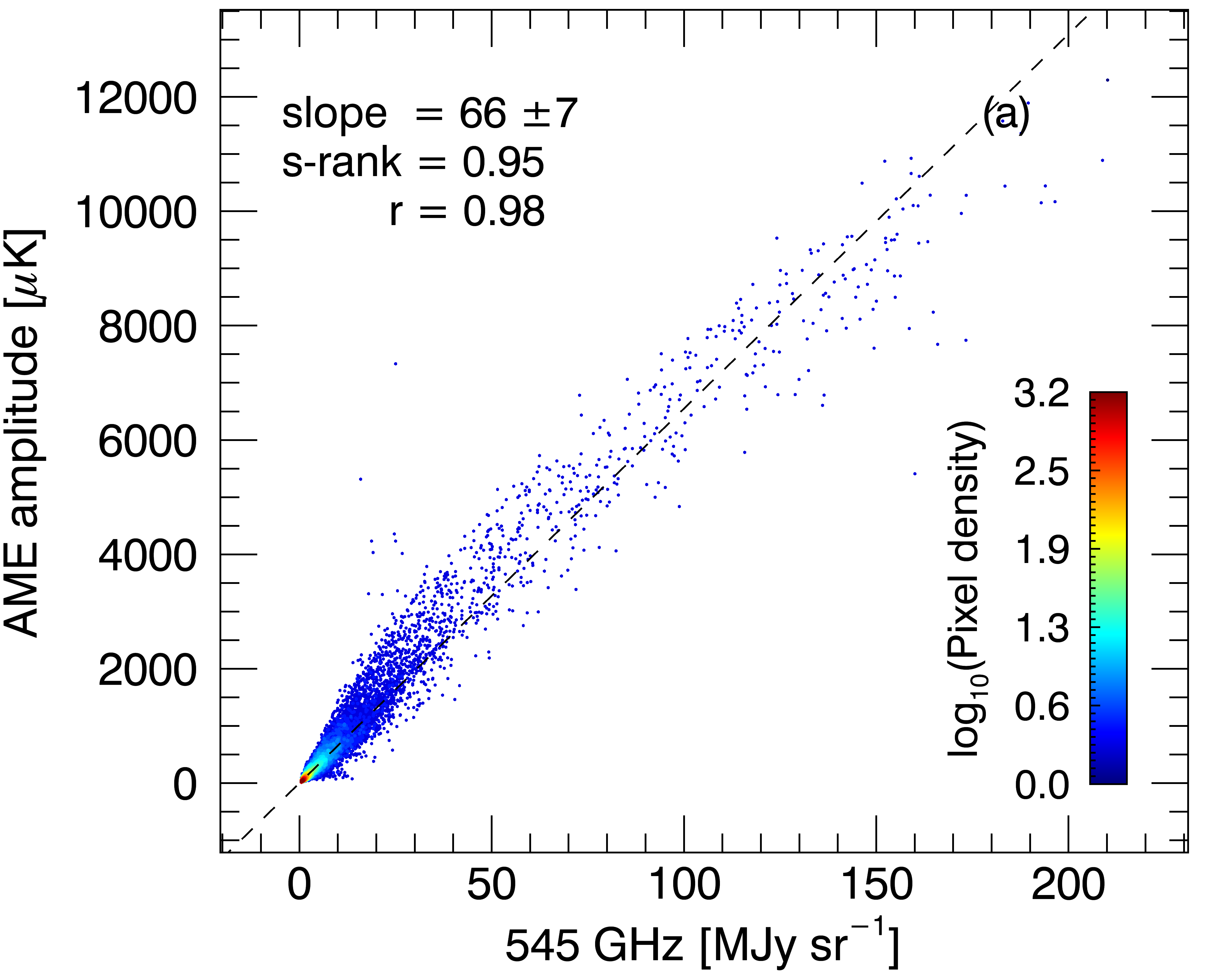}
  \includegraphics[angle=\angfig,width=\widthfig\textwidth]{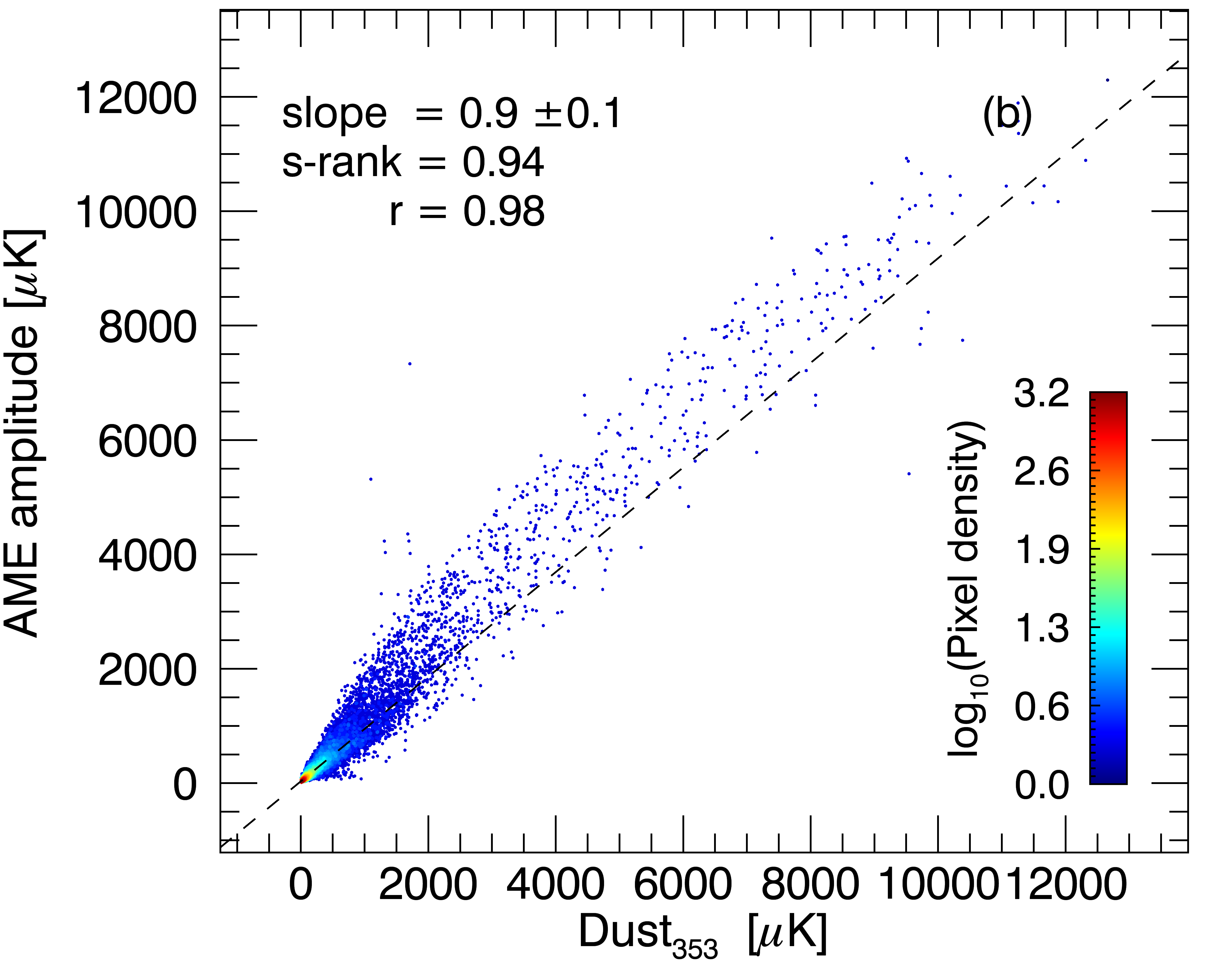}
  \includegraphics[angle=\angfig,width=\widthfig\textwidth]{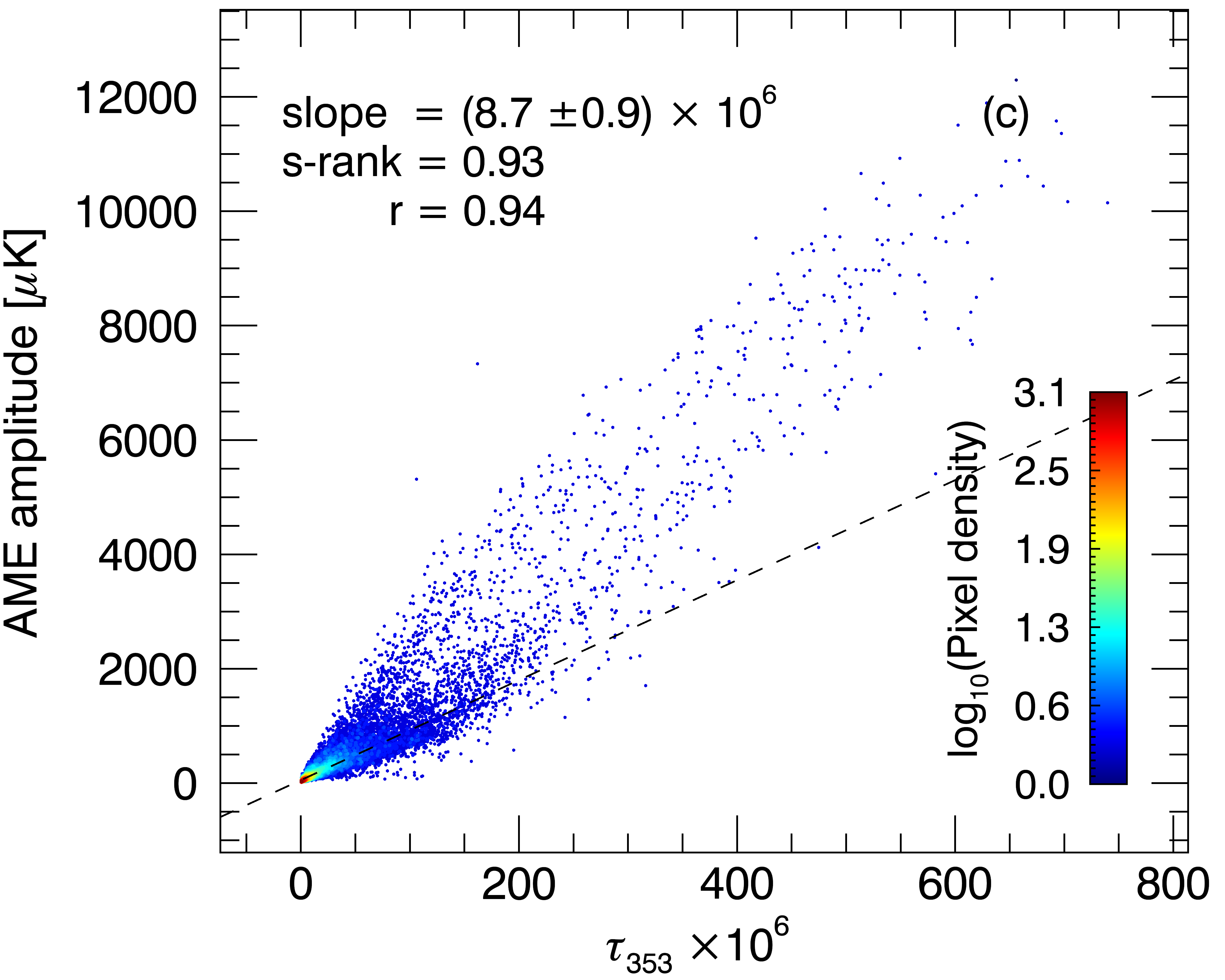}
  \includegraphics[angle=\angfig,width=\widthfig\textwidth]{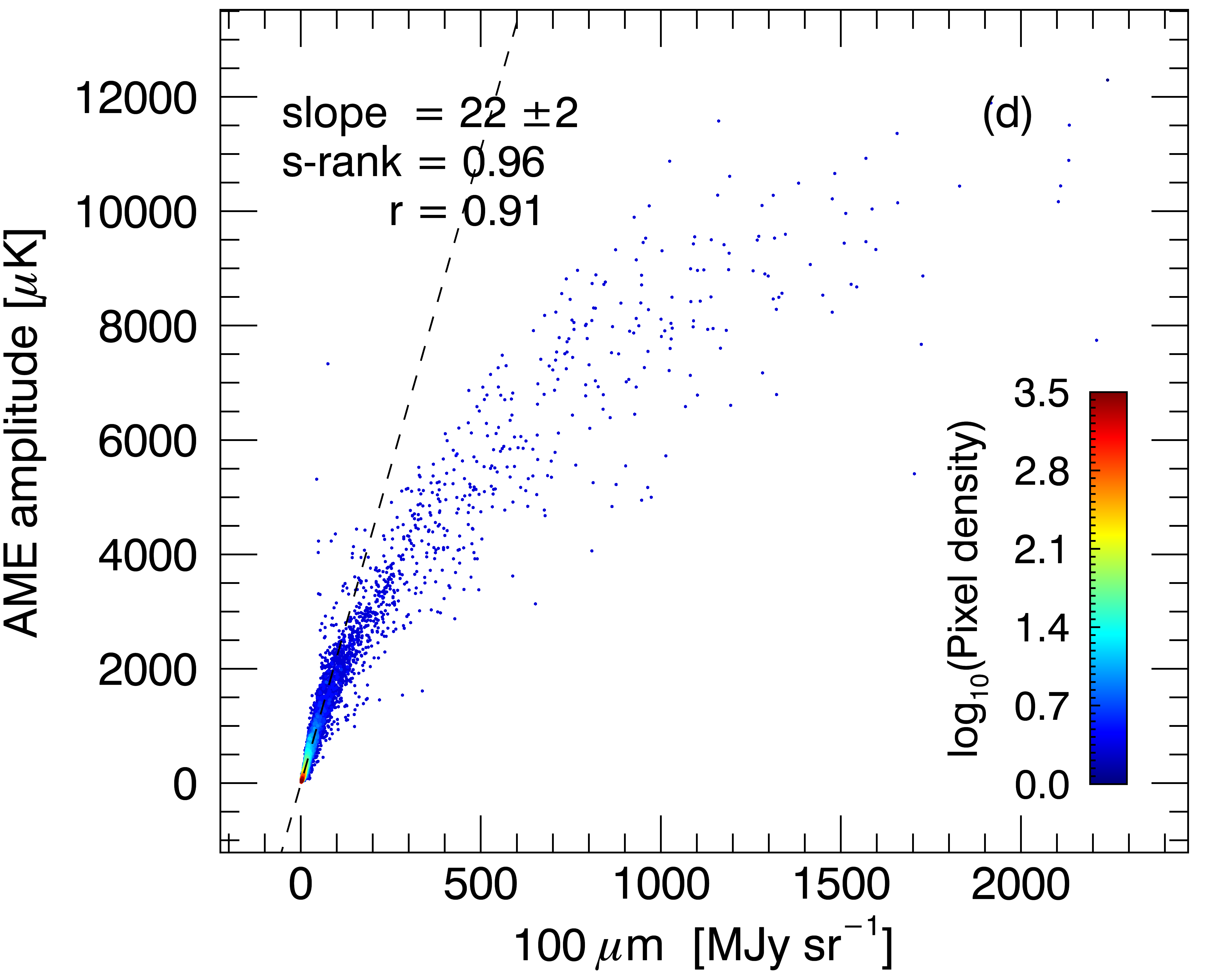}
  \includegraphics[angle=\angfig,width=\widthfig\textwidth]{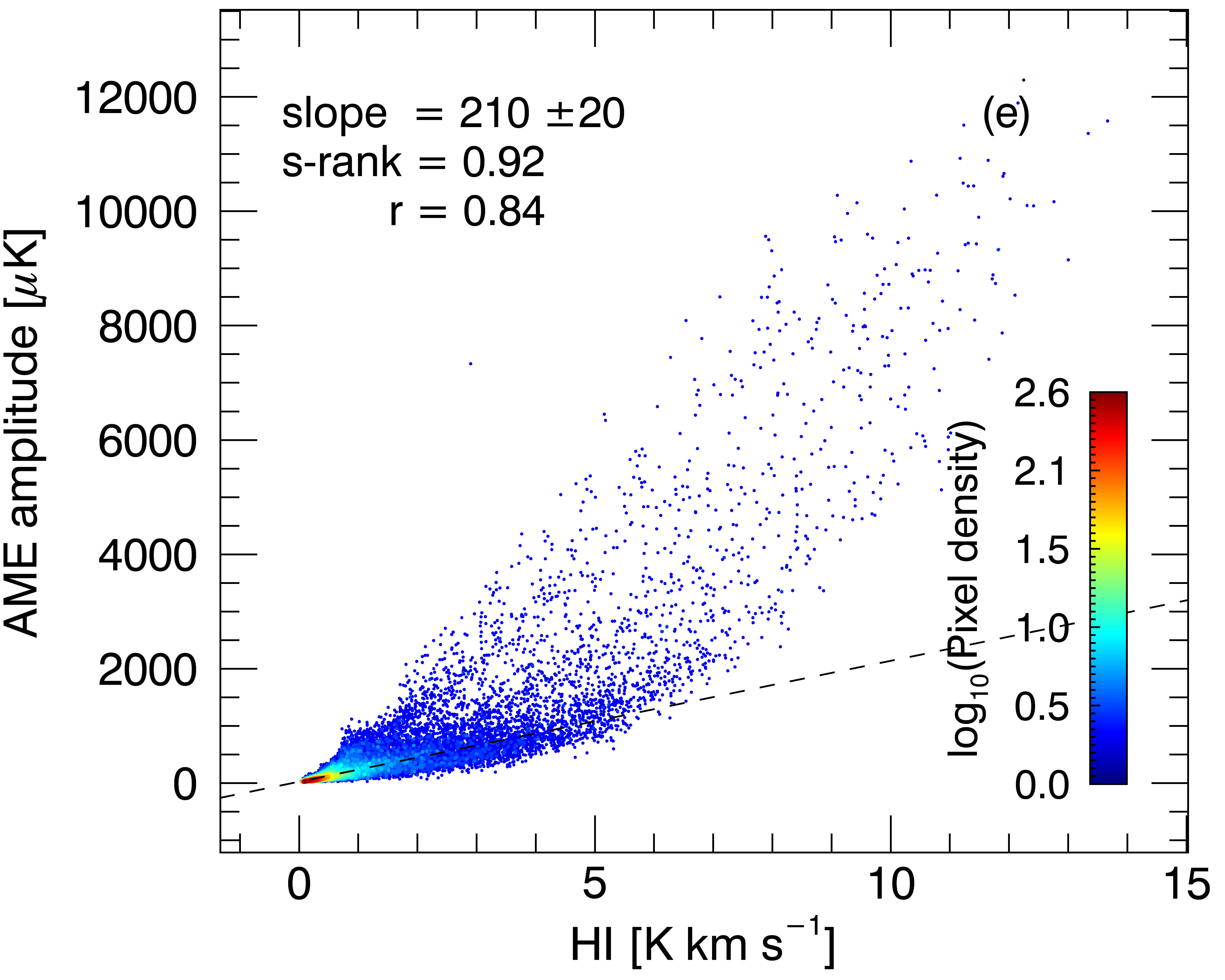}
  \includegraphics[angle=\angfig,width=\widthfig\textwidth]{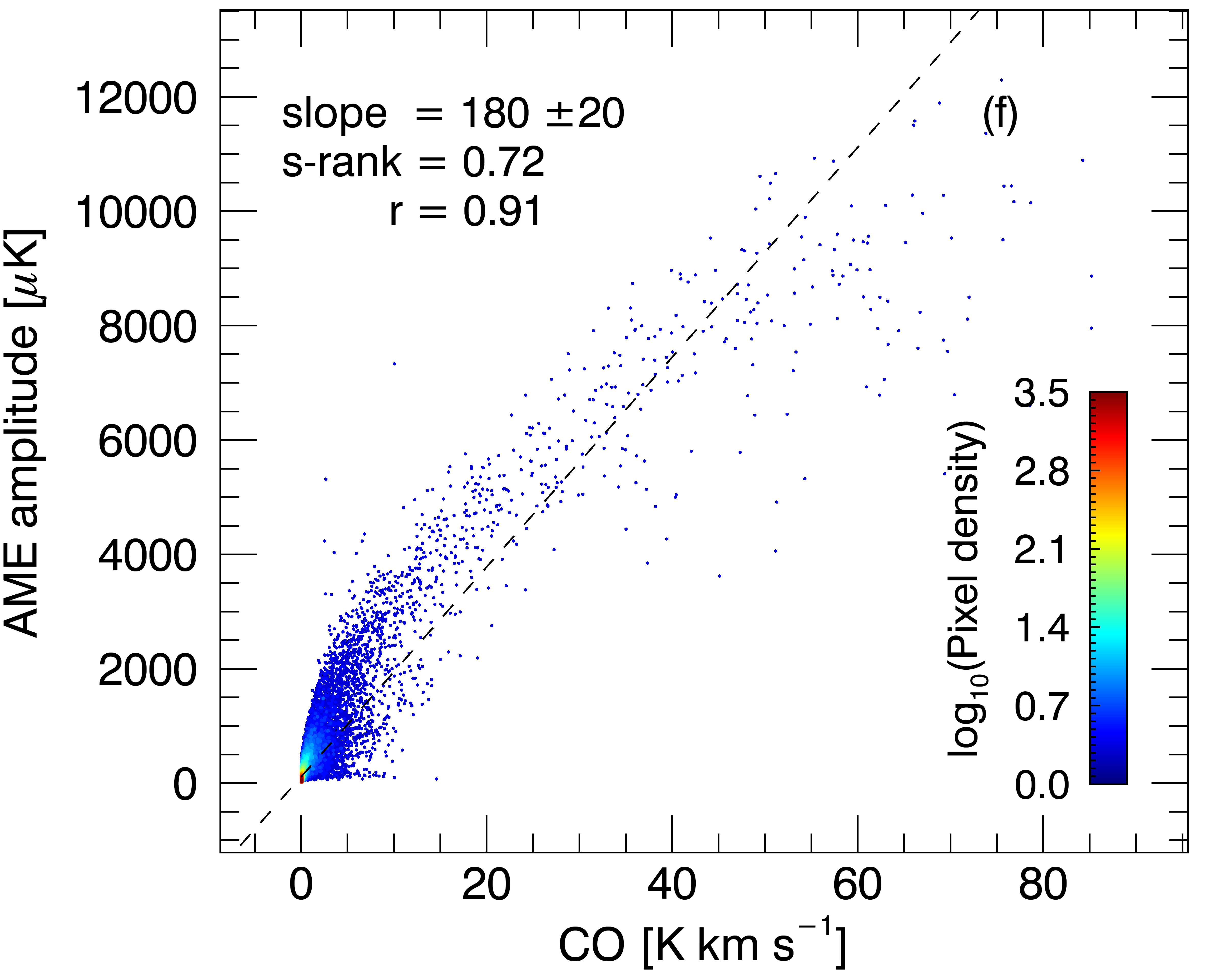}
  \includegraphics[angle=\angfig,width=\widthfig\textwidth]{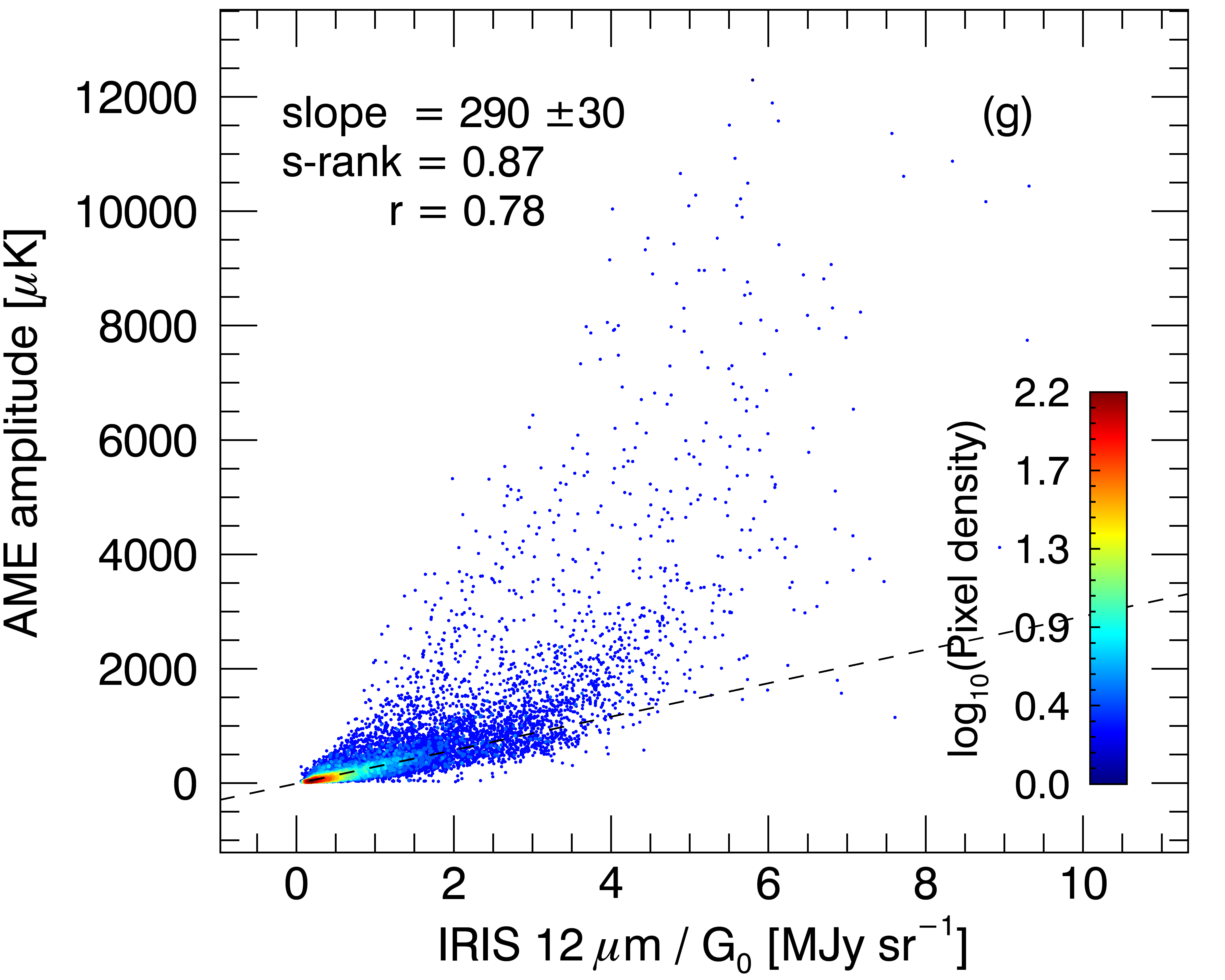}
  \includegraphics[angle=\angfig,width=\widthfig\textwidth]{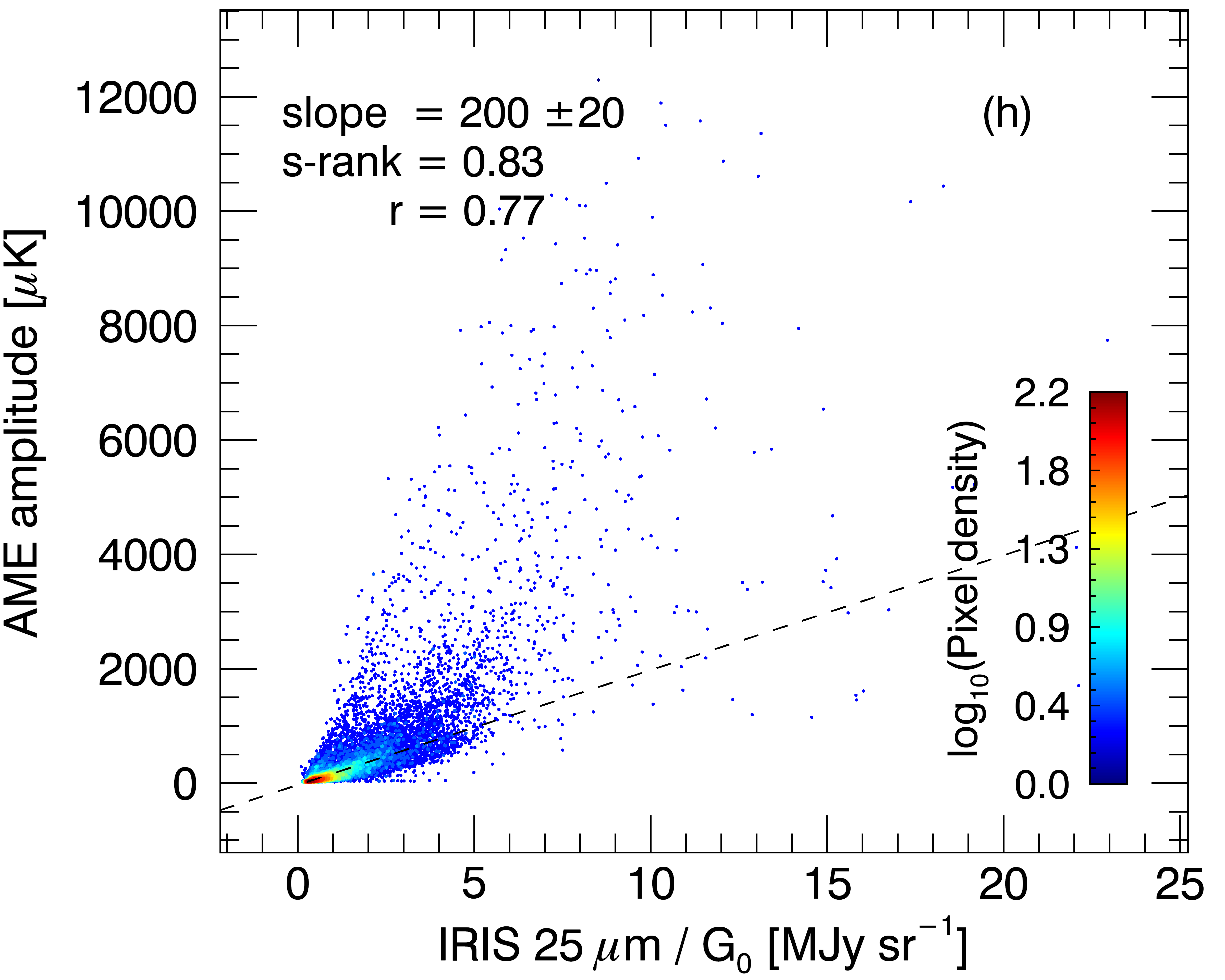}
  \includegraphics[angle=\angfig,width=\widthfig\textwidth]{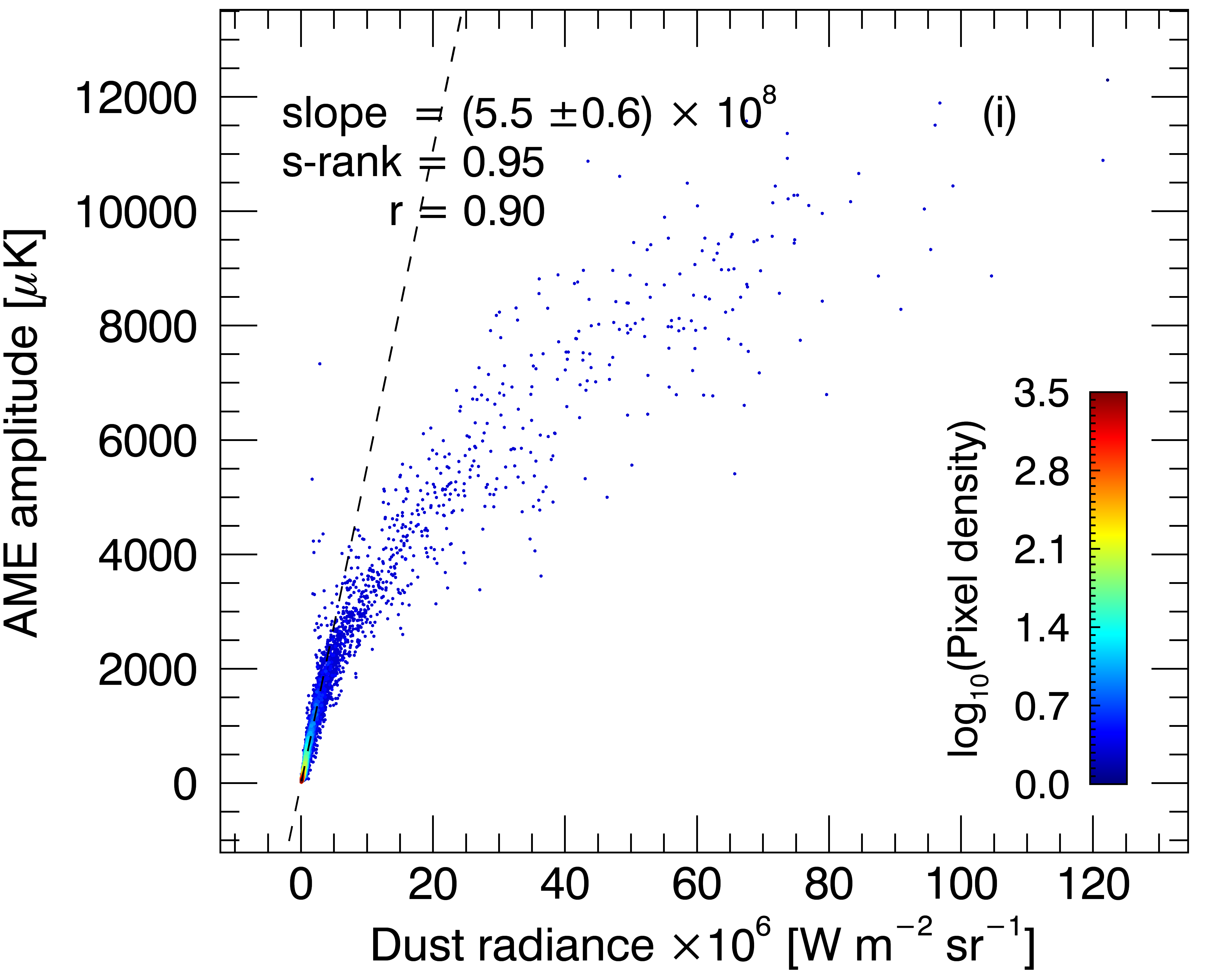}
  \caption{\ttp\ plots at $N_\mathrm{side}=64$, using the mask shown in Fig.\,\ref{fig:ame_mask}. The best-fitting lines and emissivities shown are determined where the AME amplitude is less than 1\,mK. The Spearman's rank correlation coefficient (s-rank) and the Pearson's correlation coefficient ($r$) are also shown. {\it Top row}: AME ($\upmu$K at 22.8\GHz) compared to: (a) \Planck\ 545\GHz\ (MJy\,sr$^{-1}$); (b) the \commander\ thermal dust solution at 353\GHz\ ($\upmu$K); and (c) $\tau_\mathrm{353}$ (rescaled by $10^5$). {\it Middle row}: (d) 100\um\ (MJy\,sr$^{-1}$); (e) \hi\ (K\,km\,s$^{-1}$); and (f) CO $J$$=$2\,$\rightarrow$\,1 (K\,km\,s$^{-1}$). {\it Bottom row}: (g) \IRAS\ 12\um\ (MJy\,sr$^{-1}$) divided by $G_0$; (h) \IRAS\ 25\um\ (MJy\,sr$^{-1}$) divided by $G_0$; and (i) dust radiance calculated from the \commander\ products.}
\label{fig:ame_ttplots}
\end{center}
\end{figure*}

\commander\ models the AME as spinning dust, which for uniform conditions (grain size distribution, ambient radiation) gives a relatively sharply peaked spectrum; specifically it uses the \spdust\ code \mbox{\citep{Ali-Hamoud2009,Silsbee2011}} to calculate a template spectrum using parameters typical for the diffuse cold neutral medium \citep{Draine1998}. To tolerable accuracy for present purposes, models for other phases can be approximated by scaling the template in frequency.

As shown in Sect.~\ref{sec:ilc}, the AME spectrum is quite variable, with an apparent tendency to peak at substantially higher-than-average frequencies around some \hii\ regions. Low-latitude lines-of-sight will therefore contain AME with a range of peak frequencies, which will give a broader spectrum that {\em cannot} be fitted by a frequency-shifted template. In fact, a superposition of AME spectra with a distribution of peak frequencies could approximate a power law over the frequency range of primary interest here (20--70\GHz). Such a general model could not be constrained by the available data, but we take a step in this direction by using two AME components in the \commander\ fit, each based on our \spdust\ template. One component (which generally dominates) is scaled in frequency at each pixel, with a Gaussian prior for the frequency of the intensity peak, $\nu_\mathrm{p1}$, of ($19$\,$\pm$\,$3$)\GHz. The peak frequency of the other component is assumed to be constant over the whole sky with a value to be determined globally; in this \commander\ solution the result is $\nu_\mathrm{p2}=33.35$\GHz. This second model component is there to account for the broadened spectrum of the overall AME emission, and should not be interpreted as a physically distinct AME component; in this paper (as in \citealp{planck2014-a12}) we only consider the combination of the two components, which we evaluate at 22.8\GHz\ to allow comparison with earlier studies.

The all-sky map of AME as fitted by \Commander\ is shown in Fig.~\ref{fig:ame_fullsky}. The foreground emission detected at frequencies around 20--60\GHz\ in the \wmap/\planck\ maps appears to be dominated by AME \citep[e.g.,][Sect.~\ref{sec:ilc} of this paper]{Davies2006,Ghosh2012}. It has been detected with high significance over a large area of the sky. The strongest emission is observed in the Galactic plane, particularly in the inner Galaxy ($l = 300\deg \rightarrow 0\deg \rightarrow 60\deg$). The AME in the inner plane was identified and discussed by \mbox{\citet{planck2011-7.3}} and \mbox{\citet{planck2014-XXIII}}. The map also shows clear, strong AME in many of the molecular cloud regions that are well known to exhibit AME, including the Perseus and $\rho$ Ophiuchus clouds. While previous \Planck\ AME papers \citep{planck2011-7.2,planck2013-XV} focused on compact AME regions, here we also look at diffuse AME regions away from the Galactic plane ($|b|>10$\deg). We identify strong diffuse AME in the Chameleon/Musca regions as well as the Orion complex, which is discussed later in this section. These regions are less likely to be affected by foreground degeneracies than others: regions such as the Auriga dust feature above the Galactic centre and a large structure around the North Celestial pole can also be seen in the AME map, however they overlap with significant sources of synchrotron emission.

Overall there is a high level of correlation of AME with the thermal dust maps made at higher frequencies of \Planck\ and in the infrared; however, the AME clearly cannot be accounted for by the Rayleigh-Jeans tail of the thermal dust emission, which is at least a factor of 30 below the AME component at 30\GHz. At low Galactic latitudes ($|b|\lesssim10$\deg) and in bright compact and diffuse AME regions, the complexity of separating AME from CMB, synchrotron, and free-free emission is the dominant uncertainty for the \commander\ solution. At high latitude, where the AME is weak, the \Commander\ solution is dominated by instrumental noise. The \commander\ AME map also contains residual extragalactic source contributions when their spectrum is intermediate between the assumed synchrotron and free-free models (e.g., Centaurus A is clearly visible, as are bright radio sources such as 3C\,273); to remove these we mask sources in the PCCS2 catalogue at 28.4\GHz\ that are brighter than 1\,Jy.

We evaluate the peak frequency of the combined AME spectrum for each pixel. The mean and median across the whole sky are 20.6 and 20.4\GHz, respectively, and at \mbox{$|b|>10$\deg} they are 20.5 and 20.2\GHz, respectively. Apart from bright regions, the peak frequency map is very noisy, largely constrained by our prior on $\nu_\mathrm{p1}$ around 19\GHz\ (see section~4.1 and figure~16 of \mbox{\citealp{planck2014-a12}}) and contains residual signal from point sources. Regions such as Perseus, $\rho$ Oph and Orion have higher than average peak frequencies, around 25--30\GHz, in agreement with previous analyses \mbox{\citep[e.g.,][]{planck2011-7.2}}. In some places the fits are dominated by the second AME component (notably the Sh\,2-27 and southern Gum nebulae). This is consistent with the ILC analysis of Sect.~\ref{sec:ilc}, although in reality some of these regions have even higher peak frequency, e.g., \mbox{\citet{planck2013-XV}} found a best-fitting value of $\nu_\mathrm{p}$ of 50\GHz\ for the California Nebula. Sh\,2-27 has an exceptionally high $\chi^2$ in the \commander\ fit \mbox{\citep[figure 22]{planck2014-a12}}, which we found can be alleviated by allowing a similar peak frequency (recall that in our baseline \commander\ model $\nu_\mathrm{p2}$ is fitted globally to keep the number of free parameters acceptable). \cite{Dobler2008b} previously found a bump at $\sim$50\GHz\ in the SED of H$\alpha$-correlated emission from the Warm Ionized Medium that they speculated could be a high-frequency spinning dust component, which is likely the same emission we are seeing here. We find that in bright free-free regions (primarily on the plane, but also bright high-latitude regions such as M\,42 and the California nebula) the AME peak frequency correlates with the free-free amplitude (at 22.8\GHz) with a best-fitting slope of \mbox{($0.066\pm0.012$)\,GHz\,mK$^{-1}$}, which may be an indication of the importance of the interstellar radiation field in the excitation of AME carriers \mbox{\citep[e.g.,][]{Tibbs2013}}. 

\subsubsection{All-sky correlations}

In order to look at the correlation of the AME with other components, we have degraded the AME map and other tracer maps to $N_\mathrm{side}$\,$=$\,$64$ (pixel widths of 55\arcm). We define a mask, shown in Fig.~\ref{fig:ame_mask}, to exclude pixels where the S/N for AME is less than 3.0 and point sources brighter than 1\,Jy, except at $|b|<5\deg$ to avoid masking Galactic plane sources. However, we mask five Galactic plane regions containing the brightest \hii\ complexes (e.g., Cygnus-X) in the sky. We use a 3\deg\ radius disc centred at $(l,b)$=(49\pdeg6,$-0\pdeg6$), (79\pdeg2,0\pdeg4), (268\pdeg2,$-1\pdeg3$), (287\pdeg5,$-1\pdeg1$) and (291\pdeg8,$-0\pdeg9$). These are particularly bright in the IRAS 12 and 25\um\ maps, where there could be contamination by starlight and/or line emission.  We show the \ttp\ plots for the entire sky outside of the mask for different potential tracers in Fig.~\ref{fig:ame_ttplots}. We also test the effects of additionally masking the Galactic plane ($b<10$\deg). We calculate the emissivity, i.e., the ratio of AME emission compared to AME tracers, from the slope of the best-fitting line as shown in the \ttp\ plots for pixels where the AME amplitude at 22.8\GHz\ is less than 1\,mK (shown in light blue in Fig.\,\ref{fig:ame_mask}). The difference between the \commander\ maximum-likelihood and mean solutions for the AME component is around 5--6\,\%, depending on frequency, due to the non-Gaussian shape of the probability distribution. We therefore assume a conservative 10\,\% modelling uncertainty on top of the \commander\ component uncertainty maps; we also assume a 13.5\,\% calibration uncertainty for the IRAS 100\um\ map \mbox{\citep{Miville-Deschenes2005}}.

We find an excellent correlation ($r$\,$=$\,$0.98$) with the \planck\ map at 545\GHz, shown in Fig.\,\ref{fig:ame_ttplots}a, which is predominantly thermal dust emission. A mean emissivity of ($65\pm7$)\ukmjy\ is found (($70\pm7$)\ukmjy\ when the Galactic plane is masked), which is consistent over a wide range of amplitudes. The overall emissivity against \IRAS\ 100\um\ that we find when looking at the entire sky is \mbox{($22\pm2$)\ukmjy} (Fig.~\ref{fig:ame_ttplots}d), with $r$\,$=$\,$0.91$; this is somewhat lower than the weighted average found by \mbox{\citet{planck2013-XV}} of \mbox{($32\pm4$)\ukmjy}, although the latter is for a specific set of 27 low-latitude clouds selected for significant emission, rather than an all-sky average. \citet{Davies2006} used template fitting to estimate the emissivity of AME compared with thermal dust; they calculated the emissivity in terms of the ``FDS model 8'' dust map at 94\GHz\ \citep{finkbeiner1999}, which can be multiplied by a factor of 3.3 to obtain the equivalent 100\um\ emissivity (on average). Our analysis agrees well their results, where they found ($21.8\pm1.0$)\ukmjy\ across the whole sky outside of the \WMAP\ Kp2 mask, and an average of ($25.7\pm1.3$)\ukmjy\ in the 15 regions they examined. We note that this good agreement in the overall high-latitude emissivity comes from different component-separation methods.

For thermal emission tracers there is some curvature in the relation at the brightest pixels, where there is less AME for a given amount of thermal dust emission. This is more noticeable in the 100\um\ correlation than the 545\GHz\ correlation, because the 100\um\ amplitude is much more affected by the temperature of the thermal dust grains than the 545\GHz\ emission \citep{Tibbs2012}, which increases towards the Galactic plane \mbox{\citep[e.g.,][]{planck2011-7.0}}. This effect is not seen in the $\tau_{353}$ correlation, however there is a much larger scatter in the $\tau_{353}$ correlation, as well as a trend towards there being more AME per unit $\tau_{353}$ at high optical depth.

There is a good correlation ($r$\,$=$\,$0.84$) with \hi\ emission integrated in the velocity range $[-450,+400]\,$km\,s$^{-1}$ from the LAB survey \citep{Kalberla2005}, shown in Fig.~\ref{fig:ame_ttplots}e, with a slope of \mbox{($210\pm20$)\,$\upmu$K\,K$^{-1}$} (($280\pm30$)\,$\upmu$K\,K$^{-1}$ when the Galactic plane is masked); there is also a noticeable upturn and increase in the scatter of points in this relation above an \hi\ integrated velocity intensity of around 3\,K, in agreement with previous results \citep[e.g.,][]{Lagache2003}. \mbox{\citet{planck2013-XVII}} compared \Planck\ and \WMAP\ data to \hi\ data, and found an emissivity of $(1.75\pm0.14)\times10^7$\,$\upmu$K\,$\tau_{353}^{-1}$;\footnote{\citet{planck2013-XVII} give a value of \mbox{($17\pm1$)\,$\upmu$K\,$(10^{20}$\,H\,cm$^{-2})^{-1}$} at 23\GHz, which can be converted to $\upmu$K\,$\tau_{353}^{-1}$ using $E$($B-V$)/$N_\mathrm{H}=(1.44\pm0.02)\times10^{-22}$\,mag\,cm$^2$ and $E$($B-V$)/$\tau_{353} = (1.49\pm0.03)\times10^{4}$\,mag$^{-1}$\,cm$^{-2}$ \mbox{\citep{planck2013-p06b}}.} they also considered AME models, which gave slightly lower emissivities with $\tau_{353}$ of  $1.3\times10^7$\,$\upmu$K. We find an emissivity against $\tau_{353}$ of $(8.7\pm0.8)\times10^6$\,$\upmu$K (Fig.~\ref{fig:ame_ttplots}c, $(1.1\pm0.1)\times10^7$\,$\upmu$K excluding the Galactic plane), which is lower than the values found by \mbox{\citet{planck2013-XVII}} by a factor of 1.5-2.1. However, the emissivities have not been calculated on the same areas of sky, and this difference is within the variations of $E$($B-V$)/$N_\mathrm{H}$ reported in \citet{planck2013-p06b}. We also compare AME at 22.8\GHz\ with the \commander\ thermal dust solution at 353\GHz\ and find a slope from the regions of (0.9$\pm$0.1)\,K\,K$^{-1}$ (Fig.~\ref{fig:ame_ttplots}b, $r$\,$=$\,$0.98$, (1.0$\pm$0.1)\,K\,K$^{-1}$ excluding the Galactic plane); this is also slightly lower than the value of ($1.12\pm0.03$)\,K\,K$^{-1}$ found by \mbox{\citet{planck2014-XXII}}.

The upturn in the \hi\ correlation implies that the AME is related to another phase that is not traced by \hi\ emission. We also look at CO emission, which will trace denser regions, and is a marker for \hii\ emission. The correlation with the \Commander\ CO $J$$=$2\,$\rightarrow$\,1 map is rather more complicated than the other correlations, however. The best-fitting slope is ($180\pm20$)\,$\upmu$K\,$(\mathrm{K\,km\,s}^{-1})^{-1}$ (Fig.\,\ref{fig:ame_ttplots}f), with $r$\,$=$\,$0.91$, reducing to ($150\pm20$)\,$\upmu$K\,$(\mathrm{K\,km\,s}^{-1})^{-1}$ when the Galactic plane is masked. However, at low amplitudes the emissivity is clearly a lot lower than the best-fitting value, and at high amplitudes it is clearly higher than the best-fitting value. This could be due to component separation issues; alternatively, since the curvature largely takes place at AME amplitudes above 1\,mK, it may indicate that there is less AME in the densest parts of the Galaxy than at higher latitudes.

We have also looked at the correlations with the \IRAS\ 12\um\ and 25\um\ data, which trace the very small dust grain population that could produce AME, as well as containing PAH line emission. Indications of an improved correlation with these data have been seen on smaller scales by e.g., \citet{Casassus2006}. However, these data are considerably contaminated by zodiacal light and other systematics; as such, we have masked the ecliptic plane ($|\beta|<20\deg$) before fitting these data. The resulting AME emissivity against 12\um\ emission is ($460\pm60$)\ukmjy, with Pearson correlation coefficient $r$\,$=$\,$0.93$ and Spearman's rank correlation coeffient $s$\,$=$\,$0.71$, and against 25\um\ emission is ($370\pm40$)\ukmjy, with $r$\,$=$\,$0.87$ and $s$\,$=$\,$0.52$. However, these will still include other systematic effects, such as stellar contamination and high-latitude zodiacal light. We nomalize the maps by the interstellar radiation field intensity, which we estimate by \mbox{$G_0=(T_\mathrm{d}/17.5)^{\beta+4}$} \citep{Ysard2010}, so that we are correlating against column density. Dividing by $G_0$ improves the correlation as measured by Spearman's rank, which includes the curvature in the correlation, while the Pearson's rank, which relies on a linear correlation, reduces. This improvement in correlation is as expected by the spinning dust model \citep{Ysard2010}, however the value of $G_0$, and hence the correlation coefficient, is highly dependent on the dust temperatures and spectral indices. The resulting emissivity against 12\um\ is \mbox{($290\pm30$)\ukmjy} (Fig.~\ref{fig:ame_ttplots}g, $r$\,$=$\,$0.78$, $s$\,$=$\,$0.87$) and against 25\um\ it is \mbox{($200\pm20$)\ukmjy} (Fig.~\ref{fig:ame_ttplots}h, $r$\,$=$\,$0.77$, $s$\,$=$\,$0.83$). We return to this topic in the discussion of the Musca region, below.

In a recent analysis, \citet{Hensley2015} have compared the \commander\ AME maps with different potential tracers of AME, and find a good correlation against dust radiance. This is the integrated intensity from thermal dust, and it characterises the energy absorbed and emitted by the thermal dust particles. We have also compared the AME map with dust radiance (calculated using the \Commander\ products and equation 10 in \citealp{planck2013-p06b}) in Fig.\,\ref{fig:ame_ttplots}i. We find a strong correlation at low amplitudes, with an emissivity against AME at 22.8\GHz\ of $(5.6\pm0.6)\times10^8$\,$\upmu$K, and $r$\,$=$\,$0.90$. At higher amplitudes (i.e., in the Galactic plane), the dust radiance calculated using the \commander\ products turns over, in the same way as the correlation with the \iras\ 100\um\ data does, implying that this is due to temperature effects. If we use the radiance map from \citet{planck2013-p06b}, which includes the 100\um\ map, in the analysis, this effect reduces in magnitude, but is still present.

\citet{Hensley2015} also compare the variations in AME emissivities (defined by dividing the AME map by the average emissivity against dust radiance) with variations in the fraction of the \WISE\ 12\um\ data \citep{Meisner2014} that can be attributed to polycyclic aromatic hydrocarbon (PAH) emission, in contradiction with expectations from the spinning dust model where PAHs are though to be the spinning molecules. Using the \commander\ thermal dust amplitude and $T_\mathrm{d}$ and $\beta_\mathrm{d}$ maps from \citep{planck2013-p06b}, they find no correlation between AME emissivities and PAH emission. We repeat their analysis, using the \WISE\ 12\um\ data and masking the Galactic plane (\mbox{$|b|<5\deg$}) and the Ecliptic plane ($|\beta|<20\deg$) in addition to the mask shown in Fig.\,\ref{fig:ame_mask}. We find similar results when using the dust radiance maps calculated using the products from \citep{planck2013-p06b}, with correlation coefficients of 0.10 (\citealp{planck2013-p06b} radiance map) and 0.23 (using the \commander\ dust amplitude). However, if we calculate dust radiance using only the \Commander\ products, then we find a correlation with a slope of $1.1\pm0.1$, with a correlation coefficient of 0.52.\footnote{The first preprint version of \citet{Hensley2015} used a full-sky \WISE\ map that later turned out to be contaminated by \Planck\ 857\GHz\ data on scales larger than 2\deg, which might also affect our analysis of their results. A revised version, \citet{Hensley2015b}, used an alternative approach to process the \WISE\ data to avoid this issue; they found that their conclusions were unchanged. As such, we have not revised our conclusion here.} Although the \commander\ dust radiance map will be biased low due to the absence of data points tracing warmer dust temperatures, this demonstrates the dependence of this result on the quality of the radiance map; it will also depend on the quality of the AME map. One possibility is that this correlation is affected by the large angular scale structure in the Galaxy; as such we return to this in the discussion of the Musca region, below.

In conclusion, we find the best correlation with AME at all amplitudes is from the 545\GHz\ \planck\ map, followed by the \commander\ dust solution at 353\GHz\ and the optical depth, $\tau_\mathrm{353}$. The dust radiance has a tight correlation with AME away the Galactic plane, but has a worse correlation in the plane. The correlation with 100\um\ is significantly affected by temperature effects, which the choice of lower frequency thermal dust maps avoids. We find a reasonable correlation with \hi\ emission, although this does not appear to correlate with the brightest AME emission, and with CO emission, although this is not well-fitted with a single emissivity. The correlation with dust radiance is very good at low amplitudes, but using a single emissivity would over-predict the amount of AME present in the Galactic plane. We caution that these correlations will depend on the choice of mask due to large-scale biases, contamination from other emission mechanisms (e.g., point sources, zodiacal light, etc.), and variations across the sky.

\begin{figure*}
\begin{center}
\includegraphics[width=0.18\textwidth]{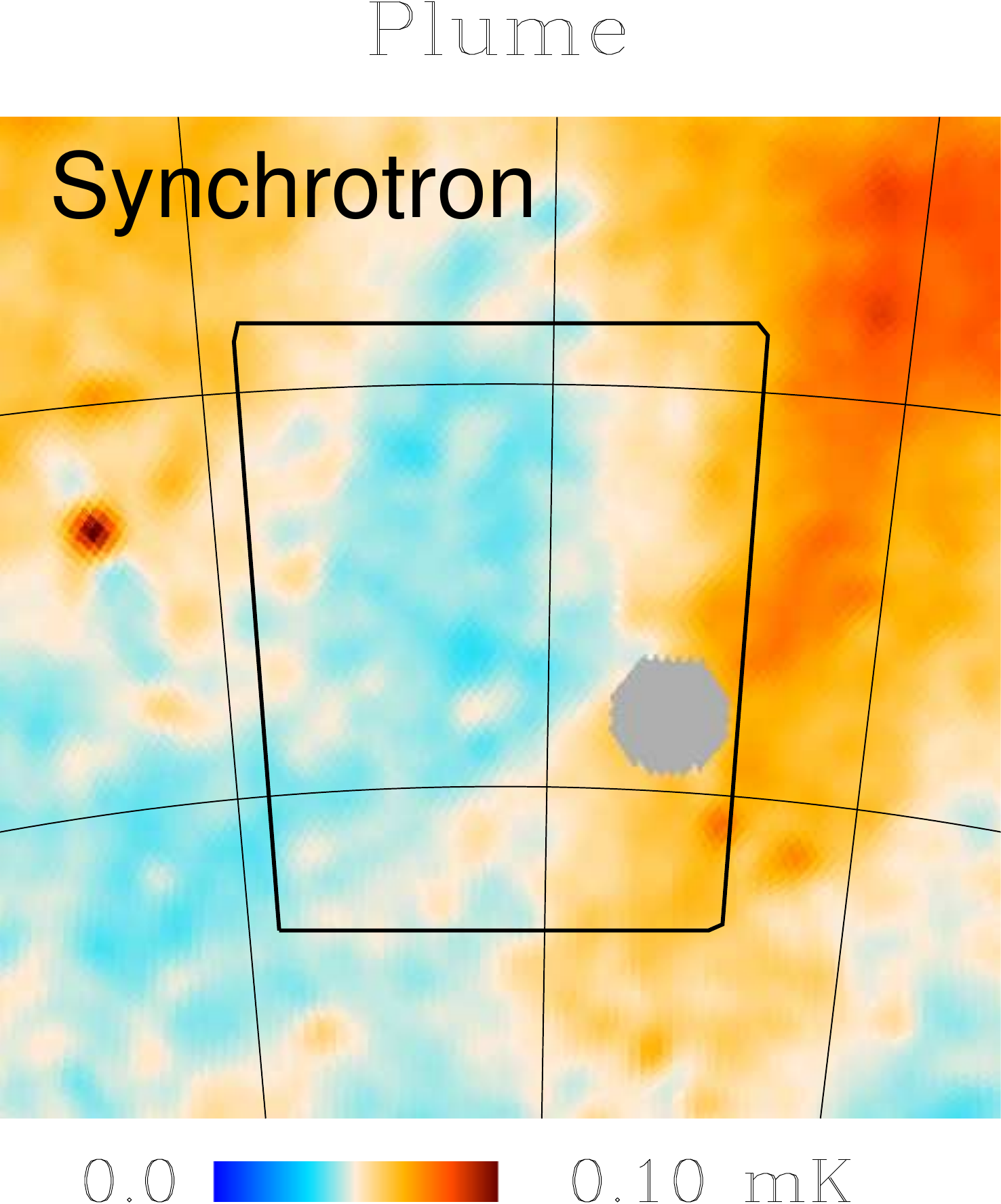}
\includegraphics[width=0.18\textwidth]{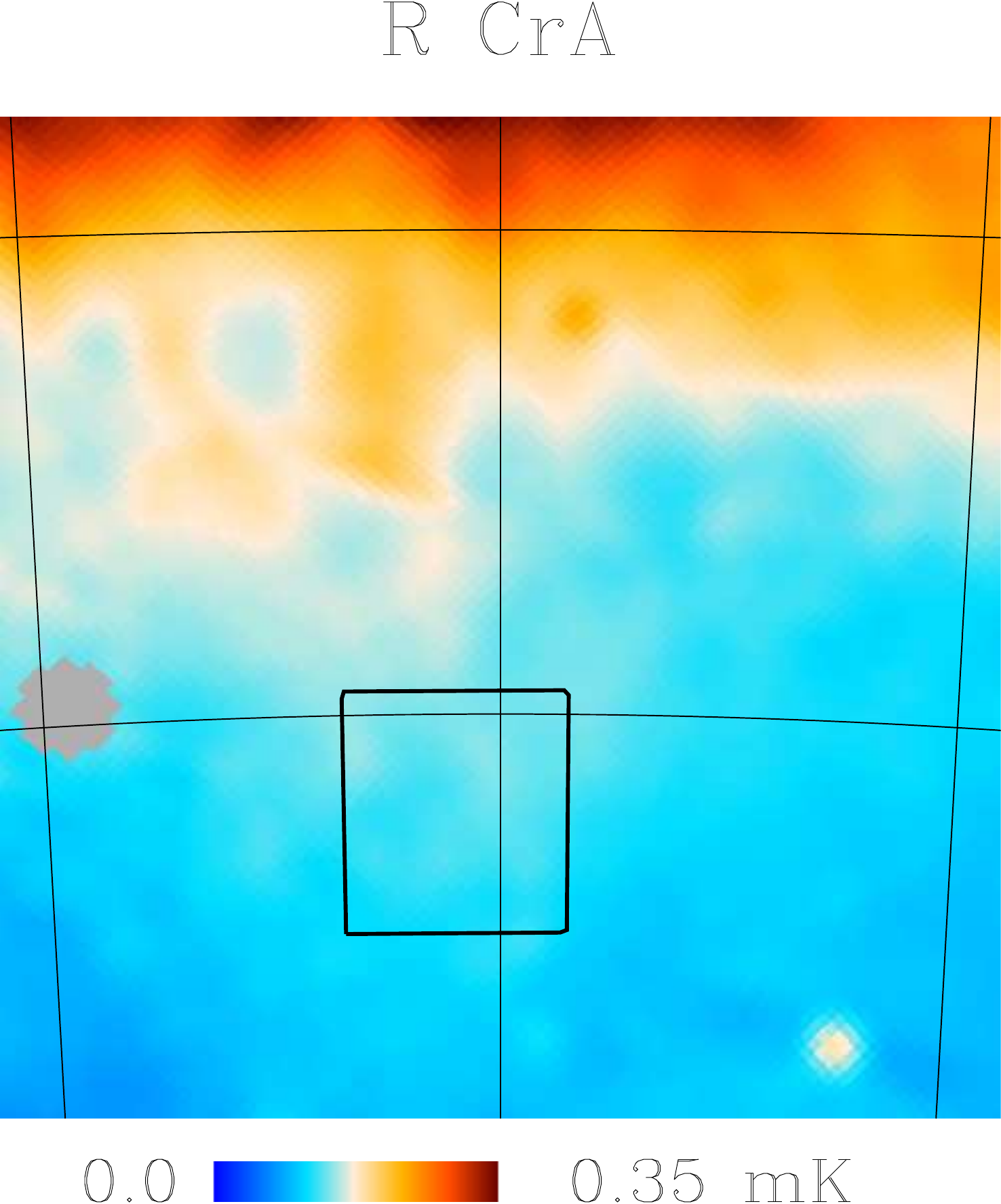}
\includegraphics[width=0.18\textwidth]{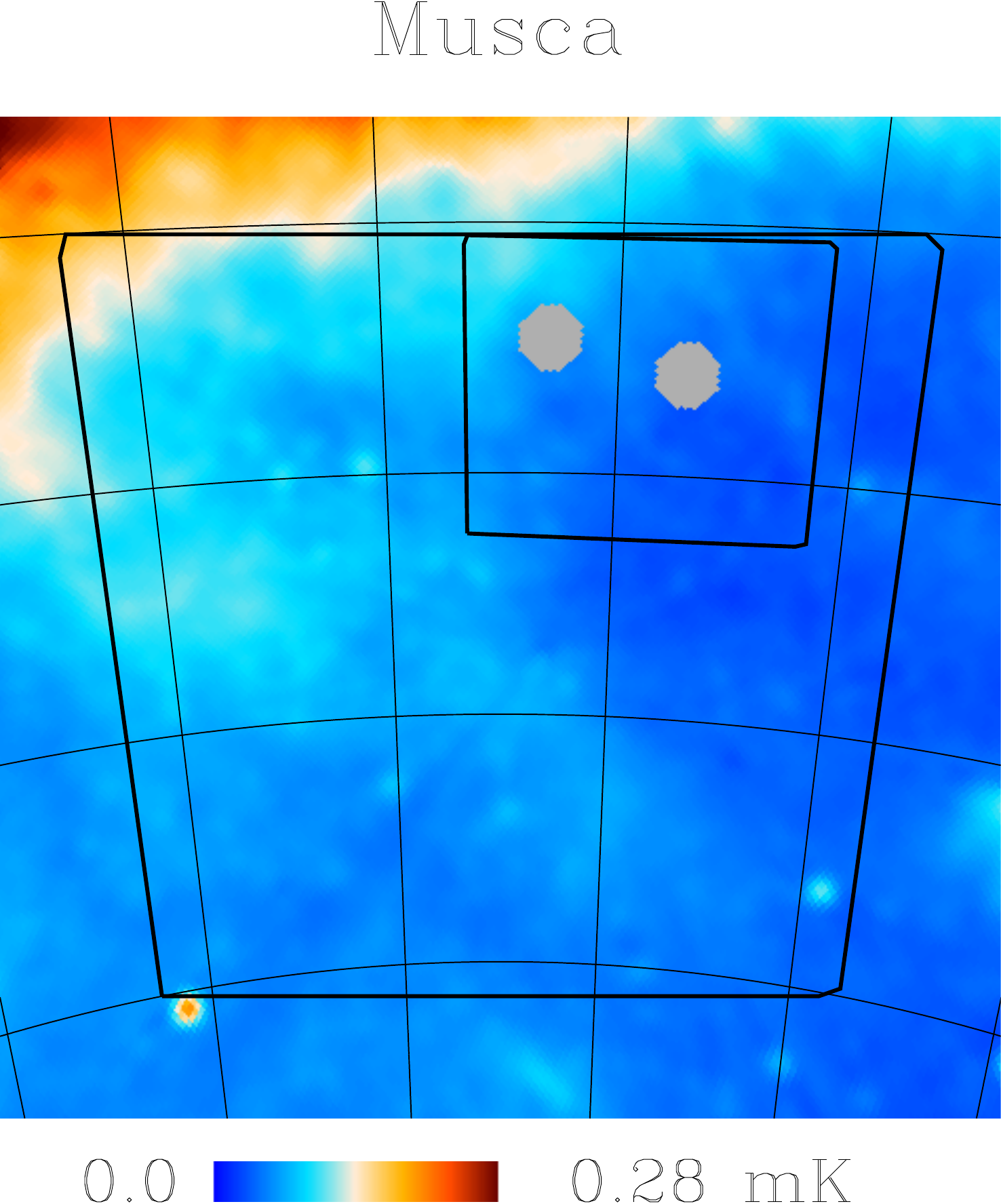}
\includegraphics[width=0.18\textwidth]{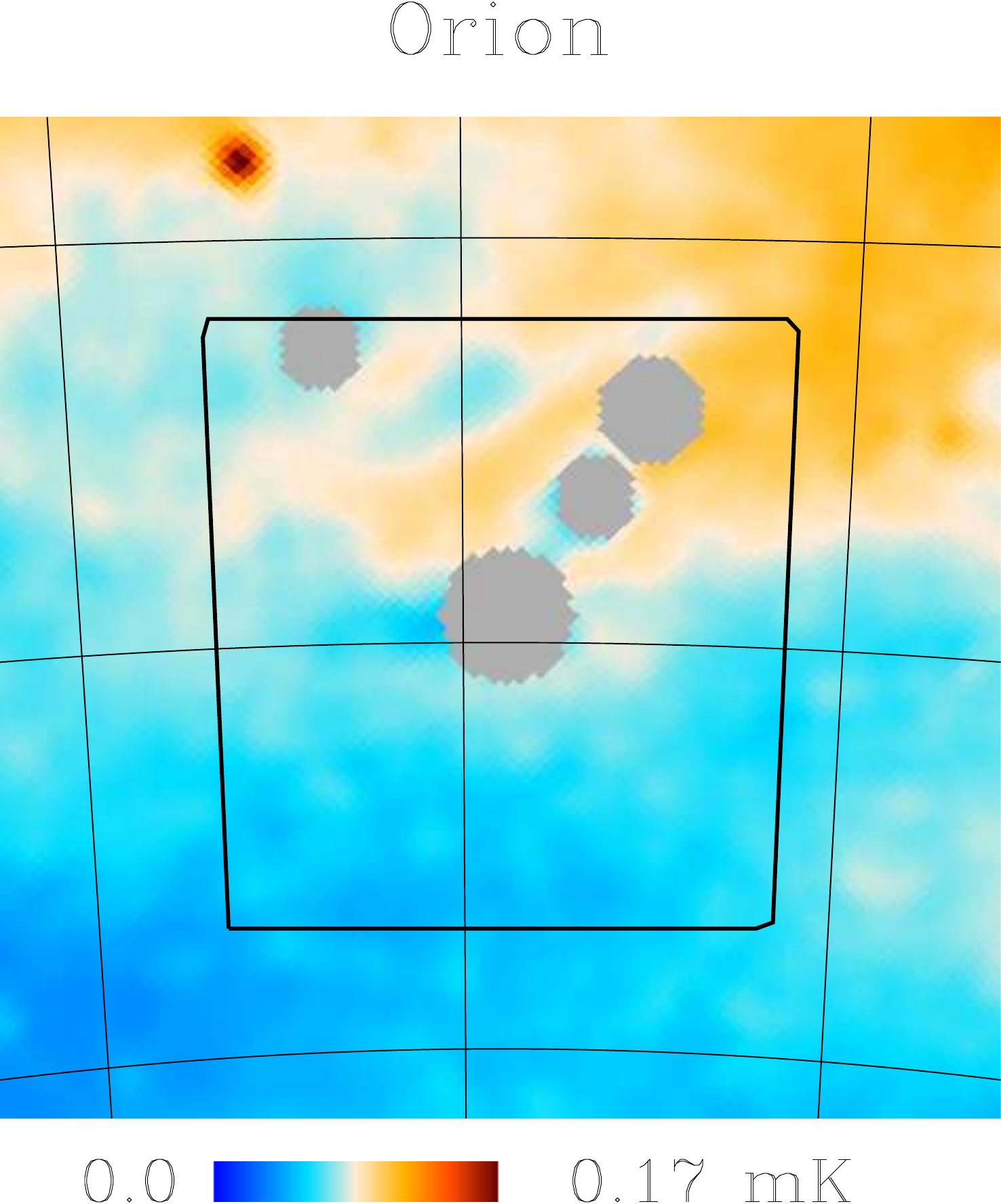}
\includegraphics[width=0.18\textwidth]{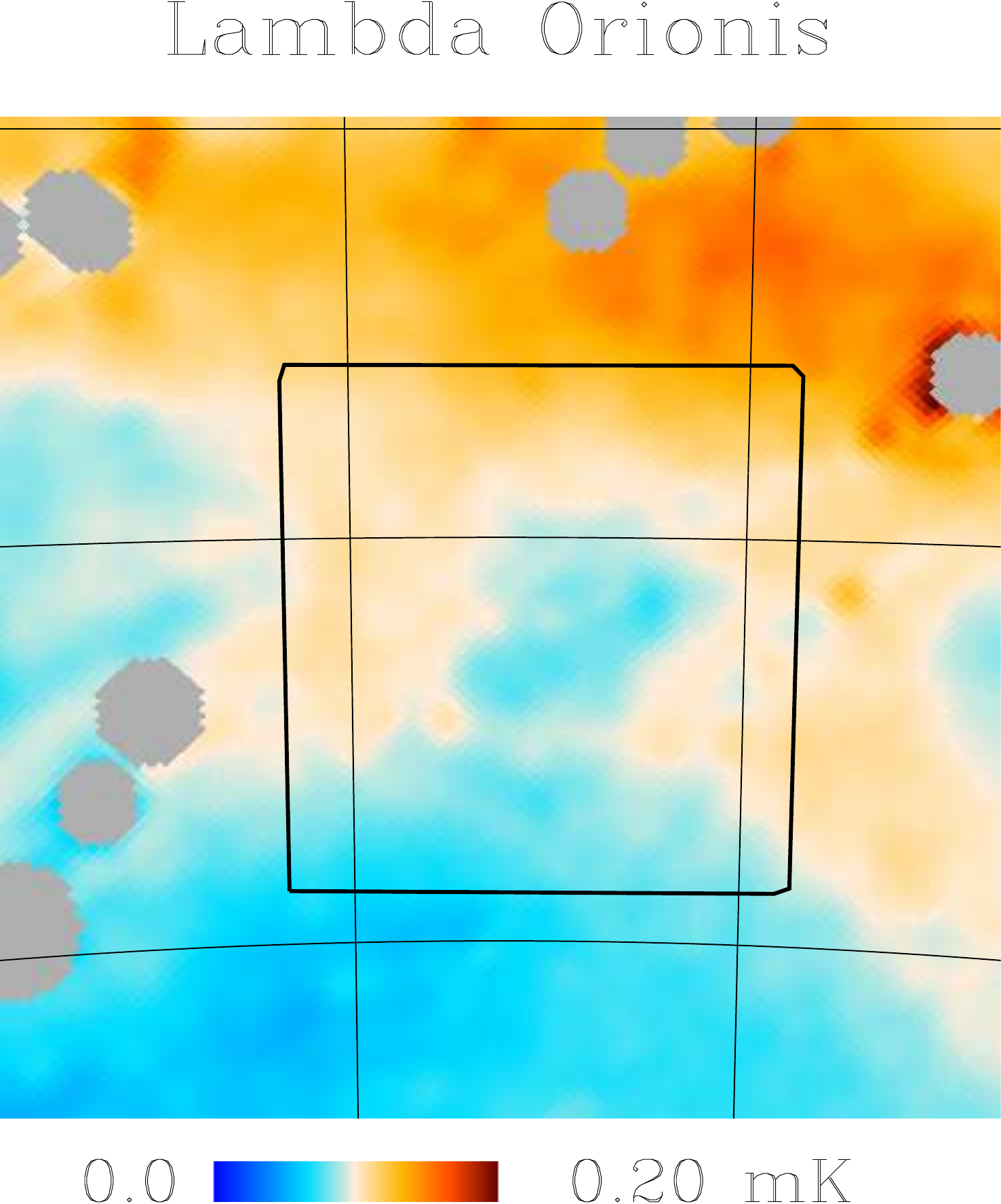}

\includegraphics[width=0.18\textwidth]{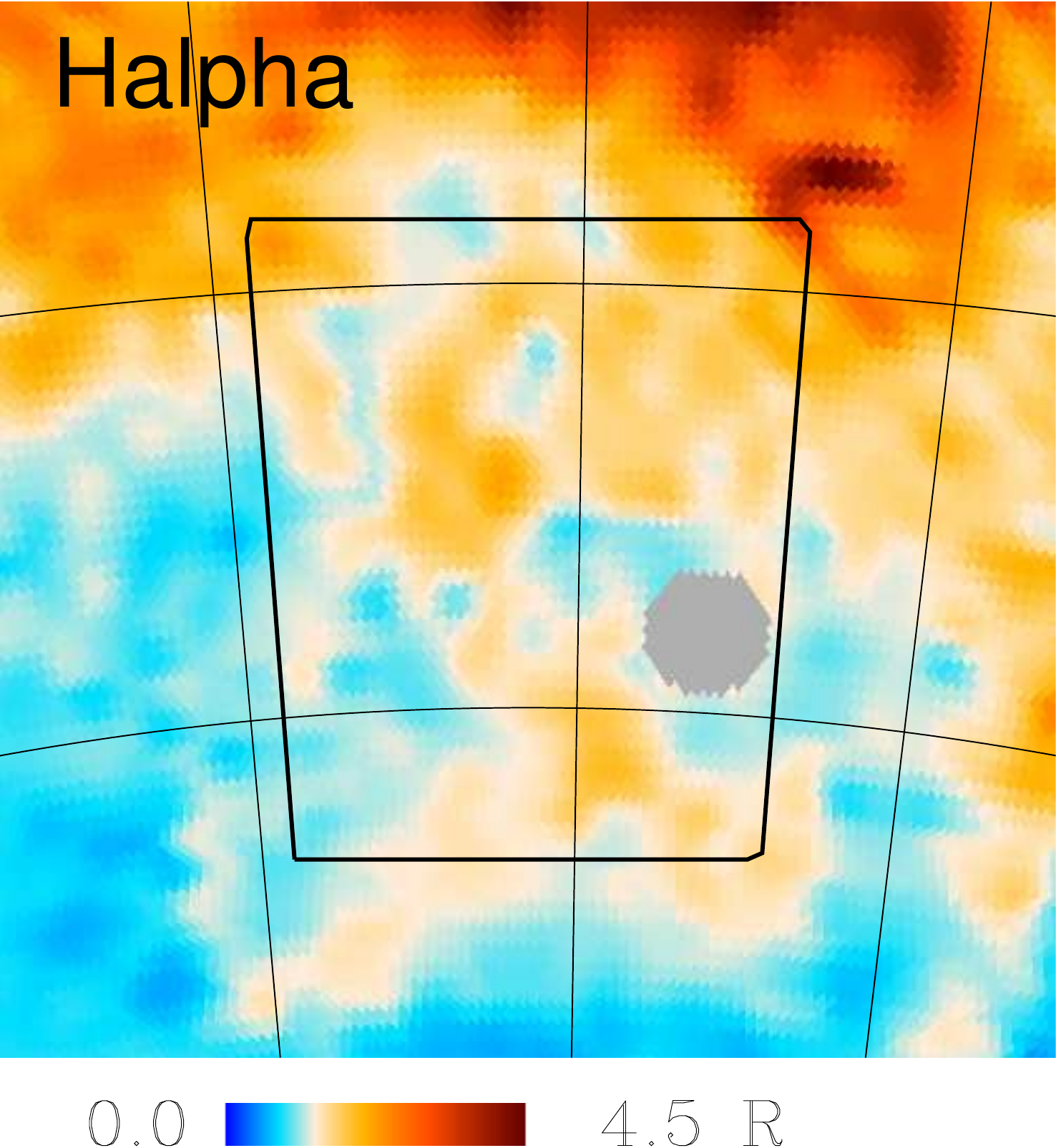}
\includegraphics[width=0.18\textwidth]{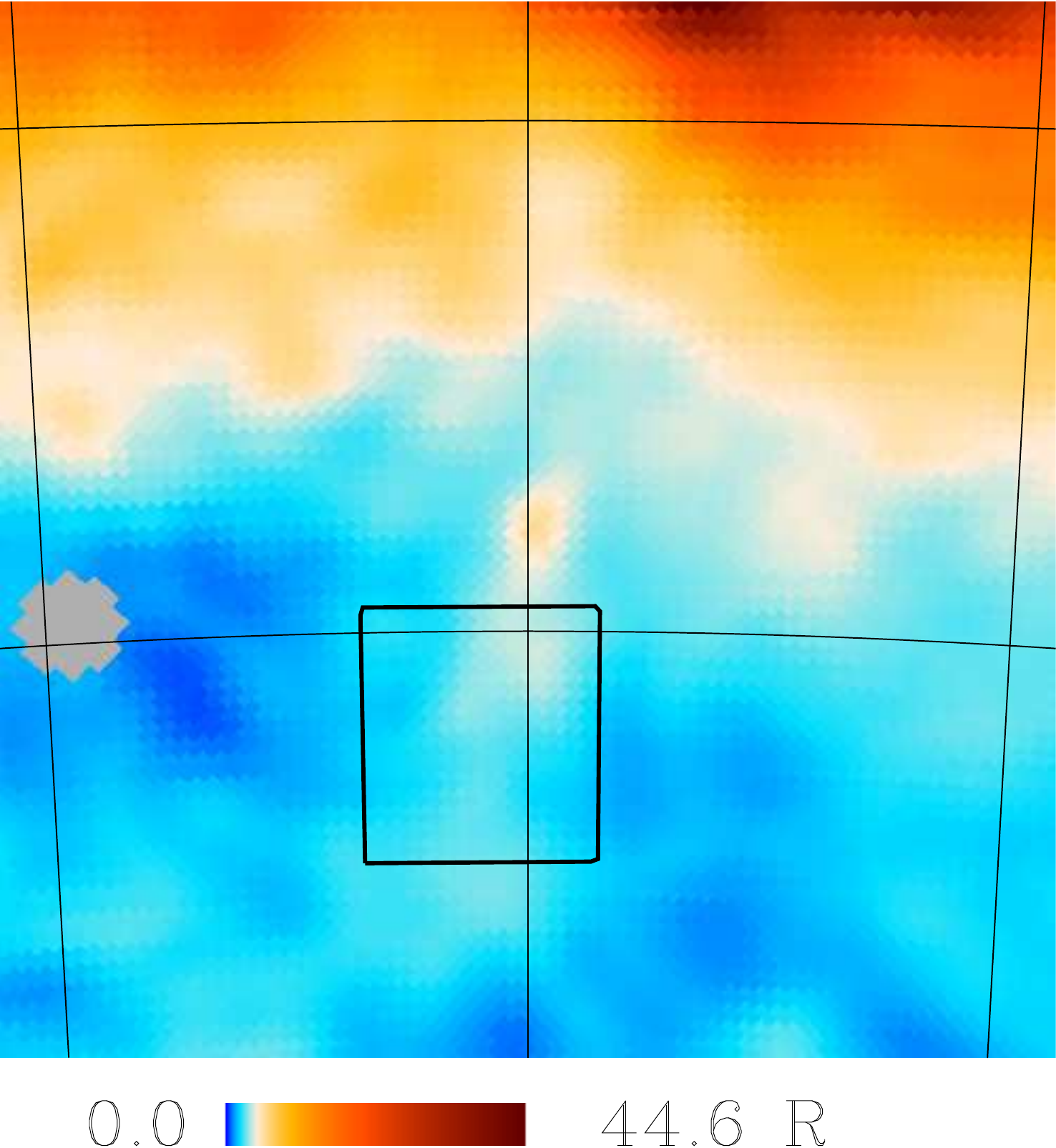}
\includegraphics[width=0.18\textwidth]{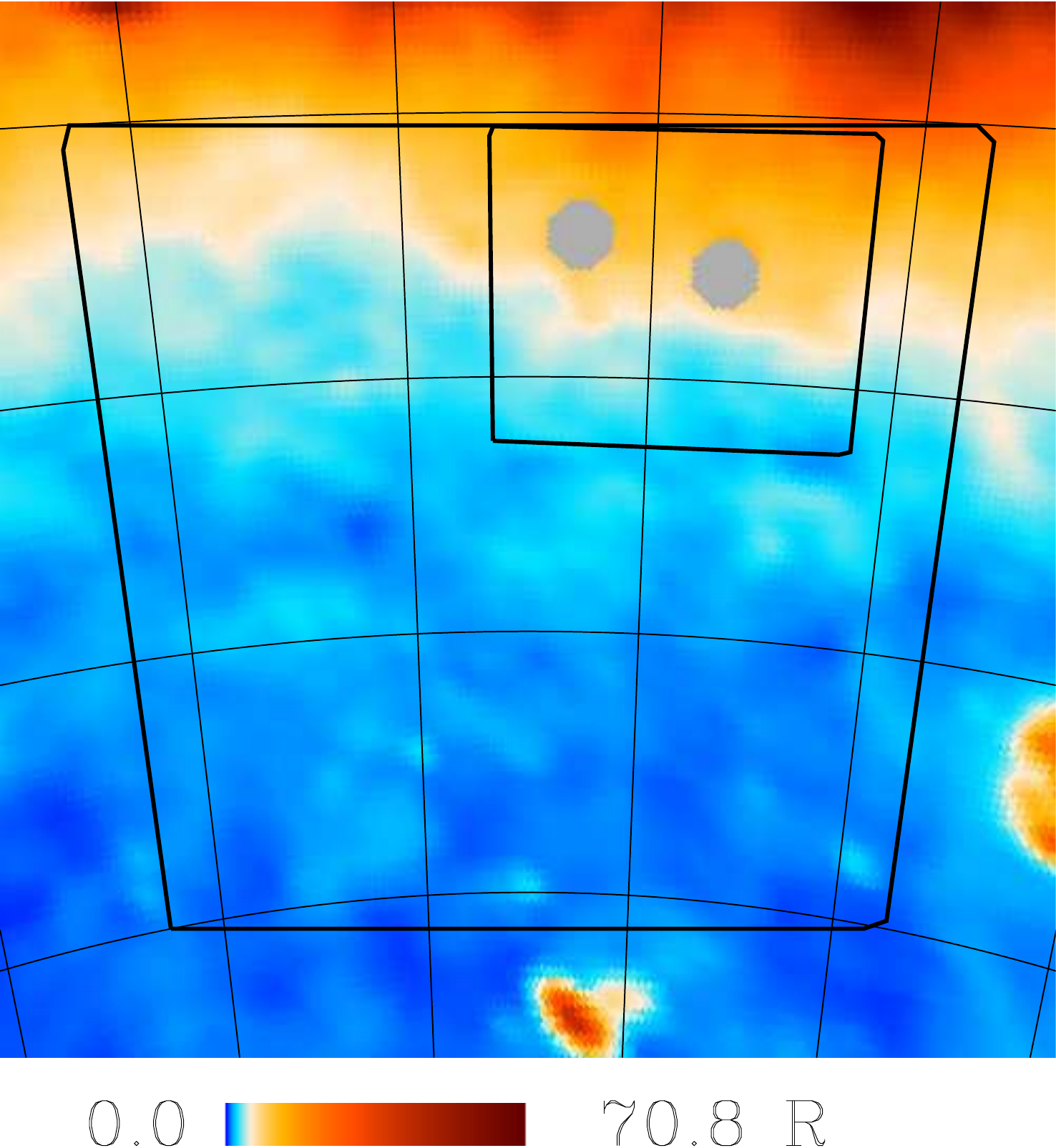}
\includegraphics[width=0.18\textwidth]{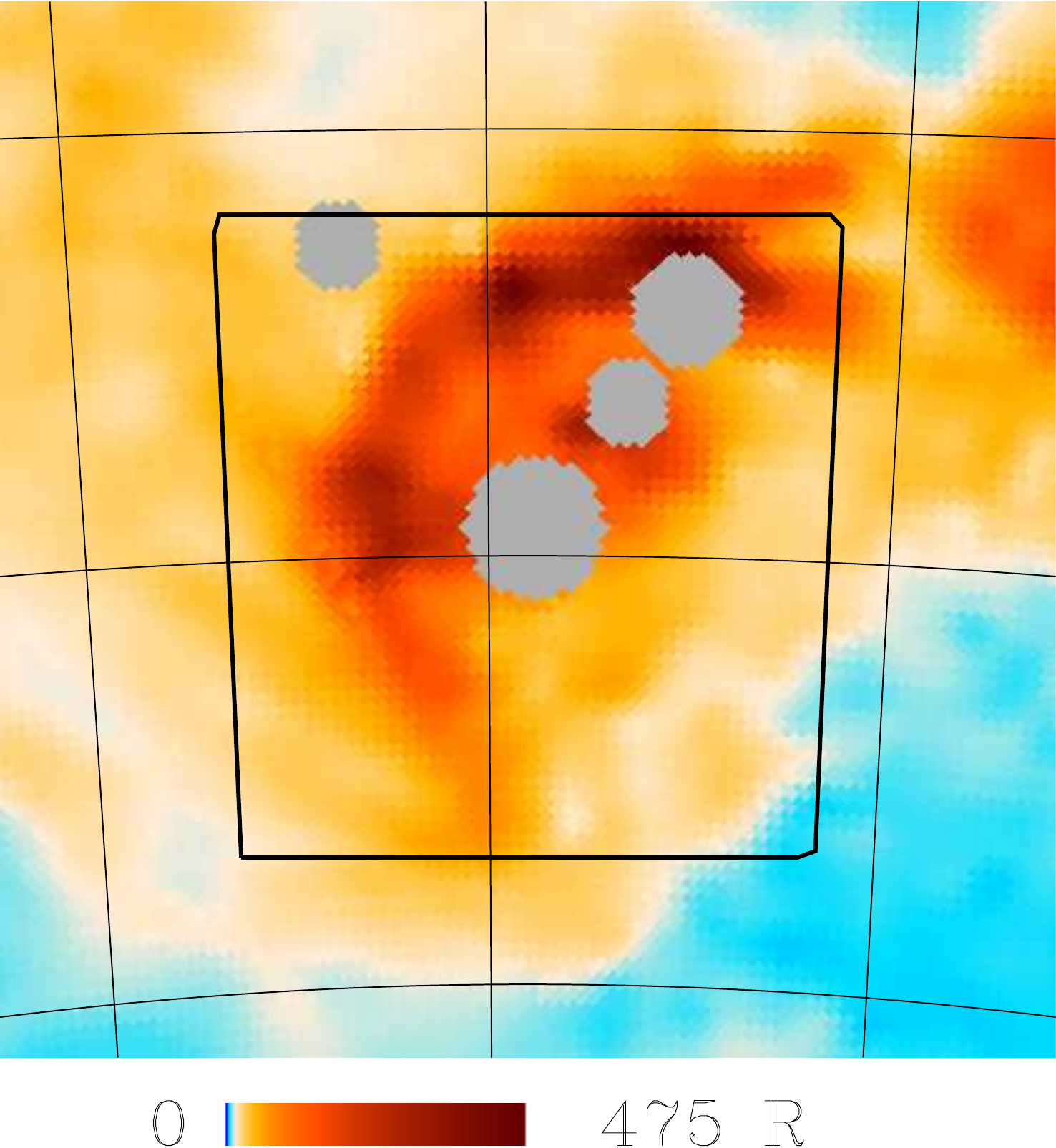}
\includegraphics[width=0.18\textwidth]{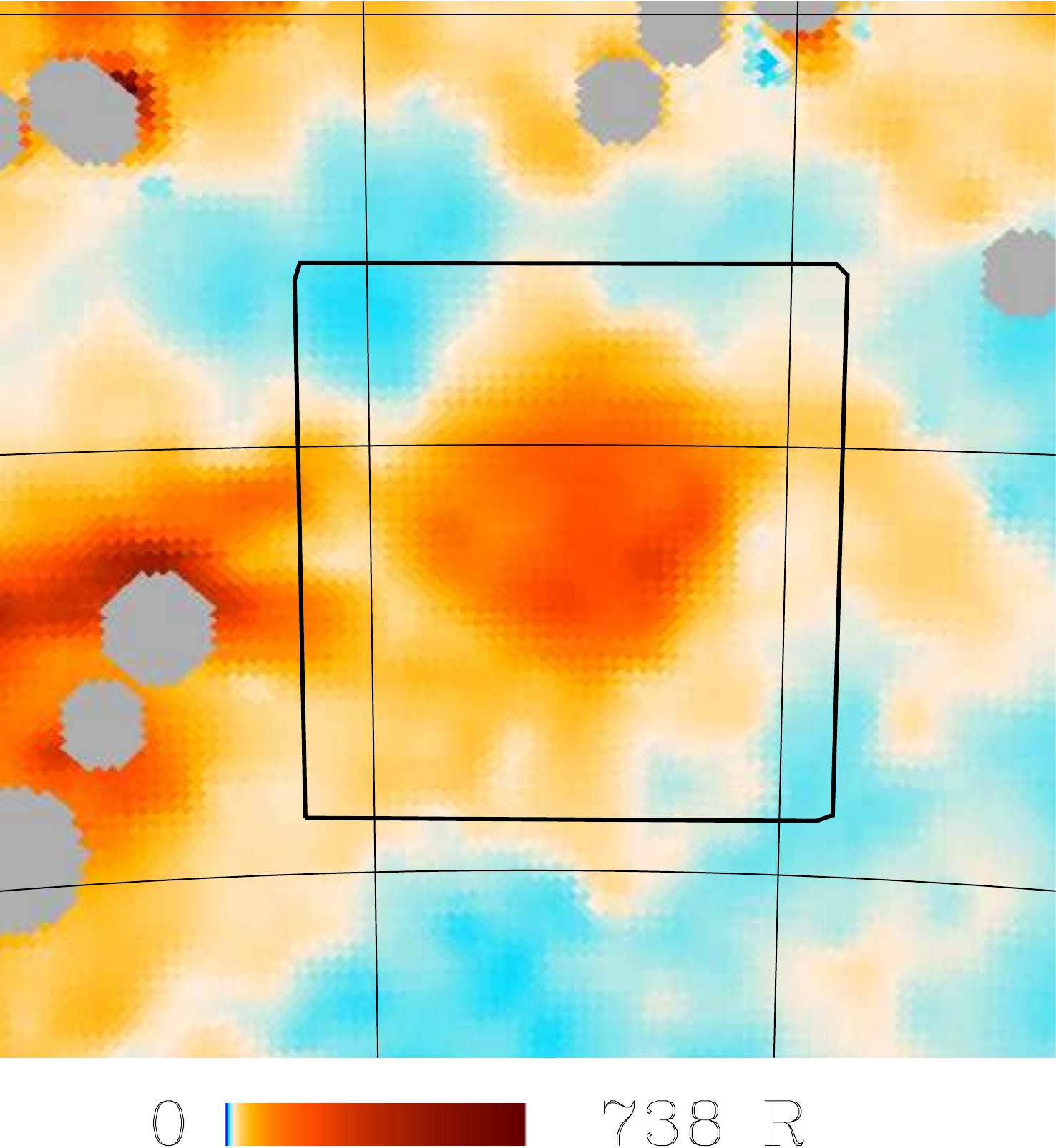}

\includegraphics[width=0.18\textwidth]{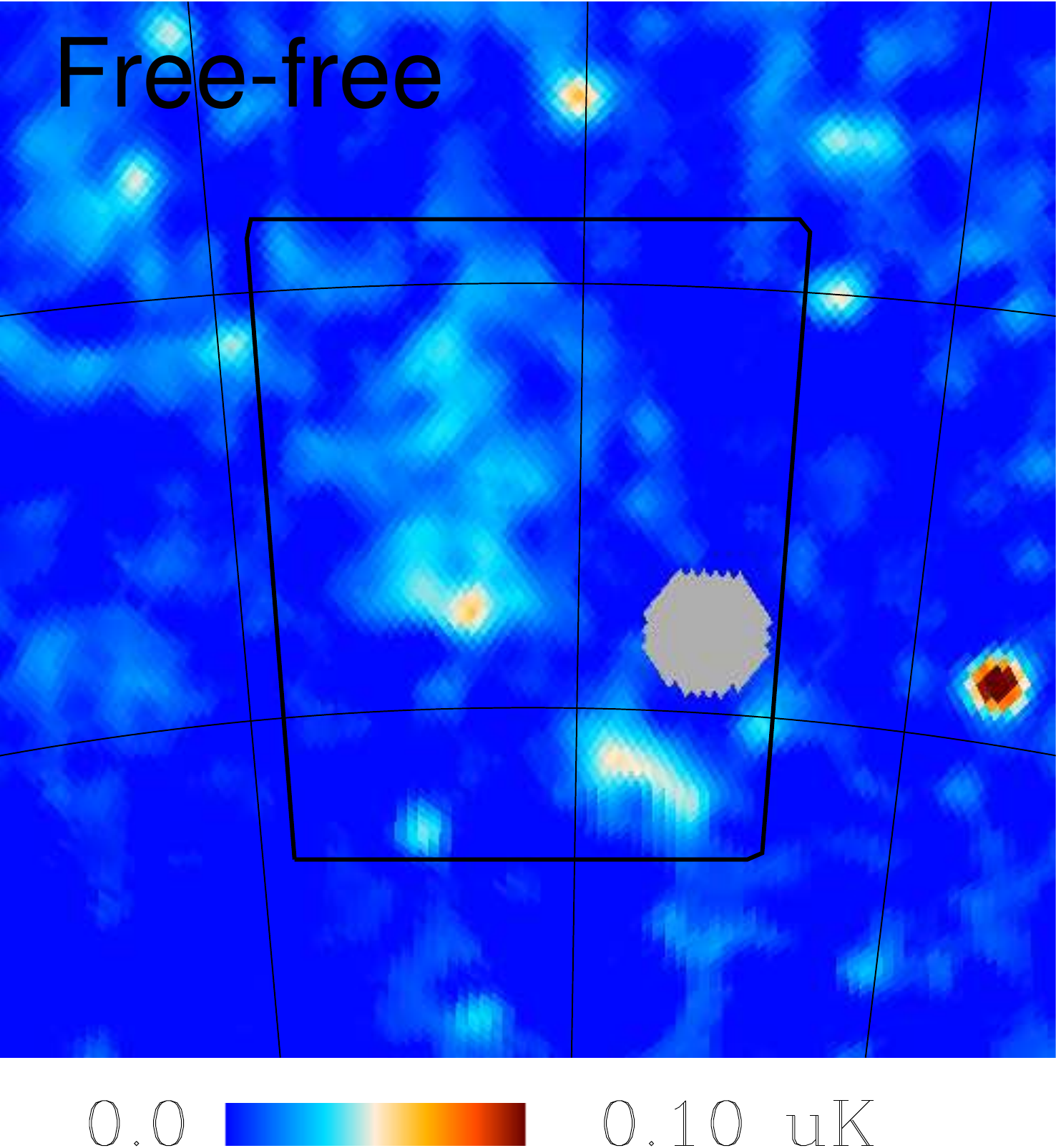}
\includegraphics[width=0.18\textwidth]{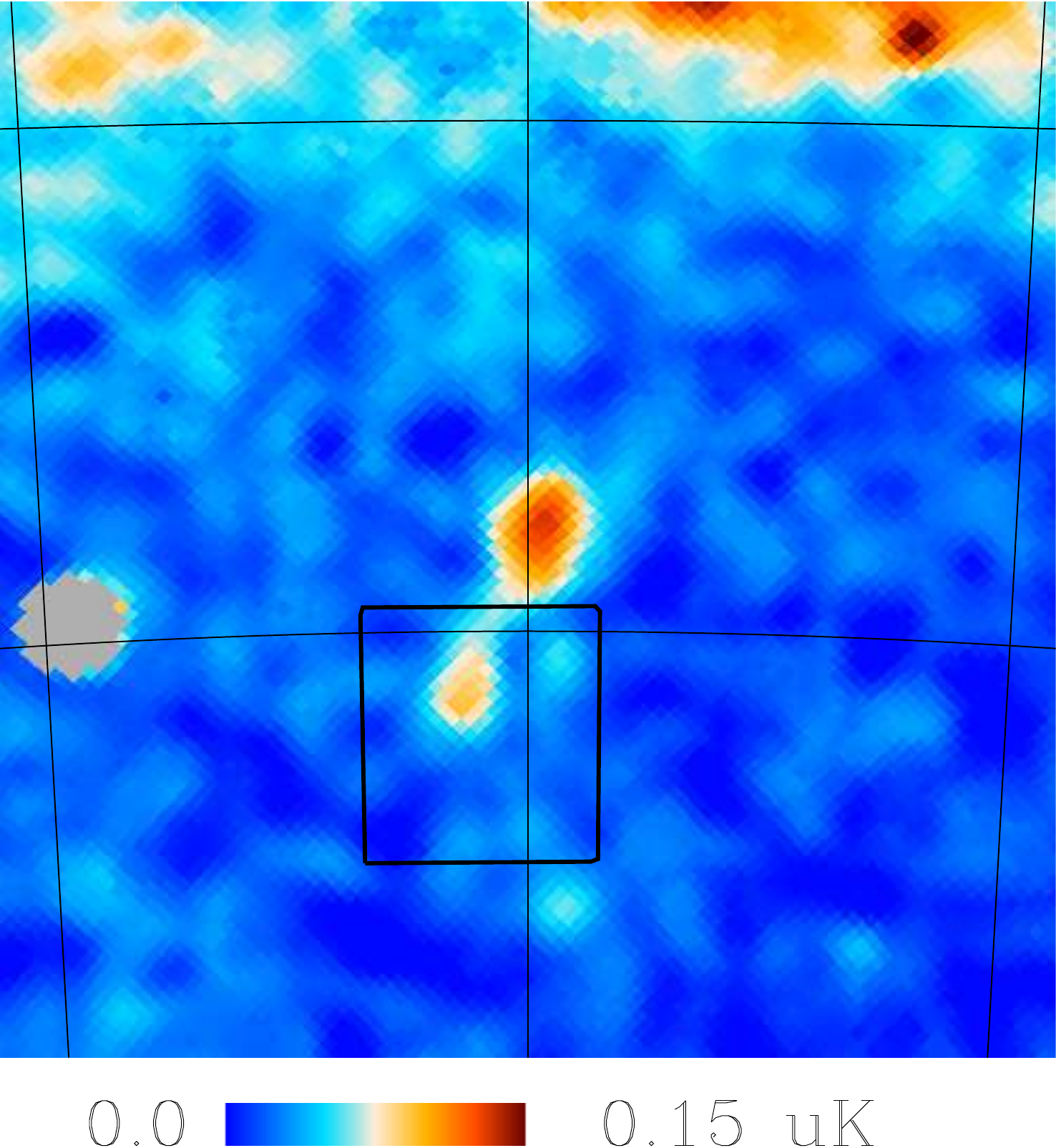}
\includegraphics[width=0.18\textwidth]{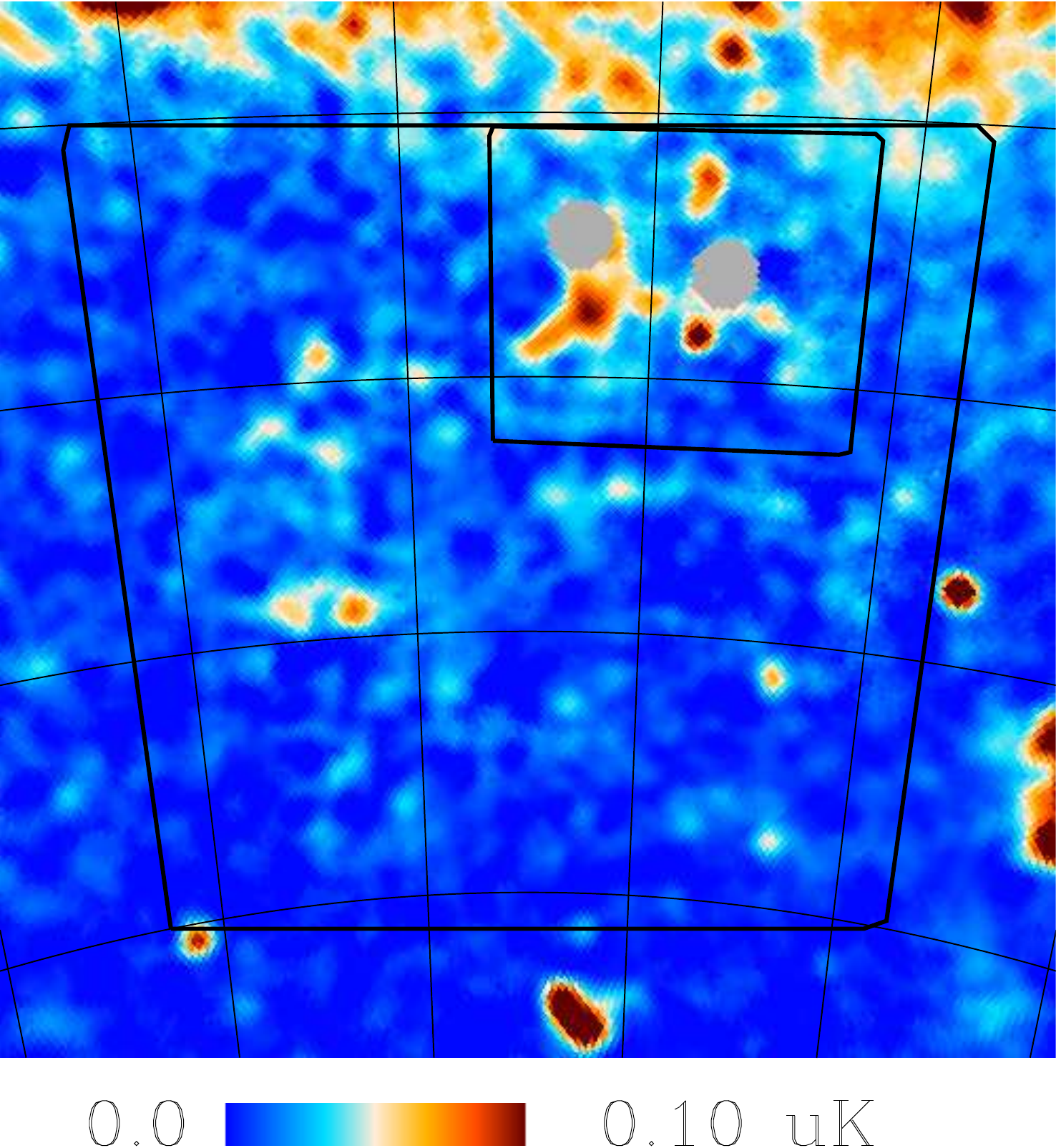}
\includegraphics[width=0.18\textwidth]{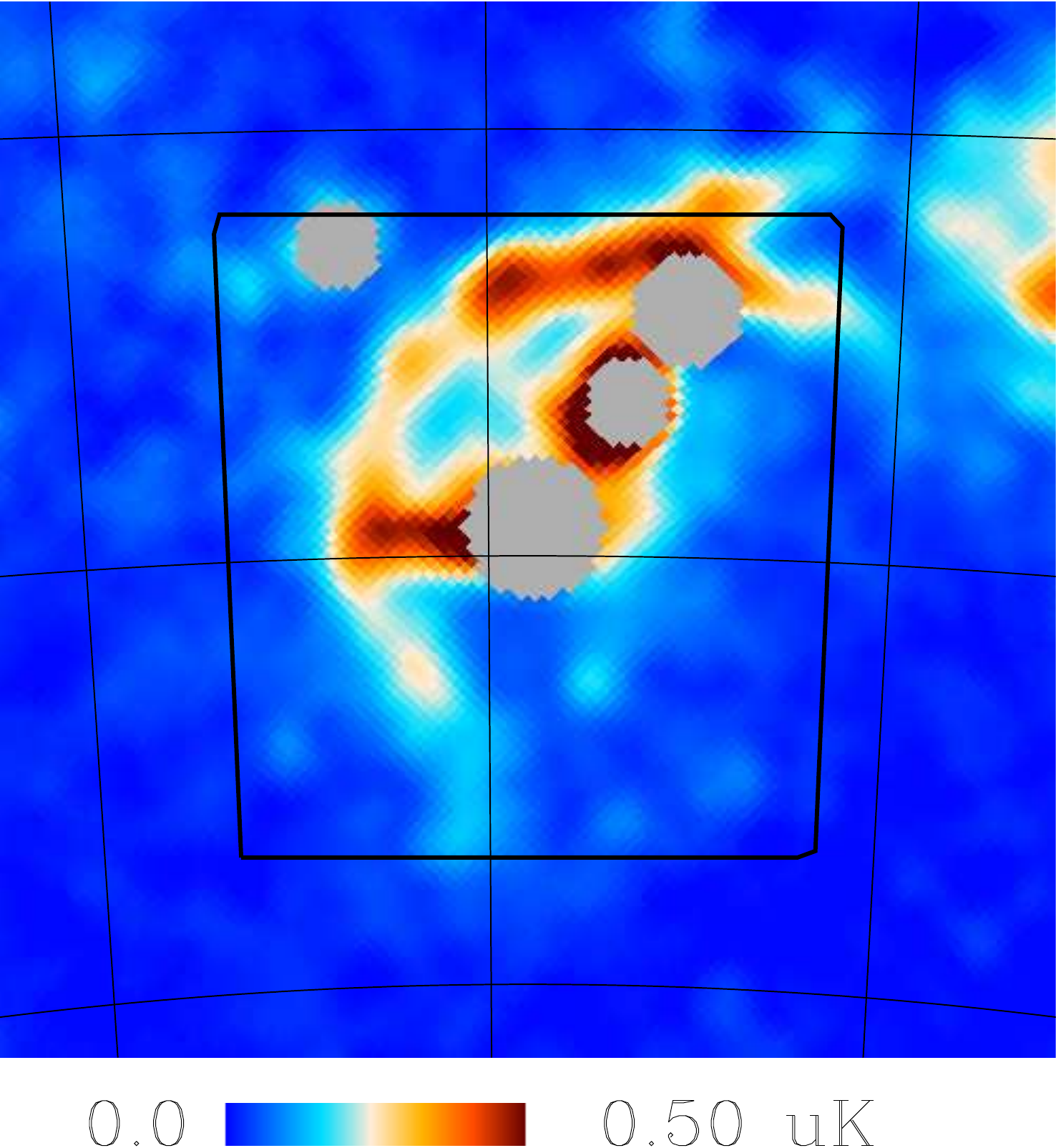}
\includegraphics[width=0.18\textwidth]{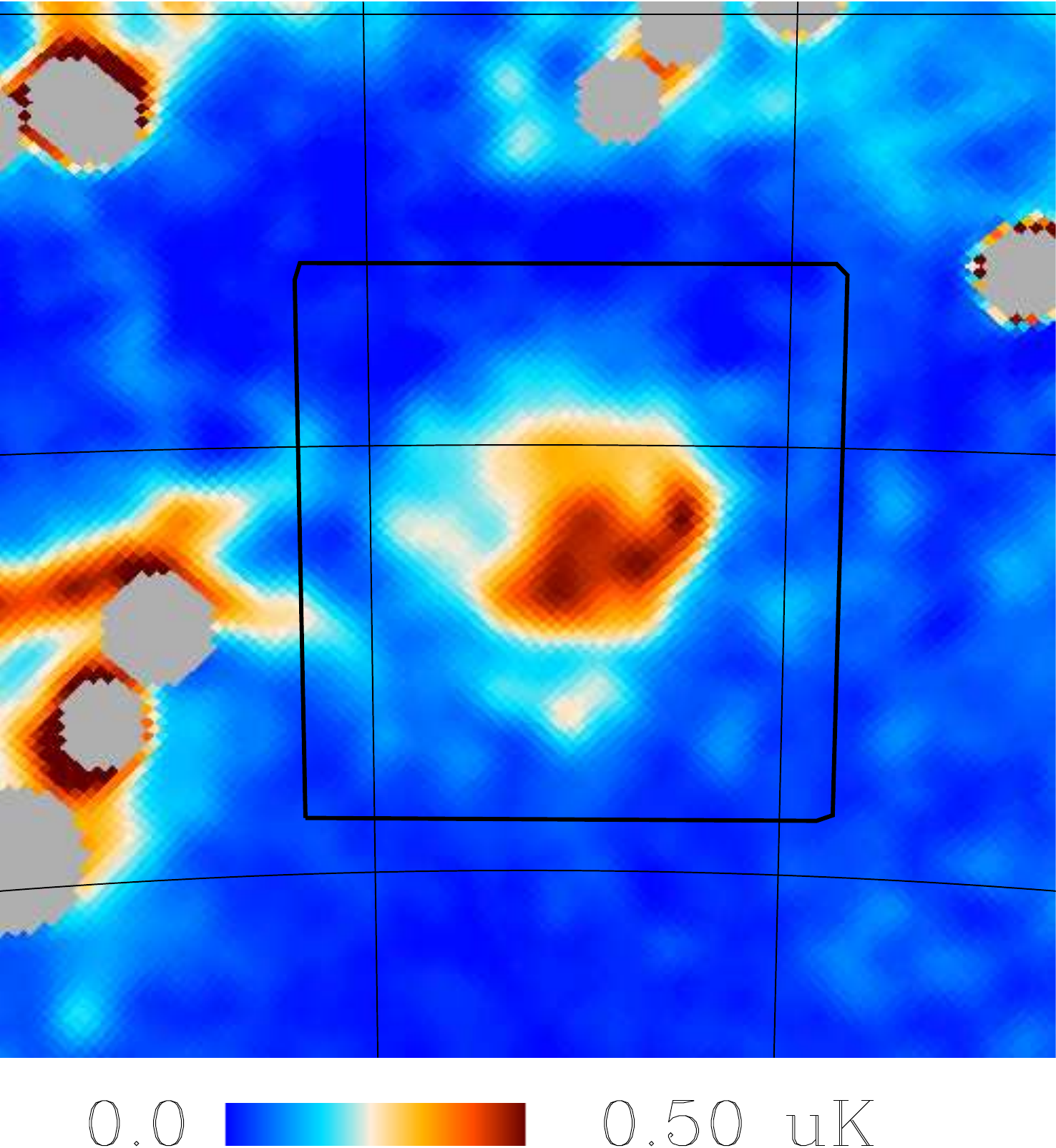}

\includegraphics[width=0.18\textwidth]{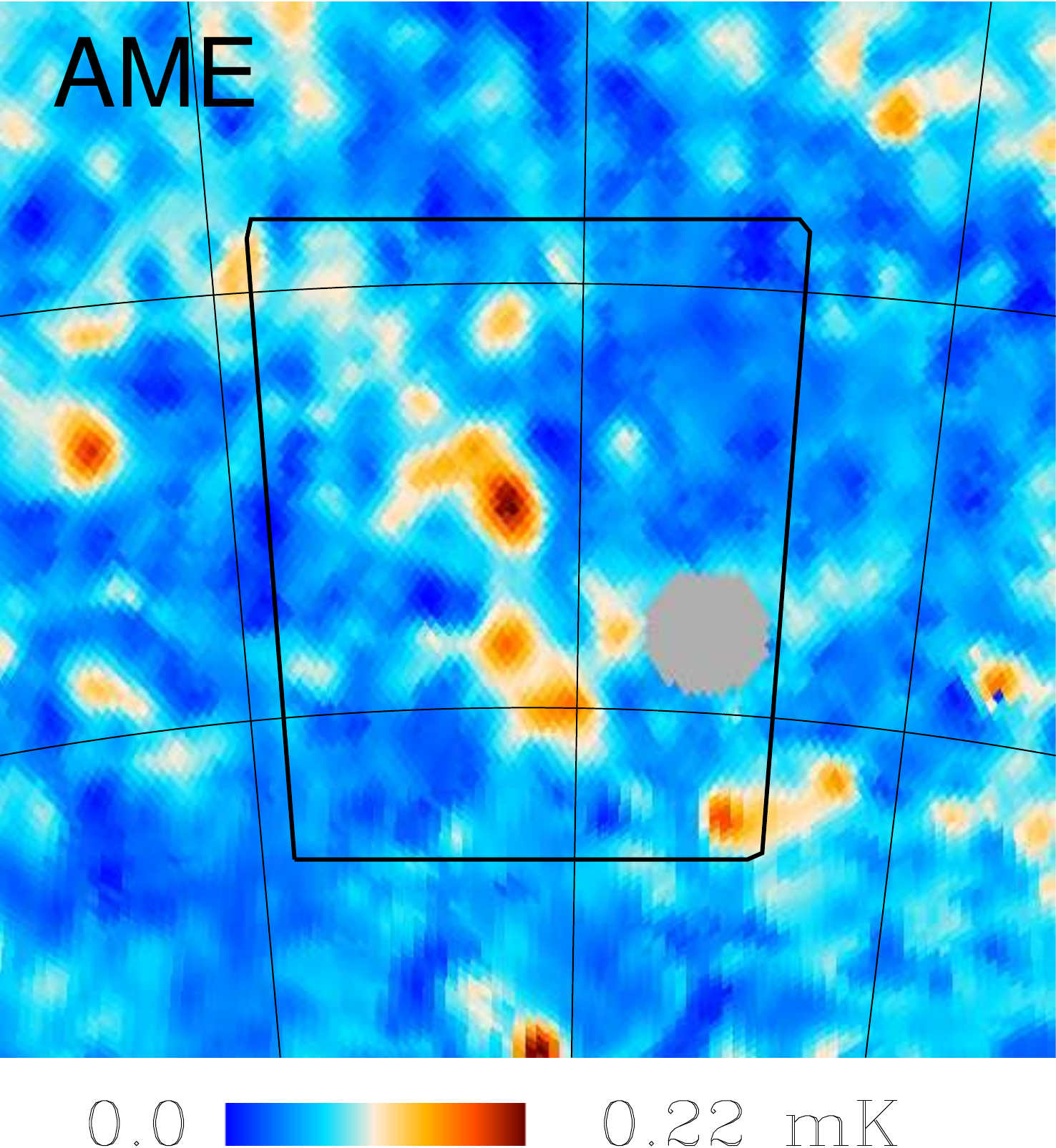}
\includegraphics[width=0.18\textwidth]{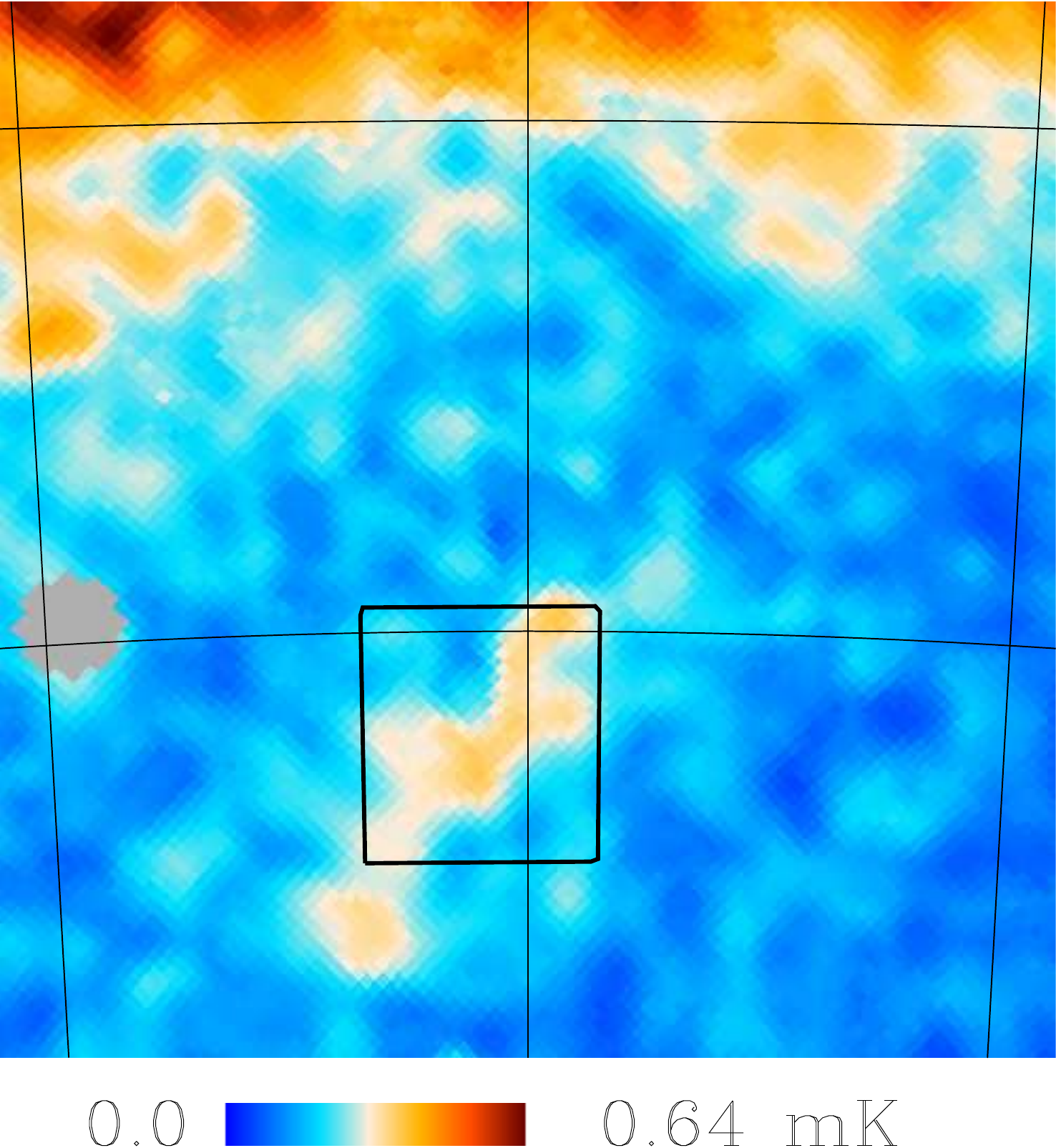}
\includegraphics[width=0.18\textwidth]{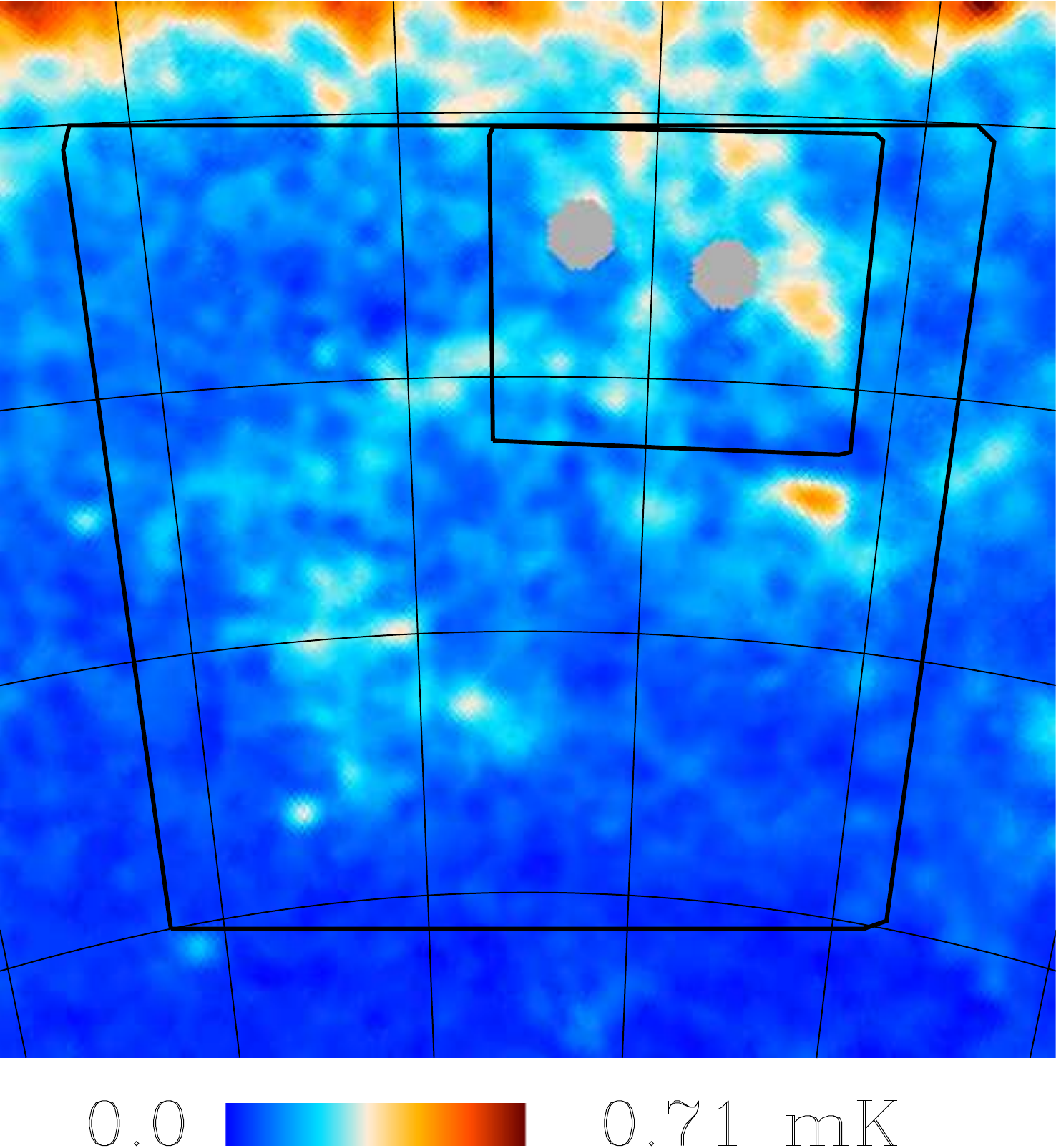}
\includegraphics[width=0.18\textwidth]{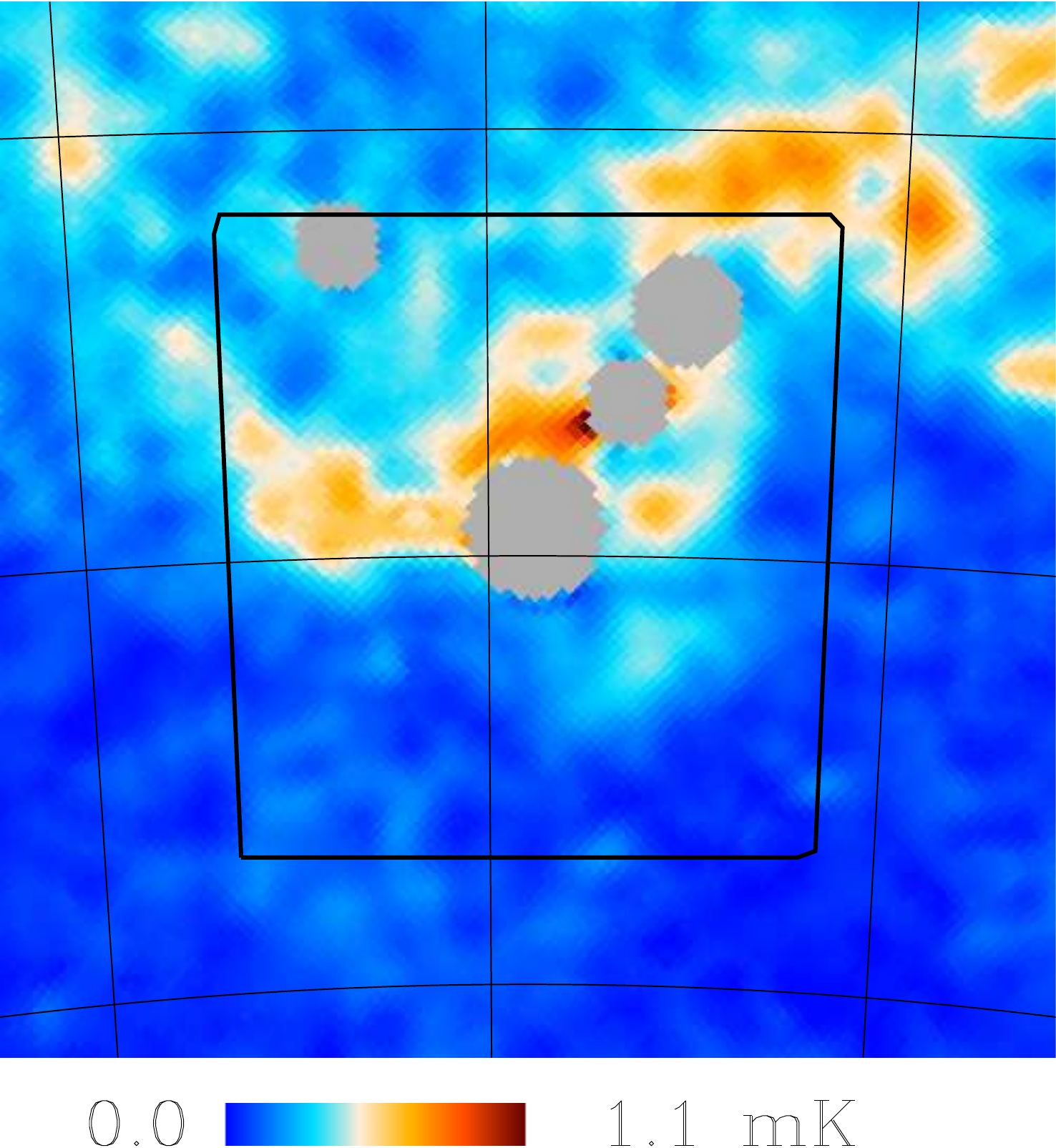}
\includegraphics[width=0.18\textwidth]{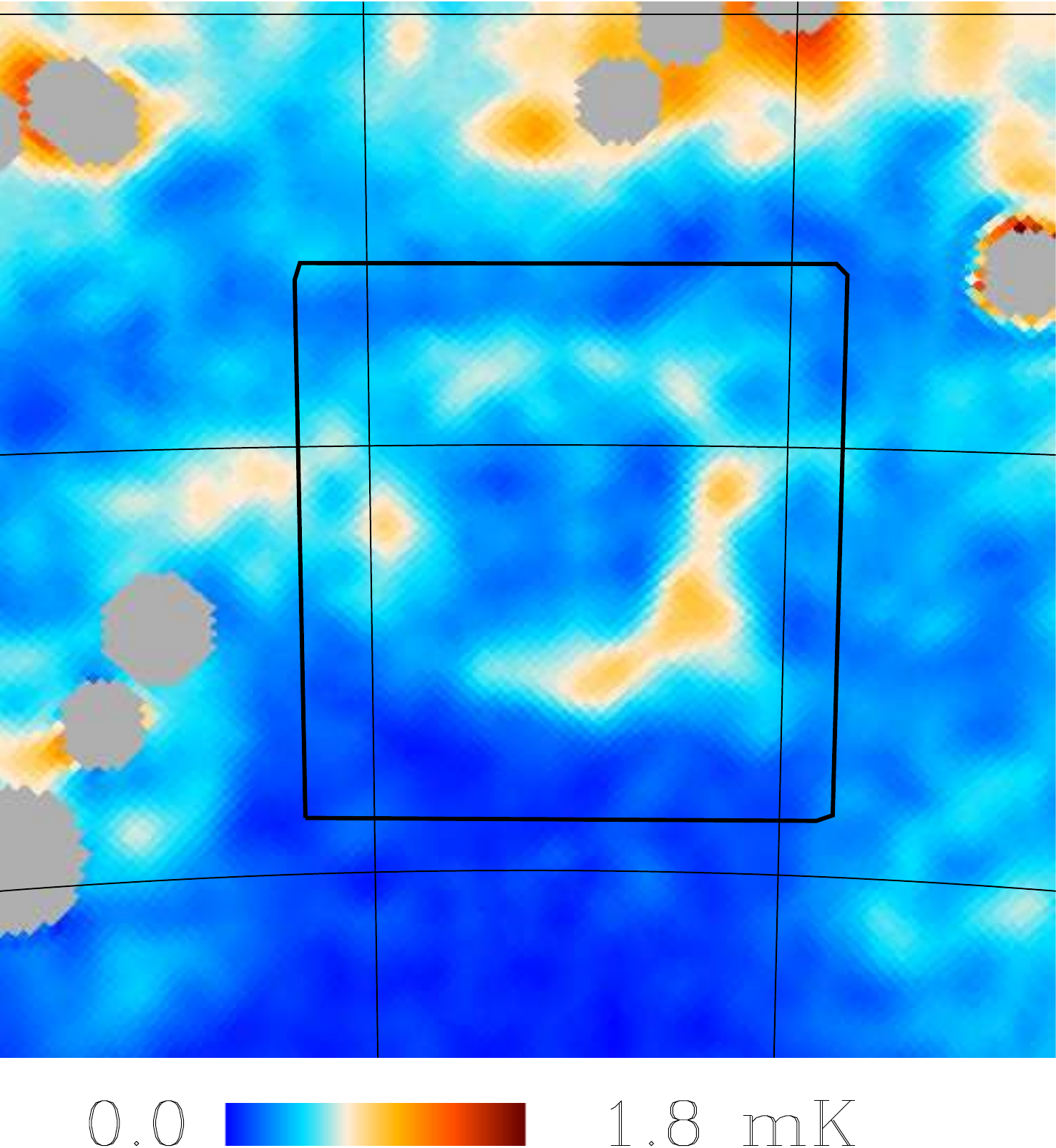}

\includegraphics[width=0.18\textwidth]{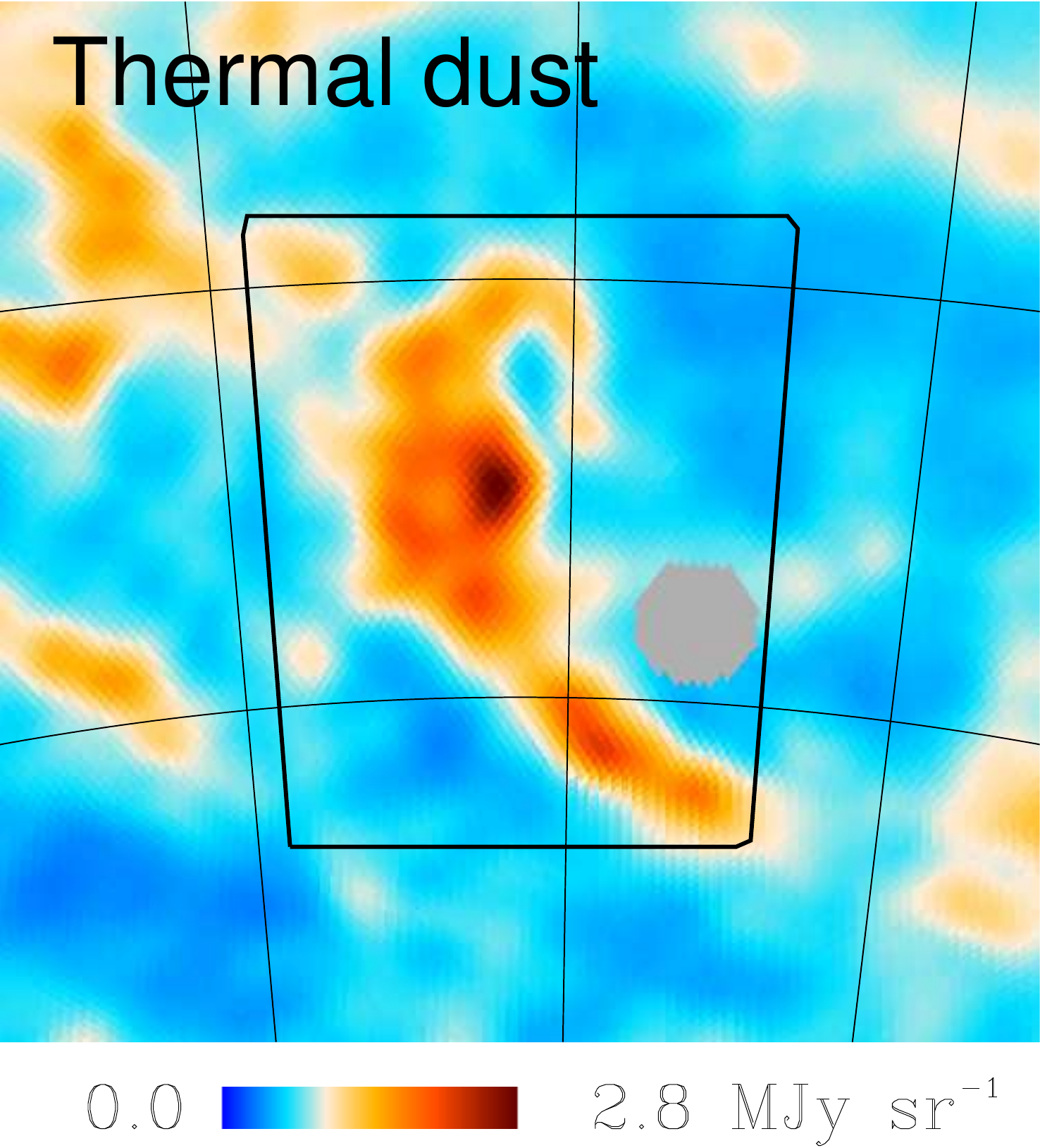}
\includegraphics[width=0.18\textwidth]{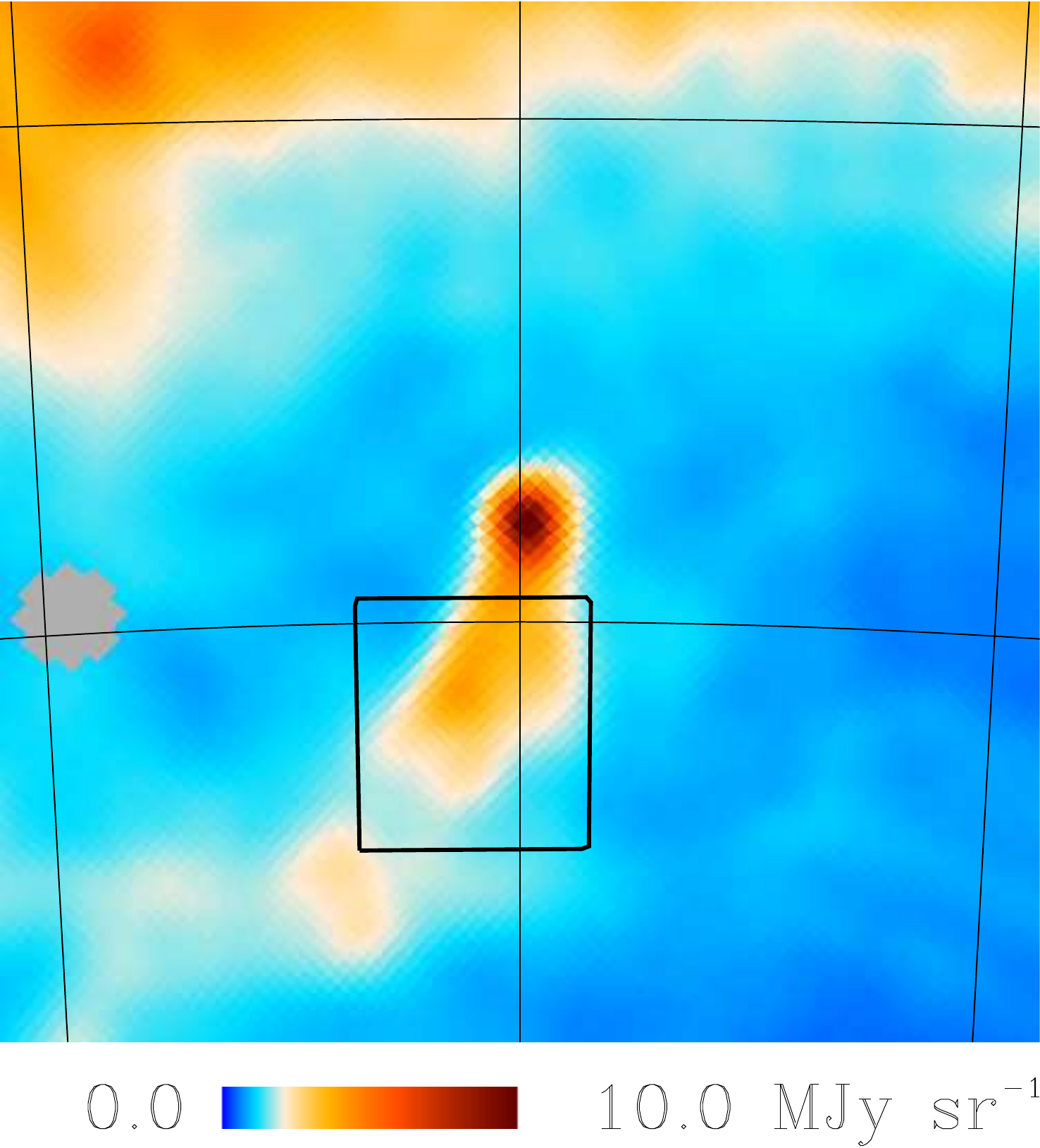}
\includegraphics[width=0.18\textwidth]{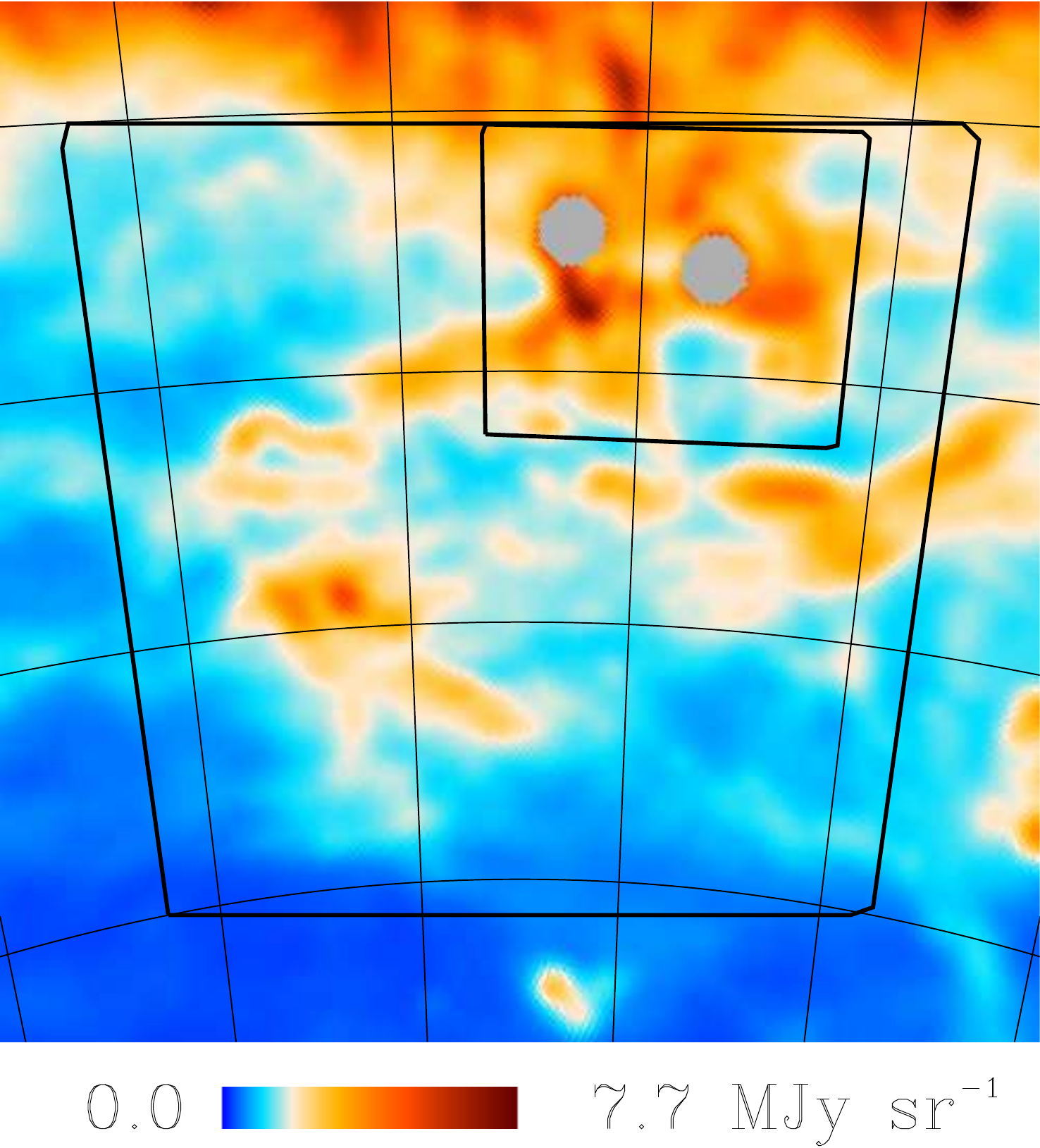}
\includegraphics[width=0.18\textwidth]{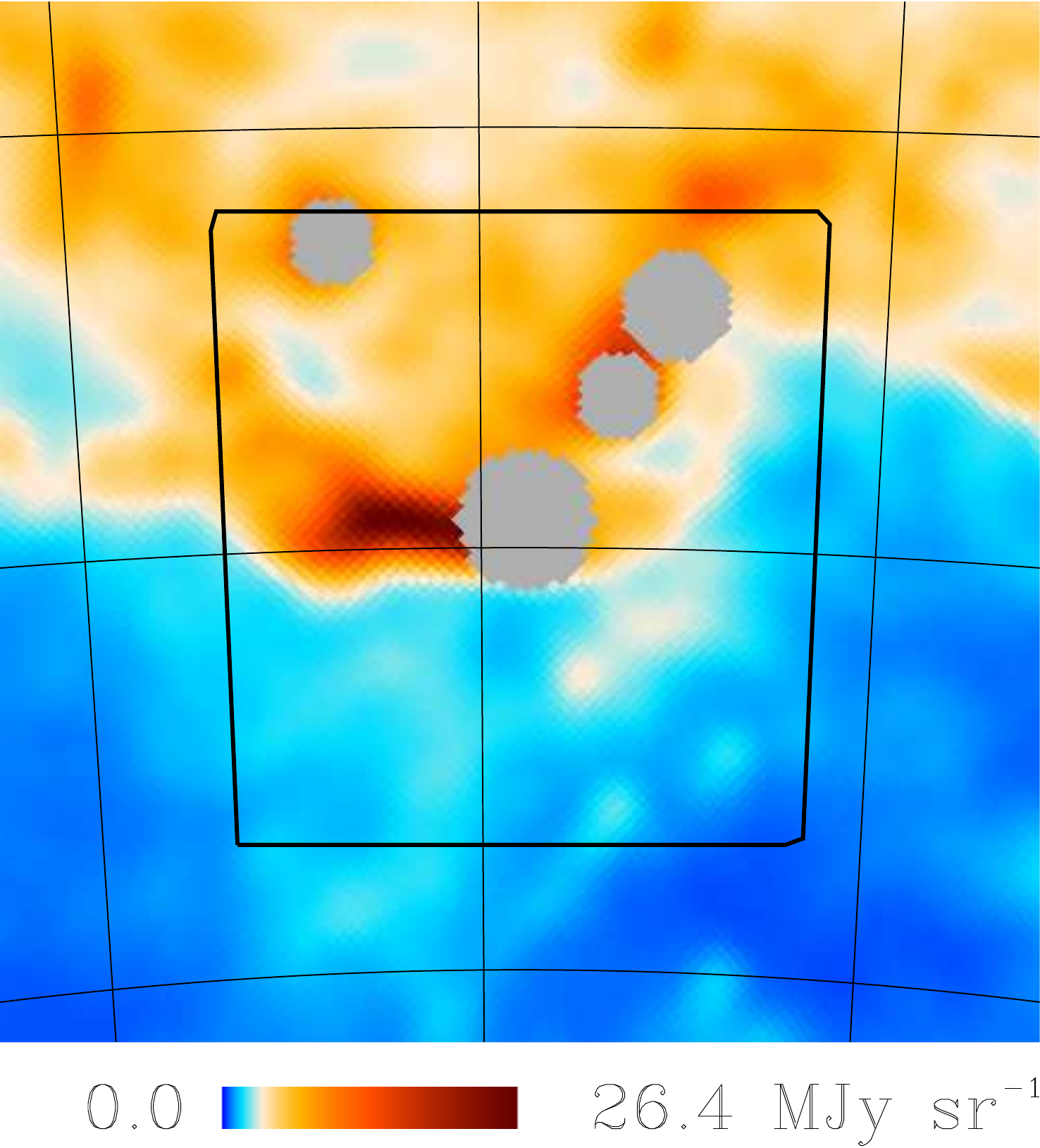}
\includegraphics[width=0.18\textwidth]{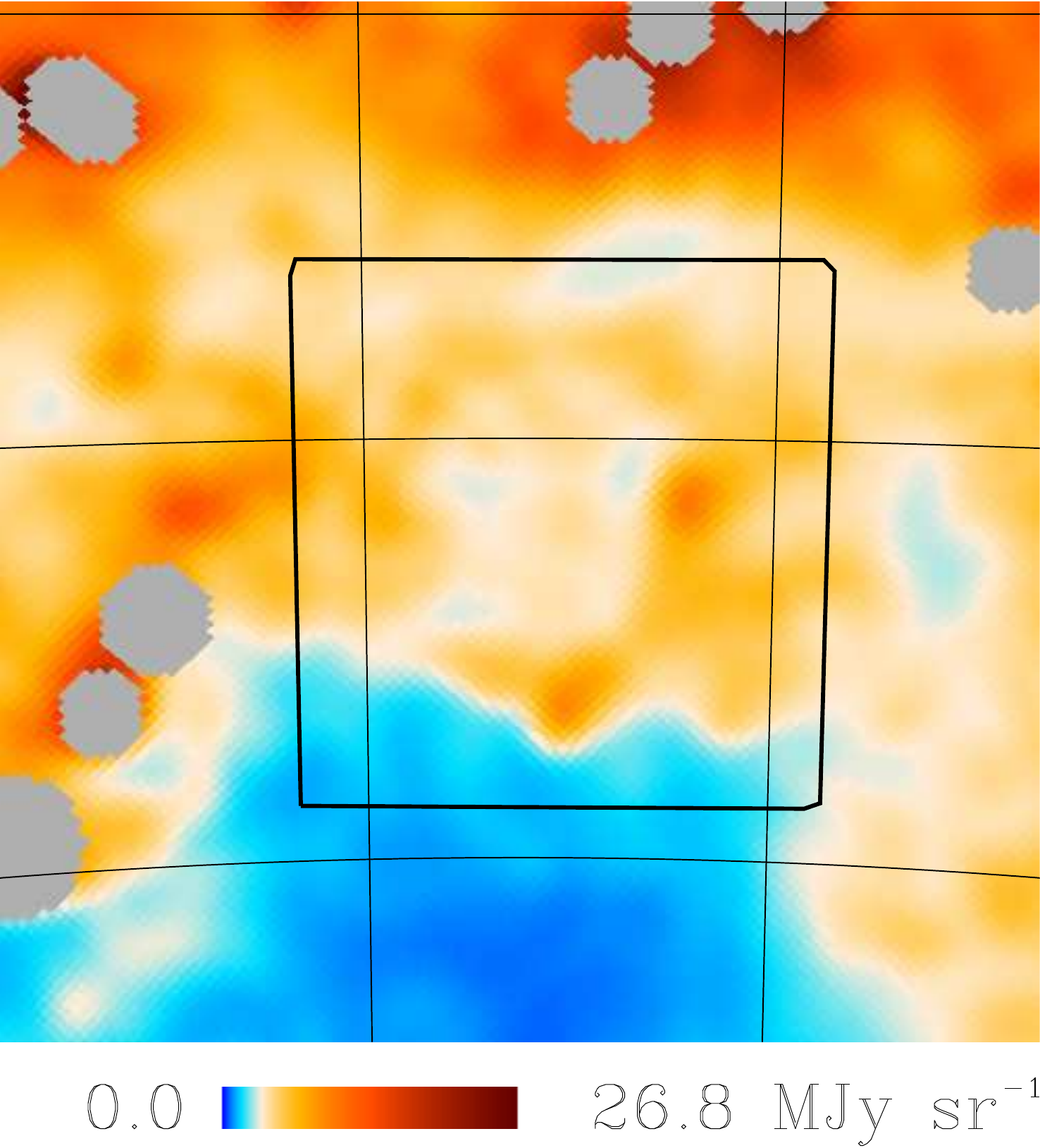}

\includegraphics[width=0.18\textwidth]{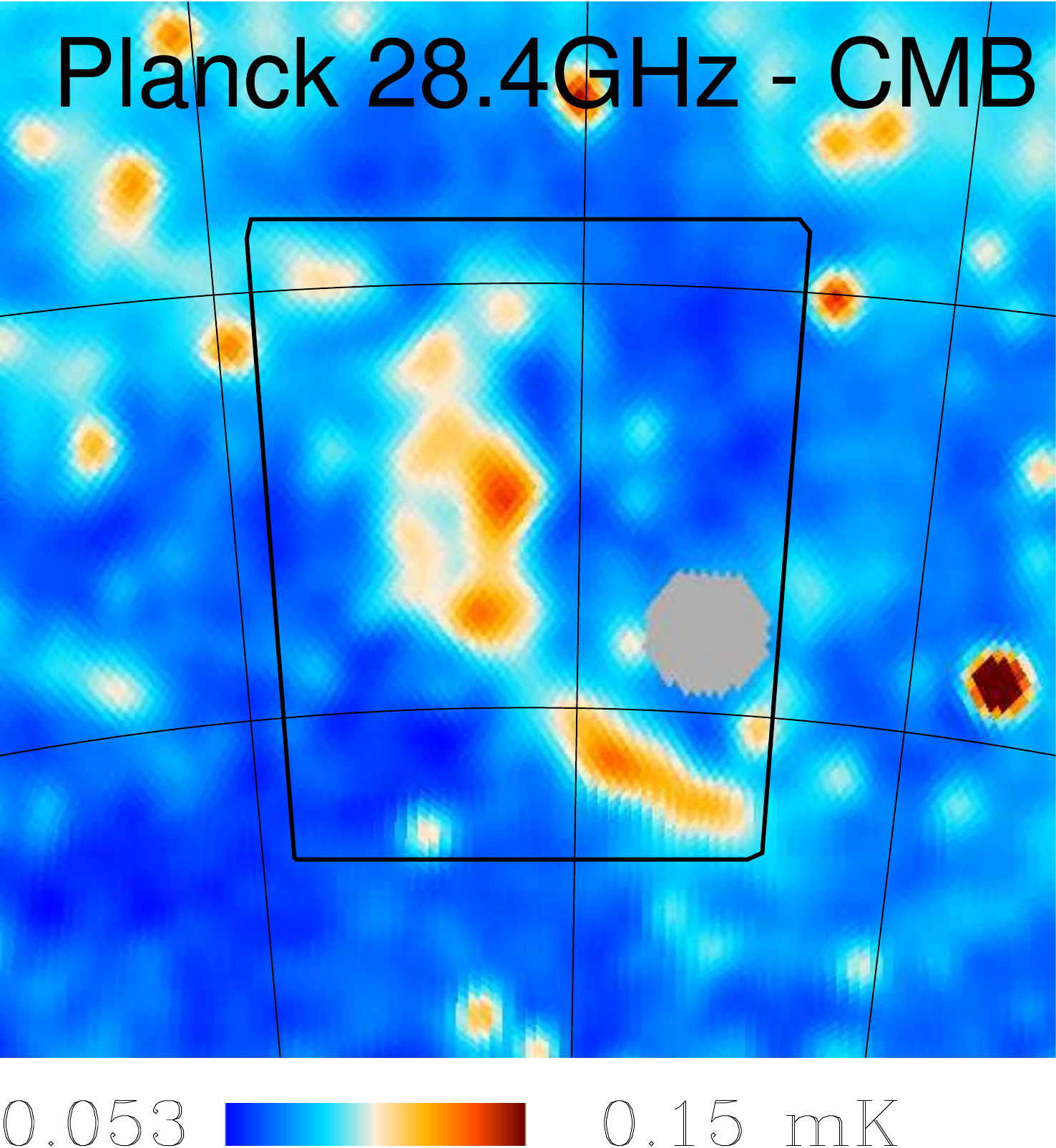}
\includegraphics[width=0.18\textwidth]{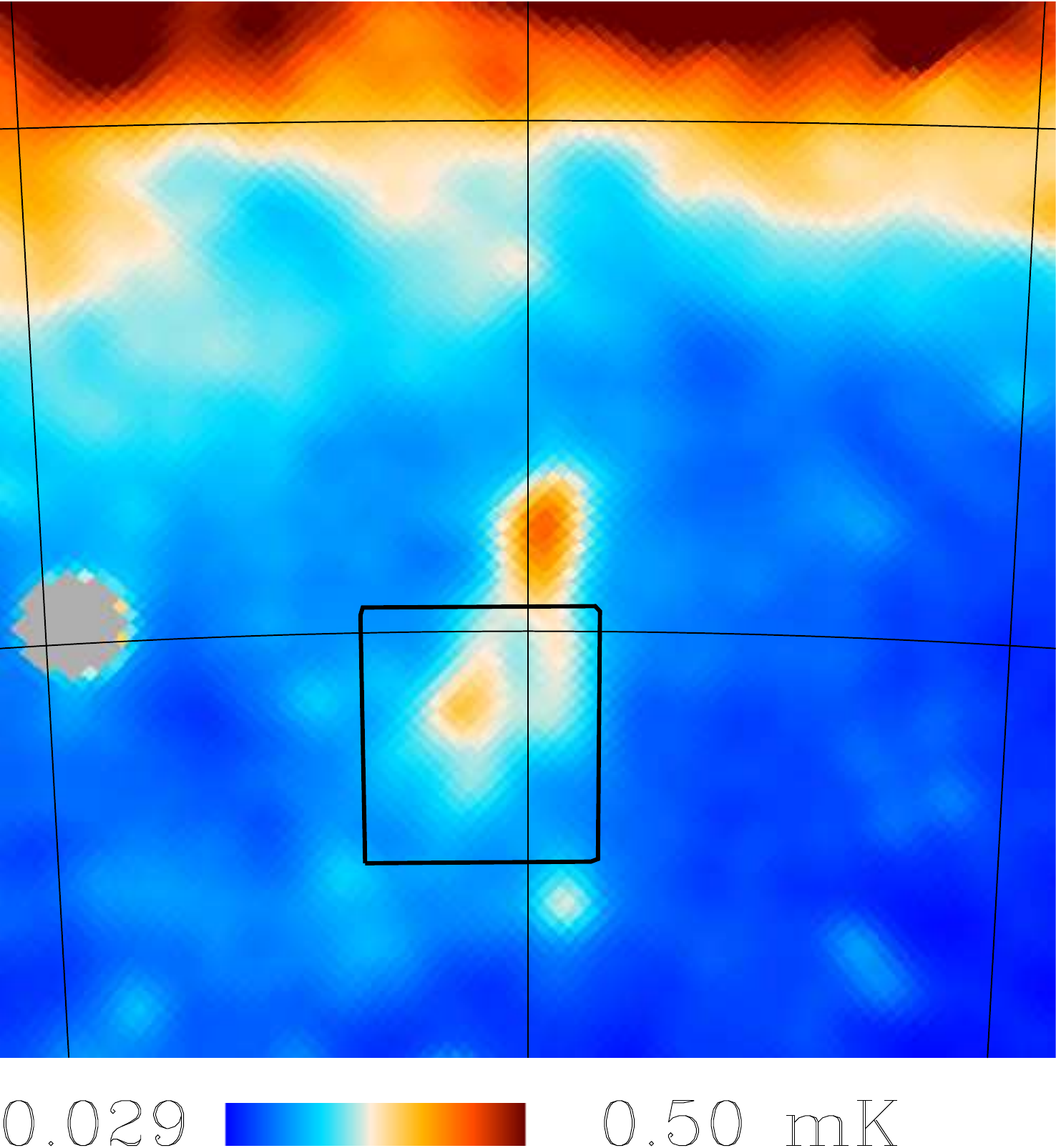}
\includegraphics[width=0.18\textwidth]{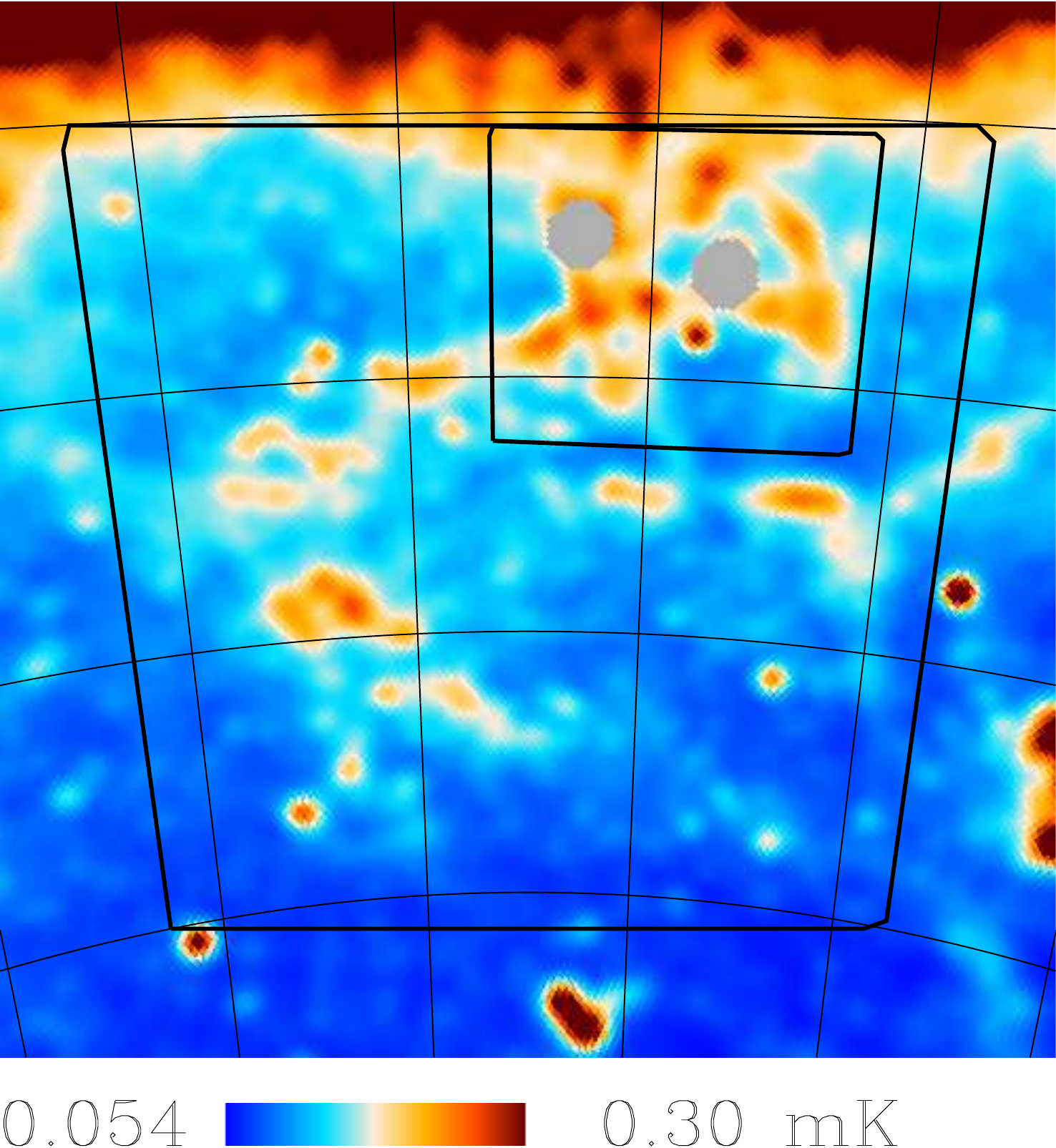}
\includegraphics[width=0.18\textwidth]{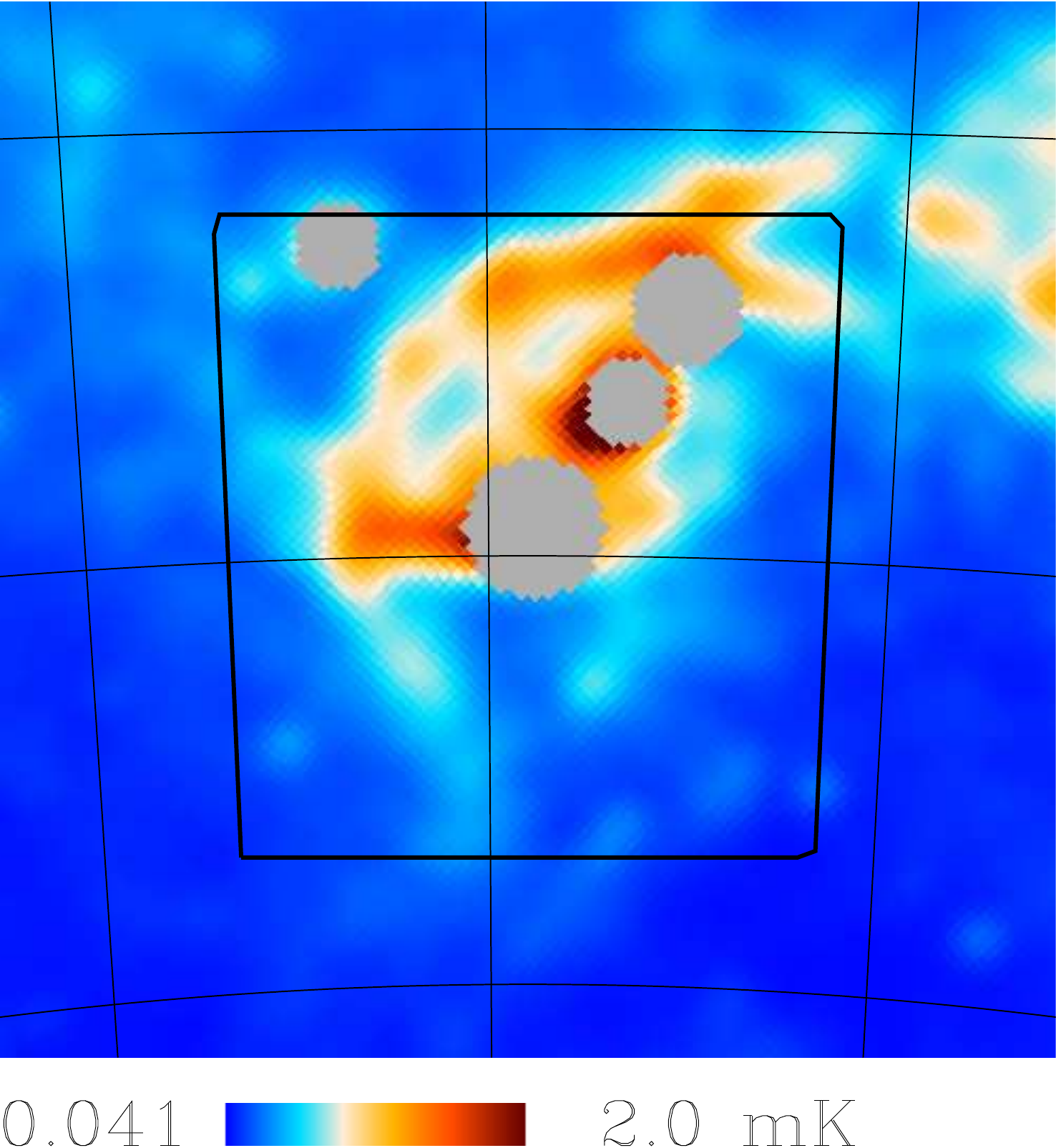}
\includegraphics[width=0.18\textwidth]{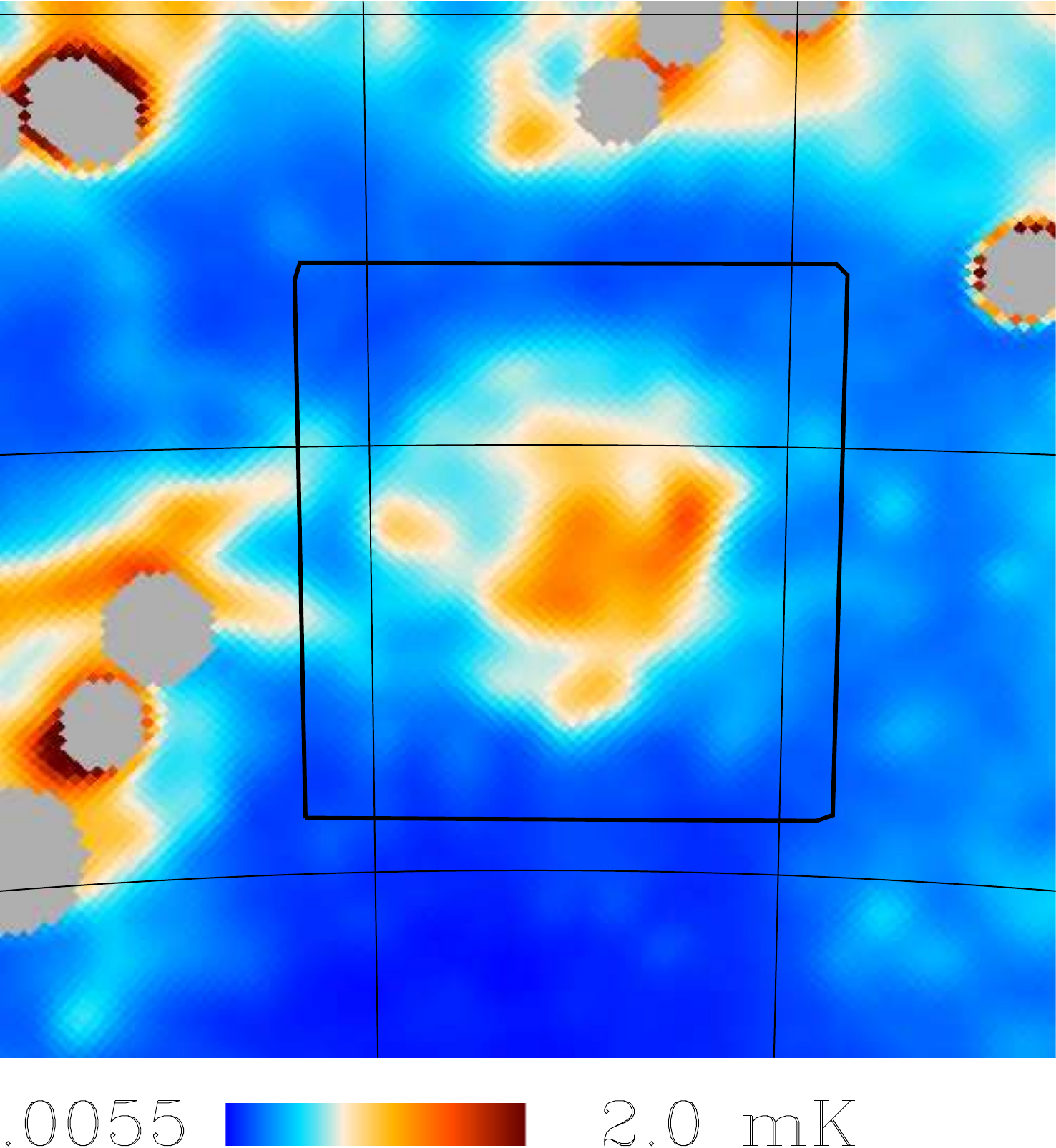}

\caption{\commander\ component maps for diffuse AME regions. From top to bottom: synchrotron amplitude (mK at 22.8\GHz), \ha\ (\citealp{Dickinson2003}, Rayleighs), free-free amplitude (mK at 22.8\GHz), AME amplitude (mK at 22.8\GHz), thermal dust amplitude (MJy\,sr$^{-1}$ at 545\GHz),  CMB-subtracted \Planck\ 28.4\GHz\ map (mK). {\it Col. 1}: a plume of emission in Pegasus, centred at $(l,b)=(91\pdeg5,-35\pdeg8)$. {\it Col. 2}: the Corona Australis region at $(l,b)=(0\deg,-18\deg)$. {\it Col. 3:} an extended region of emission in Musca/Chamaeleon, centred at $(l,b)=(305\deg,-26\deg)$ (the Large Magellanic Cloud can be seen to the bottom-right). {\it Col. 4}: the Orion (M42) region of emission centred at $(l,b)=(209\deg,-19\pdeg38)$. {\it Col. 5}: the $\lambda$ Orionis region at $(l,b)=(196\deg,-12\deg)$. The graticule separation is $10\deg$ in both directions, and the colour scales are \asinh\ (although most are close to linear). The regions looked at for the correlation analyses are shown as black rectangular boxes.}
\label{fig:amemaps}
\end{center}
\end{figure*}

\begin{figure*}[tbh]
\begin{center}
\includegraphics[width=0.32\textwidth]{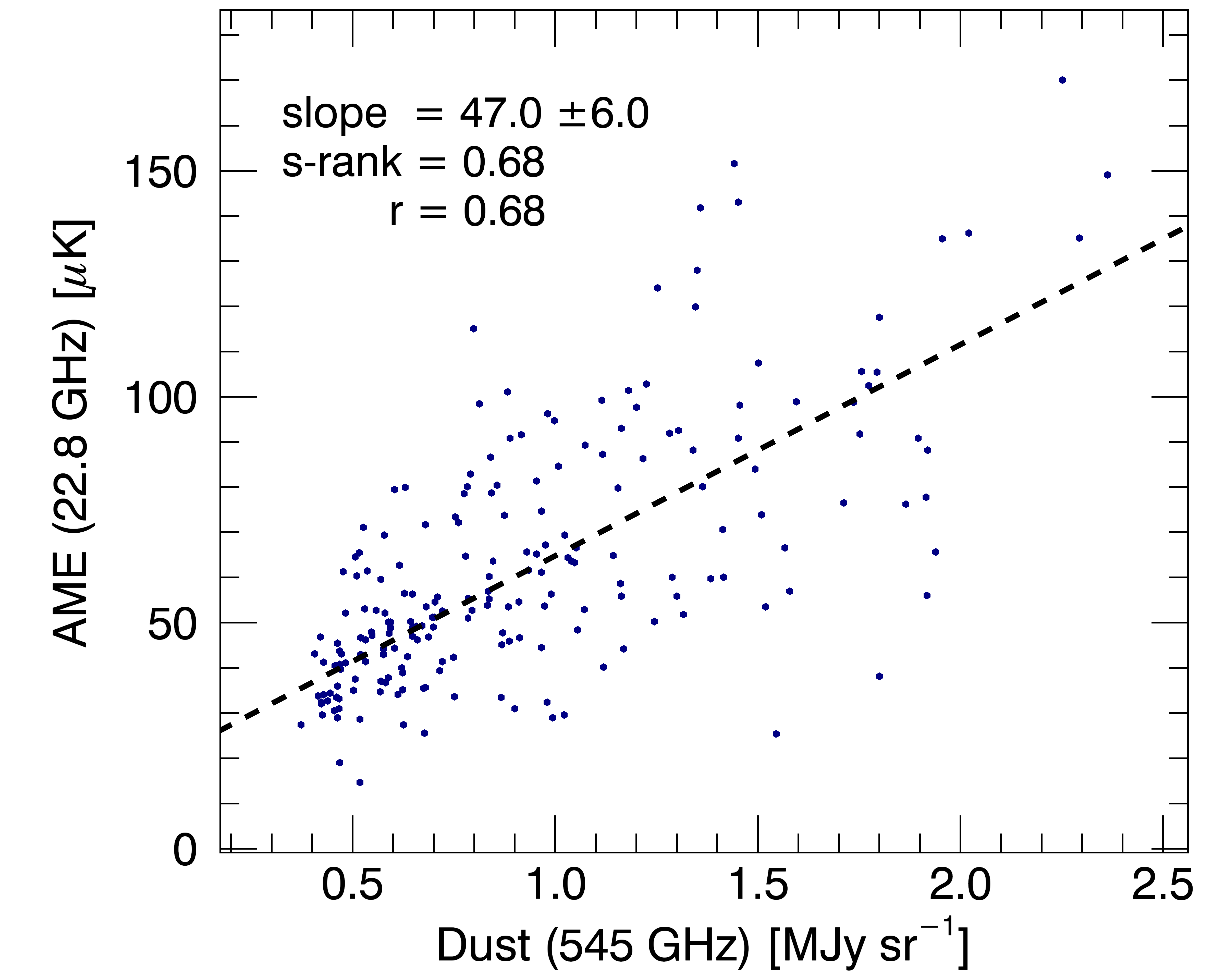}
\includegraphics[width=0.32\textwidth]{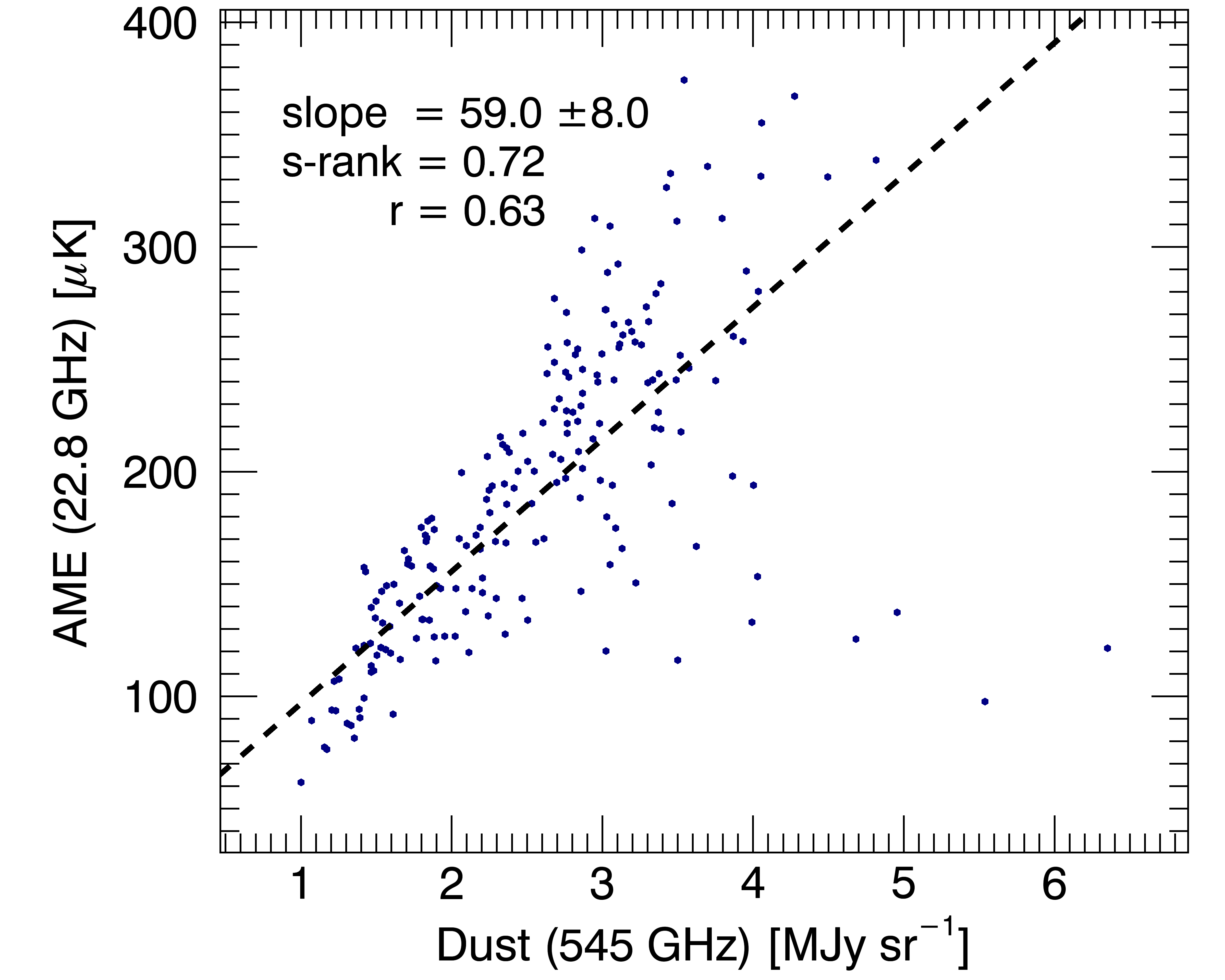}
\includegraphics[width=0.32\textwidth]{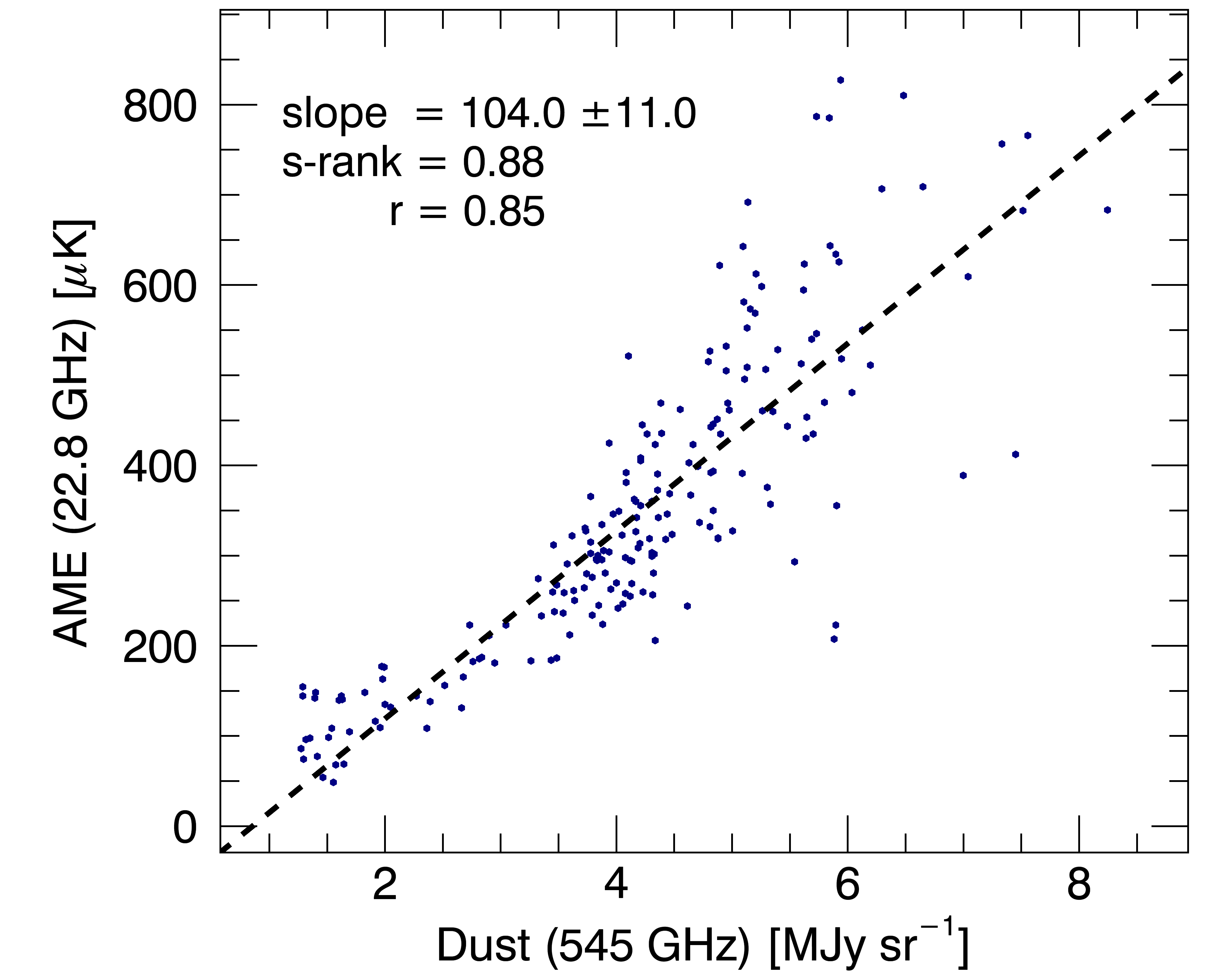}
\caption{\ttp\ plots comparing the \commander\ AME map evaluated at 22.8\GHz\ to the \commander\ dust map at 545\GHz\ in three regions, as shown in Fig.\,\ref{fig:amemaps}. The best fit is shown as a dashed line. From left to right, for the Pegasus plume, Musca, and $\lambda$ Orionis regions.
}
\label{fig:ame_ttplots2}
\end{center}
\end{figure*}

\begin{table}[tb]
\begingroup
\newdimen\tblskip \tblskip=5pt
\caption{The top section presents emissivities of AME at 22.8\GHz\ ($\upmu$K) relative to the \Commander\ thermal dust amplitude at 545\GHz\ (MJy\,sr$^{-1}$), the IRAS map at 100\um\ (MJy\,sr$^{-1}$), and the optical depth at 353\GHz, $\tau_{353}$, for the regions shown in Figs.\,\ref{fig:ame_fullsky} and \ref{fig:amemaps}. We also include emissivities for the LMC and SMC, discussed in Sect.\,\ref{sec:lmc}. The bottom part presents emissivities from \citet[D06; whole sky and region mean]{Davies2006} and \citet[XV; Perseus, $\rho$ Oph; and the unweighted region mean]{planck2013-XV} for comparison.}
\label{tab:emissivity}
\nointerlineskip
\vskip -3mm
\footnotesize
\setbox\tablebox=\vbox{
    \newdimen\digitwidth 
    \setbox0=\hbox{\rm 0} 
    \digitwidth=\wd0 
    \catcode`*=\active 
    \def*{\kern\digitwidth}
    \newdimen\signwidth 
    \setbox0=\hbox{+} 
    \signwidth=\wd0 
    \catcode`!=\active 
    \def!{\kern\signwidth}
    \newdimen\pointwidth
    \setbox0=\hbox{{.}}
    \pointwidth=\wd0
    \catcode`?=\active
    \def?{\kern\pointwidth}
    \newdimen\notewidth
    \setbox0=\hbox{$^\mathrm{a}$}
    \notewidth=\wd0
    \catcode`@=\active
    \def@{\kern\notewidth}
    \halign{\hbox to 1.0in{#\leaderfil}\tabskip 2pt&
            \hfil#\hfil\tabskip 2pt&
            \hfil#\hfil\tabskip 2pt&
            \hfil#\hfil\tabskip 0pt\cr
    \noalign{\doubleline}
    Region & AME/545\GHz\ & AME/100\um\ & AME/$\tau_{353}$ \cr
           \hbox to 1.0in{\hfil}\,\hfil & [\upkmjy] & [\upkmjy] & [\upk\,$10^{-6}$] \cr
\noalign{\vskip 3pt\hrule\vskip 5pt}
R1: Perseus           & $*24\pm*7$ & $12.3\pm*1.9$ & $*1.5\pm0.9$\cr
R2: Plume             & $*47\pm*6$ & $18?*\pm*2?*$ & $*7.7\pm1.0$\cr
R3: R\,CrA            & $*36\pm14$ & $50?*\pm12?*$ & $*4.1\pm1.8$\cr
R4: $\rho$ Oph        & $*40\pm*9$ & $*4.6\pm*0.9$ & $*2.2\pm1.2$\cr
R5: Musca             & $*59\pm*8$ & $26?*\pm*3?*$ & $*6.9\pm1.0$\cr
**?*Chamaeleon        & $*74\pm*8$ & $22?*\pm*2?*$ & $11?*\pm1.1$\cr
R6: Orion             & $*47\pm*5$ & $20?*\pm*2?*$ & $*4.7\pm0.6$\cr
R7: $\lambda$ Orionis & $104\pm11$ & $25?*\pm*3?*$ & $15?*\pm1.8$\cr
LMC                   & $*56\pm*6$ & $*8.5\pm*1.0$ & $*7.9\pm0.9$\cr
SMC                   & $*30\pm*9$ & $*4.7\pm*1.5$ & $*3.3\pm1.0$\cr
Entire sky            & $*65\pm*7$ & $22?*\pm*2?*$ & $*8.3\pm0.8$\cr
$|b|>10$\deg          & $*70\pm*7$ & $21?*\pm*2?*$ & $*9.7\pm1.0$\cr
\noalign{\vskip 4pt\hrule\vskip 4pt}
XV: Perseus           & \dots      & $24?*\pm*4?*$ & \dots\cr
XV: $\rho$ Oph        & \dots      & $*8.3\pm*1.1$ & \dots\cr
XV: Mean              & \dots      & $32?*\pm*4?*$ & \dots\cr
D06: Kp2 mask         & \dots      & $21.8\pm*1.0$ & \dots\cr
D06: Region mean      & \dots      & $25.7\pm*1.3$ & \dots\cr
\noalign{\vskip 5pt\hrule\vskip 3pt}
}}
\endPlancktable
\endgroup
\end{table}

\subsubsection{Diffuse AME regions}
We now move on to consider individual diffuse regions of AME. We have labelled seven regions in the all-sky AME map of Fig.~\ref{fig:ame_fullsky} that demonstrate diffuse AME away from the Galactic plane and are in areas with relatively high S/N ratios and clean component separation outputs. These include the well-known regions of Perseus and $\rho$ Ophiuchus, as well as five new regions. In Fig.~\ref{fig:amemaps} we show the \commander\ solutions for synchrotron, free-free, AME and thermal dust, along with the CMB-subtracted 28.4\ghz\ \planck\ data and \ha\ data, for the new regions. In each of these maps, an area has been defined to analyse the properties of the region using \ttp\ plots, and sources brighter than 5\,Jy in the PCCS2 catalogue have been masked out to a radius of 60\arcm, with some exceptions noted below. We show \ttp\ plots between AME and thermal dust for three regions in Fig.~\ref{fig:ame_ttplots2}. We use the best-fitting slope from these \ttp\ plots to determine the emissivity of the AME component compared to the other components. Results for all regions are given in Table \ref{tab:emissivity}.

\paragraph{1. Perseus.}
This molecular cloud is a well-known source of AME \mbox{\citep{Watson2005,planck2011-7.2}}. For the correlation analysis, we focus on an $8\deg\times8$\deg\ patch centred on $(l,b)=(160\pdeg26,-18\pdeg62)$. Perseus is included in the PCCS2 catalogue, so we unmask the point sources when calculating the emissivities in this region. We find a lower emissivity for Perseus in the \commander\ products than in \mbox{\citet{planck2013-XV}}; this is due to leakage of around 40\,\% of the AME emission to the free-free component, which also causes a decrement in the synchrotron map (see Fig.~\ref{fig:amepol} below). 

\paragraph{2. Pegasus plume.}
An example where the AME amplitude traces a filament-like structure at the edge of the Pegasus constellation at $(l,b)=(92\deg, -37\deg)$ and approximately 10\deg\ in length is shown in the first column of Fig.~\ref{fig:amemaps} (Region 2 in Fig.~\ref{fig:ame_fullsky}). We look at a $15\deg\times15\deg$ patch centred on this position. The plume was included in the CO catalogue of \citet{Magnani1985} as MBM\,53--55 (G92\pdeg97$-$32\pdeg15, G92\pdeg97$-$37\pdeg54, and G89\pdeg19$-$40\pdeg94, respectively), and MBM\,55 is also coincident with the \hii\ region S\,122 (G89\pdeg18$-$41\pdeg13) in \citet{Sharpless1959}. \citet{Fukui2014} have compared the CO emission to thermal dust emission as seen by \Planck\ at 353--857\GHz. The structure can be seen clearly in the CMB-subtracted \planck\ 28.4\GHz\ map, which is very closely correlated with the thermal dust amplitude at 545\GHz. There is no visible emission in the synchrotron map and thus the signal in the AME map is unlikely to be due to synchrotron emission. The feature does have associated free-free emission, as traced by \ha; however, the brightness at 22.8\GHz\ is about 10 times greater than would be predicted for free-free emission, assuming a typical electron temperature ($T_\mathrm{e} \approx 7000$\,K). The plume structure is not resolved at 1\deg\ resolution, and therefore absorption of \ha\ light by dust should be similar to other high-latitude regions ($|b| > 30\deg$) at 􏰁$\lesssim$\,0.1\,mag, with only a small fraction being absorbed in compact high-density regions. The section of the plume corresponding to S\,122 is definitely free-free emission; however, we conclude that the rest of this feature is predominantly AME, and that its appearance in the \commander\ free-free map is likely to be due to leakage from the AME component. This would be an interesting target for higher-resolution follow-up measurements, particularly in the bright region around MBM\,53.

\paragraph{3. Corona Australis.}
The second column of Fig.~\ref{fig:amemaps} shows Corona Australis below the Galactic centre (Region 3 in Fig.~\ref{fig:ame_fullsky}). The region has been studied by \citet{Harju1993} in CO emission and with \Spitzer\ by \citet{Peterson2011}, and it is reviewed in \citet{Neuhauser2008}. The central object is the R Corona Australis dark cloud (R\,CrA), which has a tail extending to the bottom-left of the maps consisting of the reflection nebula NGC\,6729. The reflection nebula is illuminated by the early-type star TY\,CrA, and exhibits an extended 3.3\um\ emission feature that could be to PAH emission \citep{Chen1993}. We focus our analysis on a $3\deg\times3\deg$ area centred on $(l,b)=(1\deg,-22\deg)$ that encompasses the reflection nebula but not the R\,CrA cloud. The R\,CrA region is dominated by free-free emission, and it appears in both the \ha\ and free-free maps. However, the reflection nebula clearly shows AME, with a small amount of free-free emission seen in \ha\ (there is a feature in the tail in the \commander\ free-free map; this does not show up in \ha, however). The structure does not appear in the synchrotron map.

\paragraph{4. $\rho$ Ophiuchus.}
This molecular cloud is another well-known source of AME (\mbox{\citealp{Casassus2008}}; \mbox{\citealp{planck2011-7.2}}). It is included in the PCCS2 catalogue, so we unmask the point sources when considering this region. We look at a $5\deg\times5\deg$ patch centred on $(l,b)=(353\pdeg05,16\pdeg9)$. We find a low AME-100\um\ emissivity of ($4.6\pm0.9$)\ukmjy, compared to ($8.3\pm1.1$)\ukmjy\ from \mbox{\citet{planck2013-XV}}. This is due to significant leakage between the AME, free-free, and synchrotron components in the \commander\ solution in this region: half of the emission that has been attributed to AME in previous analyses \citep[e.g.,][]{planck2011-7.2} is instead attributed to free-free, causing a decrement in the synchrotron map, similar to the Perseus region (see Fig.~\ref{fig:amepol} below). This issue is due to the complexity of this region, which lies within the Gould belt, where there is bright surrounding free-free and synchrotron emission.

\paragraph{5. Musca/Chamaeleon region.}
The third column of Fig.~\ref{fig:amemaps} shows a region of diffuse AME below the Galactic plane, in the Musca region close to the Magellanic clouds (Region 5 in Fig.~\ref{fig:ame_fullsky}). This region was previously studied in dust polarization by \mbox{\citet{planck2014-XX}}. Its distance is around 160--180\,pc, and it contains three dark clouds \citep{Whittet1997}. We focus on a $15\deg\times12$\deg\ region centred on $(l,b)=(299\deg,-16\pdeg5)$. The AME is clearly correlated with the dust emission, and there is a notable absence of emission in the synchrotron and \ha\ maps in this region. There is structure evident in the \commander\ free-free map; however, this correlates with some parts of the dust emission and not \ha, so it is likely to be AME rather than free-free. We mask two bright thermal dust sources in the Musca region (Cha\,I and Cha\,II) out to a radius of 80\arcm\ when calculating the emissivities.

There is also a much larger AME half-ring that extends south of the Musca feature, into the Chamaeleon constellation. We include emissivities for the Chamaeleon region in a $35\deg\times30\deg$ region centred on $(l,b)=(305\deg, -25\deg)$. We also use this region to look at the correlations between AME and \iras\ 12 and 25\um, since this part of the sky is at both high Galactic and ecliptic latitudes, and as such it is less contaminated by zodiacal or stellar light than the rest of the sky. We find an emissivity against 12\um\ of ($550\pm60$)\ukmjy, and against 25\um\ of ($570\pm70$)\ukmjy, both of which are higher than the all-sky values given earlier in this section. We also divide the 12 and 25\um\ data by $G_0$ (as described above), after which we find an emissivity against 12\um\ of ($250\pm25$)\ukmjy, and against 25\um\ of ($159\pm16$)\ukmjy. That these numbers are lower than for the whole sky is likely to be an artefact in the $G_0$ map, which has a very small value due to the low \commander\ dust temperature in this region of around 17\,K, compared to the mean temperature of 20.9\,K. We also correlate the variations in AME emissivities with the fraction of PAH emission at 12\um\ (as described above for the whole sky, following \citealp{Hensley2015}) in this region. We find similar results to above, with a slope of $3.4\pm0.4$ when using the \commander\ radiance map, but poor correlations when using the radiance maps based on \citet{planck2013-p06b}. The correlations presented here are sensitive on how the comparison maps have been made, both in terms of inputs to the radiance map and contaminating emission in the \IRAS\ maps.

\paragraph{6. Orion region.}
The fourth column of Fig.~\ref{fig:amemaps} shows the Orion region (Region 6 in Fig.~\ref{fig:ame_fullsky}). The strong free-free emission from M\,42 and Barnard's Loop (Sh\,2-276) is clearly visible. However, there is also an arc of AME emission that correlates with the high-frequency dust emission. This arc extends to lower longitude from M\,42 along the integral-shaped filament to the dark cloud L\,1641 and perpendicularly crosses over Barnard's loop (Orion Molecular Cloud A); in the other direction (Orion Molecular Cloud B) it extends up to M\,78 to the top-right of the figure. The arc also appears faintly in free-free emission and in \ha. This dusty feature was recently mapped in 3D by \citet{Schlafly2014}. We look at a $15\deg\times15\deg$ region centred on $(l,b)=(209\pdeg01,-19\pdeg38)$, and we additionally mask M\,42 to a radius of 100\arcm, and the point where the dust crosses the top part of Barnard's loop (where AME has likely leaked into the free-free map), when calculating the emissivities.

\paragraph{7. \lorionis.}
The \lorionis\ region (Region 7 in Fig.~\ref{fig:ame_fullsky}) can also be seen to the right of Barnard's Loop, and is shown in the fifth column of Fig.~\ref{fig:amemaps}. We focus on a $13\deg\times13\deg$ region centred on $(l,b)=(195\pdeg2,-12\pdeg2)$. This \hii\ region (also known as S\,264), which is illuminated by the O8 star \lorionis, exhibits a shell of AME around the outside of the free-free region coincidental with the thermal dust ring. The dust ring has been previously noted in \citet{Maddalena1987} and \citet{Zhang1989}, and it has been observed in \hi\ by \cite{Wade1957} and CO by \cite{Lang1998} and \citet{Lang2000}. The brightest AME regions in the ring are mostly located adjacent to bright thermal dust features, although the brightest AME region does not have a bright thermal dust counterpart. AME has been seen in PDRs around other \hii\ regions at higher resolution, e.g., in $\rho$ Ophiuchus \citep{Casassus2008} and Perseus \citep{Tibbs2010}. \lorionis\ has a particularly high emissivity against 545\ghz\ and $\tau_\mathrm{353}$, but the emissivity against 100\um\ is comparable to the average; this indicates that the AME is connected to the colder dust in this region. It is likely that this region has not been identified before due to the large free-free emission feature that the ring of AME surrounds.

\paragraph{}
There are significant differences between the emissivities for these seven regions. Some emissivities (particularly for Perseus and $\rho$ Ophiuchus) are clearly biased low due to component-separation issues. Four are relatively consistent with values of AME/545\GHz\ typically in the range 40--50\ukmjy. The notable exceptions are the emissivity values for \lorionis\ and the Chamaeleon region, which are higher than the other regions by a factor of 2. However, except for \lorionis\ and Chamaeleon, the emissivities are significantly lower than the average emissivity of ($65\pm7$)\ukmjy\ across the whole high-latitude sky. The differences in emissivities could be a component-separation artefact, or they could be an indicator of the dependence of environmental conditions for the AME, or dust grain size distribution. For example, in nearby molecular clouds the AME carrier grains could have attached themselves to larger dust grains more than in the diffuse medium \citep[e.g.,][]{Kim1994}. We find a good correlation between the emissivities at 545\GHz\ and the optical depth.

In conclusion, we find that the \commander\ AME map provides a good tracer of AME in our Galaxy, however, there are significant degeneracies between the free-free and AME components that present difficulties when using the map to calculate emissivities. Additional all-sky data at frequencies of 5--20\GHz\ are needed to improve the free-free tracer and so enable a cleaner separation of AME from the other low-frequency components, in order to determine accurate emissivities and comparison with other data sets. In particular, \lorionis\ is a distinctive region that warrants further study.

\subsection{Synchrotron}
\label{sec:synchrotron}
In this section we discuss constraints on synchrotron emission derived from total intensity. We discuss options for modelling the form of the synchrotron spectrum, which is not a simple power law, and specify the model used in our baseline \Commander\ analysis. The results are strongly limited by uncertainties in component separation; a much more reliable picture of the structure of the high-frequency synchrotron emission emerges from polarization data, discussed in Sect.~\ref{sec:polarization}. However, one advantage of total intensity is that we have a high S/N ratio data set in the 408\MHz\ Haslam map, whereas in polarization all available ground-based sky maps are at low frequency and hence too strongly affected by Faraday rotation to be useful for spectral analysis. Because of residual differences between \Planck\ and \WMAP\ polarization maps (see Sect.~\ref{sec:combo} and \citealp{planck2014-a12}), we do not attempt an independent fit of the spectrum of the polarized synchrotron emission.

The Galactic synchrotron spectrum curves significantly below a few GHz \citep[e.g.,][]{deOliveiraCosta2008,Strong2011}.  To generate a useful spectral constraint from just one low-frequency map, we cannot afford to fit this curvature independently at each pixel, as otherwise the low-frequency point would always fit perfectly and give no constraint on the high-frequency spectrum; instead we force the curvature to be constant across the sky, and we have tried several approaches to regularize the fit. As expected, simple power-law fits are inconsistent with our data, generating large and spurious ``gain corrections'' at 408\MHz.

\begin{figure}
\setlength{\unitlength}{1cm}
\begin{center}
\begin{picture}(9,6.5)
\put(-1.2,-6.7){\includegraphics[width=11cm,angle=0]{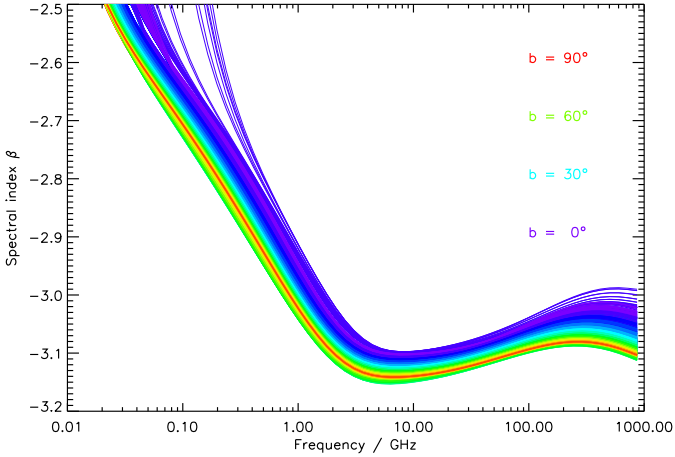}}
\end{picture}
\end{center}
\caption{Local spectral index of the synchrotron emission $\beta(\nu) = d \ln T/ d \ln \nu$ vs. frequency for a sample of pixels (one per $N_{\rm side} = 8$ super-pixel), in the \galprop\ {\tt z10LMPD\_SUNfE} model from \citet{Orlando2013}. The spectra are colour-coded by Galactic latitude: spectra at low latitudes show strong low-frequency curvature due to free-free absorption.}
\label{fig:galprop_alpha}
\end{figure}

\citet{Strong2011} and \citet{Orlando2013} model the observed synchrotron emission for given Galactic magnetic fields and cosmic-ray (CR) propagation scenarios with the \galprop\ code \citep{Strong2007}. This code solves the CR transport equation for any CR species, accounting for diffusion, reaccelerating processes, and energy losses in the interstellar medium. The CR transport properties are constrained by the local CR measurements and the observed ratio of secondary to primary nuclei, while energy losses are calculated for a given Galactic magnetic field. As implemented in the papers above, synchrotron \galprop\ models aim to reproduce the large-scale emission, but not sub-kpc scale features such as the synchrotron loops and individual supernova remnants. The modelled local interstellar energy spectrum of leptons (after propagation effects) is adjusted to reproduce the direct measurements by Fermi-LAT above 7\,GeV \citep{Abdo2009b,Ackermann2010}. At lower energies, where solar modulation makes the local spectrum hard to determine, the lepton spectrum is based on synchrotron observations. In order to reproduce the curvature of the spectrum below a few GHz, \cite{Strong2011} found that the injected electron spectrum (before propagation effects) should have a break at approximately 4\,GeV and should be harder than $p = 1.6$ below 4\,GeV. Accounting for all these constraints, the resulting injected spectral index used in those works are ($N(E) \propto E^{-p}$) with $p = 1.6/2.5/2.2$, respectively, below/between/above the breaks at 4 and 50\,GeV. Moreover, the cosmic ray lepton (CRL) spectral shape varies with Galactocentric radius due to propagation effects, CR source distribution, and the magnetic field strength. Despite these effects, the analysis in \citet{Strong2011} and \citet{Orlando2013} finds that the spectral index at any given frequency is strikingly uniform across the sky: the spectral index between 408\MHz\ and 22\GHz\ has an rms dispersion of 0.02; between 22 and 70\GHz\ the rms is 0.01 (Fig.\,\ref{fig:galprop_alpha}).

The uniformity of these model spectra over the sky contrasts with the much larger variation in $\beta(\nu)$ with frequency in the same models. For a given CRL spectrum, the frequency of synchrotron radiation scales directly with $B_\perp$, the component of the magnetic field perpendicular to the line-of-sight, and therefore the spectral index profiles shown in Fig.\,\ref{fig:galprop_alpha} will be shifted in log frequency by the change in $\ln B_\perp$. Given that $d\beta/d\ln\nu \approx -0.13$ at around 1\GHz, the spatial uniformity of the model spectral index implies that the average value of $B_\perp$ along each line-of-sight varies by at most 15\,\% (rms) from one line-of-sight to another. Of course, each model spectrum is an integral of the emission along the line-of-sight, along which $B_\perp$ varies considerably; however, the main variation of $B$ is with Galactic scale height $z$, and hence all lines of sight, except those at very low latitude, sample nearly the same range of field strengths with nearly the same relative weighting. One might expect geometric factors alone to introduce significant spectral variations, but in these models the random component of the field dominates, so to a good approximation $B_\perp \approx \sqrt{2/3}\,|B|$.

The synchrotron intensity at fixed frequency depends on $B_\perp$ as $j_\nu \propto B_\perp^{-(\beta+1)}$, which suggests that regions of enhanced field strength will disproportionately dominate the emission. In the \galprop\ models, the CRLs are supposed to diffuse efficiently through the interstellar medium, so that the local CRL density is not correlated in detail with the field strength. In reality, if the field is enhanced due to compression, e.g., by a shock wave, the CRLs will respond adiabatically if the time to diffuse out of the region is much longer than the timescale for compression. Then the particle energies will also be shifted to higher energies, giving a larger shift in the $\beta(\nu)$ profile, roughly $\propto B^2$, and $j_\nu \propto B^{-2\beta-3}$ (the exact index depending on the field geometry, see \citealp{Leahy1991} for details). Obviously this substantially increases the dominance of regions of high $B$ in the overall synchrotron emission.

The spatial uniformity of the modelled synchrotron spectrum is thus likely to be an artefact due to the omission of sub-kpc structures in the models, such as the loops and spurs that dominate the high-latitude synchrotron sky. Indeed, we know that many individual supernova remnants (SNR) have spectra significantly flatter than that of the diffuse Galactic emission, as expected from the steepening caused by energy-dependent diffusion and radiative losses. It may seem redundant to argue on theoretical grounds that the \galprop\ models show too little spectral structure, since much larger variations have been reported based on analysis of existing radio surveys \citep[e.g.,][]{Lawson1987,Reich1988,Dickinson2009}. But if one discounts features associated with free-free absorption at very low frequencies and free-free emission at higher frequencies, the remaining large variations are mostly in the regions of lowest synchrotron intensity, which are susceptible to systematic errors such as sidelobe contamination from the strong Galactic plane emission, ground spillover, zero-level errors, and (at low frequencies) pick-up of sky emission reflected from the ground. \cite{Reich1988} review previous large-area spectral index maps and conclude that all are unreliable, with the exception of the 38:408\MHz\ map of \citet{Lawson1987}, which is strikingly uniform. \citet{Reich1988} present their own map of spectral index between 408 and 1420\MHz\ in the northern hemisphere showing extensive regions at high Galactic latitude with $\beta > -2.4$, which they attribute at least in part to free-free emission uncorrelated with \ha, presumably due to extinction of the latter. They later extended their analysis to the full sky \citep{Reich2004}, finding even more extensive flat-spectrum regions in the southern hemisphere. If this were due to a free-free contribution, or even to an unusually flat synchrotron power law, the foreground temperature in the steradian or so around $(l,b) = (240\degr,-40\degr$) would be several times higher than observed
at 20--30\,GHz (e.g., Fig.~\ref{fig:overview}a). We believe that the culprit is likely to be some combination of the systematic errors mentioned above.

These considerations suggest two ways that our parameterized models can allow for larger spectral variations than found by \citet{Orlando2013}. To model emission from regions with $B$ significantly different from the average, we can shift the spectrum in $\ln\nu$, keeping the overall shape fixed. However, there is certainly more spectral freedom allowed than this: from the dispersion in the spectral indices of young SNR, we expect the injected energy spectrum to be spatially and temporally variable; moreover the propagation effects included in the synchrotron models of Fig.\,\ref{fig:galprop_alpha} induce small variations in the high-frequency spectrum, which cannot be explained by variations in $B$. If we denote by $T_{\rm GP}(\nu)$ a fiducial spectrum from that simulation, then we can steepen or flatten it by writing 
\begin{equation}
T_{\rm syn}(\nu) = (\nu/\nu_0)^{\delta\beta} T_{\rm GP}(\nu),
\label{eq:var_beta}
\end{equation}
i.e., a change of local spectral index $\beta(\nu) = d\ln T/d\ln \nu$ by a constant $\delta\beta$. Most of the variation in Fig.\,\ref{fig:galprop_alpha} can be accounted for by such a model, but a larger range of $\delta\beta$ is needed to account for the actual sky. In practice, detailed accounting for the spectral curvature below 10\GHz\ is superfluous given that we only use one observed frequency in that regime. For our baseline \Commander\ analysis our template was the spectrum of a single pixel in the {\tt z10LMPD\_SUNfE} model from \citet{Orlando2013}, chosen to be close to the all-sky median for several spectral parameters (i.e., knee frequencies, and spectral indices at the knees). As described in \cite{planck2014-a12}, this was fitted to the data along with the other foreground components allowing for a global shift in frequency (determined to be a factor of 0.26), and an amplitude that was fitted at each pixel.

The resulting synchrotron amplitude map (shown in \citealp{planck2014-a12}) is essentially determined by the high S/N 408\MHz\ data, and deviates from it only in that residual free-free emission at 408\MHz\ is corrected (or over-corrected, at bright compact \hii\ regions in which free-free absorption is significant at 408\MHz). 
The derived global frequency shift is surprisingly large and has the effect of steepening the average spectrum between 408\MHz\ and the \Planck\ bands, resulting in a relatively low amplitude for synchrotron compared to AME and free-free. This may be because the fit is most strongly constrained by the North Polar spur and the diffuse halo of the inner Galaxy, the two regions that are both strong in synchrotron and relatively free of other components that could absorb errors in the synchrotron fit. Most analyses find flatter spectra in the narrow Galactic plane \citep[e.g.,][]{planck2014-XXIII}.

\subsection{Comparison with \WMAP\ models}

Here, we discuss the main similarities and differences between the low-frequency foreground component maps from our \commander\ analysis of the combined
\Planck\ and \wmap\ datasets \citep{planck2014-a12}, and from the final \wmap\ analysis \citep{Bennett2013}.

\begin{table}[tb]
\begingroup
\newdimen\tblskip \tblskip=5pt
\caption{Key parameters of the \wmap\ MEM and MCMC foreground models (see \citealp{Bennett2013} for details).}
\label{tab:wmap_models}
\nointerlineskip
\vskip -3mm
\footnotesize
\setbox\tablebox=\vbox{
    \newdimen\digitwidth 
    \setbox0=\hbox{\rm 0} 
    \digitwidth=\wd0 
    \catcode`*=\active 
    \def*{\kern\digitwidth}
    \newdimen\signwidth 
    \setbox0=\hbox{+} 
    \signwidth=\wd0 
    \catcode`!=\active 
    \def!{\kern\signwidth}
    \newdimen\pointwidth
    \setbox0=\hbox{{.}}
    \pointwidth=\wd0
    \catcode`?=\active
    \def?{\kern\pointwidth}
    \newdimen\notewidth
    \setbox0=\hbox{$^\mathrm{a}$}
    \notewidth=\wd0
    \catcode`@=\active
    \def@{\kern\notewidth}
    \halign{#\hfil\tabskip 2.0em&
            \hfil#\hfil\tabskip 10pt&
            \hfil#\hfil\tabskip 10pt&
            #\hfil\tabskip 0pt\cr
    \noalign{\doubleline}
 Model & $\beta_{\rm ff}$ & $\beta_{\rm sync}$  & AME \cr
 \noalign{\vskip 3pt\hrule\vskip 5pt}
 MEM          & $-2.15$ & $-3.0$                & CNM, $\nu_{\rm peak }$ varies\cr 
 MCMC-c base  & $-2.16$ & $-3.0$                & \dots\cr
 MCMC-e sdcnm & $-2.16$ & $-3.0$                & CNM, $\nu_{\rm peak }$ fixed\cr
 MCMC-f fs    & $-2.16$ & varies                & CNM, $\nu_{\rm peak }$ fixed\cr
 MCMC-g fss   & $-2.16$ & varies$^\mathrm{a}$   & CNM, $\nu_{\rm peak } = 14.95$\,GHz\cr
    \noalign{\vskip 5pt\hrule\vskip 3pt}
}}
\endPlancktable
\tablenote {{\rm a}}  As described in \citet{Strong2011}.\par
\endgroup
\end{table}

\begin{table}[tb]
\begingroup
\newdimen\tblskip \tblskip=5pt
\caption{Ratios of the \WMAP\ MEM and MCMC maps to the \Commander\ maps for the high latitude sky with \mbox{$|b|>20$\deg}  (top) and for \mbox{$|b|<20$\deg} (bottom). A value $a > 1$ indicates that the \commander\ maps contain more emission than the \wmap\ maps. The Pearson correlation coefficient ($r$) is also given.}
\label{tab:ratios}
\nointerlineskip
\vskip -3mm
\footnotesize
\setbox\tablebox=\vbox{
    \newdimen\digitwidth 
    \setbox0=\hbox{\rm 0} 
    \digitwidth=\wd0 
    \catcode`*=\active 
    \def*{\kern\digitwidth}
    \newdimen\signwidth 
    \setbox0=\hbox{+} 
    \signwidth=\wd0 
    \catcode`!=\active 
    \def!{\kern\signwidth}
    \newdimen\pointwidth
    \setbox0=\hbox{{.}}
    \pointwidth=\wd0
    \catcode`?=\active
    \def?{\kern\pointwidth}
    \newdimen\notewidth
    \setbox0=\hbox{$^\mathrm{a}$}
    \notewidth=\wd0
    \catcode`@=\active
    \def@{\kern\notewidth}
    \halign{#\hfil\tabskip 2.0em&
            \hfil#\hfil\tabskip 3pt&
            \hfil#\hfil\tabskip 10pt&
            \hfil#\hfil\tabskip 3pt&
            \hfil#\hfil\tabskip 10pt&
            \hfil#\hfil\tabskip 3pt&
            \hfil#\hfil\tabskip 0pt\cr
    \noalign{\doubleline}
 Run & \multispan2\hfil Sync \hfil &  \multispan2\hfil Free-free \hfil & \multispan2\hfil AME \hfil \cr
 & $a$ & $r$  & $a$ & $r$ & $a$ & $r$   \cr
 \noalign{\vskip 3pt\hrule\vskip 5pt}
 MCMC-c base& $ 0.50$ & $ 0.62$ & $ 0.68$ & $ 0.77$ &  \dots &  \dots\cr
 MCMC-e sdcnm& $ 0.52$ & $ 0.92$ & $ 0.77$ & $ 0.87$ & $ 4.91$ & $ 0.75$\cr
    MCMC-f fs& $ 0.52$ & $ 0.62$ & $ 0.80$ & $ 0.77$ & $ 3.18$ & $ 0.67$\cr
   MCMC-g fss& $ 0.55$ & $ 0.62$ & $ 0.77$ & $ 0.78$ & $ 3.14$ & $ 0.70$\cr
        MEM&   $ 0.34$ & $ 0.84$ & $ 0.76$ & $ 0.79$ & $ 2.18$ & $ 0.86$\cr
 \noalign{\vskip 4pt\hrule\vskip 4pt}
  MCMC-c base& $ 0.12$ & $ 0.74$ & $ 0.91$ & $ 0.83$ & \dots & \dots\cr
 MCMC-e sdcnm& $ 0.61$ & $ 0.99$ & $ 0.67$ & $ 0.98$ & $ 2.71$ & $ 0.91$\cr
    MCMC-f fs& $ 0.13$ & $ 0.78$ & $ 1.09$ & $ 0.89$ & $ 4.71$ & $ 0.86$\cr
   MCMC-g fss& $ 0.15$ & $ 0.77$ & $ 1.06$ & $ 0.93$ & $ 3.22$ & $ 0.88$\cr
        MEM&   $ 0.19$ & $ 0.90$ & $ 1.01$ & $ 0.97$ & $ 2.36$ & $ 0.93$\cr
\noalign{\vskip 5pt\hrule\vskip 3pt}
}}
\endPlancktable
\endgroup
\end{table}

The \WMAP\ team have published sets of component-separated maps using a Maximum Entropy Method (MEM) and a Markov Chain Monte Carlo (MCMC) technique on a pixel basis \citep{bennett2012}. 
We have compared the synchrotron, free-free and AME maps produced by these methods with the \Commander\ maps through \mbox{\ttp\ plots} with \healpix\ $N_\mathrm{side}=64$. Table \ref{tab:wmap_models} gives the key model parameters of the \wmap\ models that we used. Point sources from the PCCS2 catalogue brighter than 1\,Jy at 28.4\GHz\ have been masked out. The best-fit ratios of the maps, derived from the \mbox{\ttp\ plots}, are given in Table~\ref{tab:ratios} for the high-latitude sky, with \mbox{$|b|>20\deg$} and for \mbox{$|b|<20\deg$}; $a$ is the best-fitting slope. Also listed in Table~\ref{tab:ratios} are the Pearson correlation coefficients ($r$) for all the comparisons. If $r \gtrsim 0.9$, the measured slope, $a$, is reliable, since there is a good linear relationship between the \commander\ and the \wmap\ templates.

A first obvious difference between the \commander\ and \wmap\ MEM/MCMC models is that the AME component is systematically higher (between 2--4 times) in the \commander\ solution, at the expense of the synchrotron emission that is lower in \commander. This is clearer on the region closer to the Galactic plane (bottom half of Table~\ref{tab:ratios}), due to the better correlations measured there.

\begin{figure}
\begin{center}          
  \includegraphics[width=0.49\textwidth,angle=0]{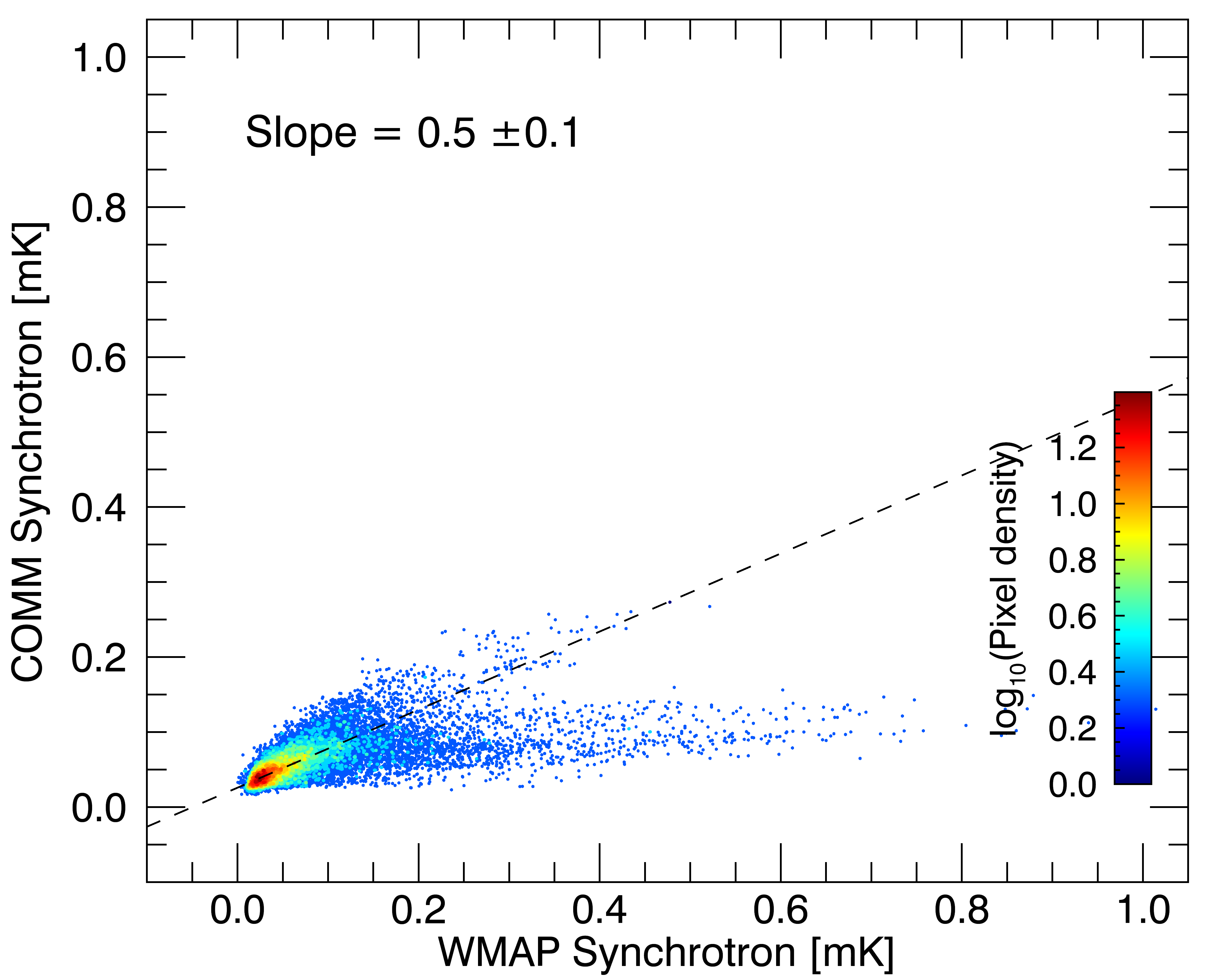}
\end{center}
\caption{Scatter plot between the \commander\ synchrotron solution evaluated at 22.8\GHz\ and the \wmap\ MCMC-f synchrotron model at the same frequency.}
\label{fig:sync_mcmc_f}
\end{figure}

When comparing the \commander\ synchrotron solution with the \wmap\ models that allow the synchrotron spectral index to vary over the sky, the data are not well-fitted with a single slope, since two populations are clearly present. Fig. \ref{fig:sync_mcmc_f} shows the scatter plot between the \commander\ synchrotron solution evaluated at 22.8\GHz\ and the \wmap\ MCMC-f synchrotron model at the same frequency. The two slopes visible in the Figure are the result of a flatter spectrum synchrotron component on the plane \citep[e.g.,][]{Kogut2007}, which is not accounted by the \commander\ synchrotron model, where $\beta_{\rm sync}$ is fixed over the sky. This is not the case for the \wmap\ MCMC-e ``sdcnm'' and MEM solutions, which use a fixed synchrotron spectral index, resulting in a much better correlation between the \commander\ and \wmap\ components (see Pearson correlation coefficients for this component in Table~\ref{tab:ratios}). We also note from Table~\ref{tab:ratios} that the \commander\ synchrotron component is always lower than the \wmap\ synchrotron models, which is due to the larger AME component in the \commander\ model.

The \commander\ free-free component is similar to most of the \WMAP\ models at low Galactic latitudes, where Pearson's $r \gtrsim 0.89$ for all but the MCMC-c model. The correlation coefficients in this case are also close to 1. At high Galactic latitudes (top half of Table~\ref{tab:ratios}), the \commander\ free-free solution is lower than the \WMAP\ models, at the expense of a higher AME contribution in \commander. The wider frequency range of \planck, specifically from 100--217\GHz, should enable a cleaner separation of the free-free component \citep{planck2013-XIV,planck2014-XXIII}.

In addition to the baseline \commander\ model, we have also compared the \WMAP\ models to several earlier \commander\ runs with different input parameters. In particular, we look at a run that used a broken power-law for the synchrotron component along with a single AME component, and another that had straight power-laws for both the synchrotron and free-free components, also with a single AME component. In these models the synchrotron spectral index is fitted locally, while the break, $\Delta\beta$, in the first is fitted globally. This parameterisation is closer to that used by the \WMAP\ team than is our baseline case with two AME
components and a \galprop\ synchrotron spectrum, and the fitted results are
also closer to the \WMAP\ results, with less AME and more synchrotron; however
some differences remain. In both models the free-free amplitude in the \commander\ outputs is still lower than in the \WMAP\ models, and in both \commander\ still finds less synchrotron emission than \WMAP\ at low latitudes 
(except for \WMAP\ ``sdcnm''). However, at high latitudes they 
find {\em more} power in the synchrotron component than does the \WMAP\ analysis. The AME components in these earlier \commander\ models are weaker than in our baseline and in much closer agreement with \WMAP, with ratios around one for both the ``fs'' and ``fss'' models at both low and high latitudes, while the \commander\ solutions find more high-latitude and less low-latitude emission than the ``sdcnm'' model. These runs were not used for the baseline product as they gave worse overall
fits and are less physically motivated than our eventual choice; 
we note them here solely to qualitatively consider the effects that 
different models can have on the component outputs.

The \wmap\ model that is closest to the baseline \commander\ solution is MCMC-e ``sdcnm'', although the AME component is brighter in the \commander\ products.  Also, the \commander\ free-free amplitude is about $30\%$ fainter than the corresponding \WMAP\ fit; this has been noted before from comparison with RRL data by \citet{Alves2010} and \citet{planck2014-XXIII}. The main limitation in obtaining accurate components maps is the precise quantification of the synchrotron component. \citet{Fuskeland2014} and \citet{Vidal2014a} have shown that the polarized spectral index of synchrotron emission varies across the sky and these variations should be taken into account to obtain an accurate separation between the synchrotron and AME components; low frequency (5--20\,GHz) data will be crucial for this. 

\subsection{LMC and SMC} \label{sec:lmc}
\begin{figure*}
\begin{center}
\includegraphics[width=0.24\textwidth]{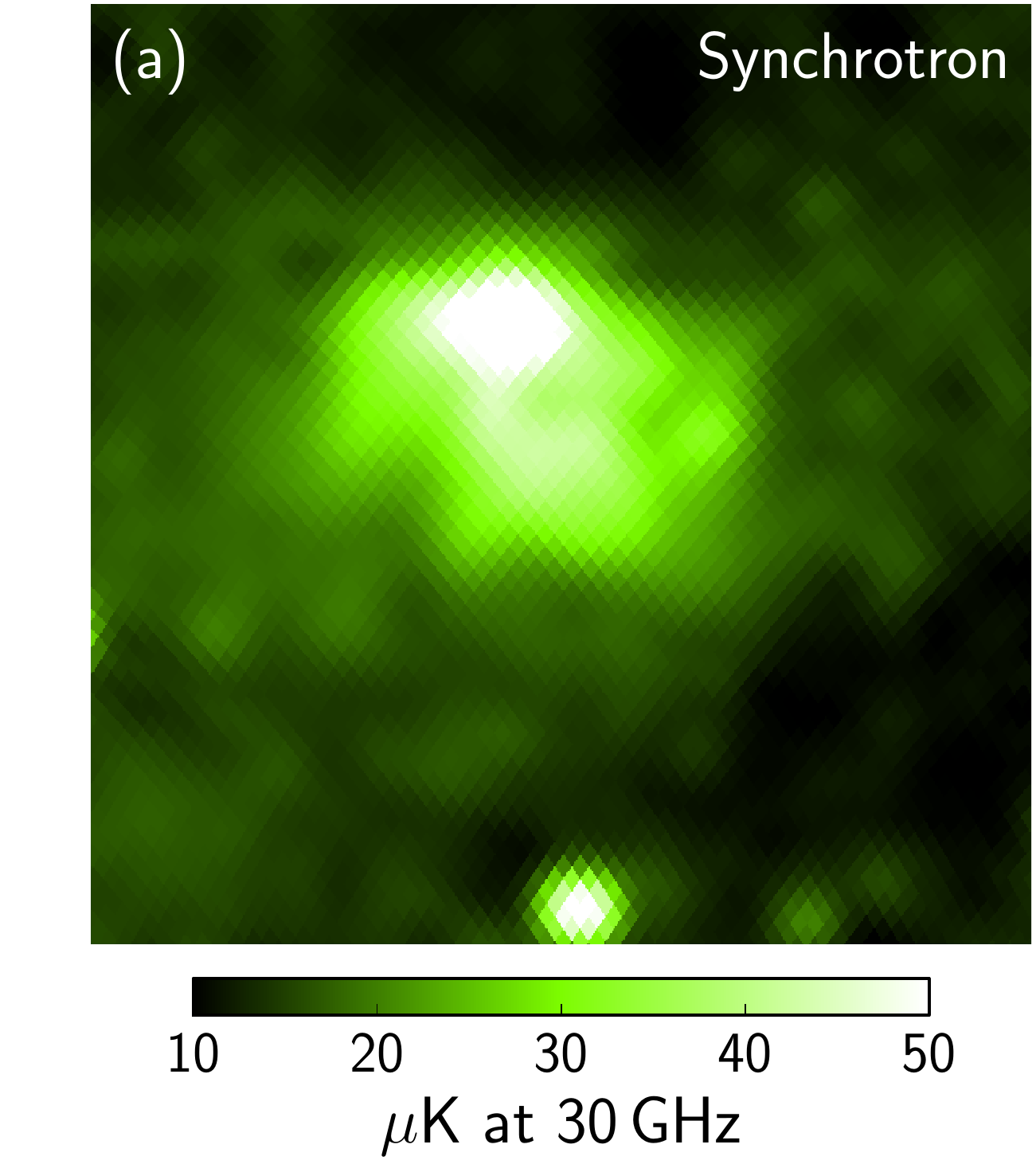}
\includegraphics[width=0.24\textwidth]{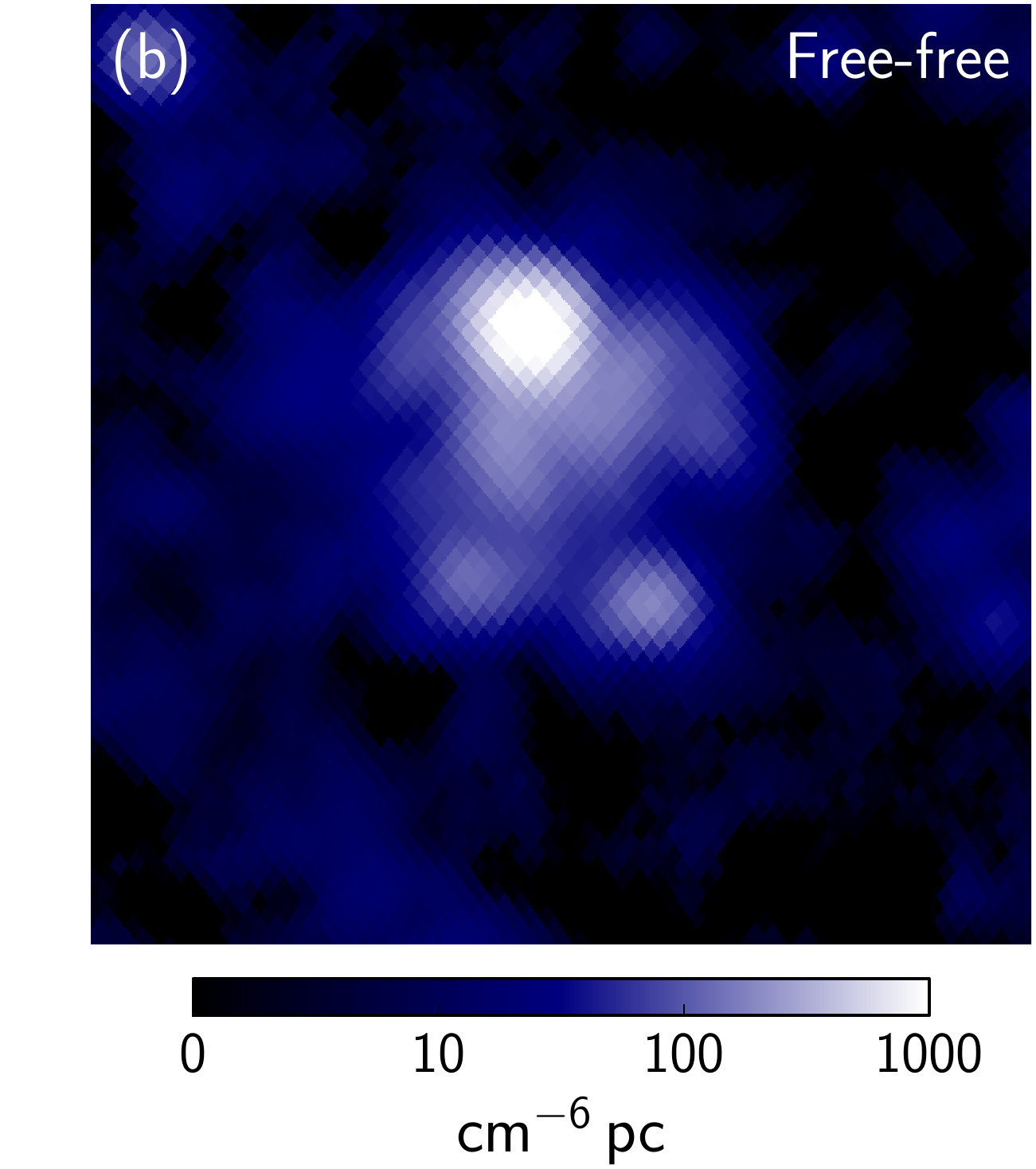}
\includegraphics[width=0.24\textwidth]{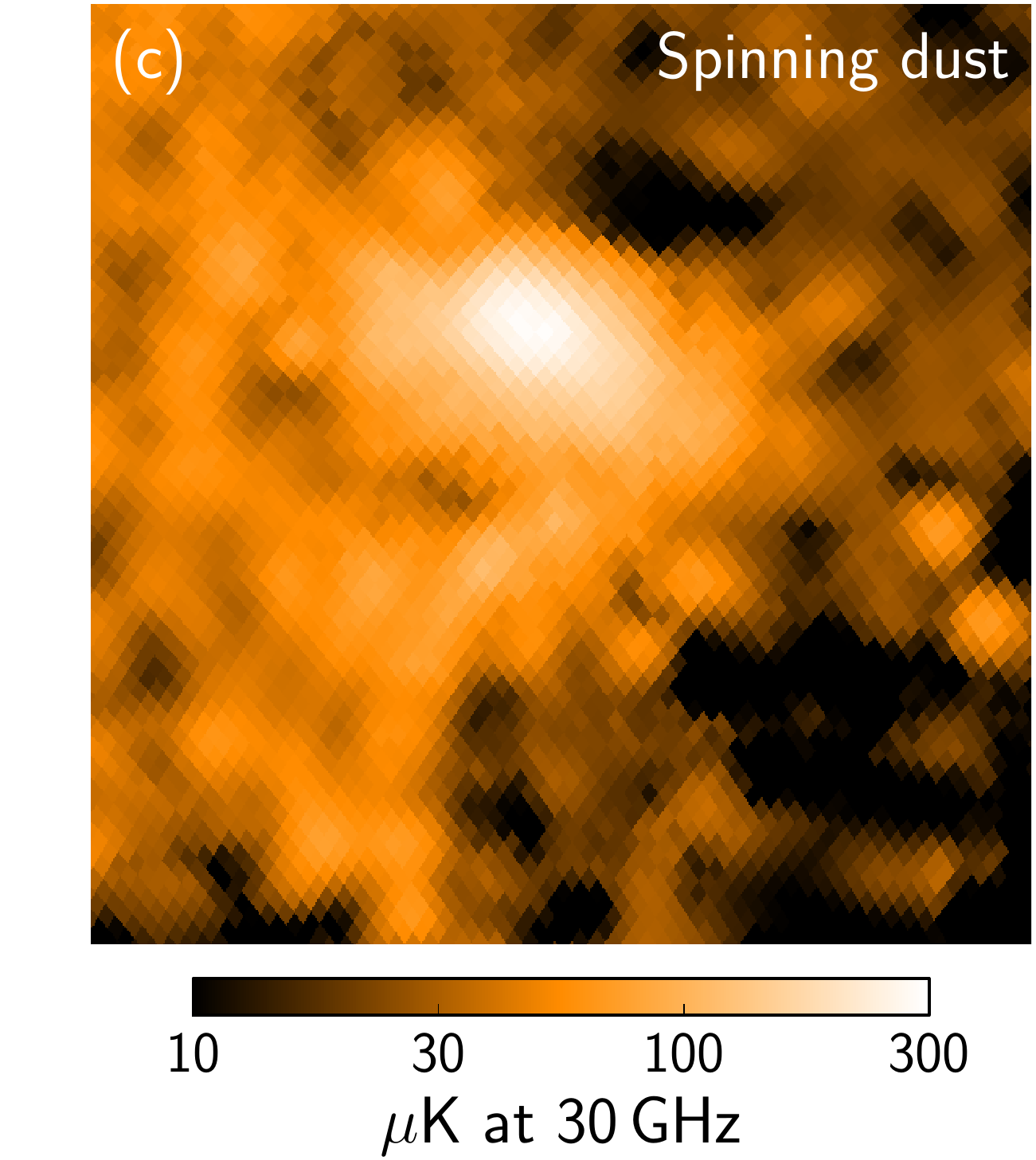}
\includegraphics[width=0.24\textwidth]{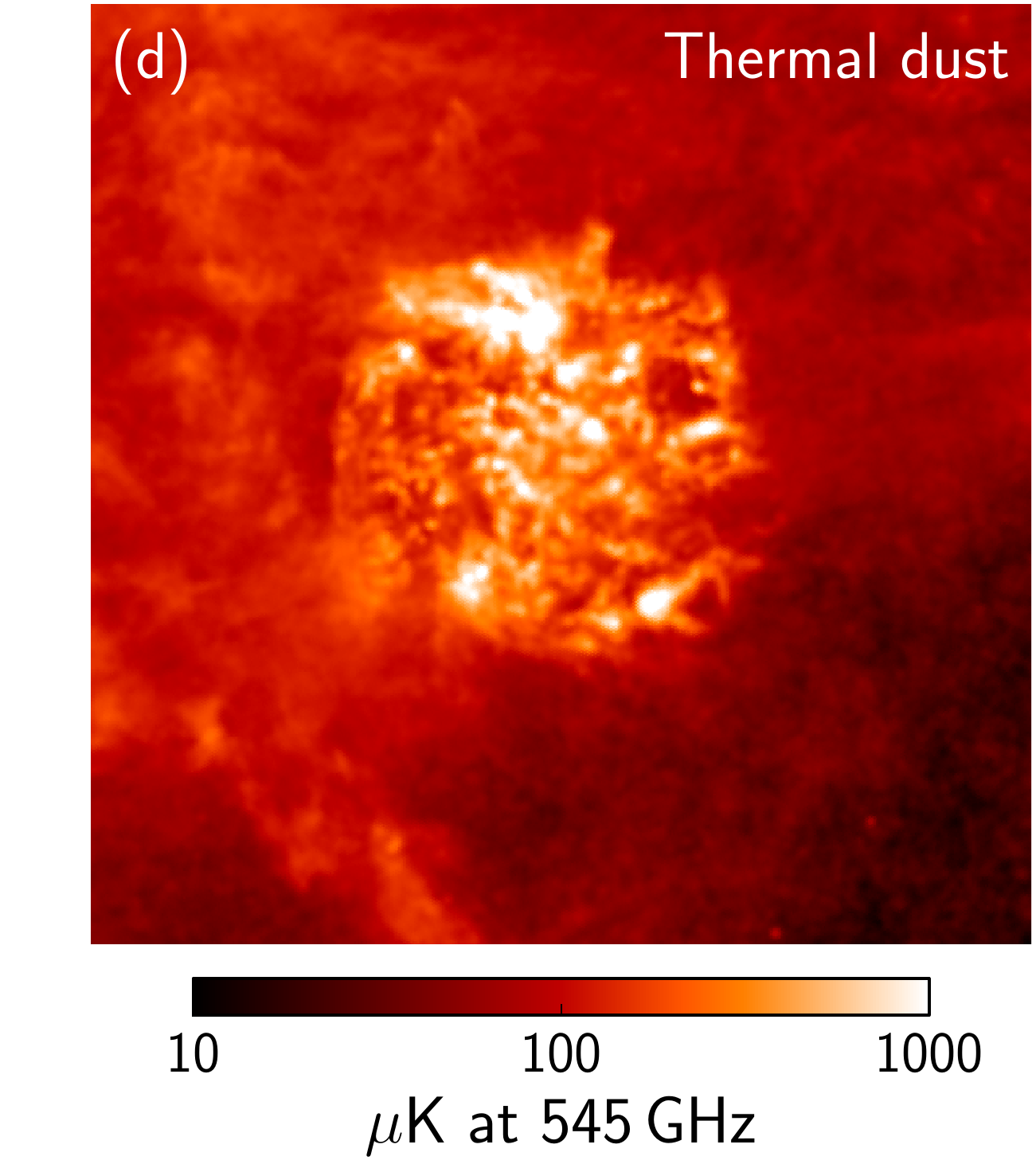}
\includegraphics[width=0.24\textwidth]{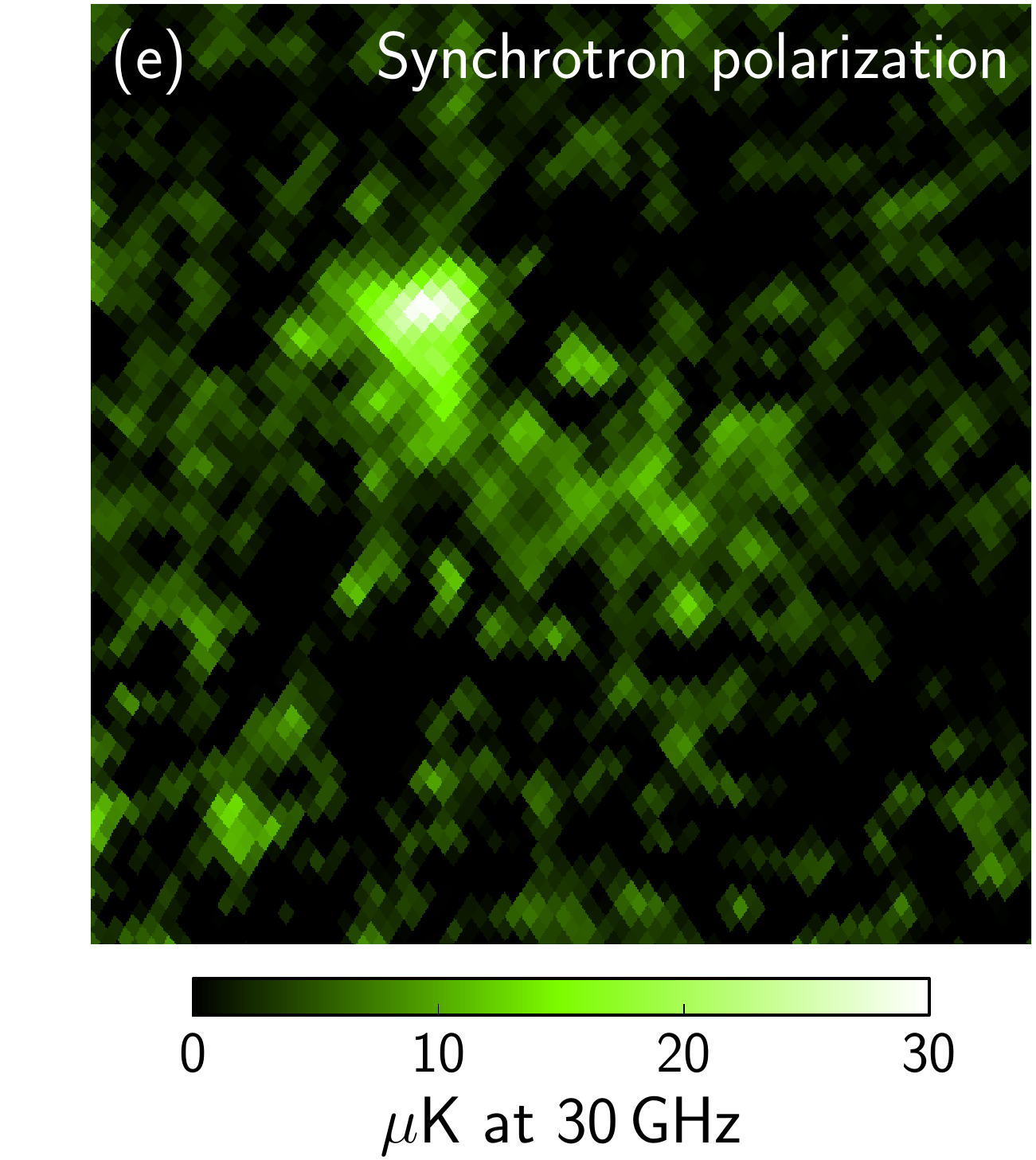}
\includegraphics[width=0.24\textwidth]{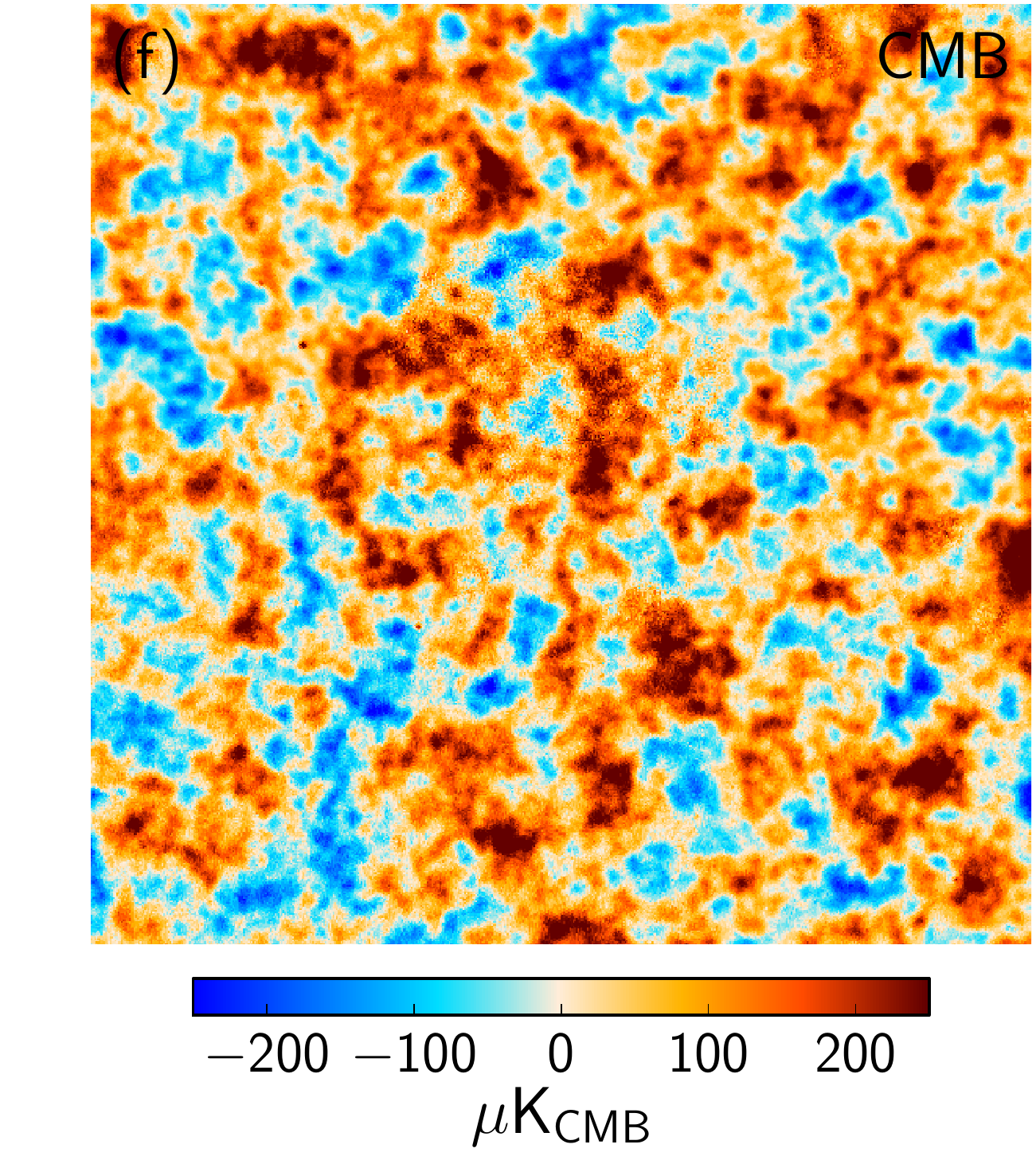}
\includegraphics[width=0.24\textwidth]{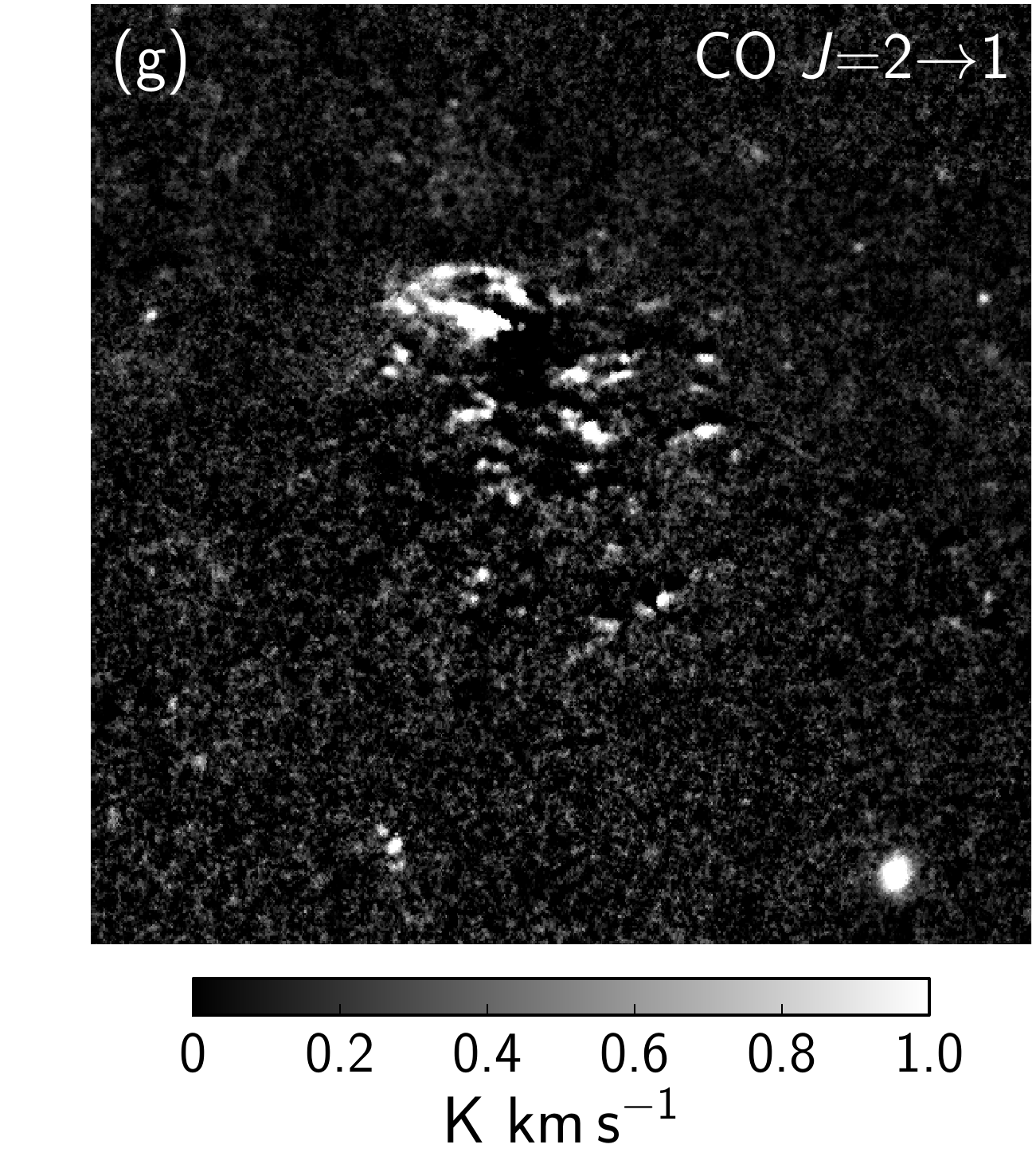}
\includegraphics[width=0.24\textwidth]{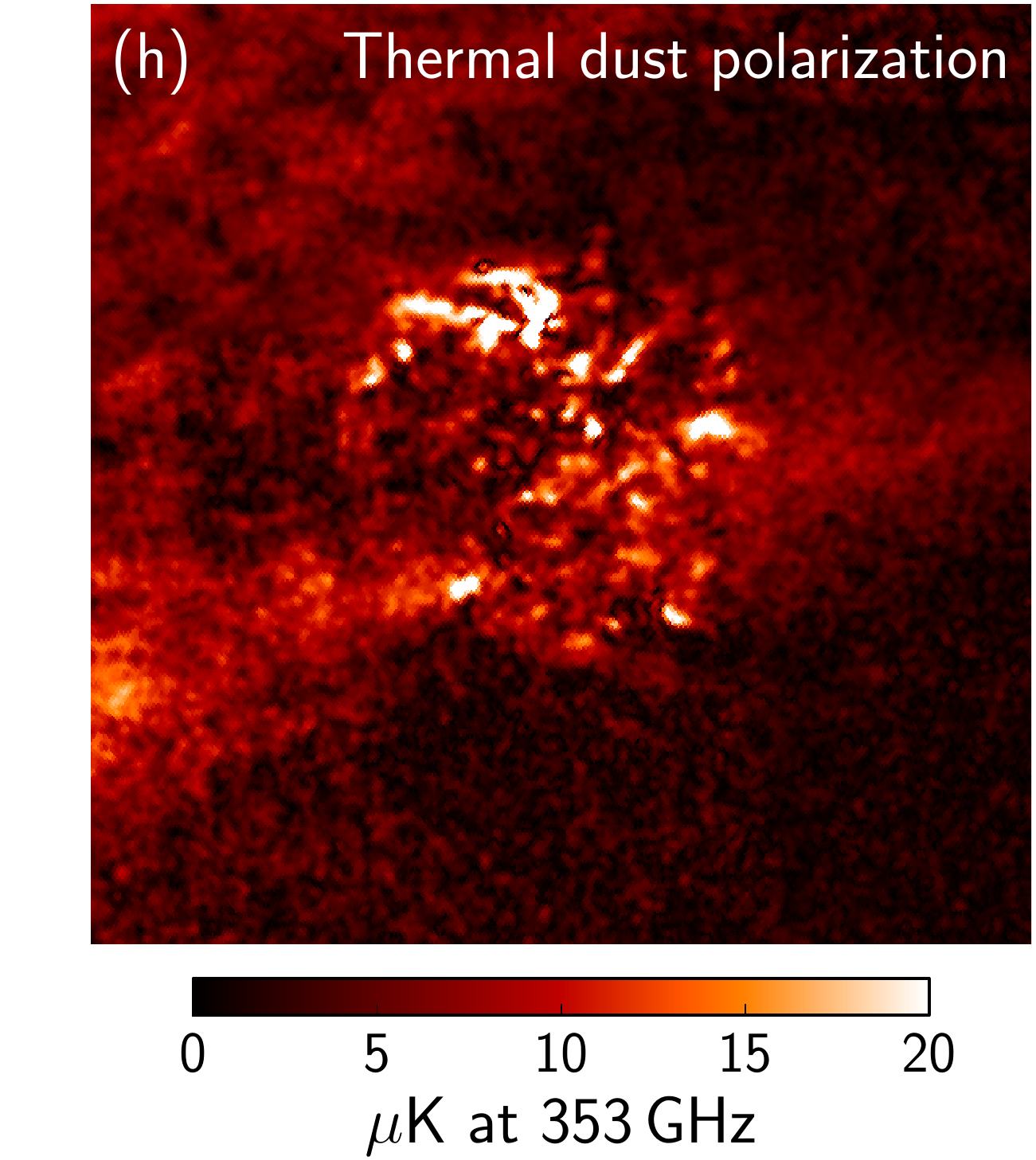}
\end{center}
\caption{\commander\ solution in the Large Magellanic Cloud region, plotted in Galactic coordinates. Panels show from left to right and top to bottom: (a) synchrotron brightness temperature at 30\GHz\ with 1\deg\ resolution (linear colour scale); (b) free-free emission measure\ with 1\deg\ resolution (logarithmic colour scale); (c) spinning dust brightness temperature at 30\GHz\ with 1\deg\ resolution (logarithmic colour scale); (d) thermal dust brightness temperature at 545\GHz\ with 5\arcm\ resolution (logarithmic colour scale); (e) synchrotron polarization amplitude, $P$, at 28.4\GHz\ with 1\deg\ resolution (corrected for polarization leakage, linear colour scale); (f) CMB temperature with 5\arcm\ resolution (linear colour scale); (g) CO $J$=2$\rightarrow$$1$ emission with 5\arcm\ resolution (linear colour scale); and (h) thermal dust polarization amplitude, $P$, at 353\GHz\ with 5\arcm\ resolution (corrected for polarization leakage, linear colour scale). Each map covers $15\deg\times15\deg$, and is centred on Galactic coordinates \mbox{$(l,b)=(279\deg,-34\deg)$}.}
\label{fig:T_highres}
\end{figure*}

The Large and Small Magellanic Clouds (LMC and SMC) are satellite galaxies of the Milky Way, close enough \mbox{($50.0\pm1.1$)\,kpc} for the LMC, \citealp{Pietrzyski2013}, and \mbox{($61\pm3$)\,kpc} for the SMC, \citealp{Hilditch2005}) that they are resolved by \planck\ both at 5\arcm\ in thermal dust and 30\arcm\ at low frequency. They have already been well studied at \planck\ and \wmap\ frequencies (\citealp{Bot2010}; \citealp{planck2011-6.4b}; \mbox{\citealp{Draine2012}}). They thus provide a good test-bed for comparing the \commander\ solution with previous results and expectations, both in the Magellanic clouds and in comparison with our own Galaxy. We show maps of the \commander\ solution in the LMC in Fig.~\ref{fig:T_highres}, which we discuss component by component below. The brightest region in all of the non-CMB component maps is the Tarantula nebula, also known as 30~Doradus, a well-known \hii\ and star-formation region located at $(l,b)=(279\pdeg5,-31\pdeg7)$.

\begin{figure*}[tbh]
\begin{center}
\includegraphics[width=0.325\textwidth]{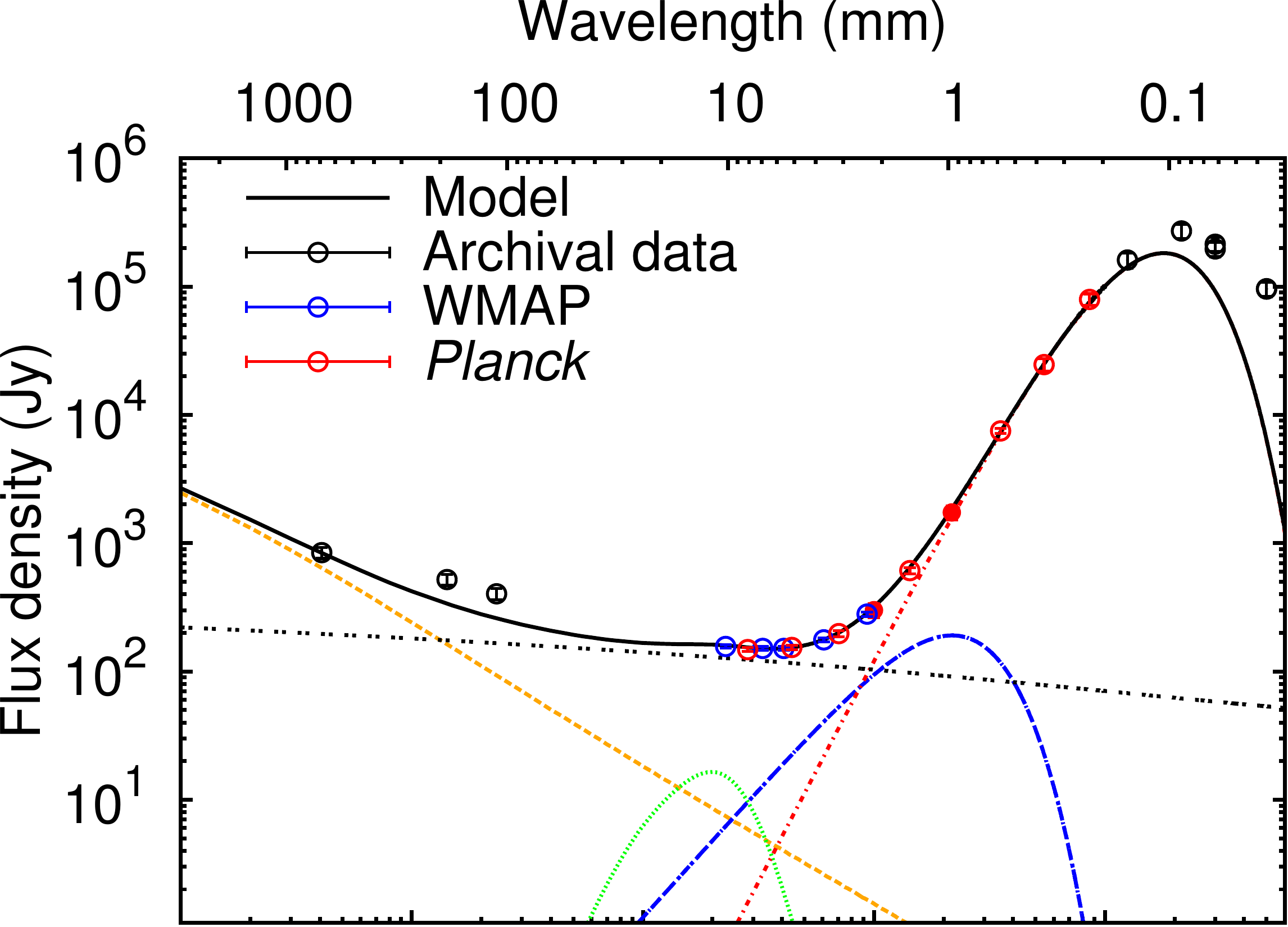}
\hspace{2px}
\includegraphics[width=0.325\textwidth]{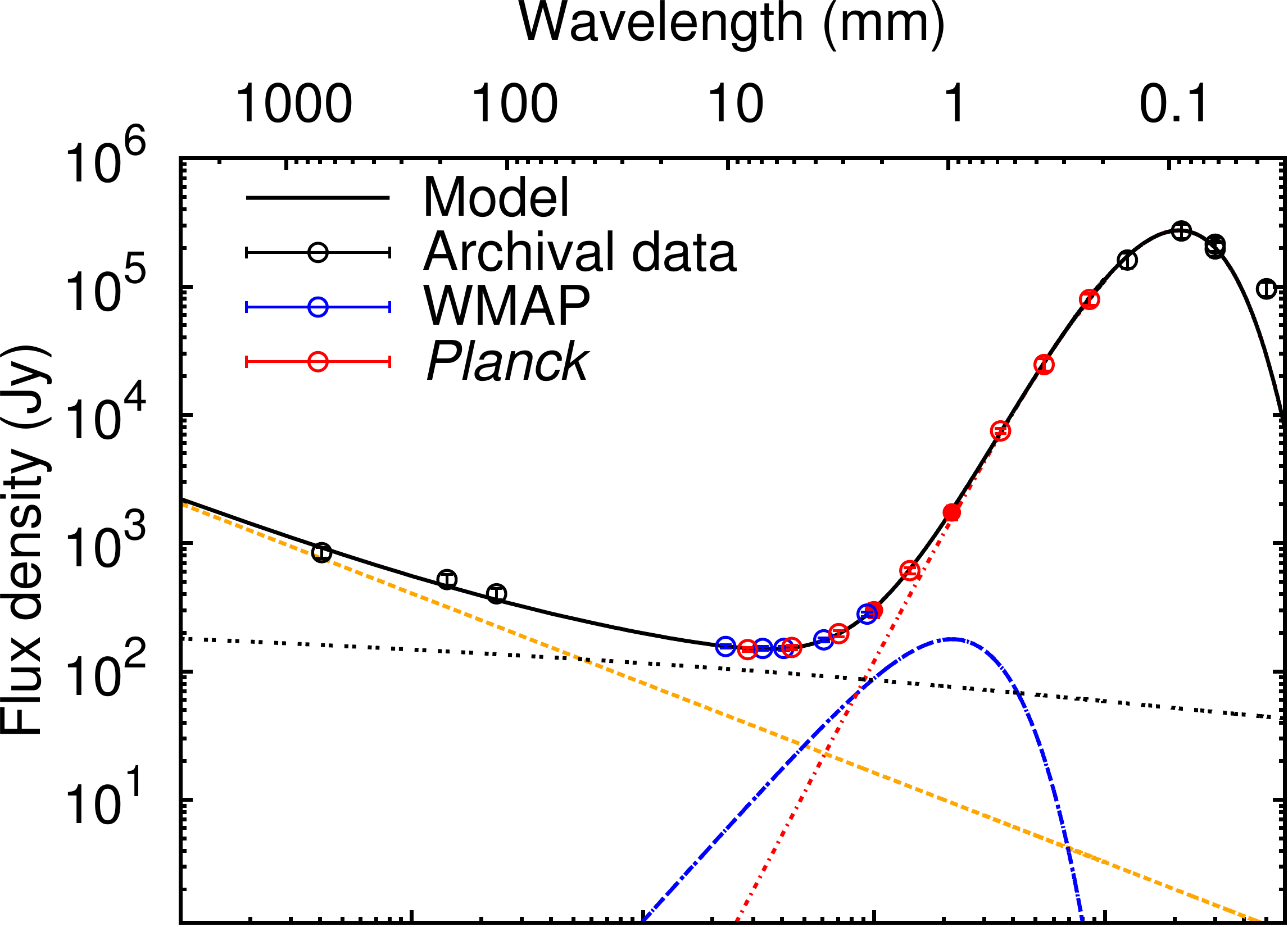}
\hspace{2px}
\includegraphics[width=0.325\textwidth]{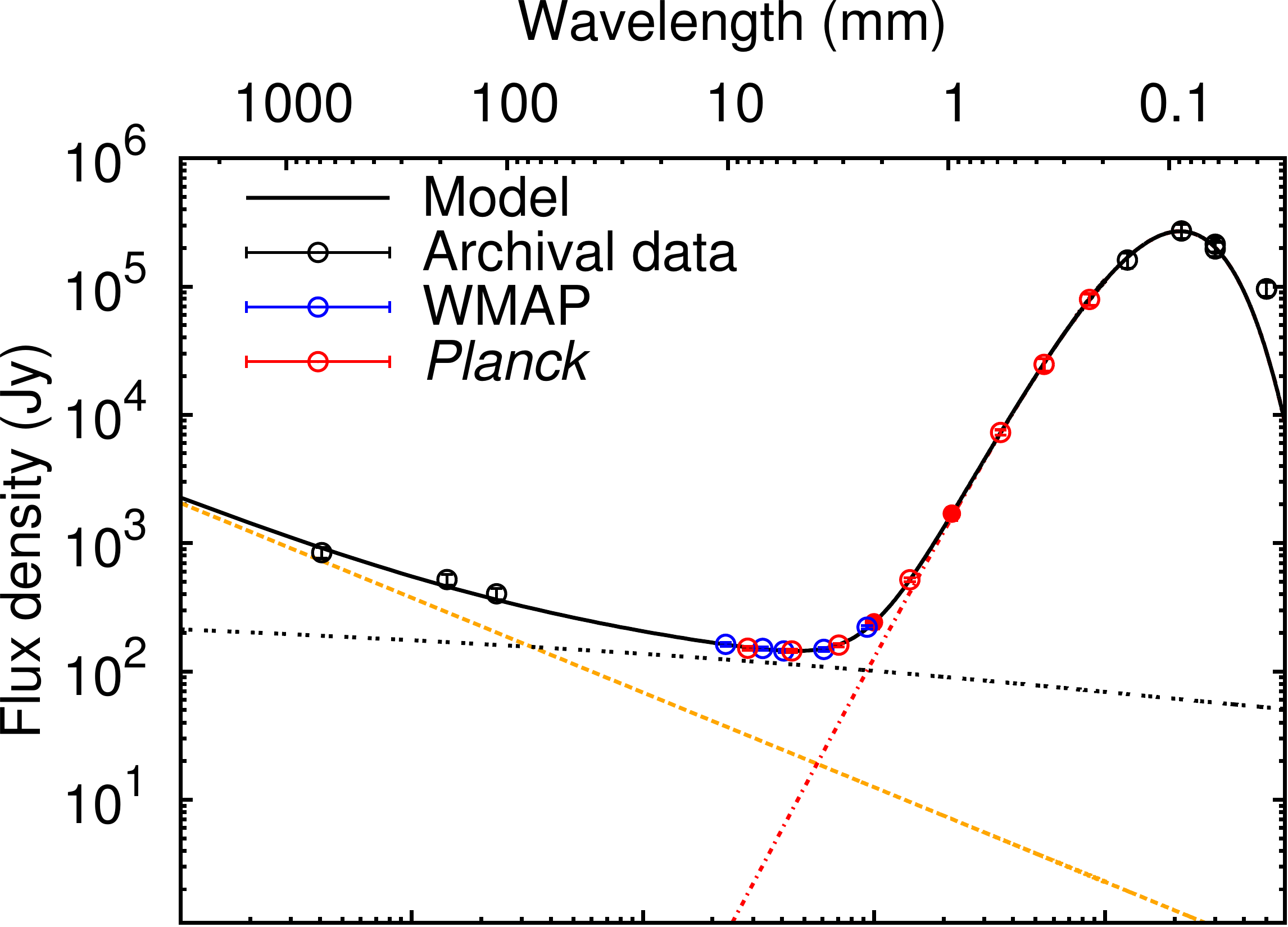}
\includegraphics[width=0.325\textwidth]{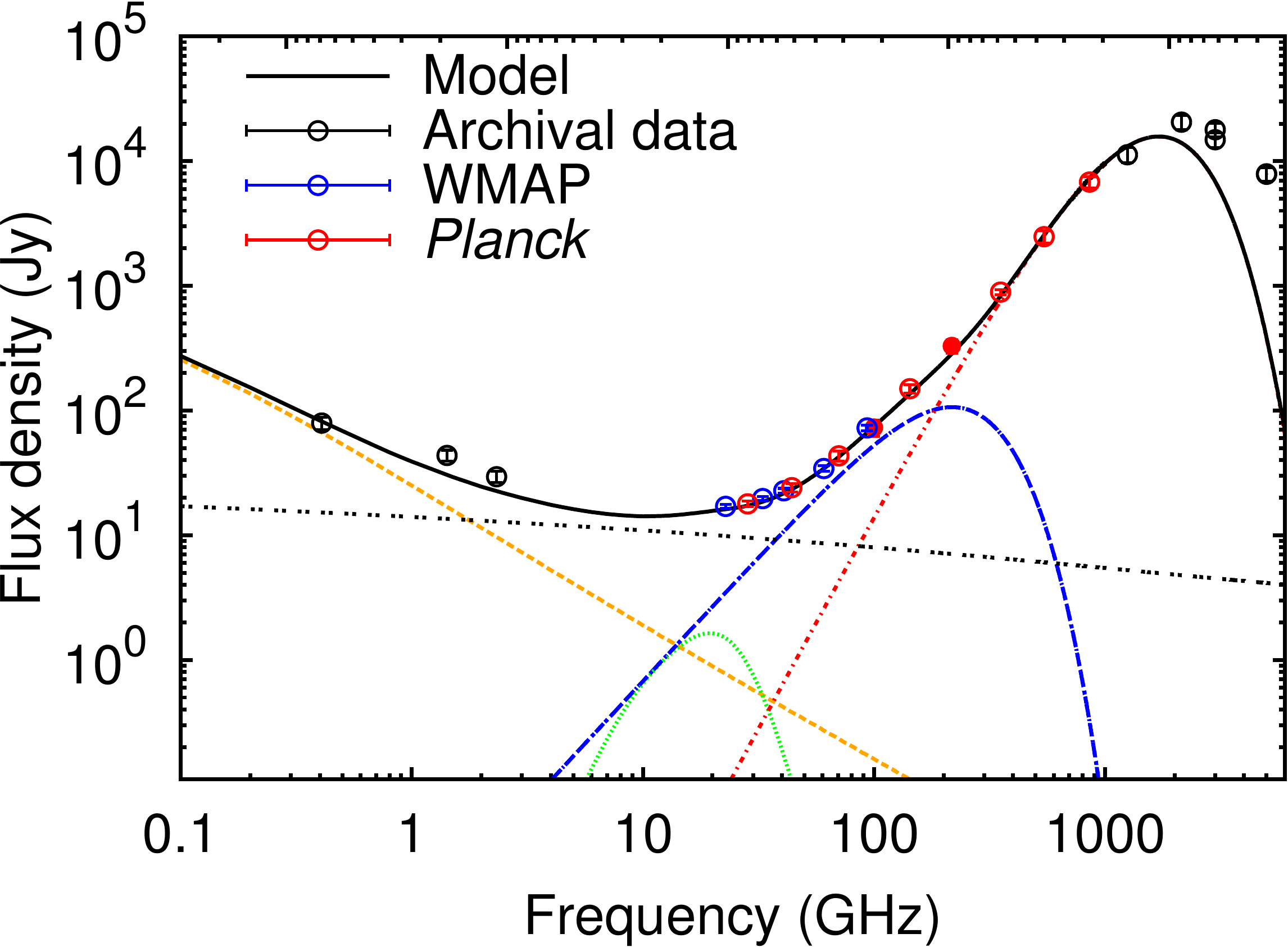}
\hspace{2px}
\includegraphics[width=0.325\textwidth]{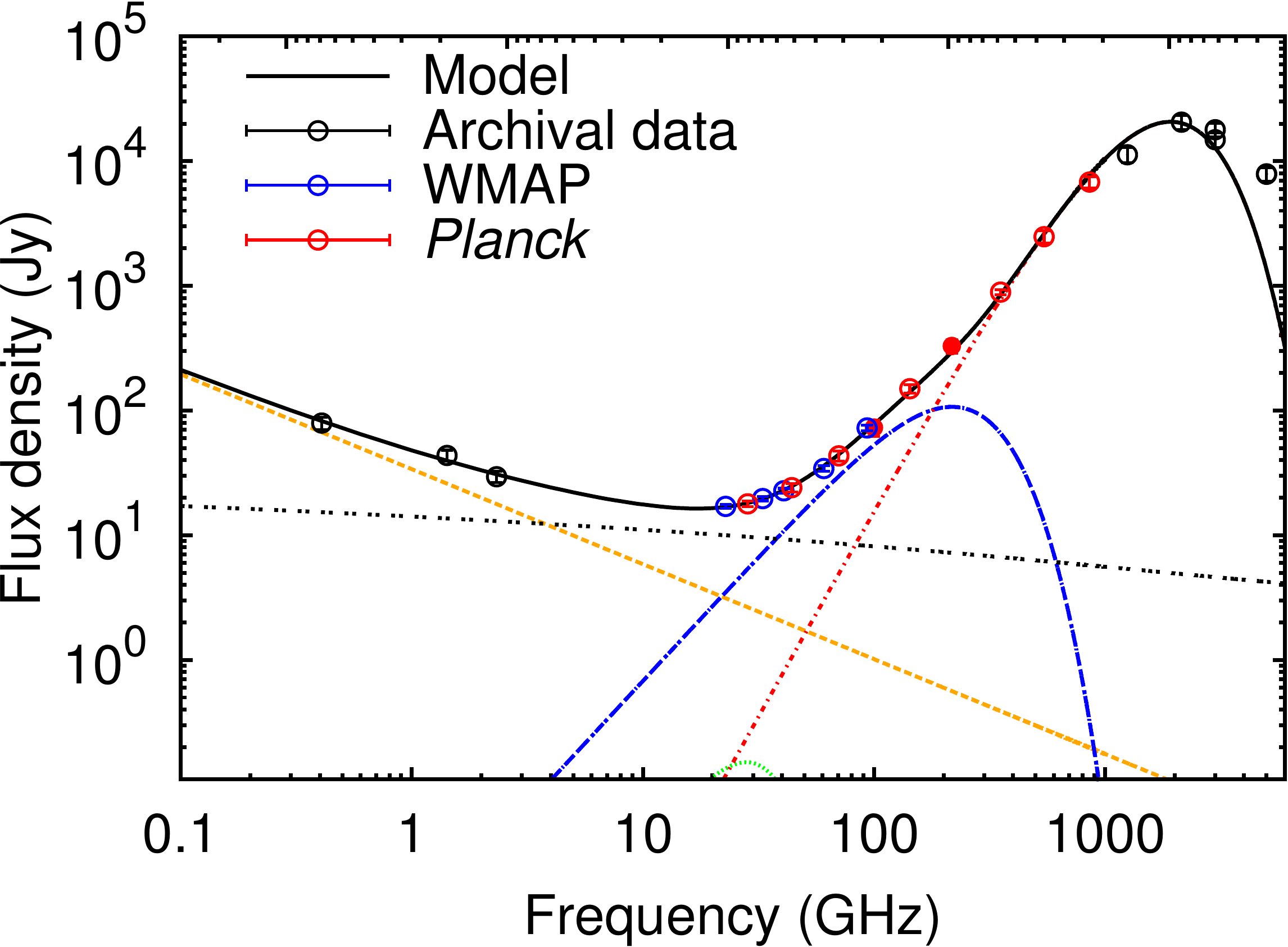}
\hspace{2px}
\includegraphics[width=0.325\textwidth]{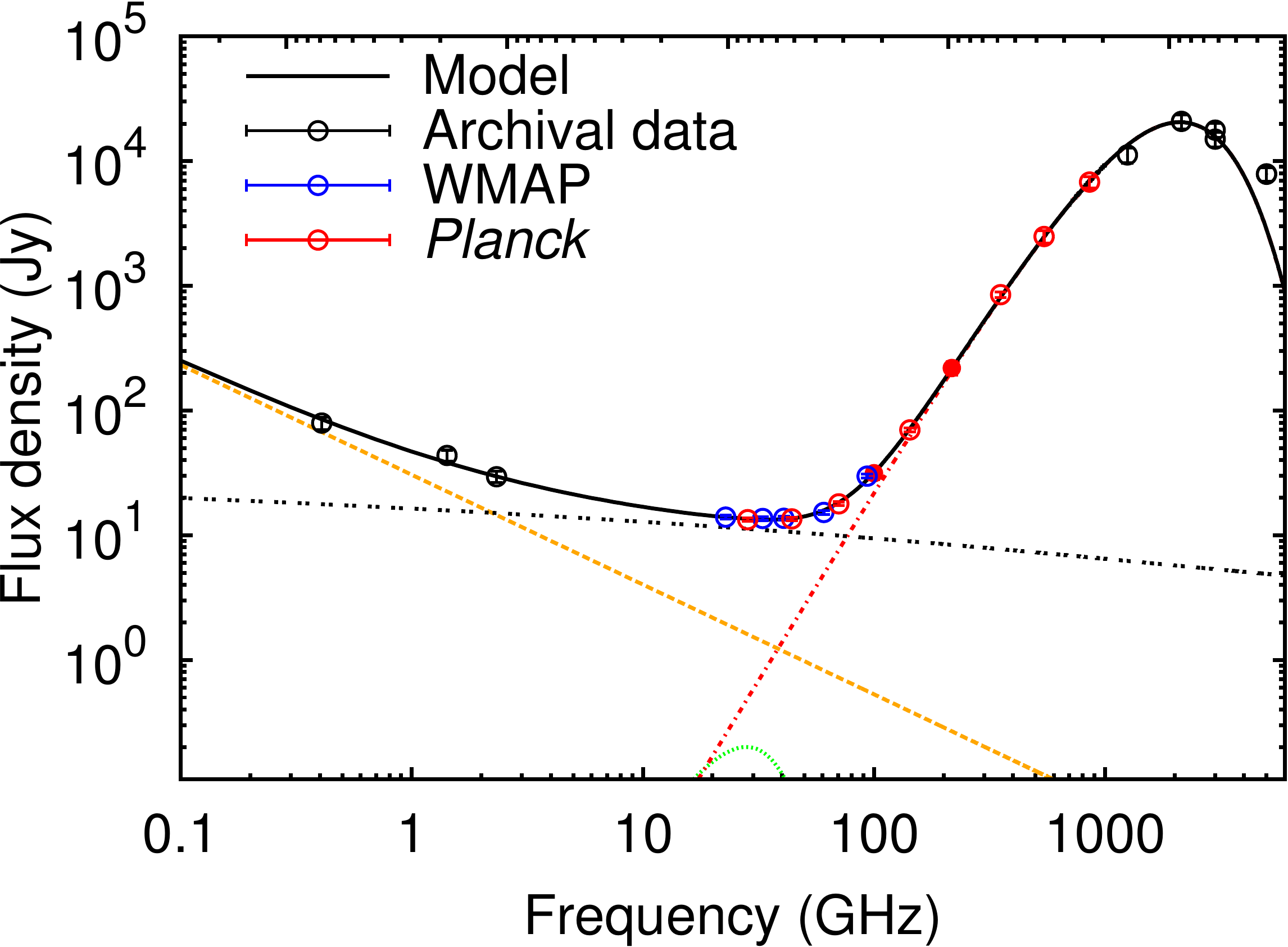}
\caption{SEDs of the LMC (top) and SMC (bottom) using aperture photometry, from the \Commander\ solution (left), and least-squares fits to the data with (middle, LSF) and without (right)) the CMB (LSF-CMB). The components are synchrotron (orange dashed lines), free-free (black double-dashed lines), AME (green dashed lines), CMB (blue dashed lines), and thermal dust (red dashed lines). The solid black line represents the sum of the model components. \planck\ data points are shown in red, \wmap\ in blue, and other ancillary data (some of which are not included in the fits, see text) in black.}
\label{fig:lmc_smc}
\end{center}
\end{figure*}

To cross-check the \commander\ solution, we perform aperture photometry on both the \commander\ maps and the latest \Planck\ and \WMAP\ maps, as well as other ancillary data, to generate spectral energy distributions (SEDs) of the LMC and the SMC. For the LMC we use central coordinates of \mbox{$(l,b)=(279\pdeg5,-33\pdeg5)$}, with a circular aperture of radius 300\arcm, and an outer annulus for background subtraction of 300--350\arcm. For the SMC we use central coordinates of \mbox{$(l,b)=(302\pdeg8,-44\pdeg3)$}, with a circular aperture of radius 150\arcm, and an outer annulus of 150--200\arcm. We do not attempt to remove Galactic foregrounds from the maps or SEDs, except through the background annulus in the photometry, since we are using the SEDs for the purpose of comparison. As a result, some Galactic foreground emission can be seen in the resulting LMC thermal dust map in Fig.~\ref{fig:T_highres}.

We use the same aperture photometry code that was developed for \cite{planck2011-7.2} and \mbox{\cite{planck2013-XV}}. We have modified the code to run on the \commander\ component maps to measure the amplitudes of the individual components in the aperture. The uncertainties are derived from the background annulus only: they do not include model or calibration uncertainties. Results are given in Table~\ref{tab:lmcparams}. We then sum those model SED components to compare with the flux densities measured directly from the Haslam, \planck, and \wmap\ maps, along with radio maps at 1.4\GHz\ by \citet{Reich2001} and 2.3\GHz\ by \citet{Jonas1998} and the infrared maps from \DIRBE\ \citep{Hauser1998} and \IRAS\ \citep{Miville-Deschenes2005}. This comparison for the LMC and SMC is shown in the left panel of Fig.~\ref{fig:lmc_smc}. We note that the \commander\ solution does not include data points above 857\GHz\ in its fit. In deriving the models, we exclude the 100 and 217\GHz\ \planck\ maps, since those bands contain CO emission. 

Using the same spectral component model described in \mbox{\citet{planck2013-XV}}, including colour corrections, we also perform least-squares fits (LSF) to the flux densities up to 3\,THz. There are two variants of the LSF model reported in Table \ref{tab:lmcparams} and Fig.~\ref{fig:lmc_smc}. One is fitted to data including the CMB contribution (henceforth referred to as ``LSF'') and the other is fitted to data post-CMB subtraction, namely flux densities from aperture photometry on CMB-subtracted maps (using the \commander\ CMB solution; henceforth referred to as ``LSF-CMB'').

\begin{table}[tb]
\begingroup
\newdimen\tblskip \tblskip=5pt
\caption{Values of the fitted parameters for the LMC (top) and the SMC (bottom) from \Commander, and the least-squares fitting before and after CMB subtraction. The \commander\ parameters are as described in \citet{planck2014-a12} except for $A_\mathrm{sync}$, which has been rescaled to a reference frequency of 1\GHz. We note that the \Commander\ uncertainties only include the standard deviation, and do not include modelling uncertainties. The LSF parameters are the same, except for $A_\mathrm{d}$, which is the optical depth at 250\um.}
\label{tab:lmcparams}
\nointerlineskip
\vskip -3mm
\footnotesize
\setbox\tablebox=\vbox{
    \newdimen\digitwidth 
    \setbox0=\hbox{\rm 0} 
    \digitwidth=\wd0 
    \catcode`*=\active 
    \def*{\kern\digitwidth}
    \newdimen\signwidth 
    \setbox0=\hbox{+} 
    \signwidth=\wd0 
    \catcode`!=\active 
    \def!{\kern\signwidth}
    \newdimen\pointwidth
    \setbox0=\hbox{{.}}
    \pointwidth=\wd0
    \catcode`?=\active
    \def?{\kern\pointwidth}
    \newdimen\notewidth
    \setbox0=\hbox{$^\mathrm{a}$}
    \notewidth=\wd0
    \catcode`@=\active
    \def@{\kern\notewidth}
    \halign{#\hfil\tabskip 2pt&
            \hfil#\hfil\tabskip 2pt&
            \hfil#\hfil\tabskip 2pt&
            \hfil#\hfil\tabskip 0pt\cr
    \noalign{\doubleline}
    Parameter & \Commander & LSF (CMB) & LSF (no CMB)\cr
\noalign{\vskip 3pt\hrule\vskip 5pt}
EM & $66.2\pm1.7$ & $57\pm4$ & $67\pm2$\cr
$T_\mathrm{e}$ [K] & 7000 & 8000 & 8000\cr
$A_\mathrm{d}$ & $24903\pm488$ & $(3.1\pm0.6)\times10^{-5}$ & $(2.82\pm0.14)\times10^{-5}$\cr
$\beta_\mathrm{d}$ & $1.516\pm0.003$ & $1.47\pm0.07$ & $1.42\pm0.03$\cr
$T_\mathrm{d}$ [K] & $19.29\pm0.09$ & $22.7\pm0.9$& $23.2\pm0.3$\cr
$A_\mathrm{AME1}$ & $15.3\pm0.3$ & \dots & \dots\cr
$A_\mathrm{AME2}$ & $3.47\pm0.14$ & \dots & \dots\cr
$A_\mathrm{CMB}$ [$\upmu$K] & $16\pm3$ & $15\pm3$ & \dots\cr
$A_\mathrm{sync}$ [Jy] & $239\pm4$ & $406\pm33$& $375\pm31$\cr
$\beta_\mathrm{sync}$ & $-3.1$ & $-2.70\pm0.05$ & $-2.74\pm0.04$\cr
\noalign{\vskip 4pt\hrule\vskip 4pt}
EM & $20.0\pm1.8$ & $21.7\pm0.9$ & $25.1\pm0.4$ \cr
$T_\mathrm{e}$ [K] & 7000 & 8000 & 8000\cr
$A_\mathrm{d}$ & $2541\pm70$ & $(1.26\pm0.07)\times10^{-5}$ & $(7.2\pm0.3)\times10^{-6}$\cr
$\beta_\mathrm{d}$ & $1.463\pm0.004$ & $1.36\pm0.05$ & $1.06\pm0.02$\cr
$T_\mathrm{d}$ [K] & $18.57\pm0.14$ & $21.5\pm0.3$ & $25.35\pm0.18$\cr
$A_\mathrm{AME1}$ & $1.497\pm0.07$ & \dots & \dots\cr
$A_\mathrm{AME2}$ & $0.57\pm0.03$ & \dots & \dots\cr
$A_\mathrm{CMB}$ [$\upmu$K] & $36\pm5$ & $37\pm2$ & \dots\cr
$A_\mathrm{sync}$ [Jy] & $25\pm2$ & $34\pm3$ & $31\pm3$\cr
$\beta_\mathrm{sync}$ & $-3.1$ & $-2.76\pm0.11$ & $-2.88\pm0.11$\cr
\noalign{\vskip 5pt\hrule\vskip 3pt}}}
\endPlancktable
\endgroup
\end{table}

We now go through each component in turn for both the LMC and SMC, considering both the morphology and the SEDs.

\paragraph{CMB.}
The CMB maps for both the LMC and SMC do not appear to be strongly contaminated by foreground emission; there is no clear correlation between the CMB map and the foreground components. The average CMB contribution agrees well in both the \commander\ and LSF estimates in the LMC and SMC. CMB contributions in both objects are positive, as was seen in \citet{planck2011-6.4b}. We thus conclude that the \commander\ CMB solution is robust at separating the CMB from foreground emission in this region.

\paragraph{Synchrotron.}
Low-frequency synchrotron emission is present in both the LMC and SMC. It is subdominant in total intensity at \planck\ and \wmap\ frequencies. In the LMC the peak of the emission is slightly offset from the Tarantula nebula, and diffuse emission is also present. The \commander\ spectral index between 408\,MHz and \WMAP/\Planck\ frequencies is assumed to be $\beta_\mathrm{sync}$\,$=$\,$-3.1$ according to the \galprop\ model (see Sect.~\ref{sec:synchrotron}); in the LSF and LSF-CMB fits for the LMC we find $\beta_\mathrm{sync}$\,$\approx$\,$-2.7$, and for the SMC $\beta_{\mathrm{sync}} \approx$\,$-2.8$, effectively between 408\,MHz and 22.8\GHz. At similar frequencies, \citet{Israel2010} found that $\beta_\mathrm{sync}$\,$=$\,$-2.70\pm0.05$ in the LMC; however, they found a steeper index of $-3.09\pm0.10$ in the SMC that agrees better with the \commander\ model. As such, \commander\ will under-estimate the synchrotron contribution at \planck\ frequencies, particularly in the LMC and to a lesser extent in the SMC. The synchrotron amplitudes at 1\GHz\ in \commander\ for both the LMC and SMC are significantly lower than that in the LSF, due to the steep spectrum assumed, with the difference absorbed by the free-free component. The LMC also appears in the synchrotron polarization map; the polarized emission is at its highest (around 30\,\% polarized, although this would be lower if the \commander\ synchrotron intensity spectral index were flatter) to the left of the peak in the synchrotron total intensity, offset from Tarantula, which corresponds to the two polarized synchrotron filaments identified by \citet{Klein1993}.

\paragraph{Free-free emission.}
The free-free component dominates the SEDs of both the LMC and the SMC at frequencies of 5--50\GHz. The majority of the emission comes from the Tarantula nebula; there is also diffuse emission closer to the centre of the LMC, and other compact ($<$\,$1\deg$) objects are present. The morphology agrees well with the \ha\ maps of individual sources in the LMC by \citet[][DEM]{Davies1976}: the main region of 30\,Dor is surrounded by a large number of smaller sources that show up in the \Commander\ map as diffuse emission. The two lower regions are groups of sources, with one comprising DEM\,4, 6, and 36 (bottom-left), and the other consisting of DEM\,27, 28, and 29 (bottom-right). The amplitude of the free-free emission can be converted to a star-formation rate (SFR) using the equations in \citet{Condon1992}. The \Commander\ free-free amplitude at 10\GHz\ is 141\,Jy, which gives an SFR of 0.10\mdotyr; the LSF model gives 0.08\mdotyr, and the LSF-CMB model gives 0.09\mdotyr. These are lower than the estimate of 0.2\mdotyr\ for the average SFR from analyses of stellar populations by \citet{Harris2009} and \mbox{\citet{Rezaeikh2014}}, but agree well with the recent star formation rate of 0.06\mdotyr\ calculated by \citet{Whitney2008} based on young stellar objects; these authors also give a range of SFR estimates of 0.05--0.25\mdotyr\ from infrared and \ha\ data. However, all of these estimates correspond to an SFR over different timescales \citep[e.g.,][]{Murphy2012}, as well as being subject to systematic effects; a more detailed study would be necessary to disentangle these effects. For the SMC, the amplitude at 10\GHz\ of 11\,Jy from the \Commander\ and LSF models gives an SFR of $1.1\times10^{-2}$\mdotyr; the LSF-CMB model gives a slightly higher value of $1.3\times10^{-2}$\mdotyr\ from a flux density of 12.8\,Jy. These estimates of the SMC free-free amplitude agree well with those from \citet{Draine2012}, who attributed 11\,Jy to the free-free component at 10\GHz, in very good agreement with the \commander\ amplitude.

\paragraph{AME.}
The \commander\ solution finds a small component of AME in both the LMC and SMC. In the LMC, the AME map has a bright region centred on 30\,Dor, where there are many sources seen in the higher-resolution \ha\ and radio maps (\mbox{\citealp{McGee1972}}, \mbox{\citealp{Davies1976}}), and where the brightest emission is seen in the LMC in thermal dust. The AME towards 30\,Dor has a higher peak frequency in the \commander\ solution than that on either side, following the pattern of \hii\ regions having higher peak frequencies, as noted in Sects.\,\ref{sec:ilc} and \ref{sec:ame}. At 20\GHz, the peak of the AME component, AME contributes about 16\,Jy to the model, compared with the free-free amplitude of around 130\,Jy. In the SMC, 1.6\,Jy is attributed to AME, compared with 10\,Jy for the free-free component. In both cases, AME is about 10\,\% of the free-free amplitude, and it is comparable to the component separation uncertainty in the free-free component, and the amplitudes of the synchrotron and CMB components at that frequency. The ratio of free-free to AME here is significantly higher than that seen in the Galaxy (\mbox{\citealp{planck2011-7.2}}; \mbox{\citealp{planck2014-XXIII}}) due to the presence of 30\,Dor, which dominates the free-free emission in the LMC. The LSF analysis includes an AME model, but is consistent with little or no AME (the AME contribution seen in the LSF SMC SED has a significance below 1$\sigma$). Nor was AME seen in the LMC and SMC SEDs in \citet{planck2011-6.4b}. \citet{Draine2012} included an AME component in their SED fits to the SMC of \mbox{3--5\,Jy} at 40\GHz; the \Commander\ solution returns much lower AME amplitudes in the SMC of 0.2\,Jy at 40\GHz. We calculate AME emissivities for the LMC for a $15\deg\times15\deg$ region centred on $(l,b)=(279\deg, -34\deg)$, and for the SMC in a $5\deg\times5\deg$ region centred on $(l,b)=(302\pdeg8, -44\pdeg3)$. The emissivities are given in Table~\ref{tab:emissivity}. For the LMC we find that the AME/545\GHz\ and AME/$\tau_\mathrm{353}$ emissivities are comparable to those from the AME regions; however, the AME/100\um\ emissivity is substantially lower, which is likely because the LMC has a higher dust temperature than the Galactic average. For the SMC, we find much lower values for all three emissivities. The LMC has comparable PAH levels to our Galaxy \citep{Bernard2008}, while the SMC has lower levels, which could explain why we find higher AME emissivities in the LMC than the SMC. The SMC has a very small grain (VSG) population that can be seen particularly clearly at 70\um\ \citep{Bernard2008}; given the emissivities here this VSG emission is unlikely to be connected to AME, as AME is typically thought to be due to PAHs. We thus conclude that the small amount of AME found by \commander\ in the LMC is at the level that would be expected from our Galaxy, although both the LMC and SMC regions may be contaminated by leakage from the free-free and synchrotron components, as well as being potentially influenced by the steep synchrotron spectral index assumed in the \commander\ model. Further work is needed to improve constraints on the AME and its properties in this region.

\paragraph{CO.}
The \commander\ CO $J$=1$\rightarrow$$0$ map detects various regions of CO emission, particularly around the 30\,Dor region, as well as in various other dusty regions in the LMC. The morphology of the map compares well with ground-based surveys, such as the Magellanic Mopra Assessment \citep{Wong2011}. This indicates that the separation of CO emission has worked well in this region. An exception is a large negative region in the \commander\ high-resolution map, which is a component separation artefact.

\paragraph{Thermal dust.}
The thermal dust seen by \Planck\ agrees well in morphology with \IRAS\ data. The dust temperatures from \Commander\ are significantly lower than those from the LSF: for the LMC the average dust temperature in the aperture is ($19.29\pm0.01$)\,K, compared with ($22.7\pm0.9$)\,K for the LSF. \mbox{\citet{planck2011-7.2}} found ($21.0\pm1.9$)\,K when fitting for a free $\beta_\mathrm{d}=1.48\pm0.25$ (comparable with the $\beta_\mathrm{d}$ found here). \commander\ only fits data up to 857\GHz\ and the fitted temperature and spectral index underestimates the flux densities at higher frequencies. The LSF fits data up to 3\,THz and so finds a higher temperature. We also note that where the fitted temperature is higher, $\beta_\mathrm{d}$ is lower. The same applies for the SMC, where the average from \Commander\ is ($18.57\pm0.14)$\,K, compared to ($21.5\pm0.3$)\,K from the LSF, and ($22.3\pm2.3$)\,K from \mbox{\citet{planck2011-7.2}}, where $\beta_\mathrm{d}=1.21\pm0.27$ is flatter than found by \commander\ but in better agreement with the LSF.\footnote{The \Commander\ solution has a Gaussian prior on the thermal dust spectral index of $\beta=1.55\pm0.1$, and the fitted values for the LMC and SMC are towards the lower end of this prior; however \commander\ would have returned a lower value for $\beta$ than the prior if the data had preferred it. The LSF indicate flatter spectral indices, however this is including data over a larger ($>857$\GHz) range of frequencies.} In the SMC, the submm excess (see e.g., \citealp{Israel2010}, \citealp{planck2011-6.4b}) has been subsumed into the dust spectral index and the lower dust temperature. The thermal dust polarization in the LMC traces the spiral structure, with the projected magnetic field running parallel to the structure, particularly in the 30\,Dor region where the \planck\ data have a higher sensitivity due to its scan strategy. This is also where the synchrotron polarization is seen.

\paragraph{}
In conclusion, we find that the \commander\ component maps in the regions of the LMC and SMC largely agree well with previous results and expectations, although the thermal dust properties are only representative up to 857\GHz.

\section{Polarized foregrounds at $\mathbf{\lambda}$\,$\mathbf{\approx}$\,1\,cm}
\label{sec:polarization}
The previous section showed that significant uncertainties remain for component separation in total intensity. Fortunately, the picture is quite different in polarization, where, between \Planck\ and \WMAP, we have twelve bands with maps of the all-sky polarization. Our current understanding is that there are three significantly polarized components: the CMB; synchrotron emission; and thermal dust. AME polarization has not been clearly detected, with upper limits of typically a few per cent (see Sect.~\ref{sec:other_pol_fg}).

The three strongly polarized components have radically different spectra, and moreover, CMB polarization is well separated in angular scale from the foregrounds: after convolution to 1\degr\ resolution, the CMB polarization predicted by the \Planck\ best-fit cosmology contributes an rms of only 0.54\,$\mu$K to the $Q$ and $U$ maps, negligible compared to the noise per beam (although, of course, detectable in the angular power spectrum at low multipoles).  The spectrum of the polarized dust is discussed in \citet{planck2014-XXII} and \citet{planck2014-XXX}. Here we are interested primarily in the synchrotron component and consider the dust polarization only to the extent that the magnetic field pattern that it traces sheds light on the synchrotron features (Sect.~\ref{sec:synch_pol}). \citet{planck2014-XIX} informs our understanding of the dust polarization features.

We begin by constructing a new map of polarized synchrotron emission by combining WMAP and \planck\ maps, which significantly increases the average S/N ratio compared with WMAP/\planck\ data alone. We then discuss some of the major features in the polarized sky, including the loops, spurs, filaments, and bubbles.

\begin{figure}[tb]
\begin{center}
\includegraphics[width=0.5\textwidth]{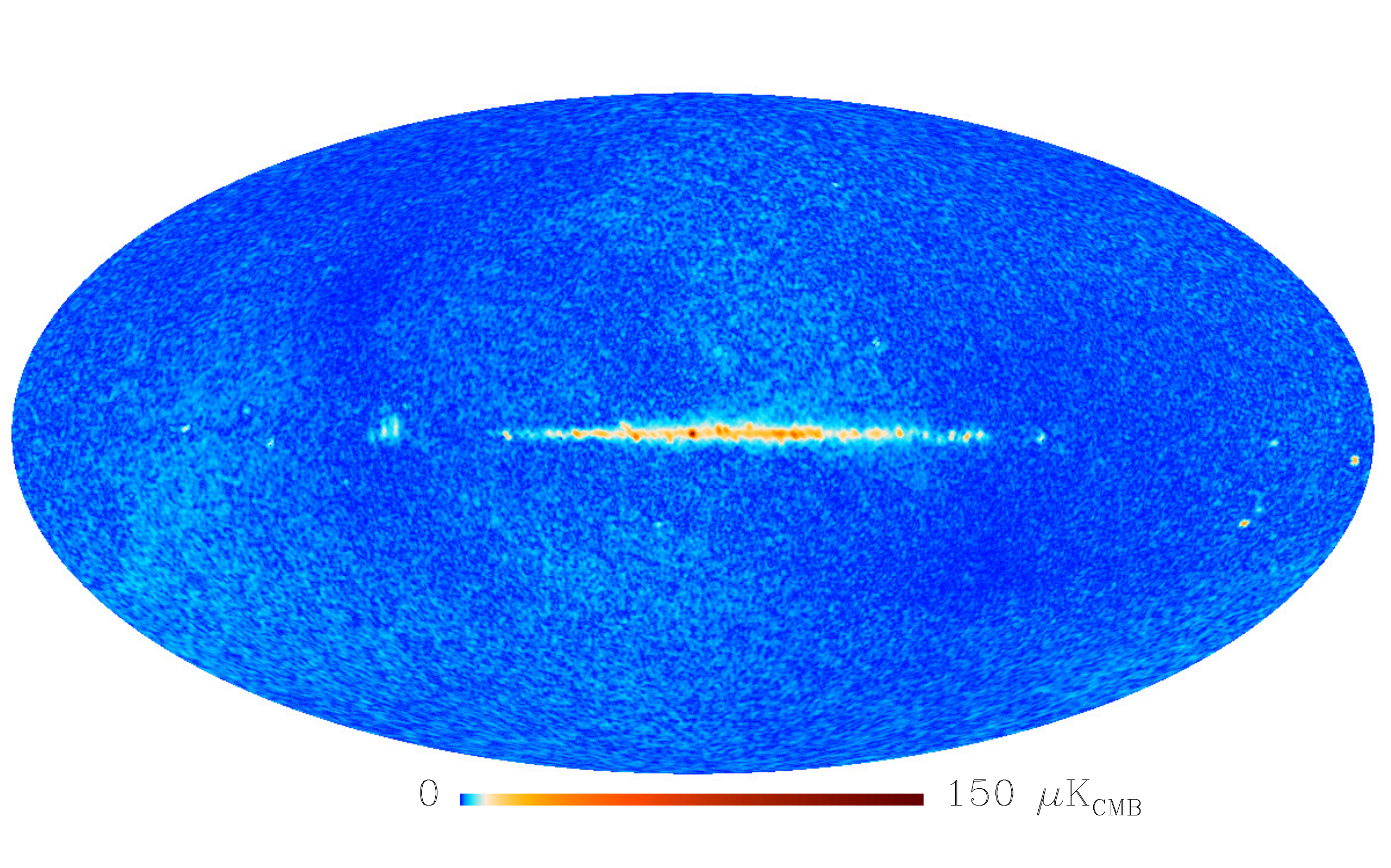}
\caption{Difference between the combined weighted polarization maps for \Planck\ and \WMAP\ at 1\degr\ resolution, defined as $\left[ (Q_\mathrm{Planck} - Q_\mathrm{WMAP})^2 + (U_\mathrm{Planck} - U_\mathrm{WMAP})^2\right]^{1/2}$.}
\label{fig:pol_comb_maps}
\end{center}
\end{figure}

\begin{figure*}[tb]
\begin{center}
\includegraphics[width=0.53\textwidth,angle=90]{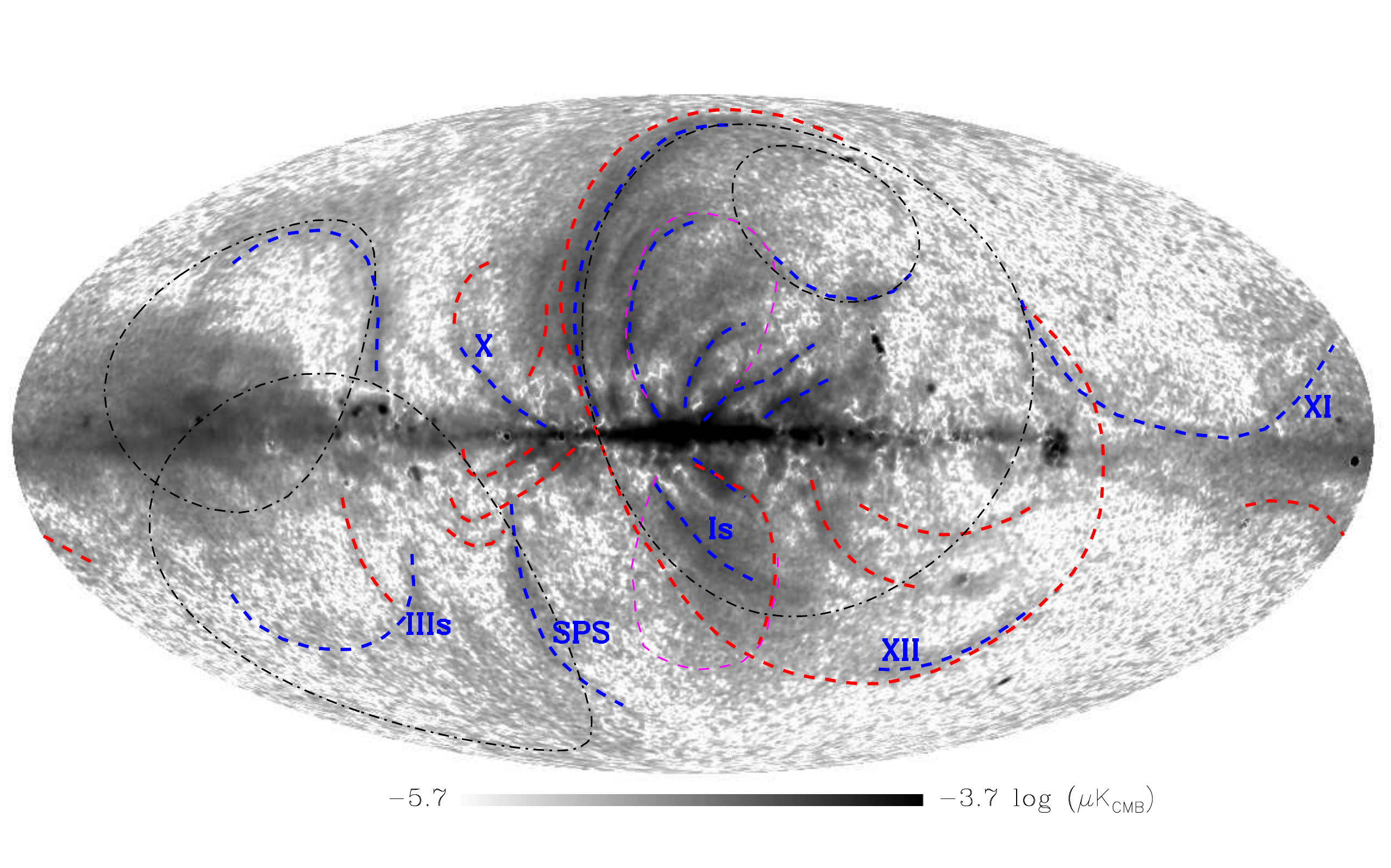}
\includegraphics[width=0.82\textwidth,angle=0]{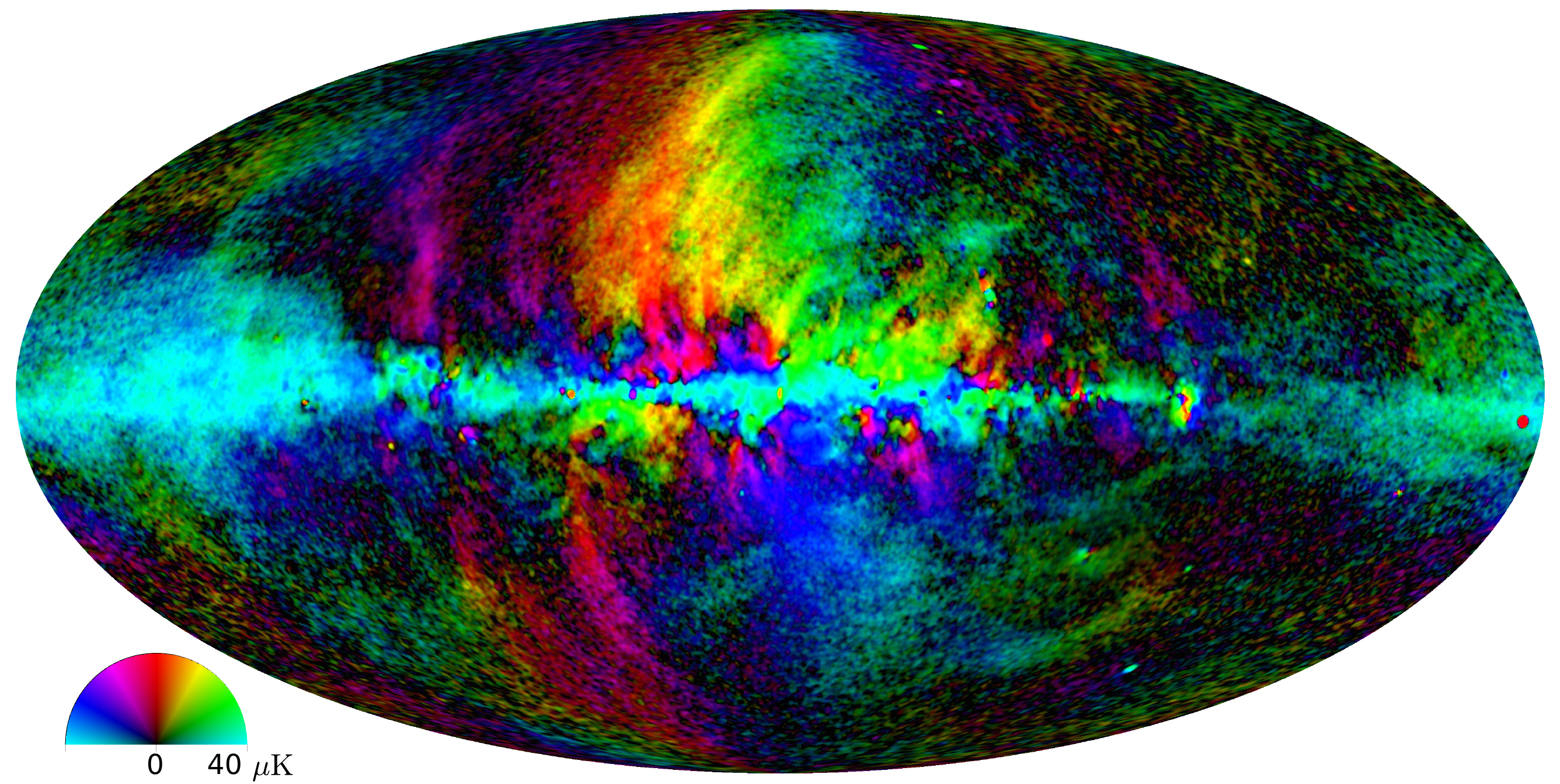}
\caption{{\it Top}: Combined weighted polarization intensity map after debiasing, with features highlighted. The black dash-dot lines show the outlines of Loops I to IV, as defined by \citet{Berkhuijsen1971}, the blue dashed lines show the filaments described by \citet{Vidal2014a} using \wmap\ polarization data, the red dashed lines show features that are visible in the new \planck\ data, and the magenta dashed lines show the outline of the \fermi\ bubbles. Filaments that are discussed in the text are labelled.  {\it Bottom}: The same combined polarization intensity map, with projected magnetic field angle (at 90\degr\ to the polarization angle) encoded in colour with asinh scaling. The coloured half-disc represents the polarization angle depicted in the map, with 0$\deg$ horizontal, while the polarization intensity is represented with the radial distance along the half-disc.}
\label{fig:pol_comb_maps_features}
\end{center}
\end{figure*}

\subsection{Combination of \Planck\ and \wmap}
\label{sec:combo}

\wmap\ \citep{Page2007,Bennett2013} provided the first clear view of the intrinsic synchrotron polarization across the sky -- previous ground-based observations \mbox{\citep[e.g.,][]{BrouwSpoelstra1976,Wolleben2006}} being strongly affected by Faraday rotation and depolarization due to their lower observing frequencies ($\lesssim 2$\GHz).  \WMAP's lowest-frequency channel, K-band, dominates the combined \wmap\ S/N ratio in polarization because of the steep spectrum of synchrotron emission. Although the full-mission \Planck-LFI maps have significantly lower noise than the final \wmap\ results, the foreground S/N ratio is very similar, since at \Planck's lowest frequency (28.4\GHz) the synchrotron brightness is half that in K-band. After smoothing to a common resolution of 1\degr, the median (mean) S/N ratio is 2.47 (3.77) for \wmap\ K-band and 2.64 (3.72) for \Planck\ 28.4\GHz. The different scan strategies of the two missions result in somewhat different sensitivity patterns, so that each map is superior in some sky regions; \Planck\ has the largest advantage in the regions around the Ecliptic poles \citep[see the hit count maps in][]{planck2014-a07}.

\begin{figure*}[!htb]
\begin{center}
  \newcommand{\widthfig}{0.32} \newcommand{\angfig}{0}
  \includegraphics[angle=\angfig,width=\widthfig\textwidth]{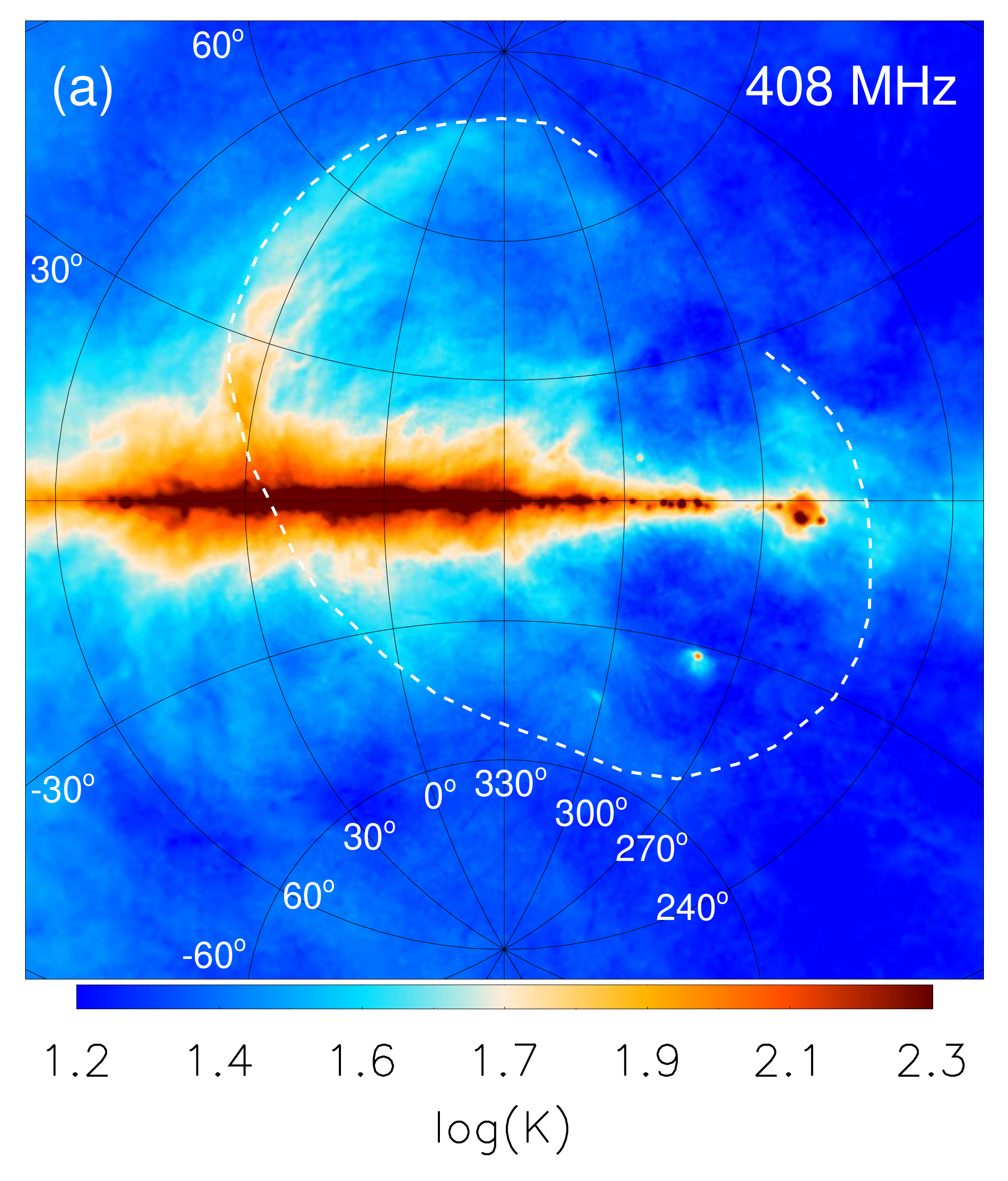}
  \includegraphics[angle=\angfig,width=\widthfig\textwidth]{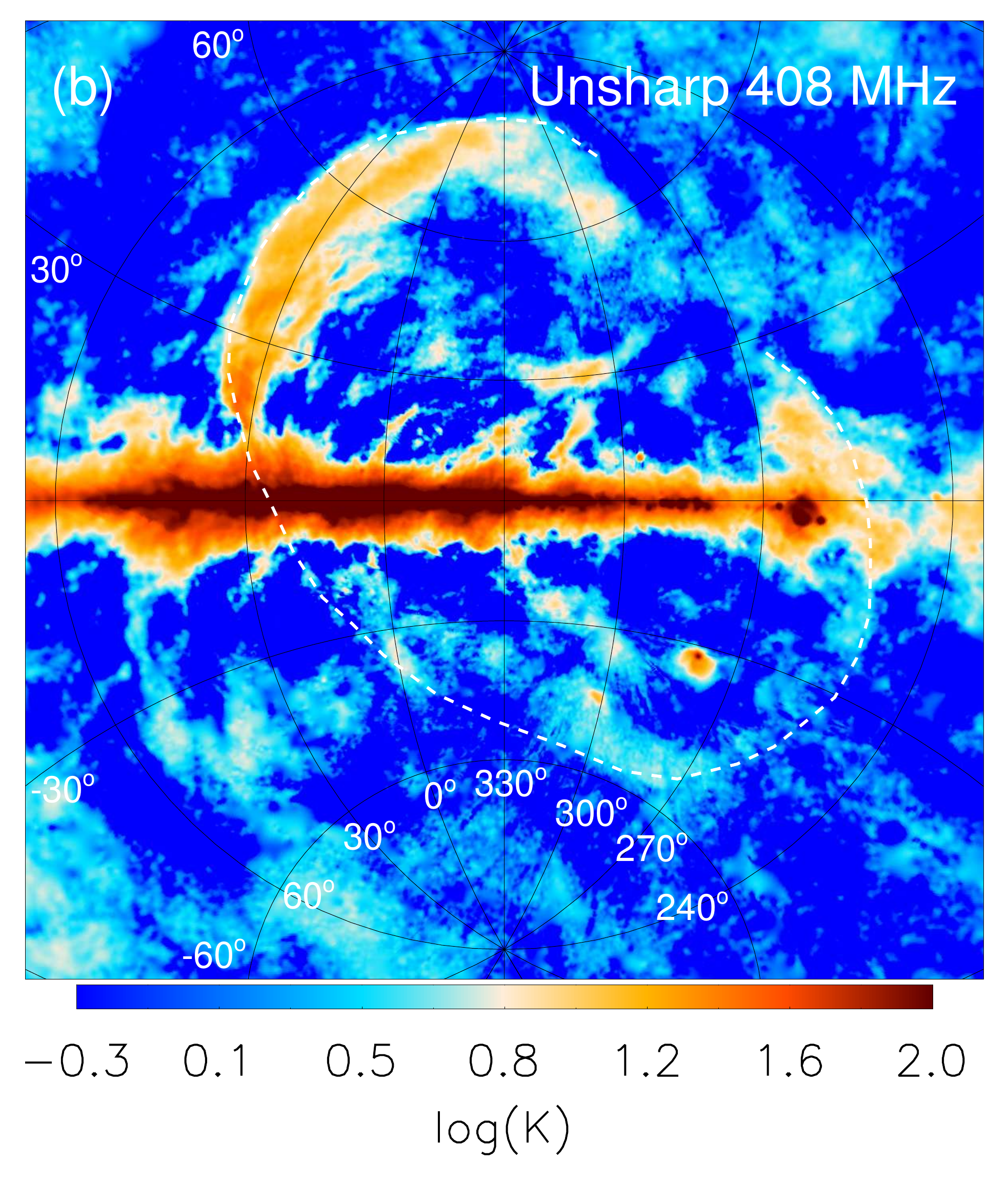}
  \includegraphics[angle=\angfig,width=\widthfig\textwidth]{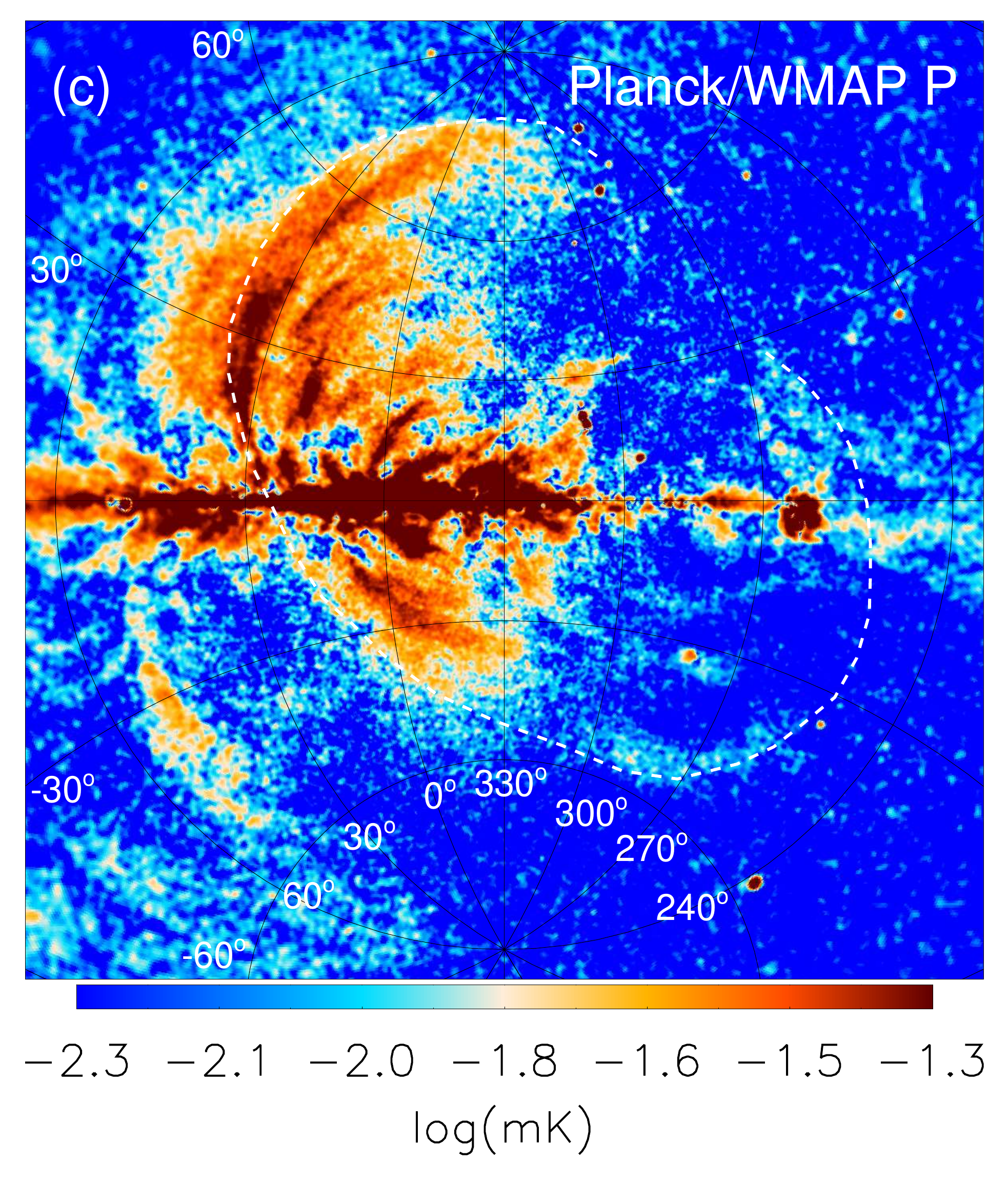}
  \includegraphics[angle=\angfig,width=\widthfig\textwidth]{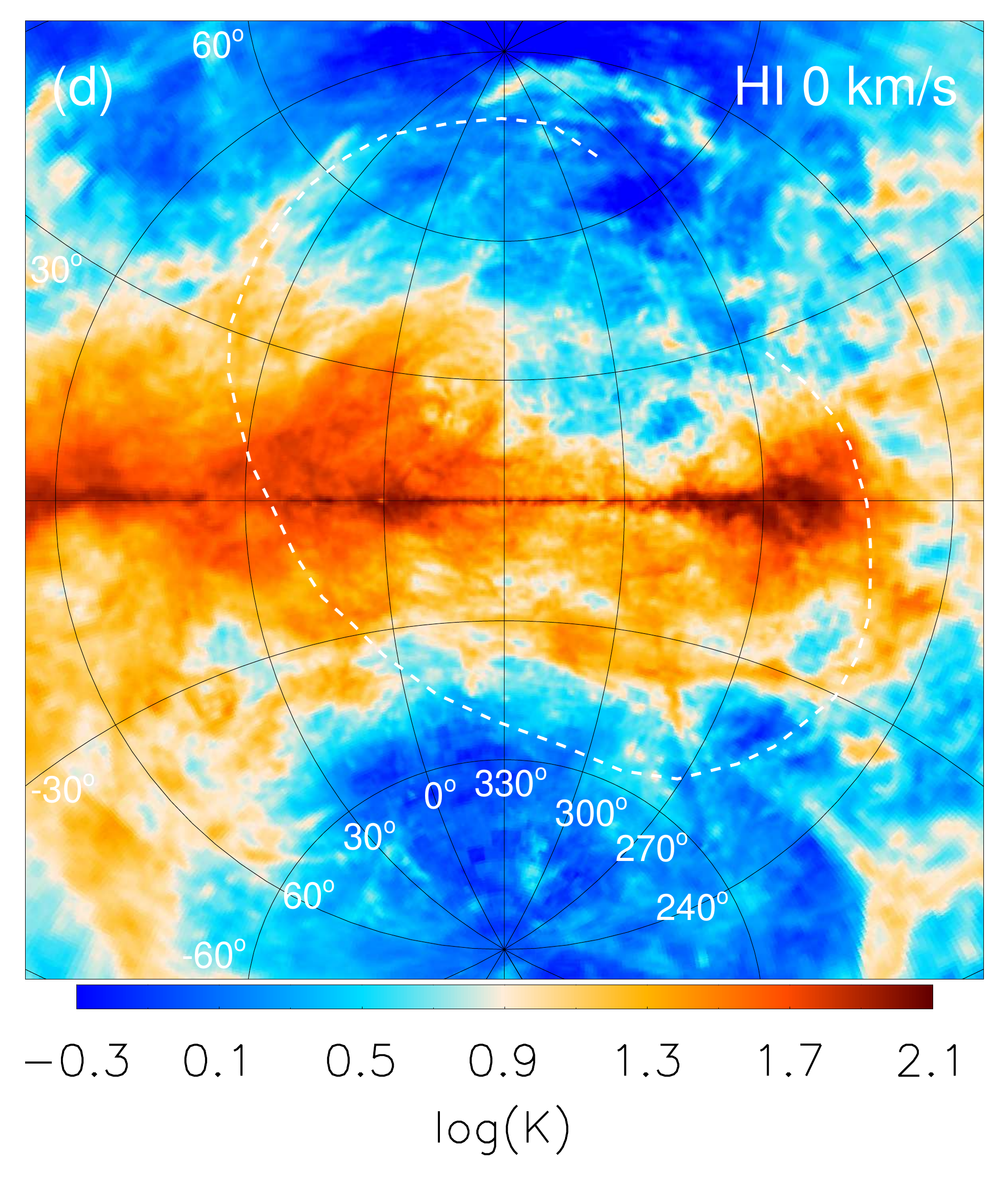}
  \includegraphics[angle=\angfig,width=\widthfig\textwidth]{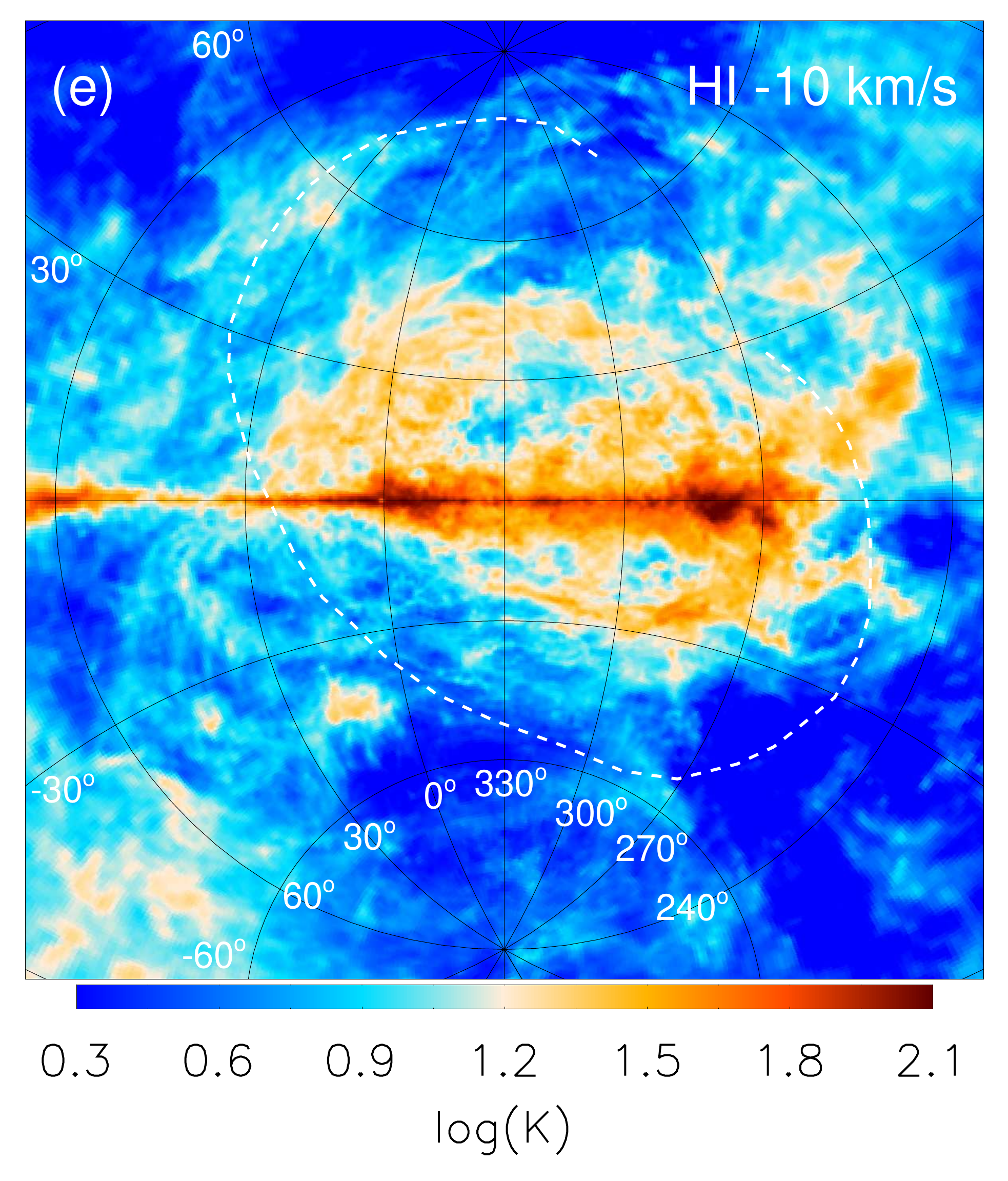}
  \includegraphics[angle=\angfig,width=\widthfig\textwidth]{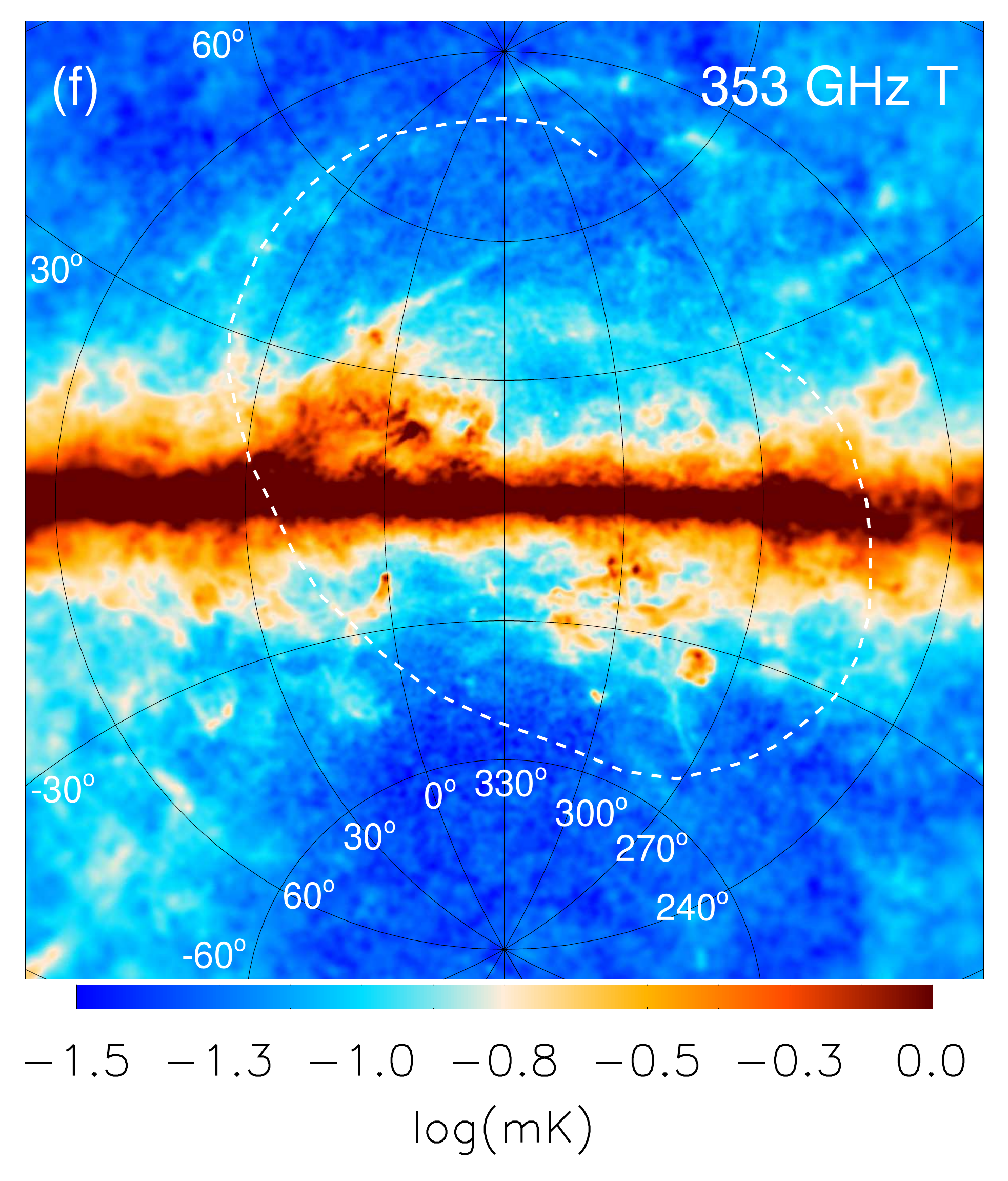}
  \includegraphics[angle=\angfig,width=\widthfig\textwidth]{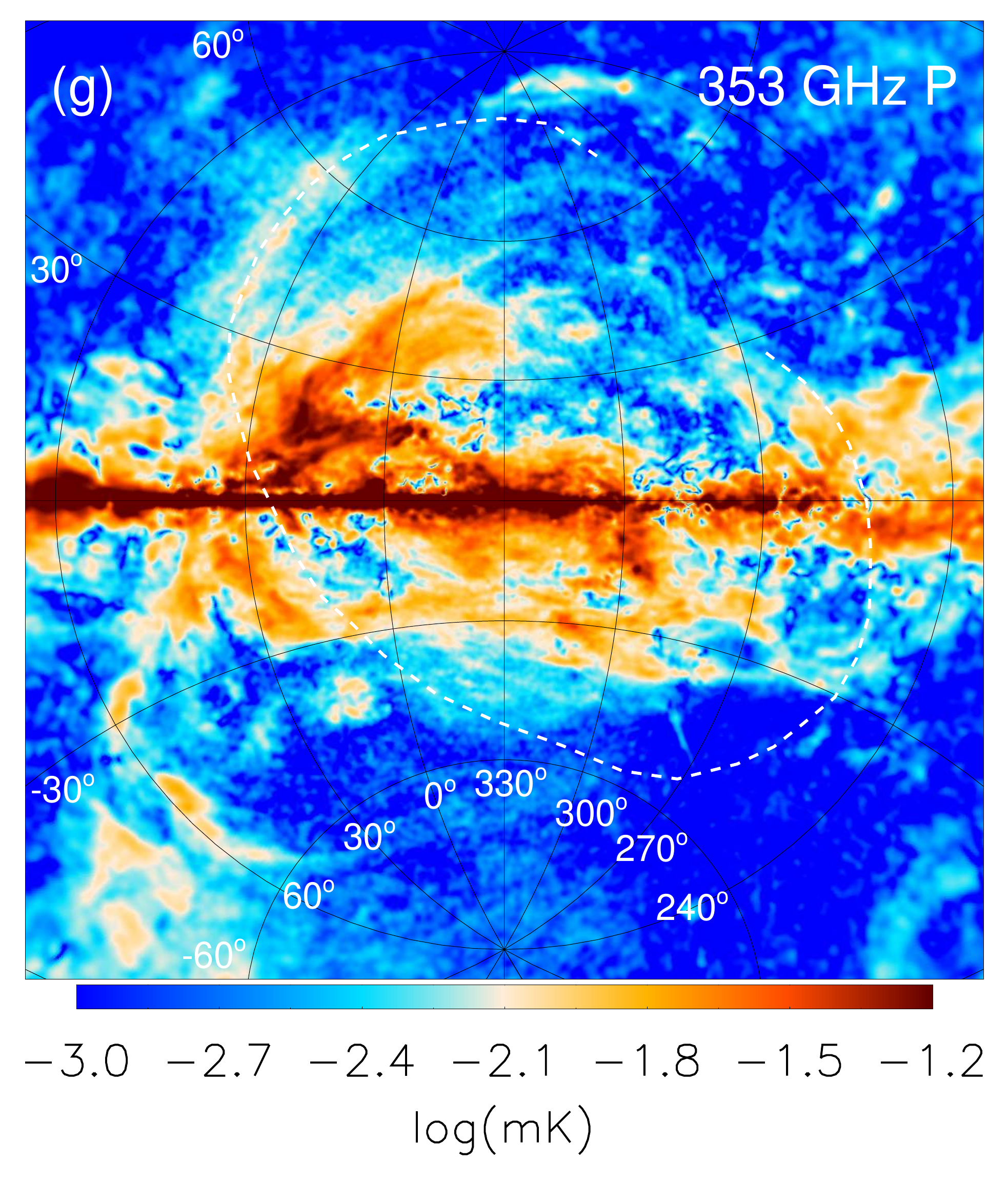}
  \includegraphics[angle=\angfig,width=\widthfig\textwidth]{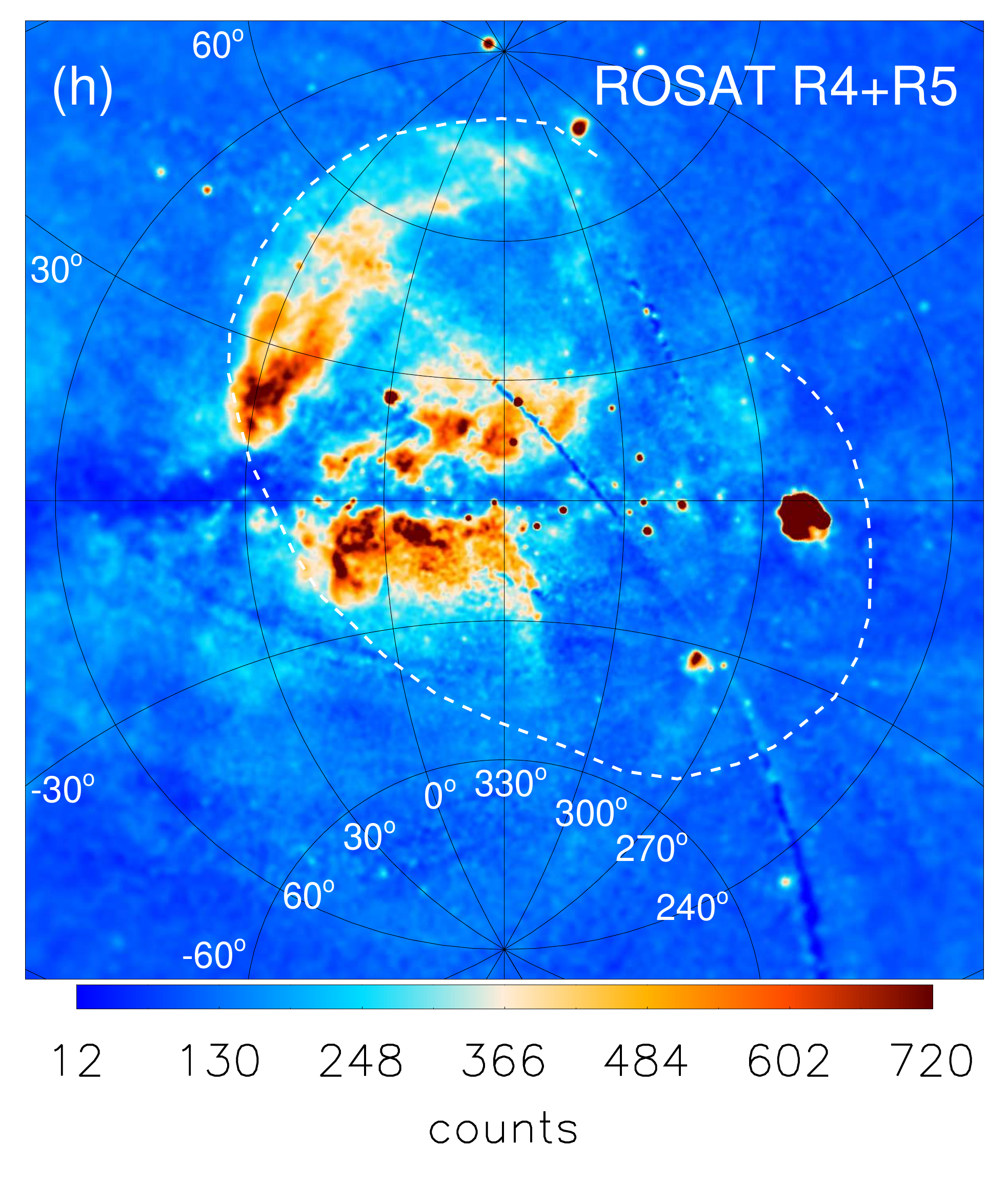}
  \includegraphics[angle=\angfig,width=\widthfig\textwidth]{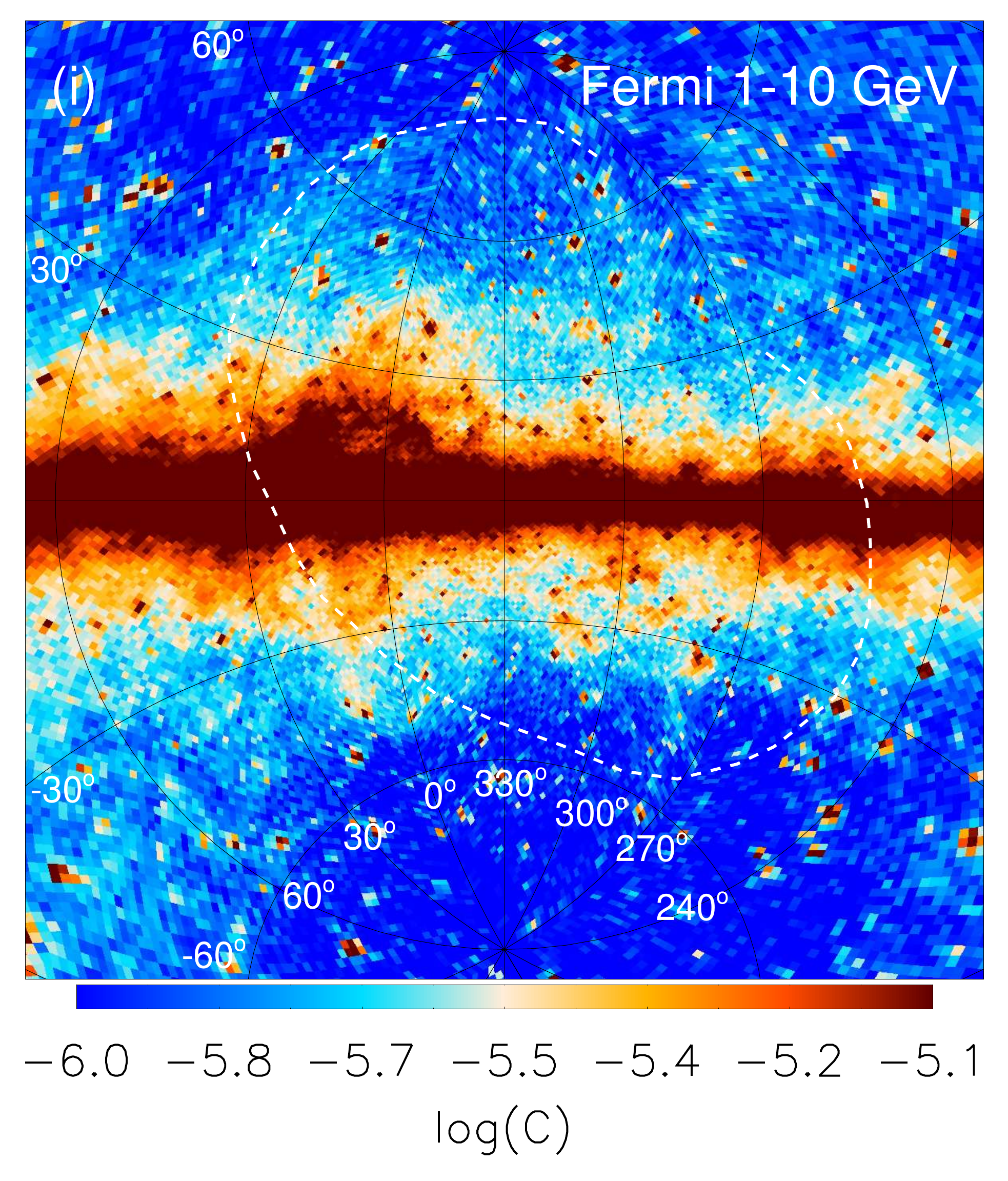}
 \caption{Various tracers of the interstellar medium in the hemisphere containing Loop I, with all maps centred at $l=330\degr$, $b=0\degr$, and using stereographic projection, so that small circles in the sky are projected as circles: (a) 408\,MHz; (b) unsharp-mask 408\,MHz; (c) combined \planck-\wmap\ polarization intensity; (d) \hi\ at $v_{\rm LSR} = 0$, velocity width 10\,km\,s$^{-1}$, from the LAB survey \citep{Kalberla2005}; (e) the same, at $-10 {\rm\, km\,s^{-1}}$; (f) \planck\ 353\GHz\ temperature; (g) \planck\ 353\GHz\ polarization intensity; (h) \rosat\ bands R4$+$R5 (0.44--1.21\,keV) from \citet{Snowden1997}; and (i) \fermi\ 1--10\,GeV from \citet{Ackermann2014}. The dashed white outline is our proposed outer boundary, derived from panels (b) and (c).}
\label{fig:stereo_grid}
\end{center}
\end{figure*}

\Planck\ and \wmap\ polarization results generally agree well, but the difference map (see Fig.~\ref{fig:pol_comb_maps}) shows significant artefacts, both on the plane and at high latitudes. Extremely large angular-scale ($\ell=3,5,7$) residuals covering much of the sky are thought to be due to the poorly-constrained modes in the \WMAP\ data as a result of the scan strategy \citep{Jarosik2011,Bennett2013}; they occur in all the \wmap\ frequency maps, with an approximately consistent pattern. However, they are most readily visible in the V- and W-bands of \WMAP\ when the foreground brightness is close to the minimum. The on-plane features are likely to be caused by inaccuracies in the \Planck\ leakage maps. These differences are further discussed in \citet{planck2014-a12}.

\begin{figure}[tb]
\begin{center}
\includegraphics[width=0.5\textwidth]{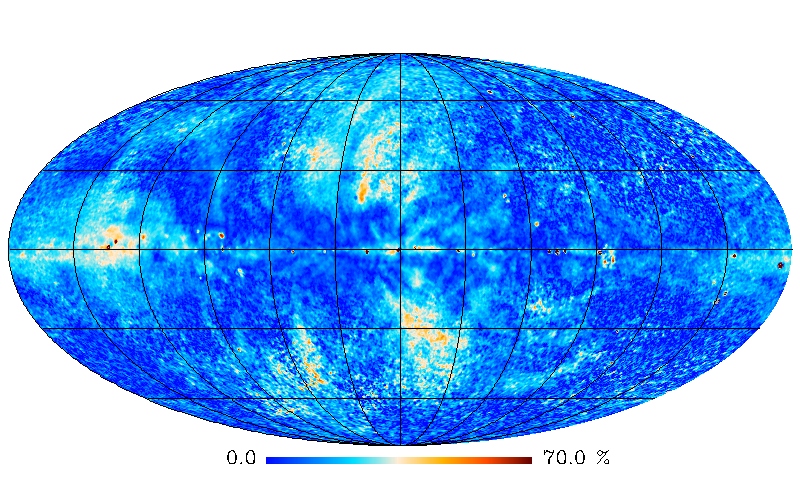}
\caption{Polarization percentage of the debiased weighted polarization intensity map against the \Commander\ synchrotron total intensity fit evaluated at 22.8\GHz. Uncertainties in the component separation discussed in the text, and also uncertainty in the absolute zero level, make this highly uncertain, except in regions of strong synchrotron emission, such as the North Polar spur and the inner Galactic halo (see also discussion in \citealp{Vidal2014a}).}
\label{fig:pol_fraction}
\end{center}
\end{figure}

Because polarization is only marginally detected in many pixels of the individual \wmap\ and \Planck\ maps, we have constructed a weighted mean image using all the maps in the range 20--50\GHz, where the polarization is dominated by optically thin synchrotron radiation, with negligible Faraday rotation \citep[except at the Galactic centre,][]{Vidal2014a}, so that the polarization angles should be consistent between the maps.  We first smooth the \Planck\ and \WMAP\ maps to a resolution of 1\deg\ and $N_\mathrm{side}=256$ in both $Q$ and $U$, and the $QQ$, $UU$, and $QU$ covariances as well. We assume that the emission has a single spectral index of $-3.0$, and we use this to scale all the maps (and their covariances) to match the \Planck\ 28.4\GHz\ image. We then use the $2\times2$ $Q$ and $U$ covariance matrices and their inverse matrix to calculate the weighted average in each pixel of the $Q$ and $U$ maps. We create three weighted maps as a result: one consisting of the three lowest \WMAP\ bands; one consisting of the two lowest \Planck\ bands; and one using both \WMAP\ and \Planck. The \Planck\ variance estimates include the uncertainty in the bandpass leakage correction. This results in a large fractional error in $Q$ and $U$ along the plane, so \wmap\ data dominate the on-plane emission in the combined map. At high latitudes, the somewhat higher sensitivity of \Planck\ down-weights the impact of the large-scale \wmap\ polarization artefacts, but we note in our discussion when these may still have an impact on our results.  The improvement in S/N ratio is significant, with: a median (mean) S/N ratio of the \WMAP\ weighted map of 2.70 (4.19); the \Planck\ weighted map of 2.68 (3.81); and the combination of both of 3.37 (5.39), where the bigger improvement comes from the combination of both \wmap\ and \planck. This corresponds to a roughly 25\,\% improvement in S/N ratio compared to using the combined WMAP or \planck\ maps separately.

\commander\ also provides a polarized intensity synchrotron map, which is calculated using only \planck\ data. As this map is the best-fit solution to a model using noisy data, it is noisier than our combined \Planck-only map (our map treats the small CMB component as noise). Moreover, the addition of \wmap\ K-band data in our map further reduces the noise in the final combined map. At 1\deg\ resolution, the \commander\ map has a median noise value over the entire sky of $\sigma_Q \approx \sigma_U $ = 4.5\,$\mu$K, while our combined map has $\sigma_Q \approx \sigma_U $ = 2.6\,$\mu$K.

Polarized intensity maps have been debiased using the asymptotic estimator \citep{Montier2015b,Vidal2014b}, which generalizes the estimator first proposed by \citet{Wardle1974} to the case of anisotropic errors in $(Q,U)$. Figure~\ref{fig:pol_comb_maps_features} shows the resulting polarization combination maps. A number of new polarized structures visible in the combined map that were unclear in the individual maps are highlighted. These are discussed in the sections that follow.

\subsection{Overview of polarized synchrotron emission}
\label{sec:synch_pol}
Synchrotron total intensity is distributed comparatively uniformly over the sky.  In the 408\,MHz map, assuming an instrumental plus extragalactic background of 8.9\,K \citep{Wehus2014}, half the total Galactic flux is contributed by 18\,\% of the sky.\footnote{The diffuse emission may be even brighter, since the background assumed may include an isotropic component of Galactic emission. Using the \citet{Lawson1987} extragalactic background estimate of 5.9\,K, half the sky flux comes from 21\,\% of the sky.}  The equivalent figure for the dust-dominated \Planck\ 545-GHz map is 4.6\,\%.  In the synchrotron intensity map, although the Galactic plane and the North Polar spur are visually prominent, they are superposed on broad diffuse emission that dominates the total flux.

This diffuse emission seems to be much weaker in polarization.  Away from the narrow Galactic plane, i.e., at $|b| \gtrsim 3\degr$, the polarized cm-wavelength emission is dominated by the synchrotron loops and spurs familiar from low-frequency radio surveys. In fact, \citet{Vidal2014a} demonstrate a close correspondence between the polarized intensity at \wmap\ K-band and an unsharp-masked version of the 408\,MHz map, in which structure on scales $\gtrsim 10\degr$ is filtered out (see Fig. \ref{fig:stereo_grid} top row a and b).

In particular, there is hardly any trace in the \Planck\ and \wmap\ polarized maps of the synchrotron halo of the inner Galaxy, which fills roughly $|l| \lesssim 60\degr$ and $5\degr \la |b| \la 15\degr$ \citep{planck2014-XXIII} and contributes 20--30\,\% of the Galactic synchrotron flux in low-frequency maps (compare panels a and c in Fig.~\ref{fig:stereo_grid}). The effect is even clearer in fractional polarization, which is very low ($\lesssim 10\,$\%) in the inner halo region (see Fig.~\ref{fig:pol_fraction}). The component separation analysis of Sect.~\ref{sec:compsep} implies that the halo is still present in the unpolarized cm-wavelength maps, because the data are well-fitted with our synchrotron model, which has a constant shape, so that the synchrotron distribution is no different at 20 or 30\GHz\ from 408\MHz. Although, as we have seen, this separation is subject to substantial uncertainties, it is corroborated by spectral analysis of ground-based surveys \citep{Reich1988,Platania2003} that find that the halo and spurs have similar spectral indices near 1\GHz.

The magnetic field in the Galactic disc follows the plane, as expected from the shearing effect of differential rotation, and one would expect the same to apply in the inner halo. Its polarization would then be roughly orthogonal to that in the spurs rising out of the plane, which all have fields roughly parallel to their axes \citep{Vidal2014a}.  Therefore, the net polarization of halo and spurs will cancel to some extent.  But the fact that the spurs are prominent in polarization, even when they are only perturbations on the inner halo in total intensity (see Fig~\ref{fig:profiles}), implies that the halo must be very weakly polarized; even between the spurs, there is hardly any sign of a field parallel to the plane in the inner halo region (a trace may be visible near $(l,b) = (34\degr,8\degr)$; see the discussion of the $l=45\degr$ feature in Sect.~\ref{sec:other_loops}).  This is a surprising contrast to the narrow plane at $|b|\lesssim 3\degr$, where the overall parallel orientation of the field is very clear in Fig. \ref{fig:pol_comb_maps_features}, despite the fact that a significant fraction of its emission is due to individual SNRs \citep{planck2014-XXIII}, whose overall polarization orientation is nearly random. The conventional analogy between the Galactic halo and the Solar corona suggests that the halo field should relax to a nearly force-free configuration with little small-scale structure, constrained primarily by the foot points where the field lines are tied to dense gas clouds in the plane. This would lead us to expect a reasonably high fractional polarization; differential rotation of the footpoints should, as with the disc field, shear the field so that it is largely parallel to the plane as viewed from Earth. Evidently the halo field is much more tangled than this naive argument would suggest.

It is worth noting that the overall fractional polarization also appears to be low ($< 15$\,\%) across the high latitude sky (Fig.~\ref{fig:pol_fraction}), although this quantity is very sensitive to the poorly-known zero level of the synchrotron total intensity \citep{Vidal2014a}. Again, the implication is that there is substantial tangling of the field even on lines of sight looking out of the disc.

\subsection{Loop I}
\subsubsection{Structure}
\label{sec:loop_I_structure}
Loop I is the nearly-circular structure of radius 58\deg\ whose top-left quadrant (as viewed in Galactic coordinates)\footnote{To avoid the non-intuitive concepts of Galactic ``East'' and ``West'', throughout this section we describe emission features in terms of left and right as projected in the figures, corresponding to the directions of increasing and decreasing longitude, respectively.} is traced by the North Polar spur (NPS).  It is also detected in soft X-rays, where the emission is dominated by thermal emission from $3 \times 10^6$\,K gas \citep{Willingale2003}, and is bordered by cold material visible via \hi\ and dust emission (both thermal and anomalous). The NPS has long been a suspected $\gamma$-ray emitter, and \citet{Ackermann2014} clearly detect Loop I at GeV energies, presumably due to inverse-Compton scattering of starlight by the CRLs in the loop, combined with pion decay emission from the cold border. Figure~\ref{fig:stereo_grid} shows the structure in several tracers in stereographic projection (chosen because circles are projected as circles and also because the angular scale increases outwards, enabling a clear display of features around the edge of this enormous structure that covers most of a hemisphere).

\begin{figure}
\begin{center}
  \includegraphics[width=0.5\textwidth,angle=0]{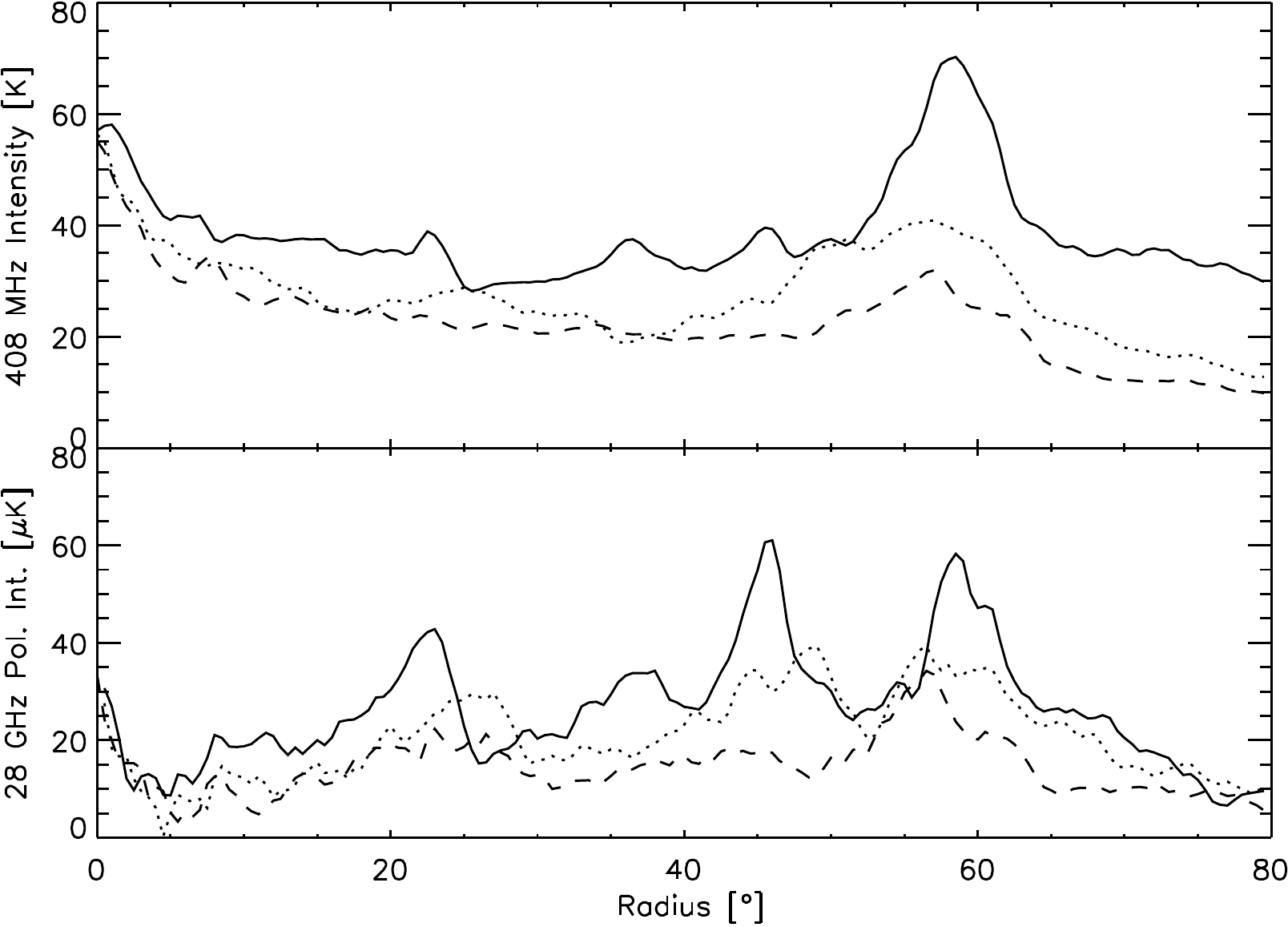}
  \caption{Profiles of Loop I in total intensity at 408\,MHz \citep{Remazeilles2014} and in our combined \Planck-\wmap\ polarized intensity image at 28.4\GHz.  Profiles are along radial cuts beginning at the nominal centre of the loop, $(l,b)=(329\degr,17\fdg5)$ and are averaged over a 10\degr -wide sector in position angle, at position angles 75\degr\ (solid line), 50\degr\ (dotted), and 25\degr\ (dashed). They cross the North Polar spur at $b = 22\degr,\, 43\degr$, and $61\degr$, respectively.}
\label{fig:profiles}
\end{center}
\end{figure}

In both Galactic hemispheres the polarization maps show a number of spurs within Loop I that parallel its outer boundary, nearly all concentrated on the left-hand side of the structure. Since Loop I covers about a third of the sky, including the inner Galaxy, some of the features projected inside it are surely unrelated, but the general coherence of the structure strongly suggests that most of the emission away from the plane has a common cause.  The internal spurs are much more obvious in polarization than in total intensity, even after unsharp masking (Figs.~\ref{fig:stereo_grid}a,b,c, \ref{fig:profiles}). These features are therefore highly polarized; however, they are projected onto diffuse extended emission that seems to be weakly polarized ($\lesssim 15$\,\%), including the inner halo of the Galaxy, so the overall fractional polarization towards the inner spurs is not much higher than towards the NPS itself ($\approx 40$--50\,\%), while between them the fractional polarization is low (Fig.~\ref{fig:pol_fraction}).

Previous measurements of the radio outline found \mbox{Loop I} to be surprisingly close to circular: the definitive study by \citet[hereafter BHS]{Berkhuijsen1971} found that 19 points on the ridge-line covering 155\degr\ around the loop (all in the north Galactic hemisphere) fitted a small circle with an rms of only 0\fdg9. The NPS conforms roughly to the brightness profile expected from a shell of emission, for which the ridge-line marks the tangent to the inner surface; the outer boundary lies several degrees beyond the ridge line, and is also clearly traceable, especially around the NPS (e.g., Fig.~\ref{fig:stereo_grid}b). Relatively bright diffuse polarized emission outside the boundary of the NPS south of $b=40\degr$ (Fig.~\ref{fig:stereo_grid}c, \ref{fig:vectors_cart}) seems to be part of a different structure (see Sect.~\ref{sec:other_loops}).

South of the plane, the loop is superimposed on the inner halo and cannot be clearly followed in total intensity, but in \wmap\ and \planck\ polarization maps (Figs.~\ref{fig:pol_comb_maps_features},\ref{fig:stereo_grid}), the pattern is clear, at least on the left-hand side of the structure. The brightest southern spur, called filament Is by \citet{Vidal2014a}, runs well inside the path of the BHS small circle. However, a fainter spur extends from $(l,b) = (22\degr,-5\degr)$, closely along the BHS circle until its ridge becomes lost in faint diffuse emission near $b=-37\degr$; at this point clearly-detectable polarization again extends several degrees outside the circle.  There is then no trace of a ridge-line close to the BHS circle along its whole bottom-right quadrant. It eventually re-aligns with the observed loop boundary at the top of the spur north of the Vela SNR $(l,b)=(268\degr,25\degr)$, near the first measured point in \mbox{\citet{Berkhuijsen1971}}. In this bottom-right quadrant three broad, faint, concentric arcs extend from the interior of Loop I, reaching some 30\degr\ beyond the BHS circle; these arcs are visible both in polarization and in the unsharp-masked 408\,MHz map. The outermost \citep[filament XII of][]{Vidal2014a} rejoins the plane near $l=250\degr$ and appears to emerge on the other side as the aforementioned spur above Vela. Although they might be unrelated, they share the general pattern of the other Loop I filaments, for instance the middle arc that passes in front of the LMC is clearly brightest on the left-hand side, as it approaches the Galactic plane near $l=300\degr$.  We consider it plausible that filament XII represents the outer boundary of Loop I, and on this basis we have marked the outline of the loop in Fig.~\ref{fig:stereo_grid}.  None of the current theories (see Sect.~\ref{sec:LI_theory}) predict a strictly spherical structure; taking the smallest plausible distance to the centre as 120\,pc, the diameter is at least twice the 100\,pc scale height of the \hi\ distribution. Therefore, even ignoring the possible interaction with the Local Cavity (LC) in which the Sun is situated, the loop is expanding into an inhomogeneous environment. Most individual SNRs, even though much smaller than Loop I, show quite large departures from circular outlines, often exceeding the level of distortion implied if filament XII traces the edge of \mbox{Loop I}.

In X-rays, the North Polar Spur peaks at a position inwards from the synchrotron ridge, but, as the outline drawn in Fig.~\ref{fig:stereo_grid} shows, the outer edge of the spur in the two tracers is coincident. This result is unexpected, because the Spur cannot be bounded by a shock front fast enough ($v \gtrsim 300 {\rm \, km\,s^{-1}}$) to heat the ambient medium to millions of kelvin, while at the same time accelerating the cold border to at most 25\,km\,s$^{-1}$ (see Sect.~\ref{sec:LI_theory} below). In young supernova remnants, co-incidence of radio and X-ray boundaries is always due to X-ray synchrotron emission from the shock, and it seems likely that in the NPS the faint X-ray emission from the outer edge is also non-thermal.

In contrast, the gas and dust tracers extend a few degrees further out, particularly obvious in the \hi\ and dust filament at the top of the loop, ($l,b)=(320\degr,84\degr$). This is just as expected, since neutral atoms and dust grains could not survive at the $3\times 10^6$\,K of the X-ray emitting gas.

\subsubsection{Distance}
\label{sec:LI_distance}
Loop I is usually associated with the Sco-Cen OB association \citep[e.g.,][]{Salter1983} at a distance $D=120$--140\,pc \citep{deZeeuw1999}.  The primary evidence is that the NPS appears to be detectable in starlight polarization for stars at distances of $\gtrsim 100$\,pc \citep{MathewsonFord1970,Santos2011}; the aligned grains must be in the cold border rather than the spur itself. \Planck-HFI maps of polarized dust (Fig.~\ref{fig:stereo_grid}g and \citealp{planck2014-XIX}) show that emission and extinction measurements of field direction agree, confirming that the extinction distances apply to the dust features seen in emission.

\begin{figure*}
\begin{center}
  \includegraphics[width=0.5\textwidth,angle=90]{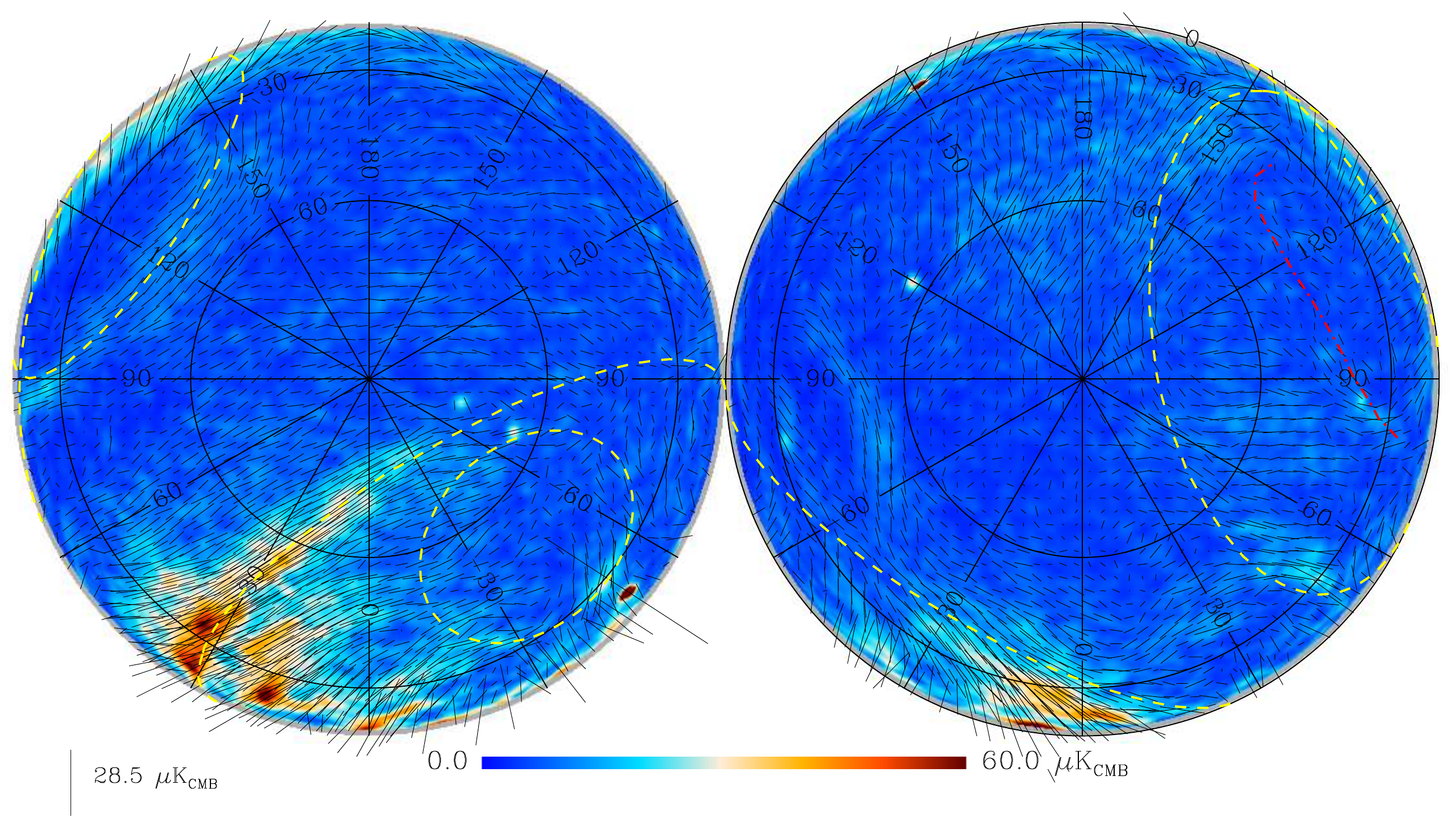}
  \caption{Line elements showing the orientation of the projected magnetic field ($90\deg$ to the synchrotron polarization angle) from the \Planck\ 28.4\GHz\ data at 2\degr\ resolution. The region within $b =\pm 10$\deg\ has been masked out (grey rings) to avoid crowding of the vectors.  The colour scale shows the polarized intensity. The maps are in orthographic projection, centred on the Galactic poles, North (left) and South (right). Light dashed circles show the outlines of Loops I to IV, as defined by \citet{Berkhuijsen1971} and shown in Fig. \ref{fig:pol_comb_maps_features}; the red dashed line is the locus of filament IIIs from \citet{Vidal2014a}. }
\label{fig:vectors_orth}
\end{center}
\end{figure*}

\citet{Sofue1977,Sofue1994} has instead suggested that the loop is due to an outburst at the Galactic centre, but in Sect.~\ref{sec:fermi} we argue that this model is now ruled out. However, evidence against the traditional distance has also been accumulating.

\citet{Iwan1980} showed that the wall of dense gas that bounds the LC would have to form a major ``dent'' in the apparently spherical structure of Loop~I, assuming $D=130$\,pc. Placing the loop several hundred parsecs away avoids the awkwardness that this dent is not reflected at all in the projected outline. If the shape is quasi-spherical, its near surface would only be about 20\,\% of the distance to the centre, so $D\approx 400$\,pc given absorption at the LC wall. It would still be plausible that stars at about 100\,pc could be polarized by dust in the border.

The interstellar medium within 200--300\,pc has now been tomographically mapped in a variety of absorption tracers against stars with \Hipparcos\ distances \mbox{\citep{Frisch2011}}. Towards the bright left side of Loop I, the LC wall is at $D=70$--80\,pc, and in the Galactic plane dense gas then extends continuously to the Ophiuchus and Lupus molecular clouds at $D=140$--160\,pc \citep[e.g.,][]{Vergely2010}. These are the remnants of the parent cloud(s) of the Sco-Cen association. Although there is a small cavity surrounding the most active part of the association \citep{Puspitarini2012}, there is no sign of a cavity with the angular size of Loop I, and this can no longer be ignored given that microwave polarization maps show that the loop crosses the Galactic plane without interruption.
\citet{Lallement2014} push the analysis out to 1 kpc by using stars with photometric distances instead of relying purely on {\sc Hipparcos}. They find a low-density cavity starting some 250 pc from the Sun, covering the longitude range of Loop I. If this is the Loop, the radius is roughly a kiloparsec, and it would stretch well into the halo. 

Given the lack of room for Loop I immediately outside the Local Cavity, Frisch has long argued \mbox{\citep[e.g.,][]{Frisch1981,Frisch2011}} that the two are continuous, and that the NPS is merely separated from us by a fold in the LC wall. However, this view is at odds with the evidence from soft X-ray absorption of the NPS. This absorption is visible in the \rosat\ map (Fig.~\ref{fig:stereo_grid}h) where the inner boundary of the X-ray NPS at $b < 40\degr$ closely follows the border of the cold material traced in panels d--g of Fig.~\ref{fig:stereo_grid}. This material can be confidently assigned to the LC wall, since it is picked up by starlight polarization at $D > 100$\,pc \citep{Santos2011}, and again extinction and emission polarization is consistent. X-ray spectra in the region are well fitted by absorbed thermal emission \citep{Puspitarini2014}. Similarly, \citet{Sofue2015} makes a strong case that the sharp truncation of the X-ray NPS below $b=10\degr$ is due to absorption in the cold clouds of the Aquila Rift. Both papers find that the hydrogen column density derived from the X-ray spectra requires a distance to the front of the spur, roughly the shell tangent-point,
of at least 200\,pc.\footnote{\citet{Sofue2015} argues for a distance of $\ga 500$\,pc, but this relies on the assumption that the Aquila Rift is at a single distance, whereas clouds at several different distances are involved, see the tomographic maps of \citet{Lallement2014}.}
The corresponding distance to the centre is $D > 400$\,pc.

Further evidence for a larger distance comes from the 21-cm polarization map of \citet{Wolleben2006}. As noted by \citet{Wolleben2007}, the emission from Loop I is strongly depolarized at $|b| \le 30\degr$. Given the cut-off in latitude, which does not correspond to any feature in the intrinsic intensity or polarization structure, the depolarization is almost certainly caused by fluctuations in the foreground Faraday depth, $\phi_F$, across the 30-arcmin beam of the Wolleben map.
The sharp cutoff in $b$ suggests that the high-latitude part of the loop projects above the dense layer of the ISM that creates the ``Faraday horizon''. At the near distance, the implied scale height, $h\approx 50$\,pc, is much too low for the most effective source of widespread Faraday rotation, the warm ionized medium (WIM) ($h = 1$--2\,kpc, \citealp{Gaensler2008}).

The Faraday depths required also imply a larger distance. Depolarization
requires fluctuations with 
$\sigma_\mathrm{F} \ga \pi/2\lambda^2 = 36 {\rm \, rad\,m^{-2}}$.  
The LC has negligible Faraday depth, due to its low density, and so the magneto-ionic medium responsible must be in the LC wall, if Loop I is directly adjacent to us, even though the wall is mainly traced by neutral gas. 
Fluctuations across a beamwidth $\theta$ requires field tangling on scales $D\theta$, and hence the Faraday depth along the line of sight of length $L$ centred at distance $D$ integrates as a random walk, with amplitude $\sigma_\mathrm{RM} = k_{\rm F} \langle n_e |B_\||\rangle \sqrt{LD\theta}$, where $k_{\rm F} =8.1\times 10^{-6} {\rm \, m\,nT^{-1}\,pc^{-1}}$. The WIM has electron density $n_\mathrm{e} \approx 1 {\rm \, cm^{-3}}$ at pressure equilibrium. 
We also assume $B=1$\,nT, with $\langle B_\|\rangle$ a factor $1/\sqrt{3}$ smaller. 
Inserting these values and taking $D = 80$\,pc, the required path length is $L \ga 70$\,pc, which places the front face of the loop beyond the the Sco-Cen association, even if the WIM has unity filling factor. Of course higher-density material would require less path and in fact the Sh\,2-27 \hii\ region (G006$+$236), ionized by $\zeta$\,Oph at $D = 112$\,pc \citep{vanLeeuwen2007}, casts a particularly deep Faraday depolarization ``shadow'' at 21\,cm, implying it is in the foreground despite a precise distance that puts it near the assumed shell centre. However, such objects are readily detected in \ha\ surveys, and cover only a small fraction of the depolarized region. The near distance can only be made compatible with these results if the polarization is dominated by the far hemisphere of the loop, as in one of the models discussed in Sect.~\ref{sec:LI_theory}.

Given all this evidence that Loop I is several times further away than the Sco-Cen OB association, we searched for an alternative
group of OB stars\footnote{We follow 
\citet{Reed2003} in defining
OB stars as main sequence stars of type B2 and earlier, and giants of type
B9 or earlier, roughly corresponding to stars massive enough to go supernova.}
to generate the Loop. Given the angle at which the NPS passes through the Galactic plane, these must be at positive latitudes; we searched  
$-45\degr < l < 15\degr$, $5\degr < b < 35\degr$. 
Apart from Sco-Cen, in this region there are no known OB associations \citep{deZeeuw1999}, but an association old enough to generate Loop I could already have lost its O stars and could easily have escaped notice.
Using SIMBAD \citep{Wenger2000}
we found 51 OB stars (mostly B giants) with parallax, $\pi$, between 
1 and 3.3\,mas; 14 with $0 < \pi < 1$\,mas; and 14 with $\pi < 0$. 
Although the individual parallaxes are barely significant, the distribution is
strongly biassed to positive values, so there certainly has been star formation
within the last 20\,Myr in the relevant volume. We will have to await 
{\em Gaia} data for a definitive assessment of stellar groupings there.

\begin{figure}
\begin{center}
\includegraphics[width=0.5\textwidth,angle=90]{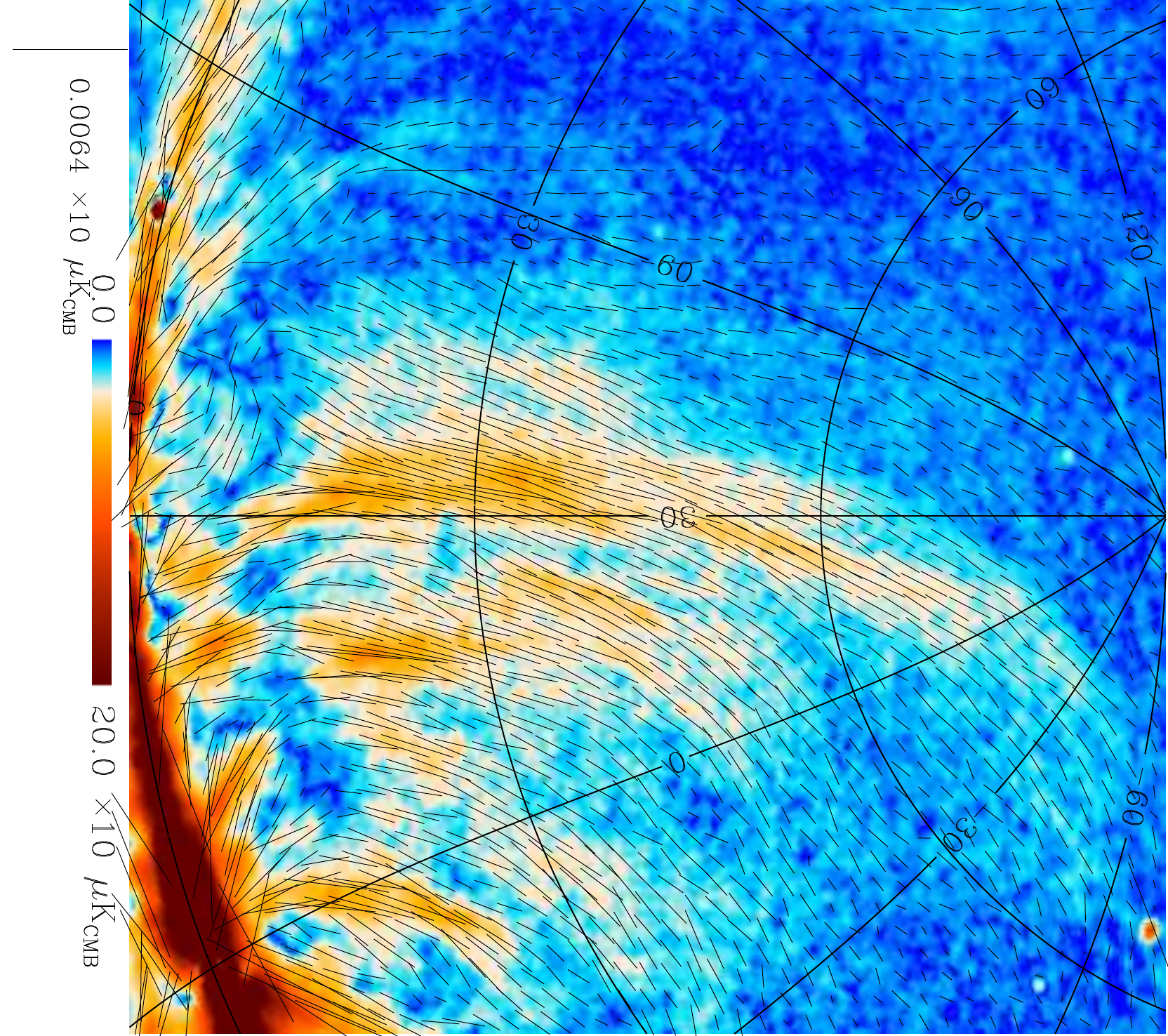}
\caption{Combined 28.4\GHz\ polarization map at 1\degr\ resolution of the North Polar spur and the region outside its perimeter, including filament X of \citet{Vidal2014a}, which leaves the Galactic plane at a shallow angle from near $l=36\degr$, and the $l=45\degr$ feature between filament X and the NPS. Colour
scale is \asinh, but the length scale for polarization line elements remains 
linear, as in all our figures.}
\label{fig:vectors_cart}
\end{center}
\end{figure}

\subsubsection{Interpretation}
\label{sec:LI_theory}

The majority view of Loop I is that it is a nearby pre-existing cavity re-energized by one or more recent supernova, in order to explain the cold neutral border seen in dust and \hi\ (see \citealp{Salter1983} for a review).

The \hi\ border appears at $V_{\rm LSR} = 0$\,km\,s$^{-1}$ (Fig.~\ref{fig:stereo_grid}d), which is uninformative, because expansion at the tangent points is perpendicular to the line of sight. \citet{Heiles1984} interpreted the border as part of a shell expanding at $\approx 25 \kms$; this implies that the obvious Sco-Cen supershell, visible most clearly at $-10 \kms$ \citep[e.g.,][Fig.~\ref{fig:stereo_grid}e]{Kalberla2005,Vidal2014a}, is merely the end-cap of Loop I itself. It is worth noting that this shell is elongated in the same direction as our proposed outline for Loop I. However, the Sco-Cen shell is more usually seen as a distinct structure with a smaller angular size than Loop I \citep[e.g.,][]{deGeus1992}. If Loop I has a much lower expansion speed, e.g., $2 \kms$ as estimated by \citet{Weaver1979}, its \hi\ emission would not be separable in velocity from the very local \hi\ surrounding the Sun. The difference between these two interpretations is dynamically significant: at 25\kms, the cold shell cannot have been accelerated by a shock, since the heating would have dissociated the gas and evaporated the dust; therefore acceleration by a pressure gradient as in the stellar wind model proposed by \citet{Weaver1979} is required. A 2\kms\ expansion is consistent with the weak remains of a shock, although of course a much faster shock would be needed to heat the interior gas above $10^6$\,K.  We also note the absence of optical emission lines characteristic of a cooling shock, which would be expected if the shock speed exceeded 10\kms. A plausible explanation is that the re-energizing blast wave has only recently hit the cavity wall, which would also explain why the NPS is so much brighter and more sharply-defined than the other large loops \citep[e.g.,][]{Borken1977}.

Figure~\ref{fig:vectors_orth} shows the projected magnetic field orientation around the Galactic polar caps. We show the \Planck\ data only, smoothed to 2\degr\ resolution, because of the suspect large-scale structure in the \wmap\ data. Figure~\ref{fig:vectors_cart} shows the pattern closer to the Galactic plane.  The quasi-parallel field pattern in the spurs has been recently discussed by \citet{Vidal2014a}. Here we wish to draw attention to the regions of organized field orientation at high latitude, {\em outside} the NPS, which parallels it quite closely over most of its length. Some of this region is occupied by the cold border, where the magnetic field may be organized by the expansion of the loop, e.g., as modelled by \mbox{\cite{Vidal2014a}}. However, Fig.~\ref{fig:vectors_orth} shows that the parallel-field region extends past longitudes of $60\degr$, well beyond the outer limit of the cold border.

The origin of this high-latitude, inter-loop polarization is unclear. \citet{Wolleben2007} sees it as emission from a ``new loop'', in whose shell the solar system is embedded, but the evidence for this loop is weak (Sect.~\ref{sec:other_loops}) and its proposed interpenetration of Loop I is, to say the least, dynamically problematic.  In a somewhat similar picture, \citet{Vidal2014a}, following \citet{Heiles1998}, place the Solar System on the surface of \mbox{Loop I}, so that loop emission fills a hemisphere: this can explain the high-latitude polarization, but it is not clear why the cold border seems to have a significantly smaller radius than 90\degr. If the loop is nearby and the polarization is from the halo, the alignment with Loop I would be an accident. However, especially if the loop is as large as suggested in Sect.~\ref{sec:LI_distance}, the emission may be from the immediate environment of the loop, suggesting that Loop I is brightest where the interstellar field is in the plane of the shell surface. This is exactly as predicted by \citet{Spoelstra1972}, who applied the model of \citet{vanderLaan1962} to the NPS. This appeals to the asymmetry of magnetic stress, which has a net tension along the field lines, so magnetic forces oppose expansion only perpendicular to the field.  The same model has been invoked to explain the occurrence of barrel-shaped SNRs aligned with the Galactic plane \citep{Gaensler1998}.

Much of the random component of the Galactic magnetic field is on scales smaller than Loop I \citep[e.g.,][]{Haverkorn2008,Brown2001}, so the external field around it is likely to show significant variation in magnitude and direction. Such large objects are therefore unlikely to show the bilateral symmetry of classic barrel-shaped remnants. Nevertheless the basic physics still operates; thus, segments of the shell expanding perpendicular to the field, along the edge of the NPS, will be more strongly impeded and so require higher internal pressure, and therefore stronger nonthermal and thermal (X-ray) emissivity. By the same token, we expect the fainter parts of the shell to expand fastest, consistent with our interpretation of the bottom-right arcs as a bulge in the loop boundary.
Of course, asymmetric densities in the ISM will also play a role, and as noted above, if the loop is at $D\approx 140$\,pc the observed density gradient is in the required sense.

In old supernova remnants the projected magnetic field is commonly aligned with the outer surface, i.e., the shock propagating through the ISM. This alignment is naturally produced by shock compression, which ensures that the field lines behind the shock tend to be parallel to the shock plane. In this picture there is no global field order in the shock plane, but the field anisotropy causes polarization at viewing angles other than face-on, reaching a maximum when the line-of-sight is in the shock plane \citep[see][]{Laing1980}. Applied to Loop I, this model implies that both the NPS and the internal spurs are tangential views of shock fronts.  There is no problem with this if the internal spurs are background features, as suggested for one of the most prominent in Sect.~\ref{sec:fermi} below.  However, if, as seems likely, at least some of the internal spurs are physically associated with Loop I, then we have to explain multiple concentric shock waves. The expected supernova rate for the Sco-Cen OB association is around 1\,Myr$^{-1}$ \citep{deGeus1992}, while remnant lifetimes are conventionally $\lesssim 30$\,kyr, and in the low-density environment of a superbubble they are likely to be shorter.  Therefore multiple active supernova remants within the superbubble are very unlikely, even if Loop I is powered by a much more massive OB association than Sco-Cen.  However, if a re-energizing blast wave has recently run into the contact discontinuity at the outer surface of the superbubble, it will separate into a transmitted and reflected shock, and for any non-ideal geometry a rather complex pattern of crossing shocks is expected in the reflected wave.  MHD simulations of this scenario would be illuminating, but are beyond the scope of this paper.

A very different model for the NPS has been proposed by \citet{Heiles1998} and is implicitly applied to the internal filaments of Loop I by \citet{Vidal2014a}. In their geometry the NPS is not a tangential view of the superbubble boundary, since the solar system is on the bubble surface. The organized field pattern is due to the wrapping of a relatively ordered interstellar field over the surface of the expanding bubble. The spurs are bundles of field lines with an enhanced density of cosmic ray electrons and therefore enhanced synchrotron emission. An attractive feature of this is that a very simple geometric model gives a good qualitative fit to the field pattern in the loop. However, the model has the peculiarity that the emission is implied to be dominated by the far side of the loop; the near side, in which the viewer is embedded, would contribute a parallel polarization over the whole hemisphere, quite unlike the observed pattern. We note that in this model, Sh\,2-27 can depolarize the loop emission even though located near the centre of the bubble.

A hybrid of the two models is possible if the solar system is moved off the loop surface in the Heiles geometry, so that the NPS resumes its usual role as the projected loop boundary. However, this would worsen the agreement of the model and observed field pattern, and is still subject to the problem that emission from the near surface of the bubble would give a much more uniform projected field.

\subsubsection{Possible contamination of the CMB by Loop I?}

\citet{Liu2014} have recently argued that emission from Loop I is contaminating current microwave background maps. Specifically, they demonstrate an alignment between the BHS small circle and positive peaks in the low-multipole ($\ell \le 20$) structure of the WMAP ILC map \citep{Bennett2013}; \citet{vonHausegger2015} confirm that the alignment is also seen in the \Planck\ CMB maps.\!\footnote{\citet{Ogburn2014}} showed that the analysis by Liu et al. underestimated the the probability, $p$, of a chance alignment, but \citet{vonHausegger2015} present a stronger statistical test that gives $p < 3 \times 10^{-3}$.  The amplitude of these peaks is about ten times larger than worst-case errors due to foreground contamination in the ILC maps estimated by \mbox{\citet{Bennett2013}}, which occur around the edge of the mask recommended for use with their ILC map at low multipoles (Liu et al. applied no mask in their analysis).  The most convincing alignment with the circle is in the southern Galactic hemisphere, including along the bottom-right section that we note above shows no sign of emission from the loop border. In general these peaks in Liu et al's low-$l$ map show little correlation with the actual synchrotron or dust emission from Loop I, especially if we mask the Galactic plane as Bennett et al. recommend.  We note that \cite{Liu2014} suggest that the contamination might be due to magnetic dipole emission from dust grains associated with Loop I. However, the securely-detected dust associated with Loop I is mostly located at larger radii than the synchrotron ridge, and would not itself give a significant signal in the analysis of \cite{Liu2014} or \citet{vonHausegger2015}.  

\begin{figure}
\centering
\includegraphics[width=0.5\textwidth,angle=0]{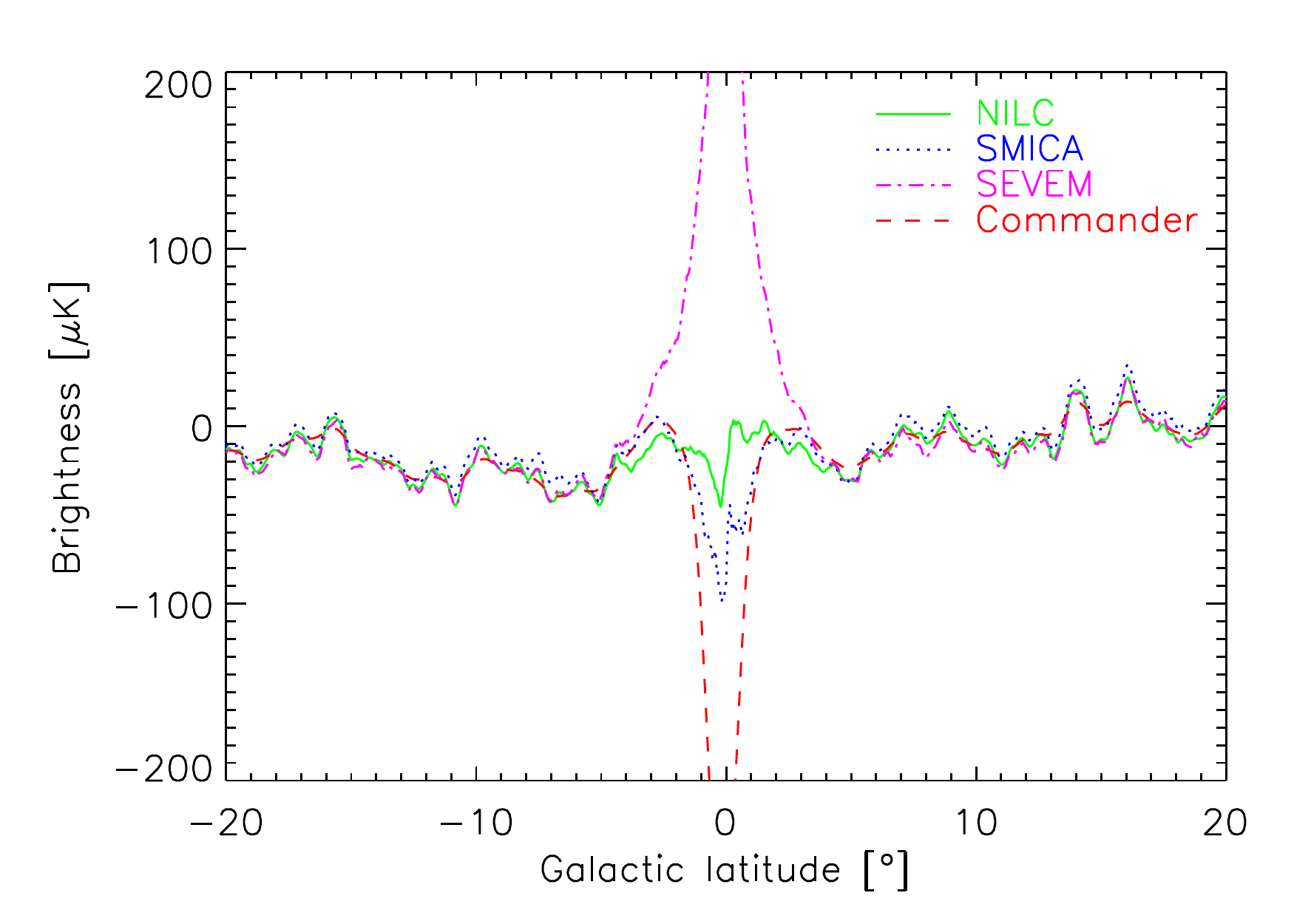}
\caption{Mean brightness vs. Galactic latitude, $b$, in the \Planck\ CMB maps 
from \citet{planck2014-a11}, in the range $|l| < 60\degr$. Solid green: {\tt NILC};
dotted blue: {\tt SMICA}; dot-dashed magenta: {\tt SEVEM} (all at 5 arcmin
resolution); dashed red: baseline 
\Commander\ model at 1\degr\ resolution \citep{planck2014-a12}.
The first three methods all use inpainting to fill masked regions around
bright compact features in some or all of the input maps: pixels
affected by this have been omitted from the averaging. High amplitudes
close to $b = 0$ are due to residual foreground contamination.}
\label{fig:lat_profile}
\end{figure}

Moreover it seems unlikely that the dust in Loop I is particularly unusual; even if we posit a unique dust type in the loops, \citet{Mertsch2013} have argued convincingly that similar structures are scattered throughout the Galactic disc and contribute a significant fraction of its synchrotron emission. By the same token, the proposed dust emission from distant loops should accumulate along the Galactic plane, but no such signal is apparent in the CMB maps. \citet{vonHausegger2015} argue that this expected signal is restricted to $|b| < 5\degr$, a region where component separation is unreliable (hence the need for a Galactic mask). While it is true that it is not possible to accurately estimate the CMB signal in this region, we can rule out foreground contamination at the level required by this model. The integrated brightness
of the distant loops is  is an order of magnitude larger than that of Loop I,
and the features responsible for the Loop I alignment are among the 
brightest in the CMB maps after filtering to $l \le 20$, with amplitudes 
50--100\,$\mu$K. Using this to scale the arbitrary units in Figure 5 of
von Hausegger et al., the brightness from distant loops on the Galactic
plane averaged over $|l| < 60\degr$ should be $\approx 900\,\mu$K, and is still over 100\,$\mu$K at $|b| = 4\degr$. Fig.~\ref{fig:lat_profile} shows profiles
of derived CMB brightness against $b$ for the four main \Planck\ component
separation methods. Except for {\tt SEVEM}, which is
the method least well adapted to removing diffuse foregrounds, substantial
contamination from Galactic foregrounds is confined to $|b| < 2\degr$. Even
the {\tt SEVEM} map has smaller Galactic residuals than predicted from the
von Hausegger et al. analysis.

In short, there is little reason to believe in the contamination suggested by Liu et al.  The alignment they suggest is with a theoretical schematic rather than the observed synchrotron loop, and is even further away from the actual dust emission in the loop. The formal improbability that the observed alignment
occurs by chance, $p \approx 0.2$\% in the best (i.e., \Planck) CMB 
maps, should be offset against the low prior probability that this schematic
accidentally aligns much better with a previously-undetected structure
than with the synchrotron loop that it was intended to model. This $p$-value
does not allow for any look-elsewhere effect such as potential alignments 
with other loops. Moreover, their proposed physical model implies additional strong contamination along the Galactic plane, which is not observed.

\subsection{Other loops and spurs}
\label{sec:other_loops}

Three other synchrotron loops that are generally agreed upon \citep[e.g.,][]{Vidal2014a}, and there are a dozen or so shorter filaments or spurs, with several mutually-incompatible proposals for joining some of them into loops. Given our emphasis on the non-circularity of Loop I, and the fact that it seems all too easy to find small circles passing plausibly close to a sequence of ridges on the sky maps, we do not find any of these newly proposed loops convincing; however, their component spurs are certainly real. We discuss the more interesting examples in turn.

\paragraph{Loop II.} Also known as the Cetus arc, Loop II is a very diffuse, 45\degr-radius structure centred at $(l,b) = (100\degr,-32\fdg 5)$. The faint, broad emission on the left-hand side of the maps in Fig.~\ref{fig:confusion} is attributed to it.  As originally defined by \citet{Large1962}, the right-hand end of the arc was considered to be the ``South Polar Spur'' (SPS), which descends from the plane at $l=45\degr$ (visible in Figs.~\ref{fig:pol_comb_maps_features} and \ref{fig:stereo_grid}). The SPS is prominent in both the unsharp masked 408\,MHz map and the synchrotron polarization map, but curves towards the right as it leaves the plane (as does its projected magnetic field), in the opposite direction from the notional path of Loop II; it therefore appears to be unrelated. The BHS small circle for Loop II just crosses the plane, but no emission from it has been traced in the northern Galactic hemisphere; the top of the loop may be obscured by Galactic plane emission, but it is also possible that the expanding shell was halted by the dense gas in the plane. Alternatively, \citet{Weaver1979} and \citet{Heiles1998} suggest that Loop III is its northern extension, although the two loops are offset by 24\degr\ in longitude.  No reliable distance information is available for Loop II.

Figure~\ref{fig:vectors_orth} shows a coherent field pattern roughly parallel to the locus of Loop II that is not visible in the \wmap\ data.  The polarized intensity essentially disappears into the noise at the bottom of the loop near the South Galactic Pole (SGP), but a coherent polarization field can be traced along a path that goes a few degrees further towards the SGP than the BHS fit, and closes a few degrees higher in longitude from the SPS.  Some of this polarization is detected at 1.4\GHz\ by \citet{Wolleben2006}, but the polarization angles are more disordered, presumably due to Faraday rotation, and so the alignment with the loop is not so clear.

\begin{figure}
\centering
\includegraphics[width=0.44\textwidth,angle=0]{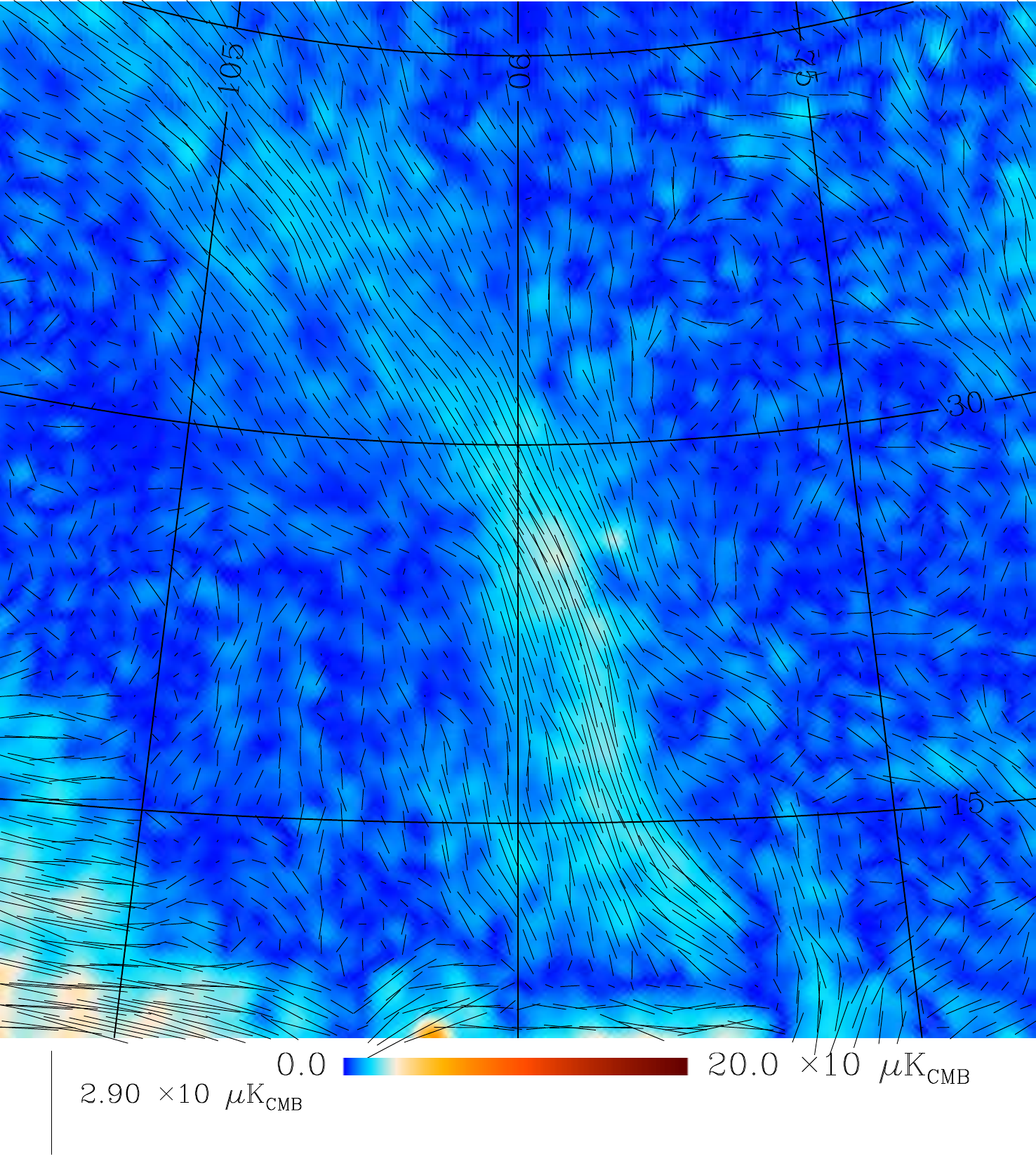}
\caption{Combined polarization map of the Cygnus spur, the right-hand end of Loop III. The map is centred at $(l,b)=(90\deg,+27\deg)$ and the graticule has a $15\deg$ spacing.}
\label{fig:cygnus_spur}
\end{figure}

\paragraph{Loop III.} Loop III is centred at $(l,b)\approx (124\degr,15\pdeg5)$;
BHS give a radius of $32\pdeg5$, although Figs.~\ref{fig:pol_comb_maps_features} and \ref{fig:vectors_orth} show that polarized emission is detected several degrees further out.  Loop III is clearly detected around most of its perimeter in the northern hemisphere in both total intensity and polarization. Its right-hand spur, rising from the plane directly above the Cyg X region, passes through the region of deep coverage near the North Ecliptic Pole $(l,b)\approx(96^{\circ},30^{\circ})$ on the \Planck\ maps, giving a particularly clean view of the well ordered magnetic field (Fig.~\ref{fig:cygnus_spur}).  This Cygnus spur is relatively narrow (FWHM $\approx 5\degr$) until it reaches $b \approx 30\degr$ after which the loop becomes much broader and fainter. As it returns to the plane its polarization is obscured by the strongly polarized Fan region ($100\degr < l < 170\degr$), in which the magnetic field is parallel to the plane and therefore orthogonal to the expected field in the loop.  BHS argue that the loop re-emerges from the fan region in the southern hemisphere at $l \approx 150\degr$, and the $B$-field pattern in our synchrotron polarization map is consistent with this, but the loop does not convincingly close in the south. \citet{Vidal2014a} imply that their filament IIIs may be associated with Loop III proper, but this is one of the regions where \wmap\ and \Planck\ disagree the most, and the existence of this filament is not confirmed in the \Planck\ maps, except for a short section near $(l,b)=(82\degr,-30\degr)$ (see Fig.~\ref{fig:vectors_orth}). \citet{Kun2007} discusses possible interactions between Loop III and the cold gas north of the plane, notably the North Celestial Loop \citep{Meyerdierks1991} that may be swept-up material.

\paragraph{Loop IV.} Loop IV is centred at $(l,b)=(315\degr,48.5\degr)$ and is projected entirely inside Loop I.  Only an arc subtending $\approx 20\degr$ along its low-latitude rim is clear in our polarization map; see \citet{Vidal2014a} for analysis. The high-latitude rim of Loop IV that parallels the top of the NPS is very noisy here but clear in the $21$\,cm polarization map of \citet{Wolleben2006}.

\paragraph{The South Polar Spur (SPS) and Filament X.} As discussed above, the SPS was once seen as the low-longitude edge of Loop II, but in fact curves in the opposite direction so the outside edge of the SPS is to its right. It also forms the most prominent segment of the S1 loop proposed by \citet{Wolleben2007}, and was labelled filament VIIb by \citet{Vidal2014a}. Wolleben's view of the SPS was obscured by Faraday rotation near the plane; as noted by \cite{Vidal2014a}, the high frequency ($\gtrsim 20$\GHz) polarization structure of the SPS is more strongly curved near the plane than expected for Wolleben's path. The other section identified by Wolleben as part of his new loop is usually seen as one of the interior ridges of Loop I and the northern part of Loop IV; thus, all components of loop S1 appear to be parts of smaller structures and there is no clear evidence that S1 exists as a coherent physical structure.

\begin{figure}
\begin{center}
  \newcommand{\widthfig}{0.24} \newcommand{\angfig}{0}
  \includegraphics[angle=\angfig,width=\widthfig\textwidth]{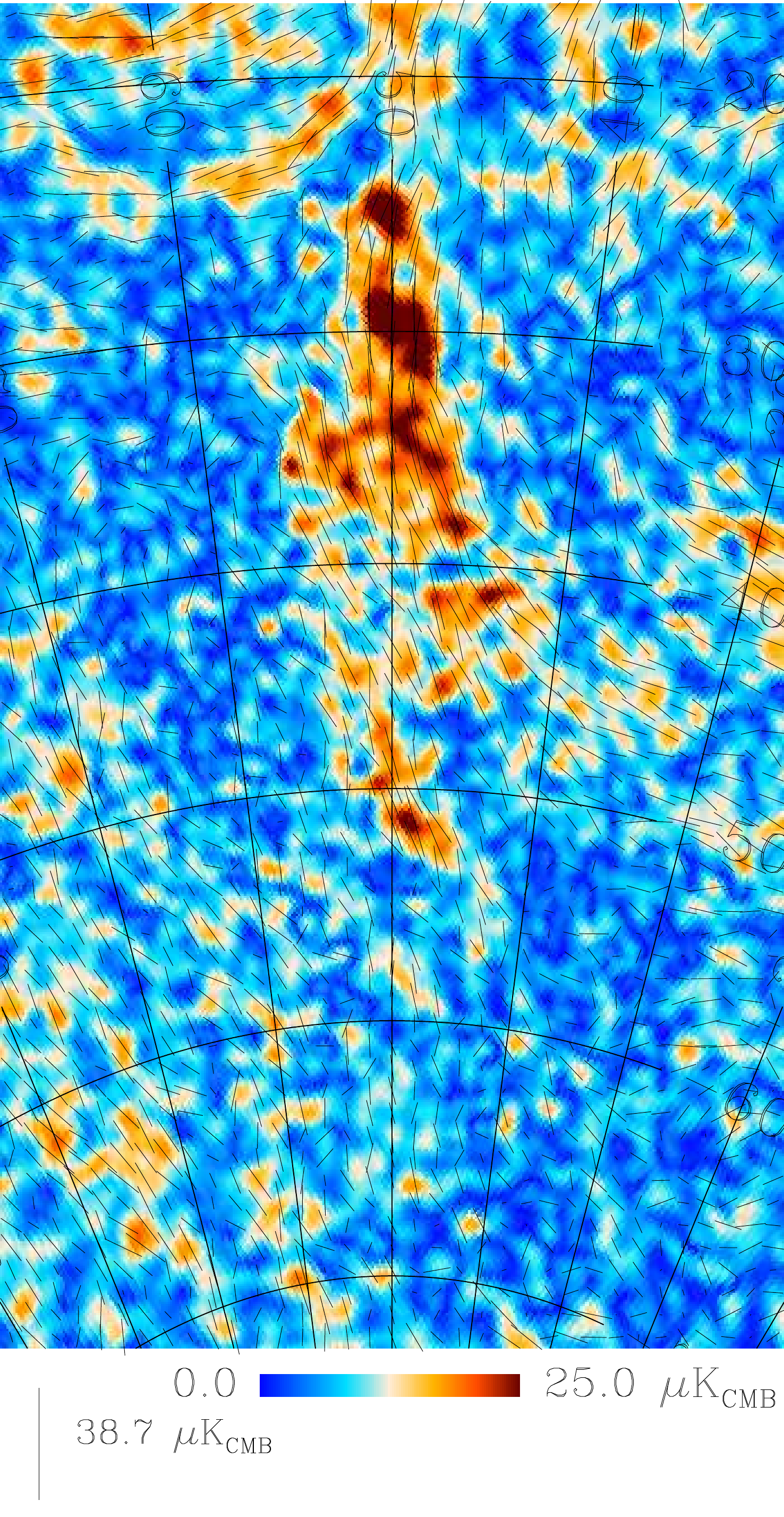}
  \includegraphics[angle=\angfig,width=\widthfig\textwidth]{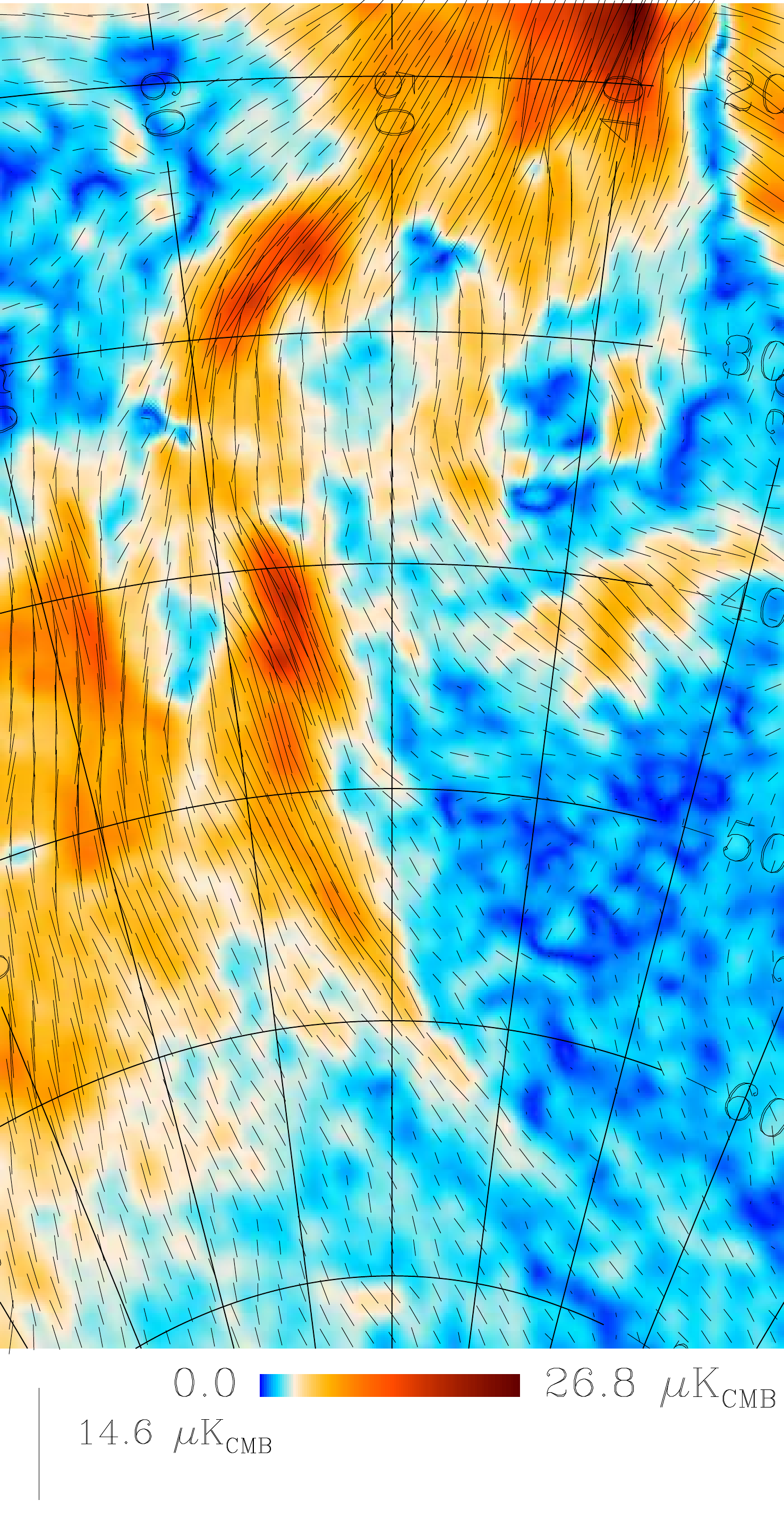}
 \caption{The South Polar Spur. {\it Left}: \Planck-\wmap\ polarization map; {\it right}: \Planck\ 353-GHz dust polarization. The maps are centred at $(l,b)=(50\deg,-45\deg)$ and the graticule has a $10\deg$ spacing. }
 \label{fig:SPS}
\end{center}
\end{figure}

The SPS has its own cold border just outside the radio ridge, at $l \approx 60\degr$, $-30\degr > b > -70\degr$, which is particularly clear in dust polarization (Fig.~\ref{fig:SPS}). The border is visible in \hi\ at $0 < v_\mathrm{LSR} < 10$\,km\,s$^{-1}$ (Fig.~\ref{fig:stereo_grid}d,e), which implies a distance of less than a few hundred parsecs.

Given our discussion above, we are reluctant to define a loop by extending a small circle from the clearly-detected arc of the SPS. In fact, a slightly fainter spur immediately north of the plane (Vidal et al.'s filament X) seems to be a reflection-symmetric twin of the SPS, suggesting a wasp-waisted cavity shaped by the dense gas in the plane (see the left-hand edge of Fig.~\ref{fig:stereo_grid}c).

Our map allows filament X to be traced further north, to \mbox{$b= 37\degr$}, where it curves over and enters the region of bright diffuse polarized emission just outside the NPS (Fig.~\ref{fig:vectors_cart}). Unusually, the field in this northern section is perpendicular to the filament. This alignment is not due to a contribution from the underlying diffuse emission; if it were, the filament would appear as a trough (due to cancellation), not a ridge in polarized intensity. A possible return section is marked in Fig.~\ref{fig:pol_comb_maps_features}.

\paragraph{The Orion-Taurus Ridge and Filament XI.} 
Unlike the other spurs, these features do not emerge from the Galactic plane but run parallel to it at closest approach, at $(l,b) \approx (220\degr, -15\degr)$, and $(l,b) \approx (230\degr,40\degr)$, respectively (see Fig.~\ref{fig:pol_comb_maps_features}). They are visible in both unsharp-masked total intensity and in polarization (see figure~2 of \citealp{Vidal2014a}). \citet{Milogradov-Turin1997} and \citet{Borka2008} propose the existence of a pair of very large, overlapping loops (V and VI in their designation) that connect a sparse collection of features including the right-hand end (at $l < 180\degr$) of the Ori-Tau ridge and Vidal et al.'s filament IIIs. However, the rest of the Ori-Tau ridge curves far away from the locus of Loops V/VI, and the magnetic field continuously follows that curvature, so these loops appear to be spurious.

Filament XI is an approximately 80\degr-radius arc that seems to curve away from the plane at both ends (Fig.~\ref{fig:pol_comb_maps_features}); however, it is unclear if features near $l = 264\degr$ belong to the left-hand end of Filament XI or the right-hand side of Loop I (or filament XII, if not part of Loop I). Extended to a full loop, Filament XI encloses most of Galactic quadrant 3 north of the Plane, which is one of the faintest regions of the polarized sky; over much of this quadrant no polarization is clearly detected by \planck\ and/or \wmap, even at $4\degr$ resolution.

\paragraph{$l=45\degr$ feature.} This is the highly-polarized patch covering \mbox{$12\degr < b < 45\degr$} and from the edge of the NPS to $l \approx 50\degr$ (Fig.~\ref{fig:vectors_cart}). The low-$b$ edge of the polarization feature coincides with a rapid brightening of the inner halo in total intensity, and when polarization re-appears at $b\approx 8\degr$ it has an orthogonal orientation, roughly parallel to the Galactic plane. Hence this minimum is due to a cancellation between the plane-parallel field in the inner halo and the near-vertical field in our feature, and does not mark the physical edge. A faint ridge runs through the top half of the feature at $l\approx 45\degr$, which is just visible in the 408\,MHz map (Fig.~\ref{fig:stereo_grid}a,b); otherwise there is little trace in total intensity. The magnetic field runs at a slight angle to the ridge of the NPS throughout; in the adjacent section of the NPS the projected field is misaligned with the ridge-line and in better agreement with that in the $l=45\degr$ feature. Plausibly the feature overlaps in projection with Loop I, but if so the current data do not allow us to trace it inside the loop.

\begin{figure}
   \centering   
  \includegraphics[width=0.45\textwidth,angle=90]{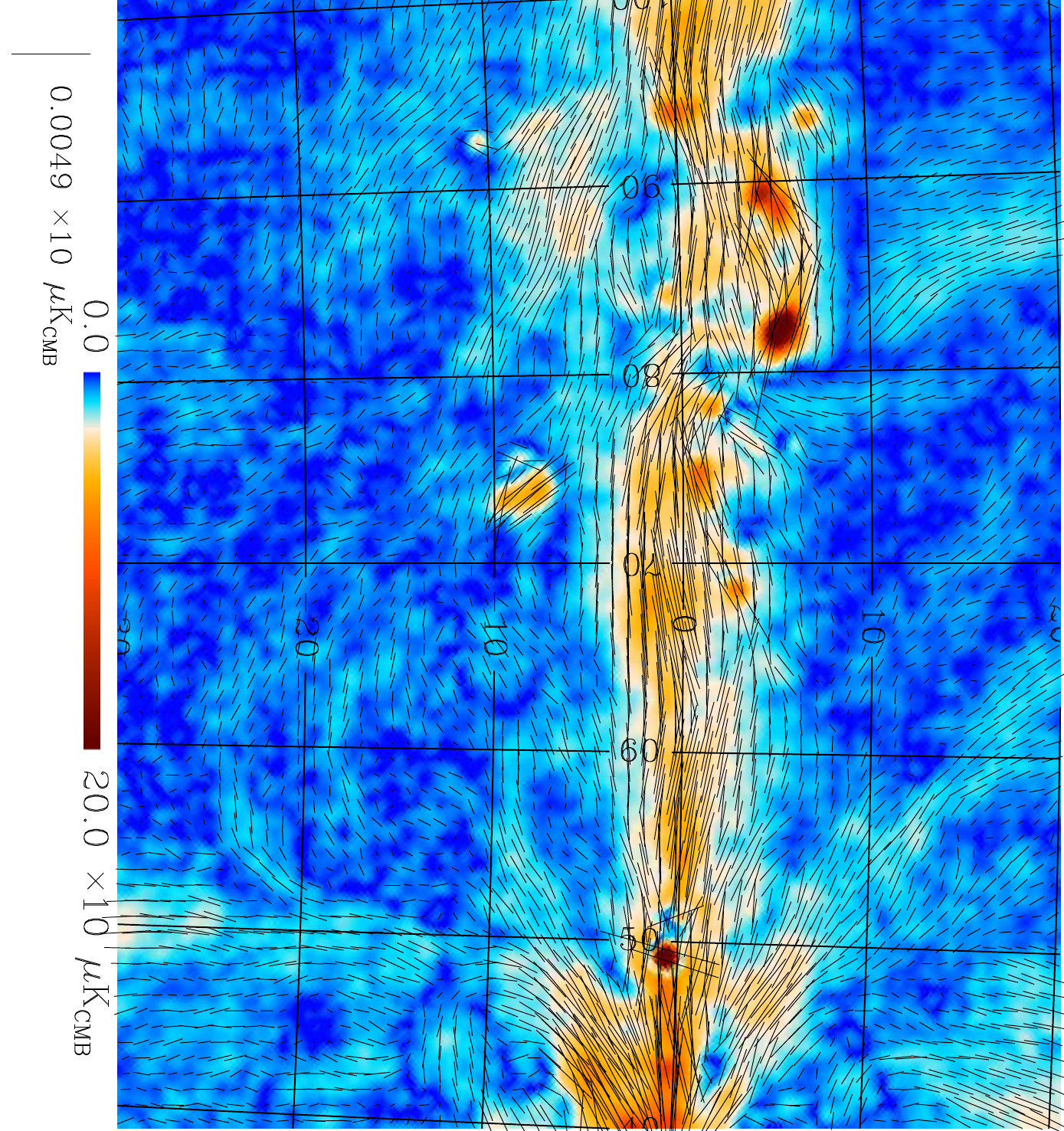}
  \includegraphics[width=0.45\textwidth,angle=90]{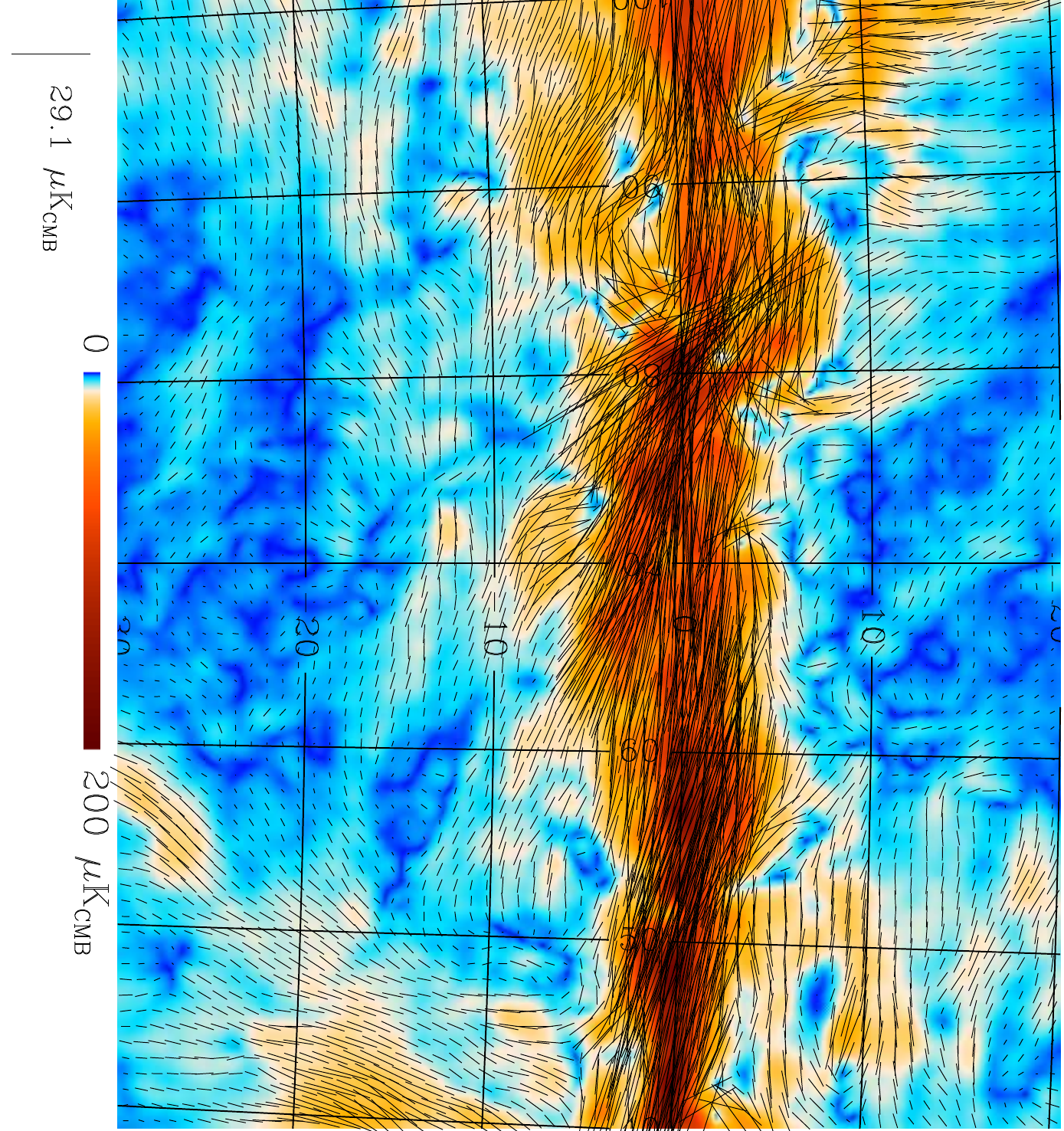}
  \caption{{\it Top:} combined polarization map with magnetic field vectors overlaid, centred on $(l,b)= (70\degr,-7\fdg5)$ showing several polarized arches south of the plane. Bright compact features along the plane are all SNRs: DA\,530 (G93.3$+$6.9); HB\,21 (G89.2$+$4.7); W\,63 (G82.2$+$5.3); CTB\,87 (G74.9$+$1.2); the Cygnus Loop (G74.0$-$8.5); CTB\,80 (G69.0$+$2.7); and W\,51C (G49.1$-$0.6). The large spur at $l\approx 50\degr$ is the SPS.  {\it Bottom:} dust polarization in the same region as seen at 353\GHz.  Maps are rectangular projections and have an \asinh\ colour scale.}
\label{fig:arches}
\end{figure}

\paragraph{Smaller loops.}
\citet{Mertsch2013} have emphasized that analogues to the large synchrotron loops should be common, and more distant and smaller examples should therefore be visible in the sky maps close to the Galactic plane. One rather clear example is the polarization structure south of the Cyg\,X region, centred at $(l,b) = (90\degr, -5\degr)$, which seems to be a superposition of a 5\degr -radius loop around this bright star-forming complex and another spur heading to the lower right (Fig.~\ref{fig:arches}).  The radius is about 120\,pc, assuming a distance of 1.4\,kpc for Cyg\,X \citep{Rygl2012}. A northern counterpart is less obvious, partly because of the two bright supernova remnants, HB\,21 and W\,63, that lie on the likely path. Figure~\ref{fig:arches} also shows another possible distant loop: the set of three arches in the southern hemisphere, centred near $l = 60\degr$;
they are also marked in Fig.~\ref{fig:pol_comb_maps_features}. The coherence of the polarization vectors allows the outer arch to be followed around to $(l,b) = (17\fdg4,-12\fdg9)$, just south of the Cygnus Loop supernova remnant.  While the polarized intensity suggests that the inner arch returns to the plane at $l \approx 61\degr$, the polarization vectors are orthogonal to the apparent ridge-line, and so this might be a different structure. Figure~\ref{fig:arches} also shows that the dust polarization has the same magnetic pattern as the synchrotron emission around Cyg\,X. There is also a polarized dust filament that runs roughly along the inner edge of the middle $l=60\degr$ arch, while the outer arch parallels the dust in polarization as it runs behind the SPS.

\subsection{Fermi bubbles in polarization}
\label{sec:fermi}

\begin{figure*}
  \begin{center}
    \includegraphics[width=0.62\textwidth,angle=90]{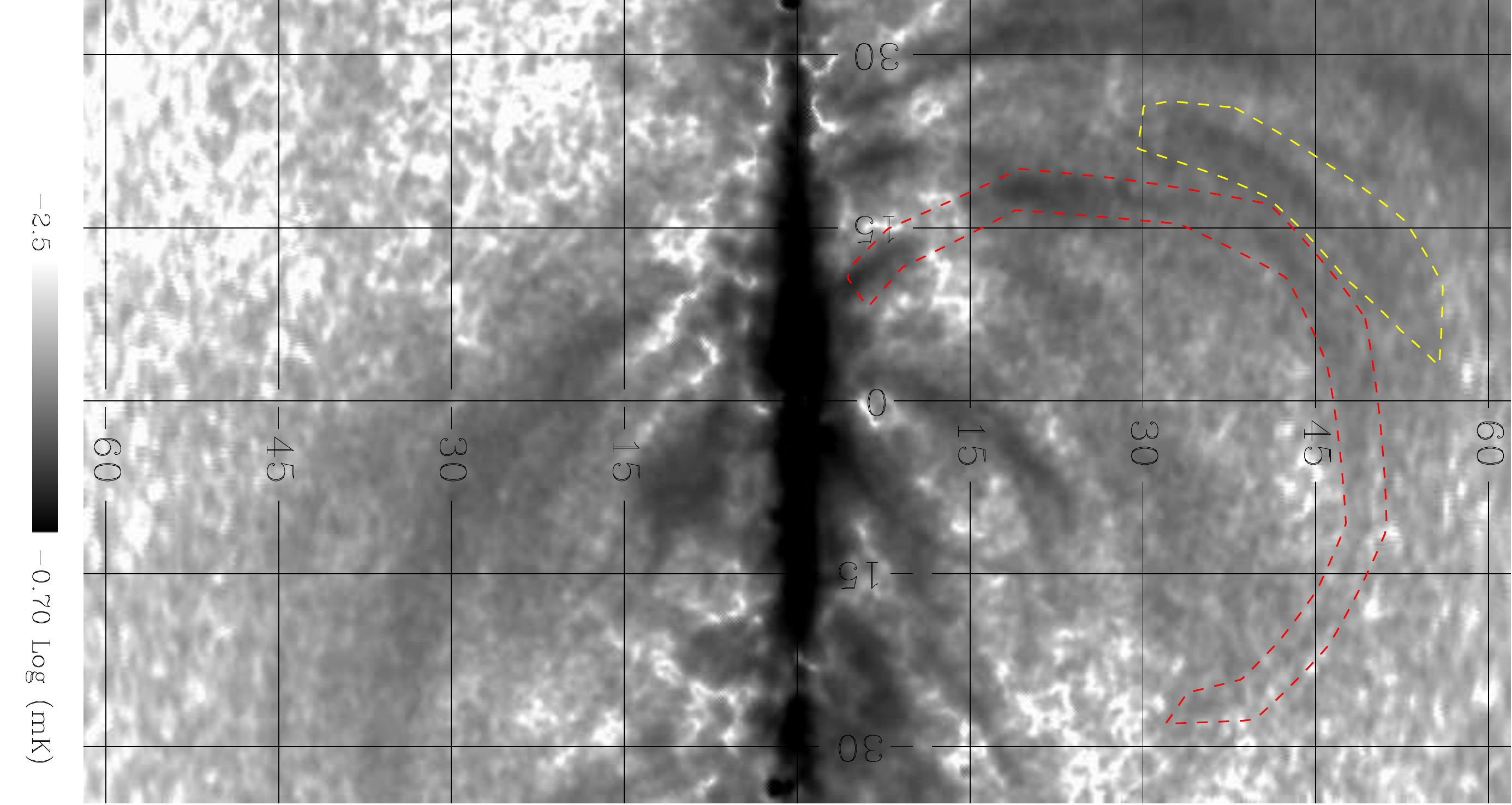}
    \includegraphics[width=0.62\textwidth,angle=90]{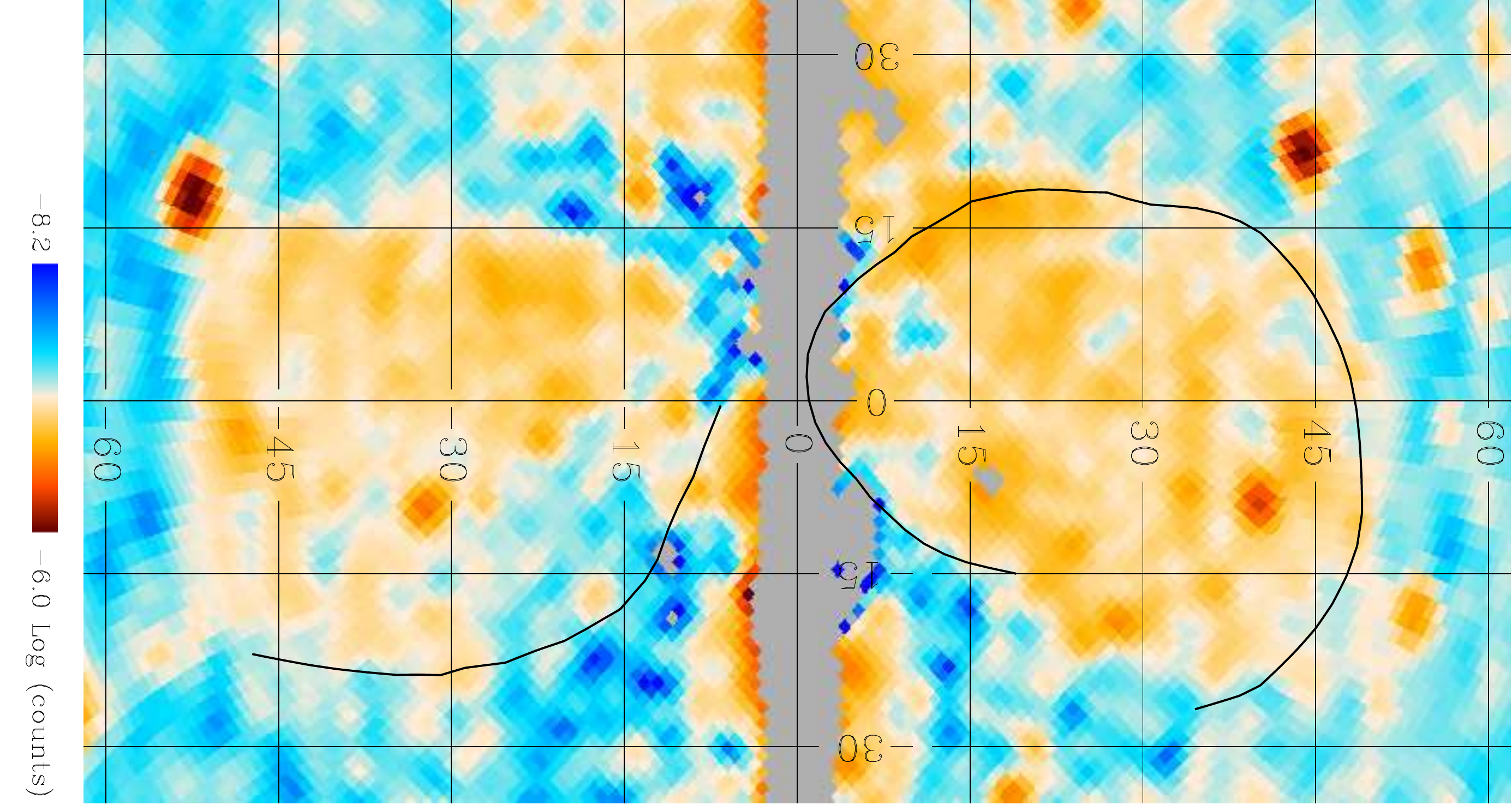}
    \includegraphics[width=0.62\textwidth,angle=90]{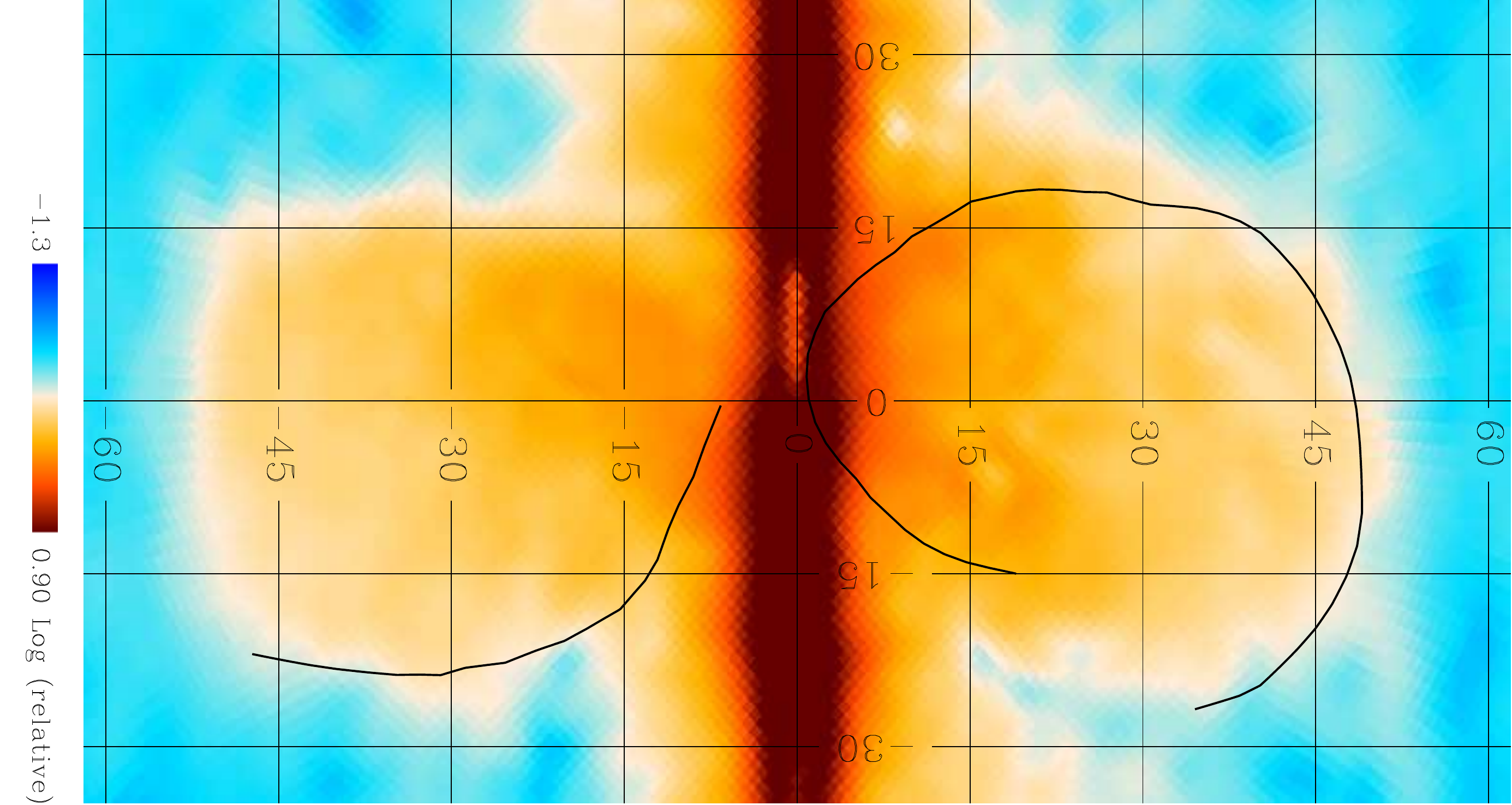}
  \end{center}
  \caption{{\it Left:} cartographic projection of the combined \planck\--\wmap\ polarization amplitude, centred at the Galactic centre, with graticule lines at 15\deg\ intervals. The regions defined by red and yellow dashed lines were selected to calculate a polarized spectral index; in red is the filament around the \fermi\ bubble and in yellow a control area. {\it Centre:} cartographic projection of the 10--500\,GeV map from \citet{Ackermann2014}, with the $\pi^0$ emission
subtracted as described in the text, showing the \fermi\ bubbles. {\it Right:} \fermi\ bubbles component from \citet{Selig2015}. The black outline corresponds to the centre of the narrow filaments visible in the polarization map on the left. }
    \label{fig:planck_fermi}

\end{figure*}

\begin{figure}
\begin{center}
  \includegraphics[width=0.49\textwidth]{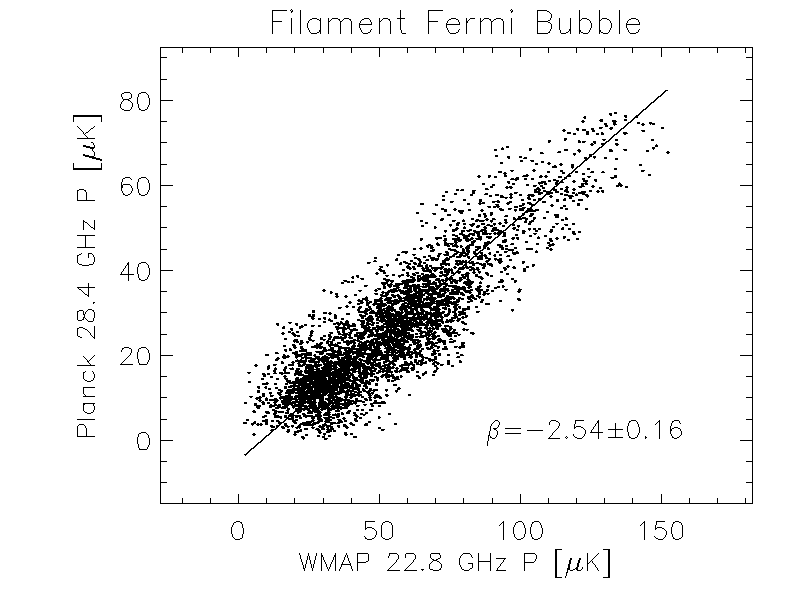}
  \includegraphics[width=0.49\textwidth]{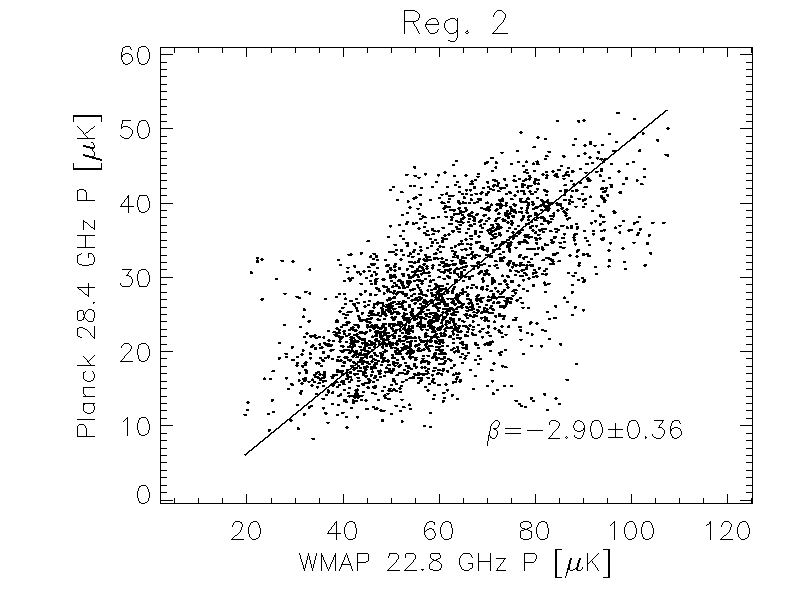}
  \caption{\ttp\ plots between the polarization amplitude maps of \wmap\ K-band and \planck\ 28.4\GHz. The two plots correspond to the regions defined in the left panel of Fig.~\ref{fig:planck_fermi}. The \fermi\ bubble filament shows a flatter spectrum than the control region. }
\label{fig:fermi_ttplots}
\end{center}
\end{figure}

The \fermi\ bubbles are two large structures extending perpendicular to the Galactic plane, up to about $\pm55$\deg. They were first discovered by \citet{Dobler2010} using spatial templates while searching for a $\gamma$-ray counterpart to the microwave haze \citep[e.g.,][]{Finkbeiner2004,planck2012-IX} in the \fermi\ data.

The origin of the bubbles is not clear, although their location and symmetry with respect to the Galactic centre suggest that they originate there. They have a $\gamma$-ray spectrum significantly harder than the inverse Comptom emission from the Galactic Halo or the one from pion decays from collisions of CR, with the ISM protons and heavier nuclei \citep{su:10}. Different models have been proposed to explain their origin, most of them relating to a recent AGN-type activity at the Galactic centre. For a recent overview on the bubbles and their spectral behaviour in $\gamma$-ray, see \citet{Dobler2012} and \citet{Ackermann2014}.

We compare our combined \planck\--\wmap\ polarization amplitude map at 28.4\GHz\ with a full-sky \fermi\ map from \citet{Ackermann2014}. The $\gamma$-ray map was produced using 50 months of \fermi\ LAT data \citep{Atwood2009}, using only the ``UltraClean'' class events, i.e., the sample of events with less contamination from misclassified interactions in the \fermi\ LAT instrument. See \citet{Ackermann2014} for a detailed description on the \fermi\ data analysis. We use the high energy map, which covers 10--500\,GeV. In this energy range, the emission from the \fermi\ bubbles appears clearly without any component separation at latitudes $b \gtrsim 20$\deg. 
The main foreground emission in gamma rays is from decay of $\pi^0$ particles produced by CR protons interacting with the ISM \citep{Ackermann2012}. If the CR density and spectrum are roughly spatially uniform over the Galaxy, then the $\pi^0$ emission will be proportional to the ISM column density \citep{su:10}. A good tracer for the ISM column density are maps of thermal emission from dust grains, since dust is well mixed in the ISM and its emission is optically thin. Here, we use the \planck\ 353\GHz\ optical depth map from \mbox{\citet{planck2013-p06b}} as a column-density proxy for $\pi^0$ emission to fit and remove it from the $\gamma$-ray map. 

In the middle panel of Fig.~\ref{fig:planck_fermi}, we show the resulting 10--500\,GeV \fermi\ map. A region of about $\pm 5$\deg along the Galactic plane, where the subtraction of the scaled \planck\ 353\GHz\ map left large residuals, has been masked out and is shown in grey. The left panel in Fig. \ref{fig:planck_fermi} shows our combined \planck\--\wmap\ polarization amplitude map, where narrow filaments are visible that correspond to the border of the \fermi\ bubbles. We note that the northern filament is much clearer than the southern one. The ridge-lines of these filaments is over-plotted in the \fermi\ map on the right to show the correspondence. It is remarkable to see how closely the polarized filament follows the border of the \fermi\ bubbles. We also show on the right panel the denoised and source-subtracted \fermi\ bubbles template constructed by \mbox{\citet{Selig2015}} using a Bayesian inference algorithm. Their reconstruction has a poorer angular resolution than the less processed version we show at the centre of Fig.~\ref{fig:planck_fermi}, but still it is clear that the synchrotron filaments outline the border of the bubbles.  The polarized filaments are unresolved in the \planck\ polarization maps (40\arcm\ FWHM beam), implying that these are very narrow synchrotron structures.  This is further confirmation of a sharp border for the \fermi\ bubbles, and models that fail to predict this feature are now less favoured (see \citealp{Dobler2012} for a description of some of these models).

\paragraph{Polarization spectral index.}
We have measured the polarization spectral index, $\beta_{\rm pol}$, between the bias-corrected \wmap\ 23\ghz\ and \planck\ 30\ghz\ using \ttp\ plots of a region that encompasses the north filament. The left panel of Fig. \ref{fig:planck_fermi} shows two regions in the polarization intensity map that we selected to measure $\beta_{\rm pol}$, one including most of the northern filament (in red) and a second region that we use as control (yellow).

The \ttp\ plots are shown in Fig. \ref{fig:fermi_ttplots}. The polarization spectral index of the \fermi\ bubble filament is $\beta_{\rm pol} = -2.54\pm0.16$, while the nearby control region has $\beta_{\rm pol} = -2.90\pm0.36$. The spectral index of the region that includes the filament is much flatter than the standard value of $-3.0$ normally assumed for diffuse synchrotron emission. It is also flatter than all the values presented in \citet{Fuskeland2014} and \mbox{\citet{Vidal2014a}}, who explored the polarized spectral indices in different regions using \wmap\ data. This flatter spectral index indicates that the energy distribution of the synchrotron radiation along the filament has more higher energy electrons than that of the diffuse component next to it. The polarization spectral index of the \fermi\ bubble filament we measure here is identical to the spectral index of the microwave haze of $\beta_{\rm Haze} = 2.54\pm 0.05$ \mbox{\citep{planck2012-IX}}. This supports the relationship between the filament and the haze/bubble emission.

\paragraph{\fermi\ bubbles and Loop I.}
A relationship between Loop I (see Sect.~\ref{sec:loop_I_structure}) and the \fermi\ bubbles has been hypothesized \citep[e.g.,][]{Kataoka2013}. This would put Loop I at the Galactic centre, with a much larger size and radio luminosity, as long argued by Sofue \mbox{\citep[e.g.,][]{Sofue1977,Sofue1994,Sofue2015}.}

With the new \planck\--\wmap\ combined polarization map, in Sect.~\ref{sec:loop_I_structure} we trace Loop I below the Galactic plane. It is clear from the top panel of Fig.~\ref{fig:pol_comb_maps_features} and from Fig.~\ref{fig:planck_fermi} that the southern part of the \fermi\ bubble extends outside Loop I, and there is no trace of an interaction with the bubble in the radio maps, given the continuity and smoothness of the Loop I polarized filaments in the southern hemisphere. The two structures must therefore be at different distances. The \fermi\ bubbles are well centred on the Galactic centre, show a pinched structure symmetric about the Galactic plane, as expected for an outburst from the nucleus \citep{Sofue1994}, and are unique in the $\gamma$-ray sky. All this makes it highly likely that they are located at the distance of the Galactic centre, as usually assumed. In contrast, Loop I is centred 35\degr\ from the nucleus and significantly above the plane, while on its left-hand side, where it can be clearly followed, it appears to extend through the plane without any sign of deviation (Sect.~\ref{sec:loop_I_structure}). This is just as expected if it is relatively nearby and embedded in the disc rather than extending far beyond it.  Of course, it also has a larger angular size than the \fermi\ bubbles, and there are at least several similar loops (Sect.~\ref{sec:other_loops}). We are therefore confident that Loop I is a foreground feature.

\subsection{An anti-correlation of an \ha\ filament and polarized intensity}
\begin{figure}
  \centering
  \newcommand{\widthfig}{0.24}
  \newcommand{\angfig}{90}
  \includegraphics[angle=\angfig,width=\widthfig\textwidth]{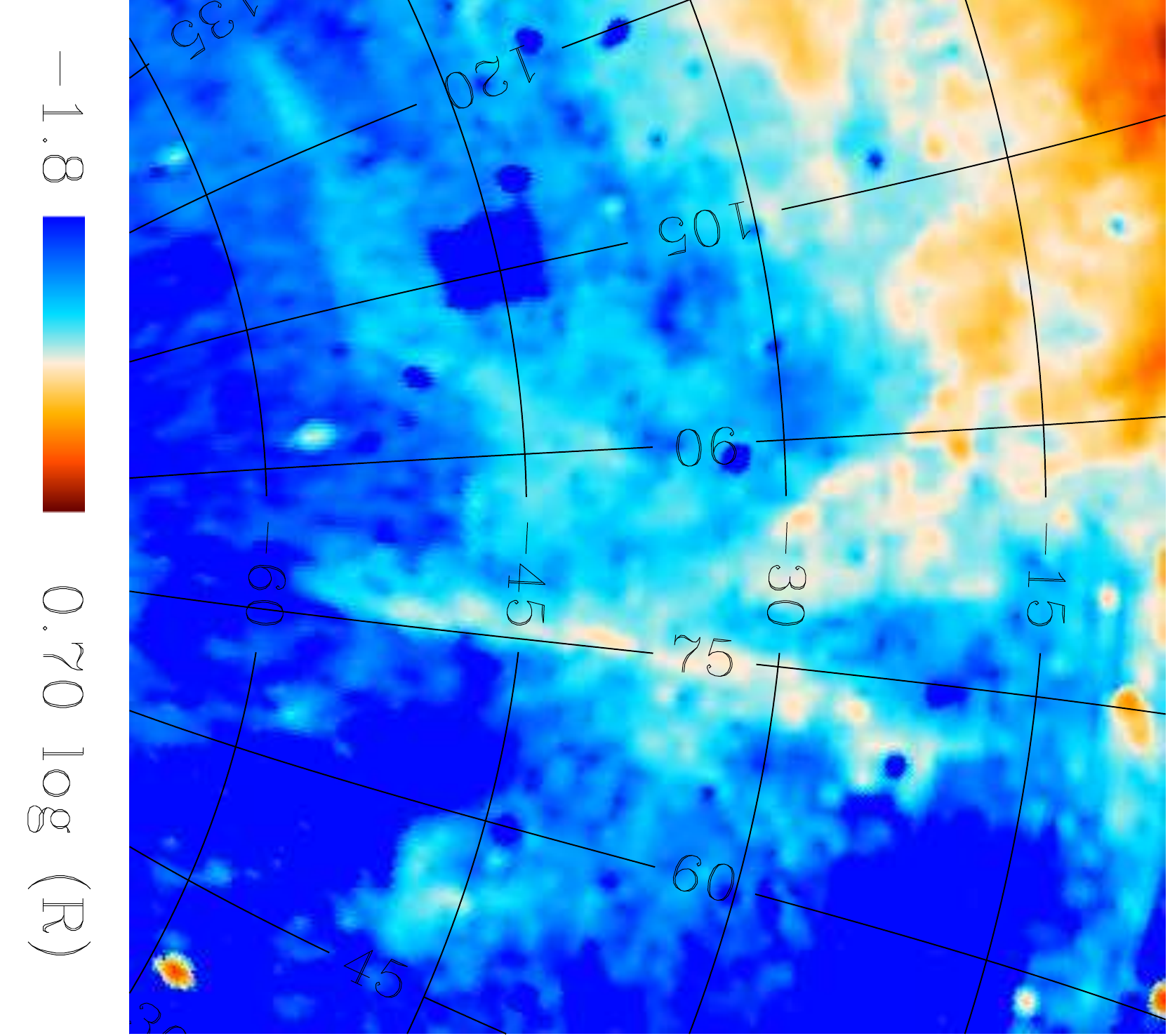}
  \includegraphics[angle=\angfig,width=\widthfig\textwidth]{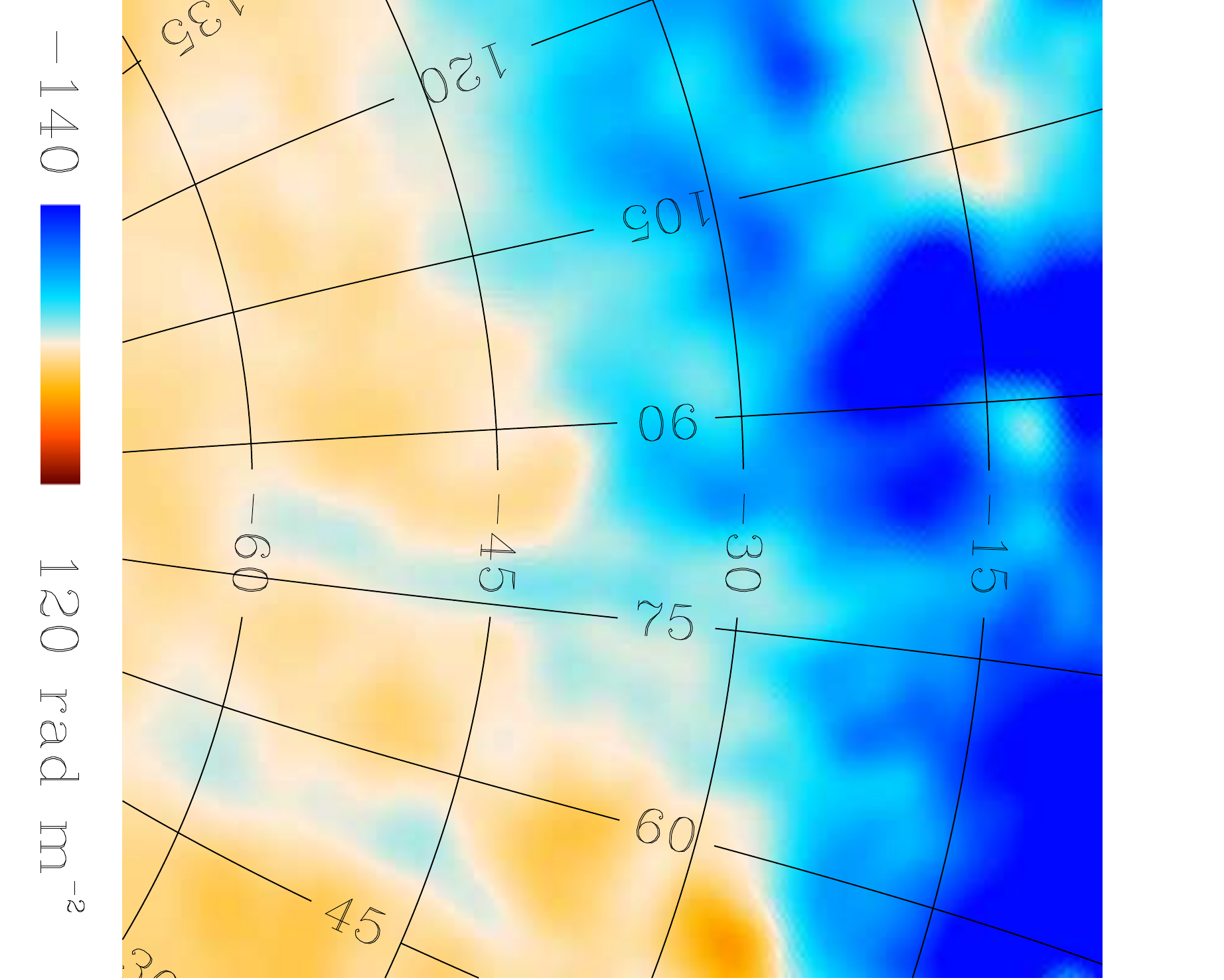}
  \includegraphics[angle=\angfig,width=\widthfig\textwidth]{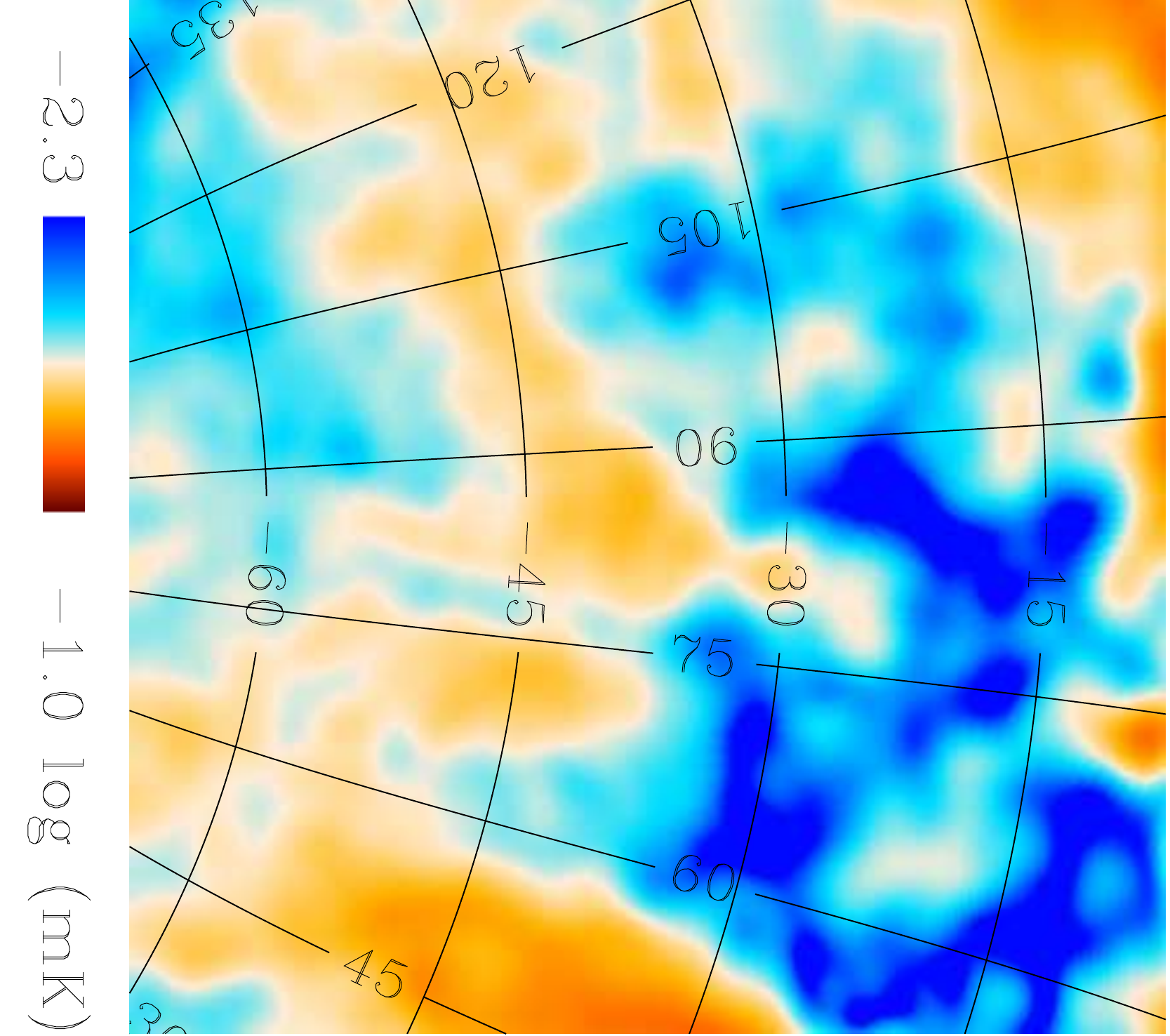}
  \includegraphics[angle=\angfig,width=\widthfig\textwidth]{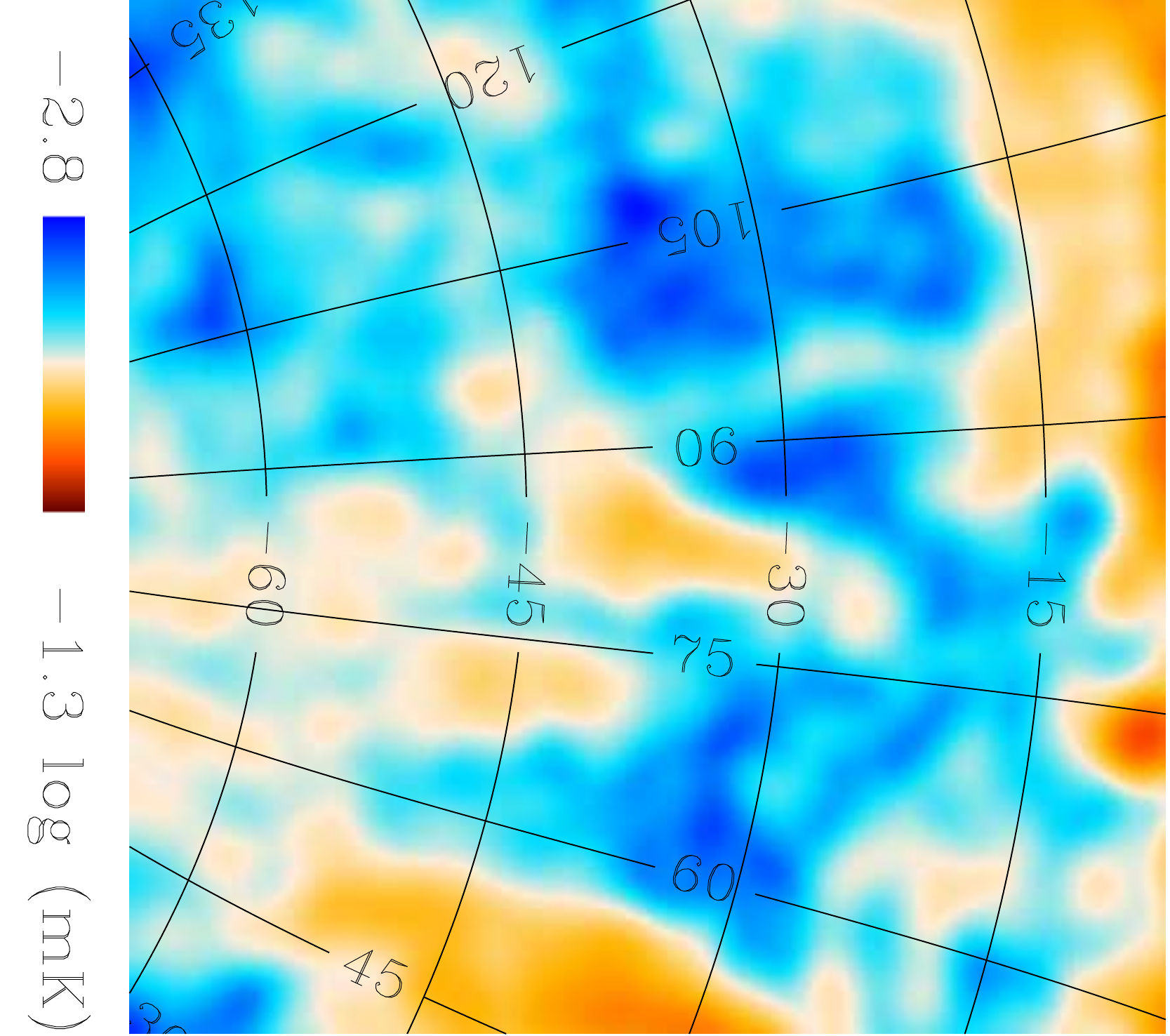}
  \caption{{\it Top left:} \ha\ map in the velocity range $-80 <
    V_\mathrm{lsr}<-40$\,km\,s$^{-1}$. Notice the vertical filament that
    runs at $l \approx 75\mdg$ and $ -60\mdg < b < -20\mdg$. {\it Top right:} Faraday depth map from \protect\cite{Oppermann2012},
    which also shows a filament at the same location, with a mean
    value of $-25$\,rad\,m$^{-2}$ along its extension. The filament
    has a counterpart in ``absorption,'' visible as a trough in
    polarization intensity maps. {\it Bottom left:} \wmap\ 23\GHz\
    polarization intensity map. {\it Bottom right:} \planck-\wmap\
    polarization intensity map. The \ha\ map on the Top left corner
    has an angular resolution of 1\deg\!, while the other three maps
    have a common resolution of 3\deg. The grid spacing is
    15\deg\!\!.}
  \label{fig:halpha_fil}
\end{figure}

Another interesting aspect that can help in the determination of the distance to some of the polarized structures is the discovery of a region that shows an anti-correlation between \ha~emission and the polarized intensity in our combined \planck\--\wmap\ map. This has not been noticed before as far as we are aware, because the filaments are located at high Galactic latitude, in the Faraday-thin regime.

In Fig.~\ref{fig:halpha_fil} we show an \ha~map, integrated over the velocity range $-80<V_\mathrm{lsr}<-40$\,km\,s$^{-1}$, from the Wisconsin H-Alpha Mapper (WHAM) survey \citep{Haffner2003}. The filament that runs along $l \approx 75\deg$ is about 40\deg\ in length.  The \wmap\ 22.8\GHz\ polarized intensity, and the \planck\--\wmap\ polarized intensity map also shown on Fig.~\ref{fig:halpha_fil}, show a trough at the same location as the \ha\ filament. This feature is also visible in the Faraday depth map from \citet{Oppermann2012}, shown on the same figure. The Faraday depth, $\phi$ is proportional to the magnetic field along the line of sight and the electron density:
\begin{equation}  
  \phi = k_F \int_{r_0}^0 \! \mathrm{d}r\, n_{\rm e}(r)B_{r}(r).
\end{equation}

We have ruled out a chance correlation since the polarized feature is also clearly visible, at higher angular resolution, in the maps from the GALFACTS \citep{Taylor2010} consortium (Jereon Stil, priv. comm.), and the correlation is still strong at angular scales of a few arcmin.

The origin of the observed anti-correlation between \ha\ intensity and polarization amplitude is not clear.  A first possibility is that the trough visible in the combined \planck-\wmap\ polarization map is a depolarized region, meaning that the \ha\ filament lies in between the polarized background emission and us.  If this is the case, the ionized gas traced by the \ha\ map could produce Faraday rotation, due to the presence of a magnetic field in the plasma, depolarizing the diffuse background emission along its extension. However, this hypothesis is not compatible with the low density of the ionized gas.  The intensity of the filament in \ha\ is 2\,R above the background at the original resolution (6\arcm) of the \citet{Finkbeiner2003} \ha\ map. This corresponds to a mean electron density of 2.0\,cm$^{-3}$ (assuming that the filament has a thickness of 1\,pc), which is very low to produce significant Faraday rotation at 23\GHz\ for typical values of the magnetic field. Moreover, the Faraday depth map from \citet{Oppermann2012} has a value of about 25\,rad\,m$^{-2}$ along the filament, which corresponds to a $0\mdgp3$ change in polarization angle at 23\GHz. Therefore, Faraday rotation of this high latitude filament is not enough to cause any major depolarization at 23\GHz.

A second alternative is that there could be a strong coherent magnetic field parallel to the line of sight along the filament. This might be the result of the dynamical processes that created the filament. If this is the case, there would be less polarized emission at the filament location than around it.

Lastly, it might well be that the synchrotron emission from the region of the \ha\ filament is intrinsically weakly polarized. This might be due to a less organized magnetic field in this region in comparison with the diffuse emission seen in the vicinity of the filament on the polarization map.

We also note the fact that the \ha\ filament is only visible at negative radial velocities. This corresponds to the velocity range of the Perseus arm of the Galaxy. If the \ha\ filament belongs to that arm, it would imply that the distance to the diffuse synchrotron background is much larger than a few hundred parsecs, lying at least 2\,kpc away from us.

\subsection{Limits on AME polarization}
\label{sec:other_pol_fg}

\begin{figure}[tb]
\begin{center}
\includegraphics[width=0.24\textwidth]{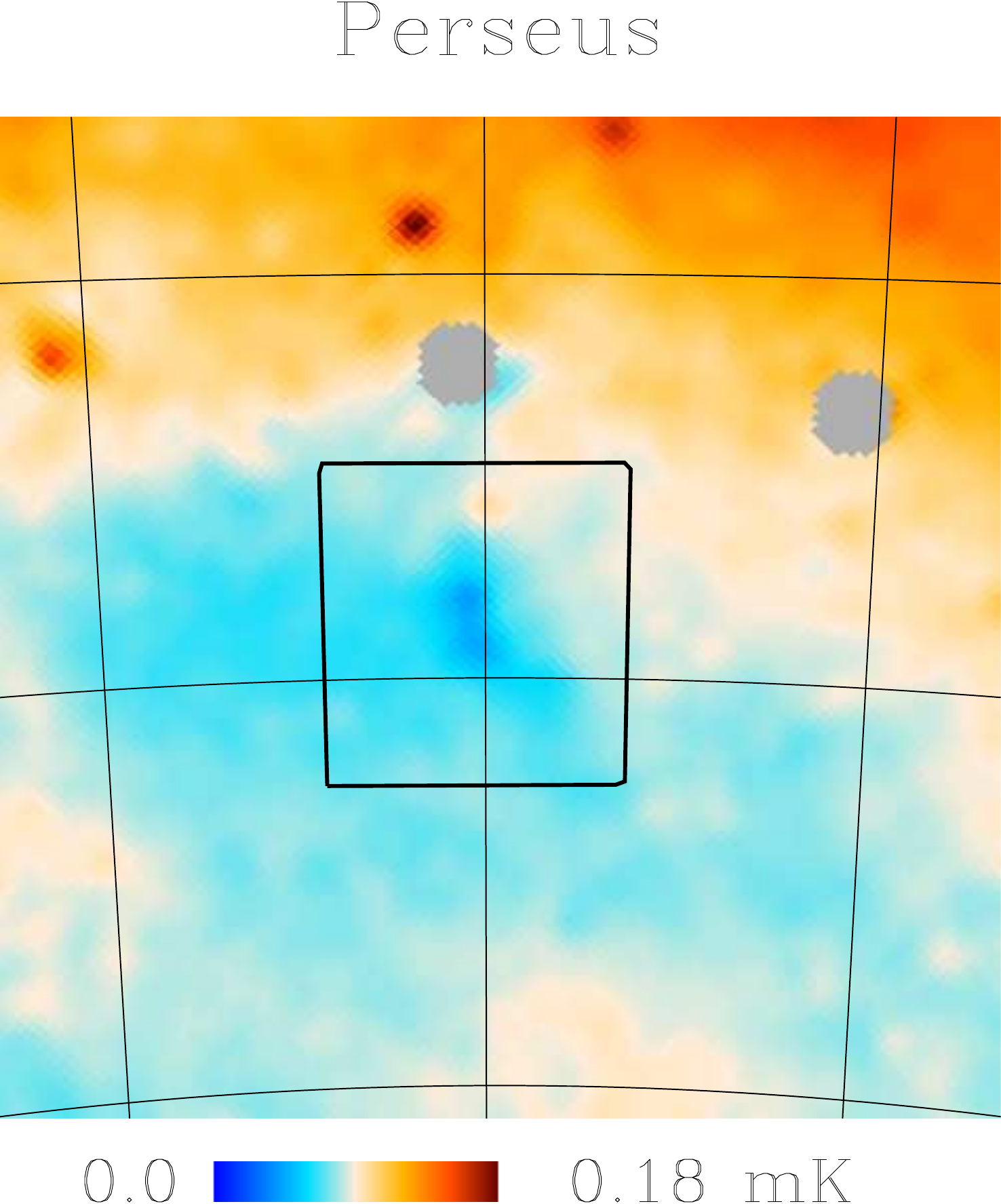}
\includegraphics[width=0.24\textwidth]{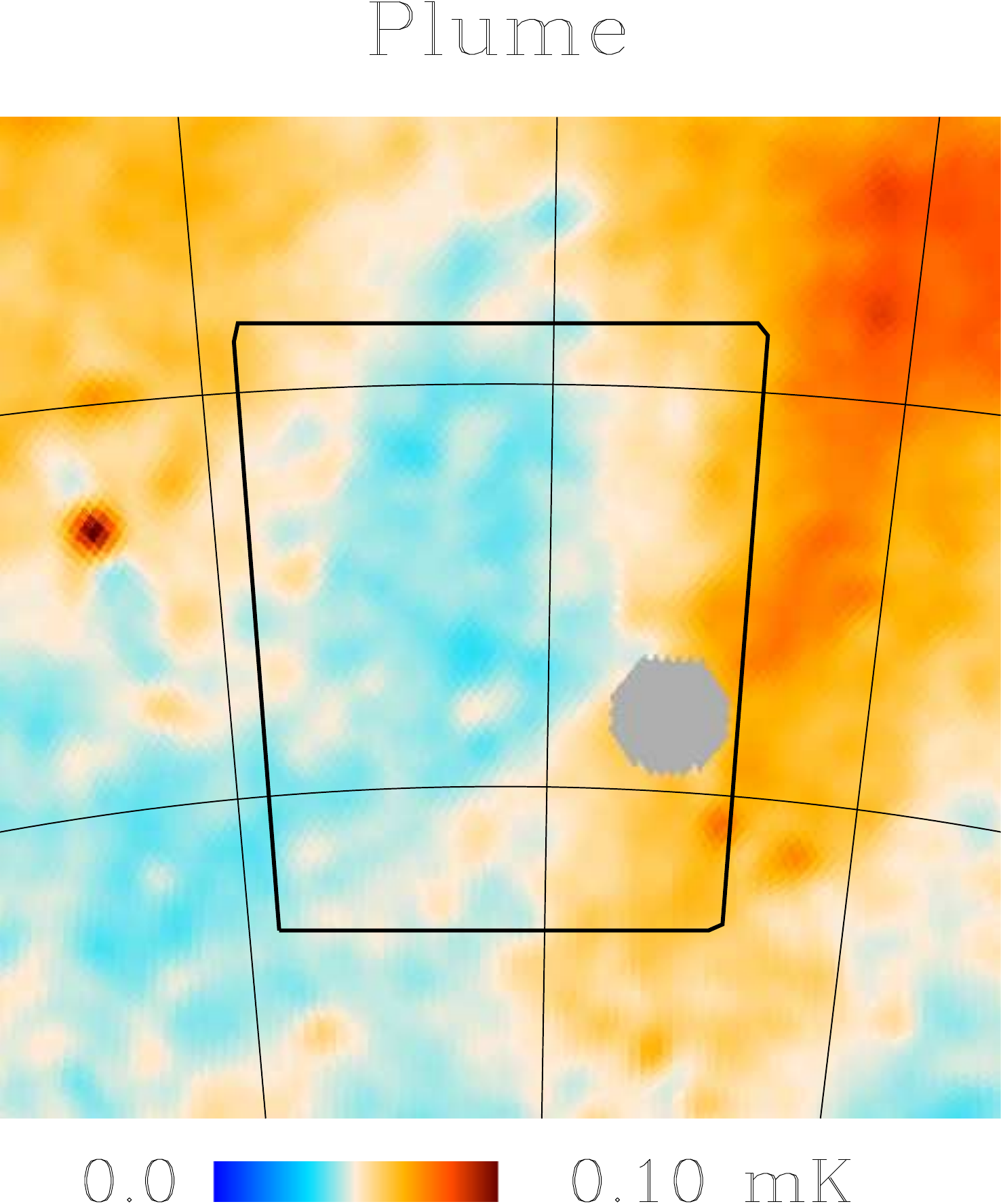}\\
\includegraphics[width=0.24\textwidth]{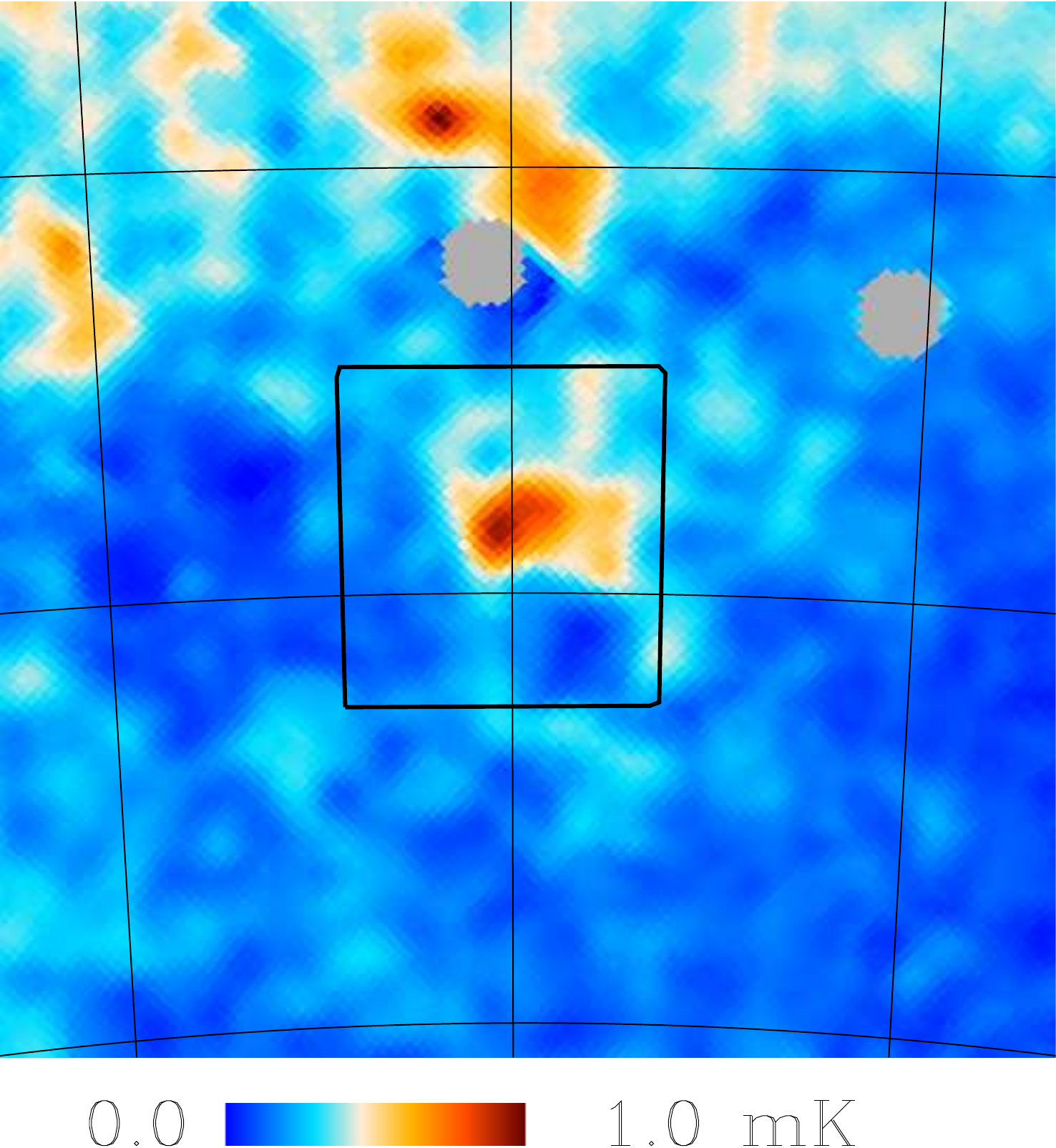}
\includegraphics[width=0.24\textwidth]{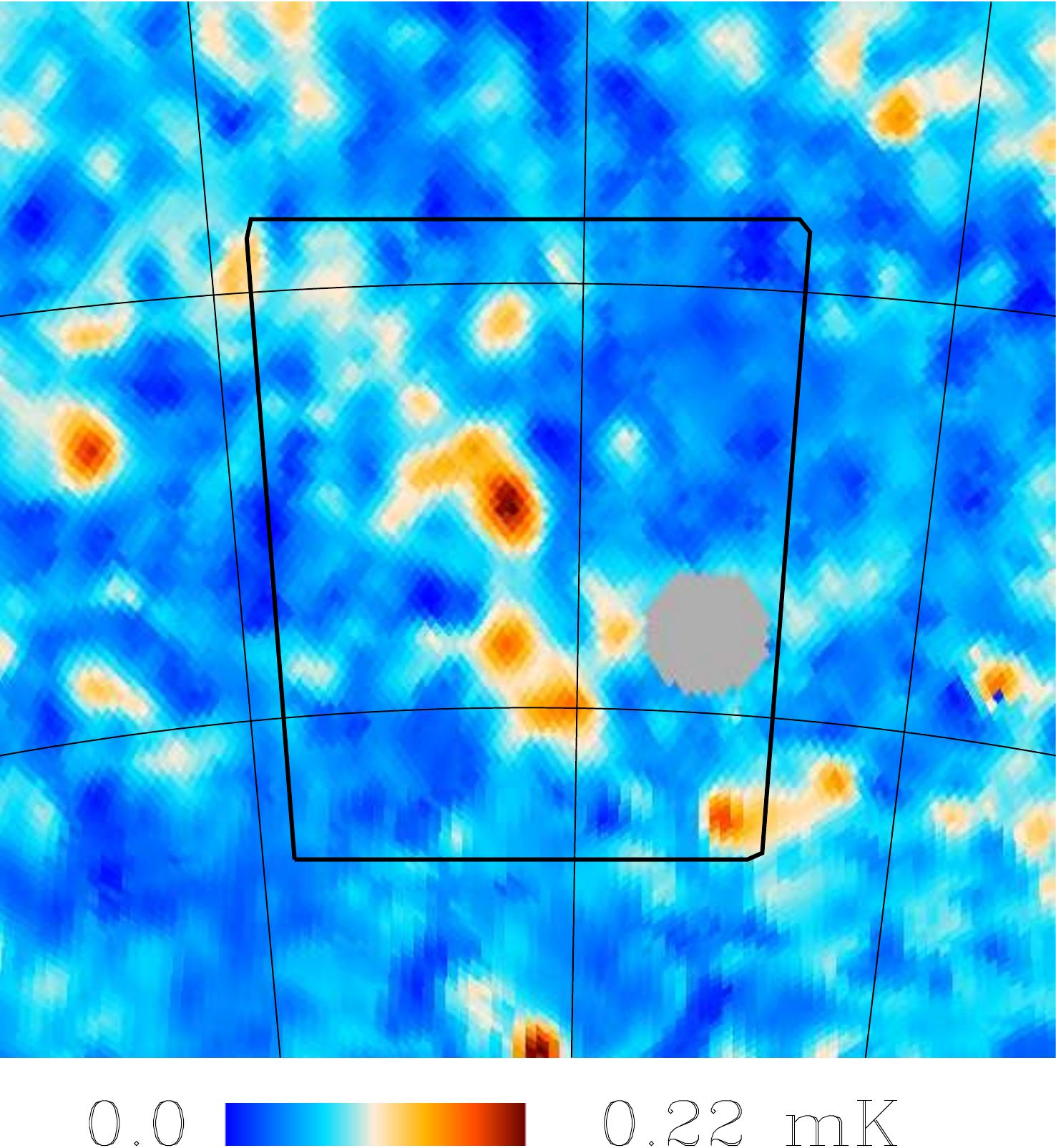}\\
\includegraphics[width=0.24\textwidth]{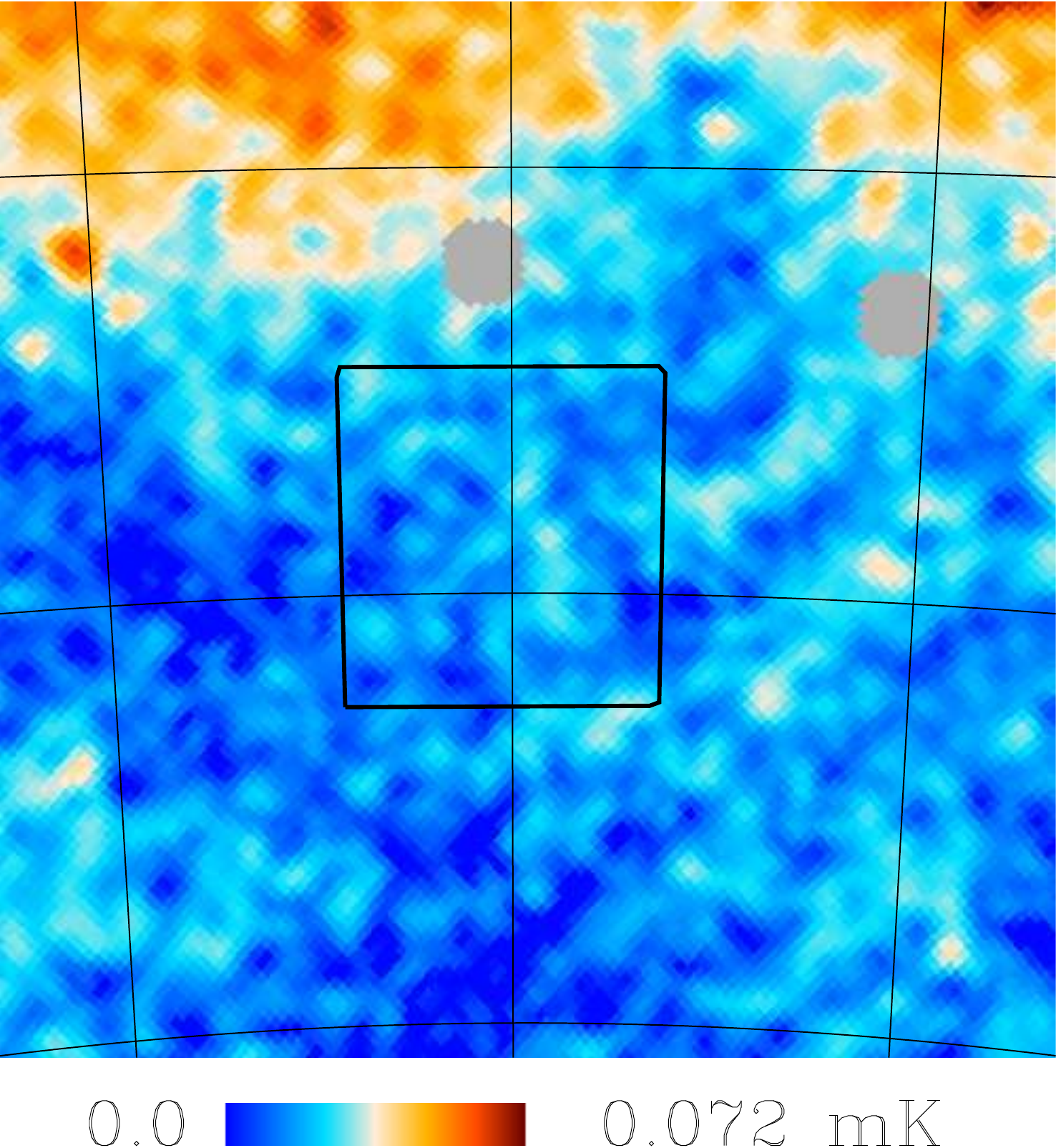}
\includegraphics[width=0.24\textwidth]{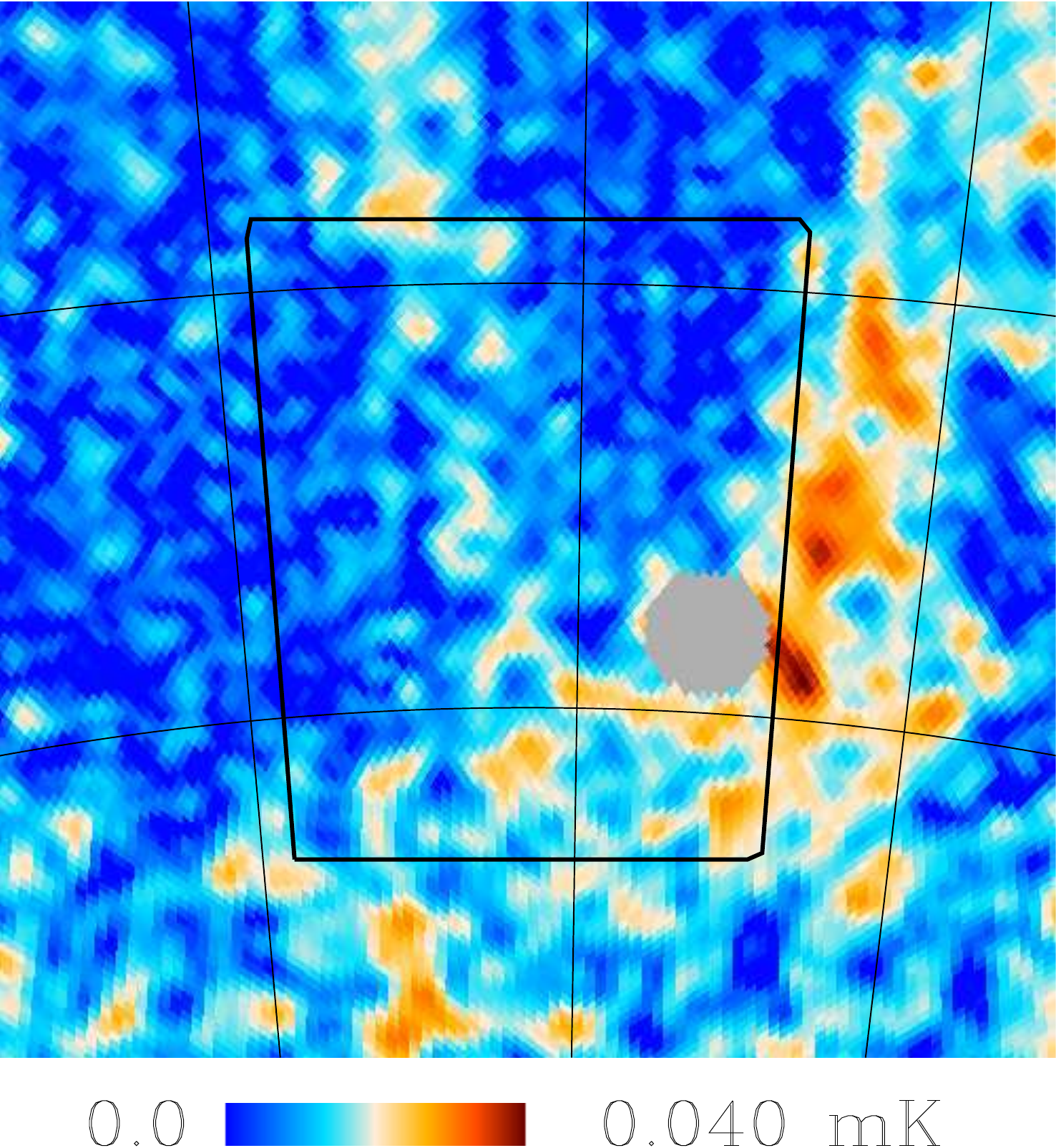}
\caption{{\it Top}: \commander\ synchrotron component maps; {\it middle}: \commander\ AME component maps; and {\it bottom}: debiased weighted polarization maps. {\it Left}: Perseus; {\it right}: the Pegasus plume. The colour scales are linear here. The regions are defined and masked as per Sect.~\ref{sec:ame}. The Perseus region does not contain significant polarized synchrotron emission, while the Pegasus plume contains several polarized synchrotron arcs.}
\label{fig:amepol}
\end{center}
\end{figure}

\begin{figure}[tb]
\begin{center}
\includegraphics[width=0.48\textwidth]{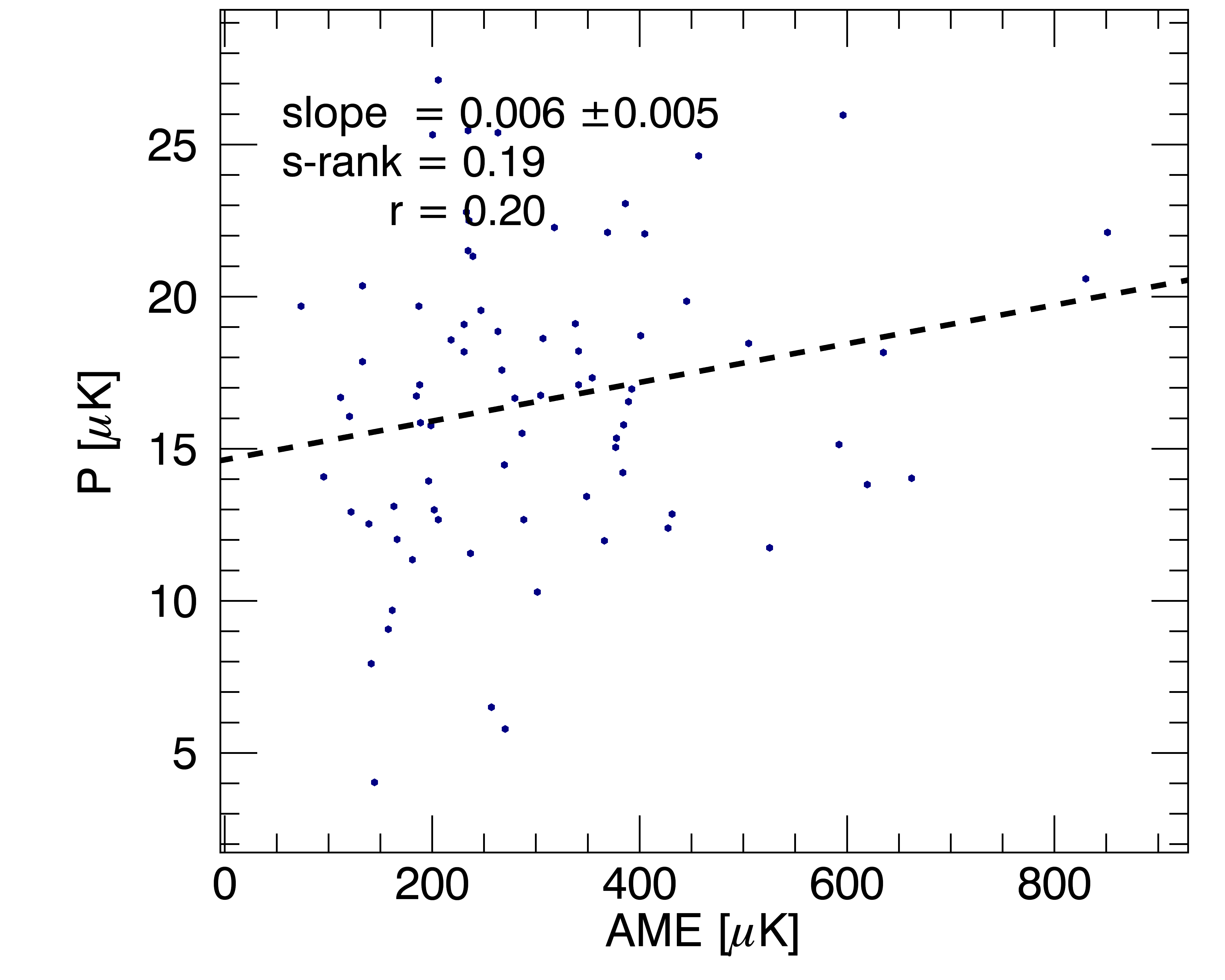}
\includegraphics[width=0.48\textwidth]{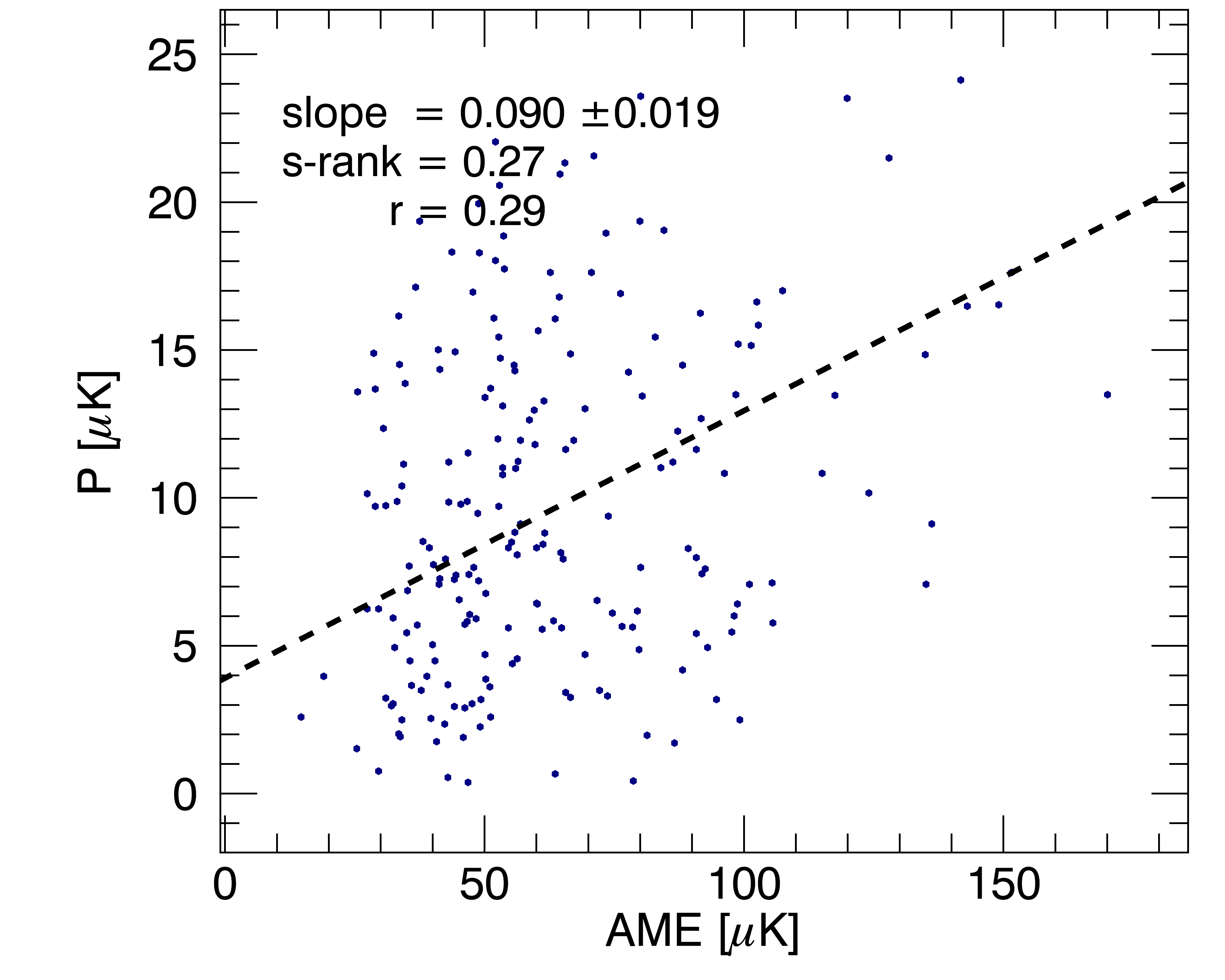}
\caption{\ttp\ plots between the polarization and AME intensities (both in units of $\upmu$K at 22.8\GHz). {\it Top}: Perseus; {\it bottom}: Pegasus plume. The scatter is mostly due to thermal noise in the polarization map (uncertainties in the polarization intensity in these regions are typically around 5\uK). The fitted slope gives the polarization fraction for the region, based on the linear fit (dashed line).}
\label{fig:amepol_tt}
\end{center}
\end{figure}

If AME is solely due to spinning dust particles, then we expect it to have a very low polarization percentage. The level of spinning dust polarization depends on the alignment efficiency of the small grains and PAHs in the interstellar magnetic field. \mbox{\cite{Lazarian2000}} considered resonance paramagnetic relaxation, which predicts $\lesssim$\,$1.5$\,\% polarization for frequencies $\gtrsim$\,$20$\GHz; \citet{Hoang2013} use constraints on the alignment of grains seen in ultraviolet polarization to predict that AME will have a polarization of $\lesssim$\,$0.9$\,\% at frequencies above 20\GHz. At lower frequencies ($\lesssim 10$\,GHz), the polarization fraction will be higher, but as the AME spectrum steeply decreases at low frequencies while other foreground components are increasing, this will be more difficult to detect than at the peak of the AME spectrum. Given this, and the fact that the observed polarization percentage would naturally be less than this due to beam and line-of-sight depolarization, we do not expect to detect significant AME polarization with \Planck, i.e., the polarization percentage should be $\lesssim$\,1\,\%.

Observational constraints on AME polarization have so far been placed using relatively compact, isolated clouds, where AME is known to be strong. \cite{Rubino-Martin2012} review the available constraints and their relation to theoretical models. The best constraint comes from the Perseus region, which has significant AME emission and relatively little contaminating synchrotron emission. \citet{Battistelli2006} reported a weak detection in Perseus at $3.4^{+1.5}_{-1.9}$\,\% at 11\,GHz, while later measurements obtained by \citet{Lopez-Caraballo2011} using \WMAP\ data found a 2$\sigma$ limit of $<$\,1\,\%; \mbox{\citet{Dickinson2011}} have also measured a 2$\sigma$ limit of $<$\,1.4\,\% for Perseus (as well as an upper limit of $<$\,1.7\,\% in $\rho$ Ophiuchus), and \citet{Genova-Santos2015} found a 2$\sigma$ limit of $<$\,2.8\,\% in Perseus at 19\GHz. Recently, \citet{Battistelli2015} have claimed to detect polarized AME emission at 21.5\GHz\ from RCW\,175, an \hii\ region, where they measured a polarization percentage of ($2.2\pm0.2$\,(random)\,$\pm$\,$0.3$\,(systematic))\,\% at 21.5\GHz. However, \cite{Battistelli2015} argue that a large fraction of this signal could be residual synchrotron radiation, leaving little or no polarized AME. There have been very few attempts to constrain the polarization from the large-scale diffuse AME; \cite{Kogut2007} find that AME accounts for less than 1\,\% of the total polarized signal, while \cite{Macellari2011} placed a limit of $<$\,5\,\% based on template fitting of \WMAP\ data. \citet{planck2014-XXII} placed upper limits by correlating \WMAP\ and \Planck\ polarization maps with the \Planck\ 353\GHz\ dust polarization map. They found that the correlation turned upwards at low frequencies, and at \WMAP\ K-band they placed an upper limit on AME polarization of 16\,\%; they also noted that the upturn could be explained by dust-correlated synchrotron polarization rather than AME polarization.

We use the debiased polarization map assembled using a weighted average of both \Planck\ and \WMAP\ data (see Sect.~\ref{sec:combo}) to look for polarized emission that is correlated with diffuse AME from the \commander\ solution. We note that the weighted polarized map assumes a spectral index of $\beta=-3.0$, rather than an AME-like spectrum. We create \ttp\ plots between the AME and polarization intensities, both rescaled to $\upmu$K at 22.8\GHz, in order to measure the percentage polarization. Assuming that any potential AME polarization has a constant percentage across the region, then this should result in a linear correlation of the data points in the \ttp\ plots. We use the uncertainty maps and assume a model uncertainty of 10\,\% for both the \commander\ AME solution and the weighted polarization map; the latter uncertainty is subdominant in these regions. We fit for the slope of the correlation, which directly gives the percentage polarization. We repixelize the maps to \mbox{$N_\mathrm{side}=64$} to ensure that the pixels are independent. We note that the debiasing method used will not perfectly remove all of the noise bias, particularly in regions of low S/N, but any residual bias should be lower than the thermal noise limits where the S/N ratio is low.

We focus on two high latitude regions from Sect.~\ref{sec:ame}, namely Perseus and the Pegasus plume. We show the AME and debiased polarization maps in Fig.~\ref{fig:amepol}, and the AME--polarization \ttp\ plot for each region in Fig.~\ref{fig:amepol_tt}. AME dominates the total emission at around 20--30\GHz\ in both of these regions. Unlike regions such as Corona Australis and Musca, which are strongly contaminated by highly polarized Galactic synchrotron structures (see Sect.~\ref{sec:synch_pol}), Perseus and the Pegasus plume are in parts of the sky with relatively little synchrotron polarization, although the Pegasus plume does have two polarized synchrotron features crossing the region; these can be seen in Fig.~\ref{fig:amepol} and are marked on Fig.~\ref{fig:pol_comb_maps_features}. One of these synchrotron arcs is somewhat aligned with the AME plume, however it is unrelated to the plume, since it extends beyond the plume towards the Galactic plane. The morphology of the polarized and total intensity synchrotron emission is quite different in the region of the plume, and the aligned arc is either highly polarized or has a flat spectrum, since it is faint in the total intensity map. Unlike the Perseus and Pegasus plume regions, the Orion and \lorionis\ regions contain strong free-free emission: while free-free emission is unpolarized there may be residual leakage from total intensity to polarization. In addition there is a broad background of polarized synchrotron emission in these regions, and due to their proximity to the Galactic plane there will be a correlation between synchrotron and thermal dust emission, simply because both components will be brighter towards the Galactic plane than they are off-plane. We apply the point source mask described in Sect.~\ref{sec:ame}: in the regions of interest this only consists of masking one point source near the Pegasus plume.

We find that the strongest constraint comes from Perseus, where we measure an AME polarization percentage (compared with the overall polarization intensity) of $0.6\pm0.5$\,\%, which gives a 2$\sigma$ limit of $<$\,1.6\,\%, matching previous constraints. The Pearson correlation coefficient is 0.20. An improved \commander\ estimate of the AME in this region (see Sect.~\ref{sec:ame}) would decrease this upper limit by around 40\,\%. In the Pegasus plume region we find a polarization percentage of $9.0\pm1.9$\,\%, with a Pearson correlation coefficient of 0.29, which, given the presence of correlated polarized synchrotron emission, we interpret to be a 2$\sigma$ limit of $<$\,12.8\,\%. We attempt to subtract some of the polarized synchrotron emission by assuming that the intensity is 10\,\% polarized, which removes some of the synchrotron emission from the region near the point source. This reduces the polarization percentage relative to the AME to $5.2\pm1.6$\,\%, or a 2$\sigma$ limit of $<$\,8.4\,\%, with a Pearson correlation coefficient of 0.24; assuming a higher level of synchrotron polarization results in significant negative pixels in the polarization map. We also compare the weighted low-frequency polarization map with the polarized dust emission at 353\GHz\ by separately comparing the $Q$ and $U$ maps at the two frequencies, at \mbox{$N_\mathrm{side}=64$}, in the regions of Perseus and the Pegasus plume. Similar analysis was performed by \citet{planck2014-XXII} across the whole sky. We find similar results, with no significant detection in Perseus, and a weak detection in the Pegasus plume, which we ascribe to dust-correlated polarized synchrotron emission, as described above.

In summary, we do not find strong evidence for diffuse polarized AME in \WMAP/\Planck\ data, with a 2$\sigma$ upper limit of 1.6\,\% in the Perseus region.  In order to improve on current constraints, we will need either reduced noise levels in regions that are free of polarized synchrotron emission (such as Perseus) or the ability to subtract accurately the synchrotron emission from less clear regions.  The forthcoming map of polarized synchrotron emission at 5\GHz\ by the C-Band All Sky Survey \citep{King2014} will assist with the latter issue.

\section{Conclusions}
\label{sec:conclusions}

In this paper we have discussed the Galactic foreground emission observed by \WMAP/\Planck\ between 20 and 100\GHz. The total intensity emission comprises at least three distinct components: synchrotron; free-free; and AME, commonly attributed to spinning dust. These components emit smooth continua that are typically decreasing with frequency. It is therefore extremely difficult to separate these components with any precision. Furthermore, these individual components are often spatially correlated (e.g., towards the Galactic plane). 

We have applied an internal linear combination technique to \WMAP/\Planck\ data to produce a map that is free of CMB, free-free, and thermal dust. This has given us some insight into the spectral form of emission at 23\GHz. For most of the intensity analysis we rely on the component-separation products from the \Commander\ code. This involved fitting a parametric model to the data. However, a number of simplifications have had to be made, given the limited number of frequencies and complexity of the foreground emission. In particular, we have not so far been able to produce a detailed synchrotron map in intensity. Instead we have had to rely on a \galprop\ spectral model to take into account the bulk of the synchrotron emission above 20\GHz. In polarization the situation is somewhat clearer. The polarization maps at 20--50\GHz\ are dominated by synchrotron radiation; free-free and AME are expected to have very low polarization fractions. 

Nevertheless, in this study we have been able to produce reasonable products for free-free and AME, from which we have inferred some physical results. Our polarization results do not rely directly on component separation.

The main conclusions from our study are as follows.
\begin{itemize}
\item{After applying an ILC procedure to remove CMB, free-free, and thermal dust emission at 22.8\GHz, we find that the residual emission has significant spectral variations. This emission is expected to be dominated by (weak) synchrotron and (stronger) AME emission. In particular, we find that for some regions the spectrum is flatter than that of free-free emission ($\beta=-2.1$).}
\item{The aforementioned areas are largely towards bright \hii\ regions such as the $\zeta$~Ophiuchi cloud (Sh\,2-27)and the California nebula. Because synchrotron emission is not observed to be flatter than about $\beta=-2.3$, the spectral changes are thought to be due to a high frequency component of spinning dust. Indeed, \cite{planck2013-XV} found evidence for a spinning dust peak at around 50\GHz\ for the California nebula.}
\item{Our free-free map appears to correspond well to previous maps made with \WMAP\ data, and also to \ha\ maps at high latitudes (where dust absorption is small). The amplitude relative to \ha\ is lower than the theoretical value for typical electron temperatures, as has been found in previous studies.}
\item{A natural explanation for the low free-free-to-\ha\ ratio is that there is significant scattered \ha\ light. We have attempted to estimate this fraction by correlating \ha\ residuals (after subtracting the mean correlation) with the dust optical depth. We do indeed find a correlation with the thermal dust optical depth. If we assume that electron temperature varies by no more than $\pm 1000$\,K, we find that the scattered \ha\ emission accounts for $28 \pm 12\,\%$, or $36 \pm 12\,$\% after making a nominal correction for absorption by dust.}
\item{We also compared the \Commander\ free-free map to the RRL Galactic plane survey of \cite{Alves2014}. Again, we find a good overall correspondence between the maps. However, there are significant differences, which might be due to variations in electron temperature.}
\item{The \Commander\ AME map is closely correlated with thermal dust emission, as traced by FIR/sub-mm maps. We have identified several new diffuse regions away from the Galactic plane that appear to emit significant AME. These regions are typically associated with large dust cloud complexes. We have compared the emissivity of AME in these regions relative to the amount of thermal dust emission. The AME emissivity against 545\GHz\ and $\tau_\mathrm{353}$ varies by a factor of approximately 2 from region to region. The \lorionis\ region has a particularly high emissivity that warrants further investigation, ideally in conjunction with additional low-frequency data to reduce leakage between the free-free and AME components.}
\item{Comparison of the \Commander\ foreground products with similar products from the \WMAP\ team shows that the amplitude of synchrotron emission is significantly lower in our analysis. This is particularly the case at low latitudes, where our fixed spectral model cannot account for the flatter index reported by \WMAP\ and in earlier \Planck\ analyses. To compensate, our AME amplitude is significantly higher by a factor of around 2--4 (depending on which model is being compared). The free-free component is in better agreement in general, but our model is typically lower by about 10--30\,\%, which has been noted before \citep{Alves2010,Alves2012}. We emphasize that our model fits the data extremely accurately, with median residuals along the Galactic plane of $<1$\,\% for \Planck\ and $<2$\,\% for \WMAP\ channels \citep{planck2014-a12}.}

\item{We have combined \WMAP/\Planck\ polarization maps to construct a higher S/N ratio polarized intensity map. This is the highest S/N ratio synchrotron polarization map available above a few GHz, where Faraday rotation can be safely ignored.}

\item{The new polarization map shows a number of large-scale polarization structures, including spurs and filaments. These new features are detected clearly only in polarization, indicating a high polarization fraction similar to the brighter well-known spurs.}

\item The halo of synchrotron emission in the inner Galaxy clearly visible at 408\,MHz appears to have a low polarization fraction ($<$\,10\,\%) compared to the high-latitude spurs with typical polarization fractions of 30--50\,\%. The diffuse high-latitude emission is also weakly (though detectably) polarized ($<$\,15\,\%).

\item We discuss in detail the characteristics of Loop I. The new polarization maps enables us to follow it further into the southern Galactic hemisphere, where there is a substantial deviation from the small circle fitted to the brighter northern part, presumably due to the inhomogenous environment. Radio depolarization, X-ray absorption, and tomographic mapping of the local ISM all suggest that the loop is at least several times more distant than the traditional 120\,pc.

\item The projected magnetic field outside the North Polar spur parallels the loop, supporting the model that the loop is in part shaped by its interaction with the ambient magnetic field.

\item The filaments projected inside Loop I are mostly associated with it, possibly a system of internal shock waves created when a re-energizing supernova shock reflected from the old boundary of the cavity. Alternatively, in accordance with the model of \citet{Vidal2014a}, they might be a system of ``illuminated'' field lines on the back hemisphere of the loop cavity.

\item On several grounds we are not convinced by the suggestion of \citet{Liu2014} that emission from the loop is seriously contaminating derived CMB maps.

\item We trace the magnetic field in Loop II for the first time. It follows the path of the loop, as in the other examples. We show that the most prominent feature in the original definition of Loop II is actually unrelated, and so the loop is somewhat smaller and fainter than previously believed.

\item The proposed extension to Loop III in the south suggested by \citet{Vidal2014a} is largely an artefact in the \wmap\ maps.

\item The South Polar Spur, previously identified as the low-$l$ edge of Loop II, is actually the high-$l$ edge of a smaller structure. Just like for Loop I, there is a filament of cold material (traced by \hi\ and thermal dust) running just outside it, with parallel magnetic field. From the \hi\ velocity the distance is at most a few hundred pc.

\item Loops V and VI \citep{Milogradov-Turin1997} and  S1 \citep{Wolleben2007} are not physical structures; our map shows that major components of the proposed loops deviate radically from the expected path. These components are independent features, with much smaller angular size than the suggested loops.

\item  We find two examples of smaller and more distant loops, one of which arcs around the Cyg X star-forming region. Both are associated with polarized dust features.

\item{We observe a clear outline of narrow highly polarized intensity around the northern Fermi bubble above the Galactic centre. The spectral index appears to be much flatter $(\beta=-2.54\pm0.16$) than other filaments, but similar to that found for the microwave haze. This is indicative of a more energetic population of electrons along these lines of sight. We note that the Southern extension of the \fermi\ bubble extends further south than Loop I, which probably rules out an association of Loop I with the \fermi\ bubbles.}

\item{We have discovered examples of anti-correlation of \ha\ and polarized synchrotron intensity. The most striking is a high latitude \ha\ filament that appears as a local minimum in polarization. Because of the low density of ionized gase we can rule out Faraday depolarization. The observed anti-correlation appears to be real in that there is no synchrotron emission emanating from the region of the \ha\ filament. The negative velocity of \ha\ and the longitude imply that the filament could be lying at a distance of at least 2\,kpc.}

\item{We correlated the polarized maps with the \commander\ AME map. We found that there was no significant correlation except for the Pegasus plume region, where a low level ($9.0\pm1.9\,\%$) of polarization was observed, consistent with contamination from surrounding synchrotron radiation.}
\end{itemize}

Although we are beginning to understand the diffuse emission at \WMAP/\Planck\ frequencies, there is still a long way to go. In intensity, the component separation is still a major difficulty. In particular, we have not been able to constrain the synchrotron spectral dependence, due to the degeneracy and confusion with free-free, AME, and CMB. New data in the frequency range 2--15\GHz\ are needed to more fully characterize all of the low-frequency foreground components. Experiments such as S-PASS at 2.3\GHz\ \citep{Carretti2011}, C-BASS at 5\GHz\ \citep{King2014}, and QUIJOTE at 11--19\GHz\ \citep{Rubino-Martin2012b} will improve the situation considerably.

In polarization, the emerging picture is that synchrotron emission is by far the dominant polarized component below 50\GHz, with very little contribution from free-free and AME. The diffuse synchrotron emission has now been measured with good S/N ratio for a large fraction of sky. However, there are still regions where the S/N ratio is low. More importantly, the detailed spectral dependence of synchrotron emission is still unknown.


\begin{acknowledgements}
This paper is dedicated to the memory of the late Professor Rodney Deane Davies CBE FRS and Professor Richard John Davis OBE, both of whom contributed greatly to {\it Planck} project. The Planck Collaboration acknowledges the support of: ESA; CNES and CNRS/INSU-IN2P3-INP (France); ASI, CNR, and INAF (Italy); NASA and DoE (USA); STFC and UKSA (UK); CSIC, MINECO, JA, and RES (Spain); Tekes, AoF, and CSC (Finland); DLR and MPG (Germany); CSA (Canada); DTU Space (Denmark); SER/SSO (Switzerland); RCN (Norway); SFI (Ireland); FCT/MCTES (Portugal); ERC and PRACE (EU). A description of the Planck Collaboration and a list of its members, indicating which technical or scientific activities they have been involved in, can be found at \href{http://www.cosmos.esa.int/web/planck/planck-collaboration}{http://www.cosmos.esa.int/web/planck/planck-collaboration}. This research was supported by an ERC Starting (Consolidator) Grant (no.~307209) and STFC Consolidated Grant (no.~ST/L000768/1). We have made extensive use of the \healpix\ package and the IDL astronomy library. This research has made use of the SIMBAD database,
operated at CDS, Strasbourg, France.
\end{acknowledgements}

\bibliographystyle{aat}
\bibliography{Planck_bib,A31_refs}

\appendix
\section{Simple ILC method}
\label{sec:ILC_method}

The well-known ILC methods return an image of a component with a known spectrum in the presence of other components whose spectra are either not explicitly considered \citep{Bennett2003b} or only known in a subset of cases \citep{Remazeilles2011a,Remazeilles2011b}. In Sect.~\ref{sec:ilc} we have the more limited aim of 
{\em eliminating} one or two of the major components with known spectra at low frequencies (CMB and free-free emission), and returning a map reasonably close to the sky at 28.4\GHz\ with these components omitted.

We model the maps as sums of components with spatially-invariant and known spectra, plus noise and other unmodelled foreground emission:
\begin{equation}
\tilde{T}_{jk} = M_{ki} A_{ij} + N_{jk} + O_{jk}
\end{equation}
where the indices are $i$ for component, $j$ for pixel and $k$ for band. $A_{ij}$ is the amplitude. The mixing matrix $M_{ki} = f_i(\nu_k)/U_k\cc_{ik}$, where $f_i(\nu_k)$ is the spectral form in Rayleigh-Jeans brightness temperature at the band reference frequency $\nu_k$, $U_k$ is the unit conversion from Rayleigh-Jeans to thermodynamic temperature, and $\cc_{ik}$ is the colour correction (\citealp{planck2013-p03d}; \citealp{planck2014-a03}).\footnote{Colour corrections and unit conversions for \WMAP\ were derived in the same way as for the LFI, using the released \WMAP\ bandpasses. Colour corrections were based on the spectral index of the model evaluated at $\nu_k$, except that we used the explicit dust colour corrections provided by the {\tt UC\_CC} code for the HFI bands \citep{planck2013-p03d}.} For the CMB, component 0, we have $M_{k0} = 1$.

Our aim is to find the weights for the frequency  maps that satisfy the conditions
\begin{equation}
\sum_k M_{ki} w_k = 0 
\label{eq:c0}
\end{equation}
for the components to be eliminated, subject to the normalization condition
\begin{equation}
\sum_k M_{k3} w_k = M_{03}
\label{eq:c2}
\end{equation}
where band $k=0$ is 30\GHz\ and component $i=3$ corresponds to a $\beta=-3$ power law, roughly appropriate for synchrotron emission, but also quite close to the effective spectrum of spinning dust. We omit maps with frequencies in the range 50--120\GHz, where the foreground components are weak and their spectrum (especially the dust components) is not well defined. The weakest point of our modelling is the assumption of a fixed spectral shape for the thermal dust, but this component is very weak below 50\GHz\ so that errors in the dust spectrum there have little effect; the spectrum is mainly used in the interpolation from 353 to 143\GHz\ needed to generate a dust-cleaned CMB template.

We solve Eqs.~\ref{eq:c0} and \ref{eq:c2} by restricting the analysis to $n$ frequency maps, where $n$ components are dealt with, to make $M_{ki}$ square, and then
\begin{equation}
\vec{w} =  M^{-1} 
\left(\begin{array}{c} 0 \\ \vdots \\ M_{03} \end{array}\right)~.
\label{eq:weight_vector}
\end{equation}

In our final application, we have more bands than constraints and so we found the weights by maximizing the S/N ratio for our reference component:
\begin{equation}
\mathrm{S/N} = \left(\sum_k w_k M_{k3} \right) \Bigg/ 
\left(\sum_k w_k^2 \sigma_k^2\right)^{1/2}, 
\end{equation}
where, to keep the weights spatially uniform, we take $\sigma_k^2$ as the median pixel variance for band $k$. We split the bands into a set $k'$ whose weights are varied by the fitting routine ({\tt AMOEBA} in IDL), and a set $k$ whose weights are found by linear regression. Our constraint is now written
\begin{equation}
\sum_k M_{ki} w_k = \delta_{i3}M_{03} -\sum_{k'} M_{k'i} w_k'. 
\label{eq:c4}
\end{equation}
Each {\tt AMOEBA} iteration provides a trial set of $w_k'$ values that are used to find $w_k$ via
\begin{equation}
w_k = \sum_i M^{-1}_{ki} \left( \delta_{i3} M_{03} - \sum_{k'} M_{k'i} w_{k'}\right), 
\end{equation}
and the full set of weights is then used to evaluate the S/N ratio.


\end{document}